\documentclass[twocolumn,traditabstract,longauth,a4]{aa}
\usepackage{graphicx,amsmath}
\usepackage{epsf}
\usepackage{txfonts}
\usepackage[hyphens]{url}
\usepackage{longtable}
\usepackage{lscape}
\usepackage[hyperindex,breaklinks=true, colorlinks, citecolor=blue]{hyperref}  
\usepackage{breakurl}
\usepackage{comment}
\usepackage{natbib}
\usepackage{upgreek}
\usepackage{float}
\usepackage{fixltx2e}

\def\setsymbol#1#2{\expandafter\def\csname #1\endcsname{#2}}
\def\getsymbol#1{\csname #1\endcsname}

\def\Planck{\textit{Planck}}

\def\all2013resultspapers{\nocite{planck2013-p01, planck2013-p02, planck2013-p02a, planck2013-p02d, planck2013-p02b, planck2013-p03, planck2013-p03c, planck2013-p03f, planck2013-p03d, planck2013-p03e, planck2013-p01a, planck2013-p06, planck2013-p03a, planck2013-pip88, planck2013-p08, planck2013-p11, planck2013-p12, planck2013-p13, planck2013-p14, planck2013-p15, planck2013-p05b, planck2013-p17, planck2013-p09, planck2013-p09a, planck2013-p20, planck2013-p19, planck2013-pipaberration, planck2013-p05, planck2013-p05a, planck2013-pip56, planck2013-p06b}}

\newbox\tablebox    \newdimen\tablewidth
\def\leaderfil{\leaders\hbox to 5pt{\hss.\hss}\hfil}
\def\endPlancktable{\tablewidth=\columnwidth 
    $$\hss\copy\tablebox\hss$$
    \vskip-\lastskip\vskip -2pt}
\def\endPlancktablewide{\tablewidth=\textwidth 
    $$\hss\copy\tablebox\hss$$
    \vskip-\lastskip\vskip -2pt}
\def\tablenote#1 #2\par{\begingroup \parindent=0.8em
    \abovedisplayshortskip=0pt\belowdisplayshortskip=0pt
    \noindent
    $$\hss\vbox{\hsize\tablewidth \hangindent=\parindent \hangafter=1 \noindent
    \hbox to \parindent{$^#1$\hss}\strut#2\strut\par}\hss$$
    \endgroup}
\def\doubleline{\vskip 3pt\hrule \vskip 1.5pt \hrule \vskip 5pt}

\def\L2{\ifmmode L_2\else $L_2$\fi}

\def\DeltaT{\ifmmode \Delta T\else $\Delta T$\fi}
\def\deltat{\ifmmode \Delta t\else $\Delta t$\fi}
\def\fknee{\ifmmode f_{\rm knee}\else $f_{\rm knee}$\fi}
\def\Fmax{\ifmmode F_{\rm max}\else $F_{\rm max}$\fi}
\def\solar{\ifmmode{\rm M}_{\mathord\odot}\else${\rm M}_{\mathord\odot}$\fi}
\def\Msolar{\ifmmode{\rm M}_{\mathord\odot}\else${\rm M}_{\mathord\odot}$\fi}
\def\Lsolar{\ifmmode{\rm L}_{\mathord\odot}\else${\rm L}_{\mathord\odot}$\fi}

\def\inv{\ifmmode^{-1}\else$^{-1}$\fi}
\def\mo{\ifmmode^{-1}\else$^{-1}$\fi}
\def\sup#1{\ifmmode ^{\rm #1}\else $^{\rm #1}$\fi}
\def\expo#1{\ifmmode \times 10^{#1}\else $\times 10^{#1}$\fi}
\def\,{\thinspace}
\def\lsim{\mathrel{\raise .4ex\hbox{\rlap{$<$}\lower 1.2ex\hbox{$\sim$}}}}
\def\gsim{\mathrel{\raise .4ex\hbox{\rlap{$>$}\lower 1.2ex\hbox{$\sim$}}}}

\def\simprop{\mathrel{\raise .4ex\hbox{\rlap{$\propto$}\lower 1.2ex\hbox{$\sim$}}}}
\def\deg{\ifmmode^\circ\else$^\circ$\fi}
\def\pdeg{\ifmmode $\setbox0=\hbox{$^{\circ}$}\rlap{\hskip.11\wd0 .}$^{\circ}
          \else \setbox0=\hbox{$^{\circ}$}\rlap{\hskip.11\wd0 .}$^{\circ}$\fi}
\def\arcs{\ifmmode {^{\scriptstyle\prime\prime}}
          \else $^{\scriptstyle\prime\prime}$\fi}
\def\arcm{\ifmmode {^{\scriptstyle\prime}}
          \else $^{\scriptstyle\prime}$\fi}
\newdimen\sa  \newdimen\sb
\def\parcs{\sa=.07em \sb=.03em
     \ifmmode \hbox{\rlap{.}}^{\scriptstyle\prime\kern -\sb\prime}\hbox{\kern -\sa}
     \else \rlap{.}$^{\scriptstyle\prime\kern -\sb\prime}$\kern -\sa\fi}
\def\parcm{\sa=.08em \sb=.03em
     \ifmmode \hbox{\rlap{.}\kern\sa}^{\scriptstyle\prime}\hbox{\kern-\sb}
     \else \rlap{.}\kern\sa$^{\scriptstyle\prime}$\kern-\sb\fi}
\def\ra[#1 #2 #3.#4]{#1\sup{h}#2\sup{m}#3\sup{s}\llap.#4}
\def\dec[#1 #2 #3.#4]{#1\deg#2\arcm#3\arcs\llap.#4}
\def\deco[#1 #2 #3]{#1\deg#2\arcm#3\arcs}
\def\rra[#1 #2]{#1\sup{h}#2\sup{m}}

\def\dots{\relax\ifmmode \ldots\else $\ldots$\fi}
\def\WHzsr{\ifmmode $W\,Hz\mo\,sr\mo$\else W\,Hz\mo\,sr\mo\fi}
\def\mHz{\ifmmode $\,mHz$\else \,mHz\fi}
\def\GHz{\ifmmode $\,GHz$\else \,GHz\fi}
\def\mKs{\ifmmode $\,mK\,s$^{1/2}\else \,mK\,s$^{1/2}$\fi}
\def\muKs{\ifmmode \,\mu$K\,s$^{1/2}\else \,$\mu$K\,s$^{1/2}$\fi}
\def\muKRJs{\ifmmode \,\mu$K$_{\rm RJ}$\,s$^{1/2}\else \,$\mu$K$_{\rm RJ}$\,s$^{1/2}$\fi}
\def\muKHz{\ifmmode \,\mu$K\,Hz$^{-1/2}\else \,$\mu$K\,Hz$^{-1/2}$\fi}
\def\MJysr{\ifmmode \,$MJy\,sr\mo$\else \,MJy\,sr\mo\fi}
\def\MJysrmK{\ifmmode \,$MJy\,sr\mo$\,mK$_{\rm CMB}\mo\else \,MJy\,sr\mo\,mK$_{\rm CMB}\mo$\fi}
\def\microns{\ifmmode \,\mu$m$\else \,$\mu$m\fi}

\def\muK{\ifmmode \,\mu$K$\else \,$\mu$\hbox{K}\fi}
\def\microK{\ifmmode \,\mu$K$\else \,$\mu$\hbox{K}\fi}
\def\muW{\ifmmode \,\mu$W$\else \,$\mu$\hbox{W}\fi}
\def\kms{\ifmmode $\,km\,s$^{-1}\else \,km\,s$^{-1}$\fi}
\def\kmsMpc{\ifmmode $\,\kms\,Mpc\mo$\else \,\kms\,Mpc\mo\fi}
\setsymbol{LFI:center:frequency:70GHz:units}{70.3\,GHz}
\setsymbol{LFI:center:frequency:44GHz:units}{44.1\,GHz}
\setsymbol{LFI:center:frequency:30GHz:units}{28.5\,GHz}

\setsymbol{LFI:center:frequency:70GHz}{70.3}
\setsymbol{LFI:center:frequency:44GHz}{44.1}
\setsymbol{LFI:center:frequency:30GHz}{28.5}

\setsymbol{LFI:center:frequency:LFI18:Rad:M:units}{71.7\GHz}
\setsymbol{LFI:center:frequency:LFI19:Rad:M:units}{67.5\GHz}
\setsymbol{LFI:center:frequency:LFI20:Rad:M:units}{69.2\GHz}
\setsymbol{LFI:center:frequency:LFI21:Rad:M:units}{70.4\GHz}
\setsymbol{LFI:center:frequency:LFI22:Rad:M:units}{71.5\GHz}
\setsymbol{LFI:center:frequency:LFI23:Rad:M:units}{70.8\GHz}
\setsymbol{LFI:center:frequency:LFI24:Rad:M:units}{44.4\GHz}
\setsymbol{LFI:center:frequency:LFI25:Rad:M:units}{44.0\GHz}
\setsymbol{LFI:center:frequency:LFI26:Rad:M:units}{43.9\GHz}
\setsymbol{LFI:center:frequency:LFI27:Rad:M:units}{28.3\GHz}
\setsymbol{LFI:center:frequency:LFI28:Rad:M:units}{28.8\GHz}
\setsymbol{LFI:center:frequency:LFI18:Rad:S:units}{70.1\GHz}
\setsymbol{LFI:center:frequency:LFI19:Rad:S:units}{69.6\GHz}
\setsymbol{LFI:center:frequency:LFI20:Rad:S:units}{69.5\GHz}
\setsymbol{LFI:center:frequency:LFI21:Rad:S:units}{69.5\GHz}
\setsymbol{LFI:center:frequency:LFI22:Rad:S:units}{72.8\GHz}
\setsymbol{LFI:center:frequency:LFI23:Rad:S:units}{71.3\GHz}
\setsymbol{LFI:center:frequency:LFI24:Rad:S:units}{44.1\GHz}
\setsymbol{LFI:center:frequency:LFI25:Rad:S:units}{44.1\GHz}
\setsymbol{LFI:center:frequency:LFI26:Rad:S:units}{44.1\GHz}
\setsymbol{LFI:center:frequency:LFI27:Rad:S:units}{28.5\GHz}
\setsymbol{LFI:center:frequency:LFI28:Rad:S:units}{28.2\GHz}

\setsymbol{LFI:center:frequency:LFI18:Rad:M}{71.7}
\setsymbol{LFI:center:frequency:LFI19:Rad:M}{67.5}
\setsymbol{LFI:center:frequency:LFI20:Rad:M}{69.2}
\setsymbol{LFI:center:frequency:LFI21:Rad:M}{70.4}
\setsymbol{LFI:center:frequency:LFI22:Rad:M}{71.5}
\setsymbol{LFI:center:frequency:LFI23:Rad:M}{70.8}
\setsymbol{LFI:center:frequency:LFI24:Rad:M}{44.4}
\setsymbol{LFI:center:frequency:LFI25:Rad:M}{44.0}
\setsymbol{LFI:center:frequency:LFI26:Rad:M}{43.9}
\setsymbol{LFI:center:frequency:LFI27:Rad:M}{28.3}
\setsymbol{LFI:center:frequency:LFI28:Rad:M}{28.8}
\setsymbol{LFI:center:frequency:LFI18:Rad:S}{70.1}
\setsymbol{LFI:center:frequency:LFI19:Rad:S}{69.6}
\setsymbol{LFI:center:frequency:LFI20:Rad:S}{69.5}
\setsymbol{LFI:center:frequency:LFI21:Rad:S}{69.5}
\setsymbol{LFI:center:frequency:LFI22:Rad:S}{72.8}
\setsymbol{LFI:center:frequency:LFI23:Rad:S}{71.3}
\setsymbol{LFI:center:frequency:LFI24:Rad:S}{44.1}
\setsymbol{LFI:center:frequency:LFI25:Rad:S}{44.1}
\setsymbol{LFI:center:frequency:LFI26:Rad:S}{44.1}
\setsymbol{LFI:center:frequency:LFI27:Rad:S}{28.5}
\setsymbol{LFI:center:frequency:LFI28:Rad:S}{28.2}

\setsymbol{LFI:white:noise:sensitivity:70GHz:units}{134.7\muKs}
\setsymbol{LFI:white:noise:sensitivity:44GHz:units}{164.7\muKs}
\setsymbol{LFI:white:noise:sensitivity:30GHz:units}{143.4\muKs}

\setsymbol{LFI:white:noise:sensitivity:70GHz}{134.7}
\setsymbol{LFI:white:noise:sensitivity:44GHz}{164.7}
\setsymbol{LFI:white:noise:sensitivity:30GHz}{143.4}

\setsymbol{LFI:white:noise:sensitivity:LFI18:Rad:M:units}{512.0\muKs}
\setsymbol{LFI:white:noise:sensitivity:LFI19:Rad:M:units}{581.4\muKs}
\setsymbol{LFI:white:noise:sensitivity:LFI20:Rad:M:units}{590.8\muKs}
\setsymbol{LFI:white:noise:sensitivity:LFI21:Rad:M:units}{455.2\muKs}
\setsymbol{LFI:white:noise:sensitivity:LFI22:Rad:M:units}{492.0\muKs}
\setsymbol{LFI:white:noise:sensitivity:LFI23:Rad:M:units}{507.7\muKs}
\setsymbol{LFI:white:noise:sensitivity:LFI24:Rad:M:units}{462.2\muKs}
\setsymbol{LFI:white:noise:sensitivity:LFI25:Rad:M:units}{413.6\muKs}
\setsymbol{LFI:white:noise:sensitivity:LFI26:Rad:M:units}{478.6\muKs}
\setsymbol{LFI:white:noise:sensitivity:LFI27:Rad:M:units}{277.7\muKs}
\setsymbol{LFI:white:noise:sensitivity:LFI28:Rad:M:units}{312.3\muKs}
\setsymbol{LFI:white:noise:sensitivity:LFI18:Rad:S:units}{465.7\muKs}
\setsymbol{LFI:white:noise:sensitivity:LFI19:Rad:S:units}{555.6\muKs}
\setsymbol{LFI:white:noise:sensitivity:LFI20:Rad:S:units}{623.2\muKs}
\setsymbol{LFI:white:noise:sensitivity:LFI21:Rad:S:units}{564.1\muKs}
\setsymbol{LFI:white:noise:sensitivity:LFI22:Rad:S:units}{534.4\muKs}
\setsymbol{LFI:white:noise:sensitivity:LFI23:Rad:S:units}{542.4\muKs}
\setsymbol{LFI:white:noise:sensitivity:LFI24:Rad:S:units}{399.2\muKs}
\setsymbol{LFI:white:noise:sensitivity:LFI25:Rad:S:units}{392.6\muKs}
\setsymbol{LFI:white:noise:sensitivity:LFI26:Rad:S:units}{418.6\muKs}
\setsymbol{LFI:white:noise:sensitivity:LFI27:Rad:S:units}{302.9\muKs}
\setsymbol{LFI:white:noise:sensitivity:LFI28:Rad:S:units}{285.3\muKs}

\setsymbol{LFI:white:noise:sensitivity:LFI18:Rad:M}{512.0}
\setsymbol{LFI:white:noise:sensitivity:LFI19:Rad:M}{581.4}
\setsymbol{LFI:white:noise:sensitivity:LFI20:Rad:M}{590.8}
\setsymbol{LFI:white:noise:sensitivity:LFI21:Rad:M}{455.2}
\setsymbol{LFI:white:noise:sensitivity:LFI22:Rad:M}{492.0}
\setsymbol{LFI:white:noise:sensitivity:LFI23:Rad:M}{507.7}
\setsymbol{LFI:white:noise:sensitivity:LFI24:Rad:M}{462.2}
\setsymbol{LFI:white:noise:sensitivity:LFI25:Rad:M}{413.6}
\setsymbol{LFI:white:noise:sensitivity:LFI26:Rad:M}{478.6}
\setsymbol{LFI:white:noise:sensitivity:LFI27:Rad:M}{277.7}
\setsymbol{LFI:white:noise:sensitivity:LFI28:Rad:M}{312.3}
\setsymbol{LFI:white:noise:sensitivity:LFI18:Rad:S}{465.7}
\setsymbol{LFI:white:noise:sensitivity:LFI19:Rad:S}{555.6}
\setsymbol{LFI:white:noise:sensitivity:LFI20:Rad:S}{623.2}
\setsymbol{LFI:white:noise:sensitivity:LFI21:Rad:S}{564.1}
\setsymbol{LFI:white:noise:sensitivity:LFI22:Rad:S}{534.4}
\setsymbol{LFI:white:noise:sensitivity:LFI23:Rad:S}{542.4}
\setsymbol{LFI:white:noise:sensitivity:LFI24:Rad:S}{399.2}
\setsymbol{LFI:white:noise:sensitivity:LFI25:Rad:S}{392.6}
\setsymbol{LFI:white:noise:sensitivity:LFI26:Rad:S}{418.6}
\setsymbol{LFI:white:noise:sensitivity:LFI27:Rad:S}{302.9}
\setsymbol{LFI:white:noise:sensitivity:LFI28:Rad:S}{285.3}

\setsymbol{LFI:knee:frequency:70GHz:units}{29.5\mHz}
\setsymbol{LFI:knee:frequency:44GHz:units}{56.2\mHz}
\setsymbol{LFI:knee:frequency:30GHz:units}{113.7\mHz}

\setsymbol{LFI:knee:frequency:70GHz}{29.5}
\setsymbol{LFI:knee:frequency:44GHz}{56.2}
\setsymbol{LFI:knee:frequency:30GHz}{113.7}

\setsymbol{LFI:knee:frequency:LFI18:Rad:M:units}{16.3\mHz}
\setsymbol{LFI:knee:frequency:LFI19:Rad:M:units}{15.1\mHz}
\setsymbol{LFI:knee:frequency:LFI20:Rad:M:units}{18.7\mHz}
\setsymbol{LFI:knee:frequency:LFI21:Rad:M:units}{37.2\mHz}
\setsymbol{LFI:knee:frequency:LFI22:Rad:M:units}{12.7\mHz}
\setsymbol{LFI:knee:frequency:LFI23:Rad:M:units}{34.6\mHz}
\setsymbol{LFI:knee:frequency:LFI24:Rad:M:units}{46.2\mHz}
\setsymbol{LFI:knee:frequency:LFI25:Rad:M:units}{24.9\mHz}
\setsymbol{LFI:knee:frequency:LFI26:Rad:M:units}{67.6\mHz}
\setsymbol{LFI:knee:frequency:LFI27:Rad:M:units}{187.4\mHz}
\setsymbol{LFI:knee:frequency:LFI28:Rad:M:units}{122.2\mHz}
\setsymbol{LFI:knee:frequency:LFI18:Rad:S:units}{17.7\mHz}
\setsymbol{LFI:knee:frequency:LFI19:Rad:S:units}{22.0\mHz}
\setsymbol{LFI:knee:frequency:LFI20:Rad:S:units}{8.7\mHz}
\setsymbol{LFI:knee:frequency:LFI21:Rad:S:units}{25.9\mHz}
\setsymbol{LFI:knee:frequency:LFI22:Rad:S:units}{15.8\mHz}
\setsymbol{LFI:knee:frequency:LFI23:Rad:S:units}{129.8\mHz}
\setsymbol{LFI:knee:frequency:LFI24:Rad:S:units}{100.9\mHz}
\setsymbol{LFI:knee:frequency:LFI25:Rad:S:units}{38.9\mHz}
\setsymbol{LFI:knee:frequency:LFI26:Rad:S:units}{58.9\mHz}
\setsymbol{LFI:knee:frequency:LFI27:Rad:S:units}{104.4\mHz}
\setsymbol{LFI:knee:frequency:LFI28:Rad:S:units}{40.7\mHz}

\setsymbol{LFI:knee:frequency:LFI18:Rad:M}{16.3}
\setsymbol{LFI:knee:frequency:LFI19:Rad:M}{15.1}
\setsymbol{LFI:knee:frequency:LFI20:Rad:M}{18.7}
\setsymbol{LFI:knee:frequency:LFI21:Rad:M}{37.2}
\setsymbol{LFI:knee:frequency:LFI22:Rad:M}{12.7}
\setsymbol{LFI:knee:frequency:LFI23:Rad:M}{34.6}
\setsymbol{LFI:knee:frequency:LFI24:Rad:M}{46.2}
\setsymbol{LFI:knee:frequency:LFI25:Rad:M}{24.9}
\setsymbol{LFI:knee:frequency:LFI26:Rad:M}{67.6}
\setsymbol{LFI:knee:frequency:LFI27:Rad:M}{187.4}
\setsymbol{LFI:knee:frequency:LFI28:Rad:M}{122.2}
\setsymbol{LFI:knee:frequency:LFI18:Rad:S}{17.7}
\setsymbol{LFI:knee:frequency:LFI19:Rad:S}{22.0}
\setsymbol{LFI:knee:frequency:LFI20:Rad:S}{8.7}
\setsymbol{LFI:knee:frequency:LFI21:Rad:S}{25.9}
\setsymbol{LFI:knee:frequency:LFI22:Rad:S}{15.8}
\setsymbol{LFI:knee:frequency:LFI23:Rad:S}{129.8}
\setsymbol{LFI:knee:frequency:LFI24:Rad:S}{100.9}
\setsymbol{LFI:knee:frequency:LFI25:Rad:S}{38.9}
\setsymbol{LFI:knee:frequency:LFI26:Rad:S}{58.9}
\setsymbol{LFI:knee:frequency:LFI27:Rad:S}{104.4}
\setsymbol{LFI:knee:frequency:LFI28:Rad:S}{40.7}

\setsymbol{LFI:slope:70GHz:units}{$-1.03$\mHz}
\setsymbol{LFI:slope:44GHz:units}{$-0.89$\mHz}
\setsymbol{LFI:slope:30GHz:units}{$-0.87$\mHz}

\setsymbol{LFI:slope:70GHz}{$-1.03$}
\setsymbol{LFI:slope:44GHz}{$-0.89$}
\setsymbol{LFI:slope:30GHz}{$-0.87$}

\setsymbol{LFI:slope:LFI18:Rad:M:units}{$-1.04$\mHz}
\setsymbol{LFI:slope:LFI19:Rad:M:units}{$-1.09$\mHz}
\setsymbol{LFI:slope:LFI20:Rad:M:units}{$-0.69$\mHz}
\setsymbol{LFI:slope:LFI21:Rad:M:units}{$-1.56$\mHz}
\setsymbol{LFI:slope:LFI22:Rad:M:units}{$-1.01$\mHz}
\setsymbol{LFI:slope:LFI23:Rad:M:units}{$-0.96$\mHz}
\setsymbol{LFI:slope:LFI24:Rad:M:units}{$-0.83$\mHz}
\setsymbol{LFI:slope:LFI25:Rad:M:units}{$-0.91$\mHz}
\setsymbol{LFI:slope:LFI26:Rad:M:units}{$-0.95$\mHz}
\setsymbol{LFI:slope:LFI27:Rad:M:units}{$-0.87$\mHz}
\setsymbol{LFI:slope:LFI28:Rad:M:units}{$-0.88$\mHz}
\setsymbol{LFI:slope:LFI18:Rad:S:units}{$-1.15$\mHz}
\setsymbol{LFI:slope:LFI19:Rad:S:units}{$-1.00$\mHz}
\setsymbol{LFI:slope:LFI20:Rad:S:units}{$-0.95$\mHz}
\setsymbol{LFI:slope:LFI21:Rad:S:units}{$-0.92$\mHz}
\setsymbol{LFI:slope:LFI22:Rad:S:units}{$-1.01$\mHz}
\setsymbol{LFI:slope:LFI23:Rad:S:units}{$-0.95$\mHz}
\setsymbol{LFI:slope:LFI24:Rad:S:units}{$-0.73$\mHz}
\setsymbol{LFI:slope:LFI25:Rad:S:units}{$-1.16$\mHz}
\setsymbol{LFI:slope:LFI26:Rad:S:units}{$-0.79$\mHz}
\setsymbol{LFI:slope:LFI27:Rad:S:units}{$-0.82$\mHz}
\setsymbol{LFI:slope:LFI28:Rad:S:units}{$-0.91$\mHz}

\setsymbol{LFI:slope:LFI18:Rad:M}{$-1.04$}
\setsymbol{LFI:slope:LFI19:Rad:M}{$-1.09$}
\setsymbol{LFI:slope:LFI20:Rad:M}{$-0.69$}
\setsymbol{LFI:slope:LFI21:Rad:M}{$-1.56$}
\setsymbol{LFI:slope:LFI22:Rad:M}{$-1.01$}
\setsymbol{LFI:slope:LFI23:Rad:M}{$-0.96$}
\setsymbol{LFI:slope:LFI24:Rad:M}{$-0.83$}
\setsymbol{LFI:slope:LFI25:Rad:M}{$-0.91$}
\setsymbol{LFI:slope:LFI26:Rad:M}{$-0.95$}
\setsymbol{LFI:slope:LFI27:Rad:M}{$-0.87$}
\setsymbol{LFI:slope:LFI28:Rad:M}{$-0.88$}
\setsymbol{LFI:slope:LFI18:Rad:S}{$-1.15$}
\setsymbol{LFI:slope:LFI19:Rad:S}{$-1.00$}
\setsymbol{LFI:slope:LFI20:Rad:S}{$-0.95$}
\setsymbol{LFI:slope:LFI21:Rad:S}{$-0.92$}
\setsymbol{LFI:slope:LFI22:Rad:S}{$-1.01$}
\setsymbol{LFI:slope:LFI23:Rad:S}{$-0.95$}
\setsymbol{LFI:slope:LFI24:Rad:S}{$-0.73$}
\setsymbol{LFI:slope:LFI25:Rad:S}{$-1.16$}
\setsymbol{LFI:slope:LFI26:Rad:S}{$-0.79$}
\setsymbol{LFI:slope:LFI27:Rad:S}{$-0.82$}
\setsymbol{LFI:slope:LFI28:Rad:S}{$-0.91$}

\setsymbol{LFI:FWHM:70GHz:units}{13\parcm01}
\setsymbol{LFI:FWHM:44GHz:units}{27\parcm92}
\setsymbol{LFI:FWHM:30GHz:units}{32\parcm65}

\setsymbol{LFI:FWHM:70GHz}{13.01}
\setsymbol{LFI:FWHM:44GHz}{27.92}
\setsymbol{LFI:FWHM:30GHz}{32.65}

\setsymbol{LFI:FWHM:LFI18:units}{13\parcm39}
\setsymbol{LFI:FWHM:LFI19:units}{13\parcm01}
\setsymbol{LFI:FWHM:LFI20:units}{12\parcm75}
\setsymbol{LFI:FWHM:LFI21:units}{12\parcm74}
\setsymbol{LFI:FWHM:LFI22:units}{12\parcm87}
\setsymbol{LFI:FWHM:LFI23:units}{13\parcm27}
\setsymbol{LFI:FWHM:LFI24:units}{22\parcm98}
\setsymbol{LFI:FWHM:LFI25:units}{30\parcm46}
\setsymbol{LFI:FWHM:LFI26:units}{30\parcm31}
\setsymbol{LFI:FWHM:LFI27:units}{32\parcm65}
\setsymbol{LFI:FWHM:LFI28:units}{32\parcm66}

\setsymbol{LFI:FWHM:LFI18}{13.39}
\setsymbol{LFI:FWHM:LFI19}{13.01}
\setsymbol{LFI:FWHM:LFI20}{12.75}
\setsymbol{LFI:FWHM:LFI21}{12.74}
\setsymbol{LFI:FWHM:LFI22}{12.87}
\setsymbol{LFI:FWHM:LFI23}{13.27}
\setsymbol{LFI:FWHM:LFI24}{22.98}
\setsymbol{LFI:FWHM:LFI25}{30.46}
\setsymbol{LFI:FWHM:LFI26}{30.31}
\setsymbol{LFI:FWHM:LFI27}{32.65}
\setsymbol{LFI:FWHM:LFI28}{32.66}

\setsymbol{LFI:FWHM:uncertainty:LFI18:units}{0.170\arcm}
\setsymbol{LFI:FWHM:uncertainty:LFI19:units}{0.174\arcm}
\setsymbol{LFI:FWHM:uncertainty:LFI20:units}{0.170\arcm}
\setsymbol{LFI:FWHM:uncertainty:LFI21:units}{0.156\arcm}
\setsymbol{LFI:FWHM:uncertainty:LFI22:units}{0.164\arcm}
\setsymbol{LFI:FWHM:uncertainty:LFI23:units}{0.171\arcm}
\setsymbol{LFI:FWHM:uncertainty:LFI24:units}{0.652\arcm}
\setsymbol{LFI:FWHM:uncertainty:LFI25:units}{1.075\arcm}
\setsymbol{LFI:FWHM:uncertainty:LFI26:units}{1.131\arcm}
\setsymbol{LFI:FWHM:uncertainty:LFI27:units}{1.266\arcm}
\setsymbol{LFI:FWHM:uncertainty:LFI28:units}{1.287\arcm}

\setsymbol{LFI:FWHM:uncertainty:LFI18}{0.170}
\setsymbol{LFI:FWHM:uncertainty:LFI19}{0.174}
\setsymbol{LFI:FWHM:uncertainty:LFI20}{0.170}
\setsymbol{LFI:FWHM:uncertainty:LFI21}{0.156}
\setsymbol{LFI:FWHM:uncertainty:LFI22}{0.164}
\setsymbol{LFI:FWHM:uncertainty:LFI23}{0.171}
\setsymbol{LFI:FWHM:uncertainty:LFI24}{0.652}
\setsymbol{LFI:FWHM:uncertainty:LFI25}{1.075}
\setsymbol{LFI:FWHM:uncertainty:LFI26}{1.131}
\setsymbol{LFI:FWHM:uncertainty:LFI27}{1.266}
\setsymbol{LFI:FWHM:uncertainty:LFI28}{1.287}

\setsymbol{HFI:center:frequency:100GHz:units}{100\,GHz}
\setsymbol{HFI:center:frequency:143GHz:units}{143\,GHz}
\setsymbol{HFI:center:frequency:217GHz:units}{217\,GHz}
\setsymbol{HFI:center:frequency:353GHz:units}{353\,GHz}
\setsymbol{HFI:center:frequency:545GHz:units}{545\,GHz}
\setsymbol{HFI:center:frequency:857GHz:units}{857\,GHz}

\setsymbol{HFI:center:frequency:100GHz}{100}
\setsymbol{HFI:center:frequency:143GHz}{143}
\setsymbol{HFI:center:frequency:217GHz}{217}
\setsymbol{HFI:center:frequency:353GHz}{353}
\setsymbol{HFI:center:frequency:545GHz}{545}
\setsymbol{HFI:center:frequency:857GHz}{857}

\setsymbol{HFI:Ndetectors:100GHz}{8}
\setsymbol{HFI:Ndetectors:143GHz}{11}
\setsymbol{HFI:Ndetectors:217GHz}{12}
\setsymbol{HFI:Ndetectors:353GHz}{12}
\setsymbol{HFI:Ndetectors:545GHz}{3}
\setsymbol{HFI:Ndetectors:857GHz}{4}

\setsymbol{HFI:FWHM:Maps:100GHz:units}{9\parcm88}
\setsymbol{HFI:FWHM:Maps:143GHz:units}{7\parcm18}
\setsymbol{HFI:FWHM:Maps:217GHz:units}{4\parcm87}
\setsymbol{HFI:FWHM:Maps:353GHz:units}{4\parcm65}
\setsymbol{HFI:FWHM:Maps:545GHz:units}{4\parcm72}
\setsymbol{HFI:FWHM:Maps:857GHz:units}{4\parcm39}
\setsymbol{HFI:FWHM:Maps:100GHz}{9.88}
\setsymbol{HFI:FWHM:Maps:143GHz}{7.18}
\setsymbol{HFI:FWHM:Maps:217GHz}{4.87}
\setsymbol{HFI:FWHM:Maps:353GHz}{4.65}
\setsymbol{HFI:FWHM:Maps:545GHz}{4.72}
\setsymbol{HFI:FWHM:Maps:857GHz}{4.39}

\setsymbol{HFI:beam:ellipticity:Maps:100GHz}{1.15}
\setsymbol{HFI:beam:ellipticity:Maps:143GHz}{1.01}
\setsymbol{HFI:beam:ellipticity:Maps:217GHz}{1.06}
\setsymbol{HFI:beam:ellipticity:Maps:353GHz}{1.05}
\setsymbol{HFI:beam:ellipticity:Maps:545GHz}{1.14}
\setsymbol{HFI:beam:ellipticity:Maps:857GHz}{1.19}

\setsymbol{HFI:FWHM:Mars:100GHz:units}{9\parcm37}
\setsymbol{HFI:FWHM:Mars:143GHz:units}{7\parcm04}
\setsymbol{HFI:FWHM:Mars:217GHz:units}{4\parcm68}
\setsymbol{HFI:FWHM:Mars:353GHz:units}{4\parcm43}
\setsymbol{HFI:FWHM:Mars:545GHz:units}{3\parcm80}
\setsymbol{HFI:FWHM:Mars:857GHz:units}{3\parcm67}

\setsymbol{HFI:FWHM:Mars:100GHz}{9.37}
\setsymbol{HFI:FWHM:Mars:143GHz}{7.04}
\setsymbol{HFI:FWHM:Mars:217GHz}{4.68}
\setsymbol{HFI:FWHM:Mars:353GHz}{4.43}
\setsymbol{HFI:FWHM:Mars:545GHz}{3.80}
\setsymbol{HFI:FWHM:Mars:857GHz}{3.67}

\setsymbol{HFI:beam:ellipticity:Mars:100GHz}{1.18}
\setsymbol{HFI:beam:ellipticity:Mars:143GHz}{1.03}
\setsymbol{HFI:beam:ellipticity:Mars:217GHz}{1.14}
\setsymbol{HFI:beam:ellipticity:Mars:353GHz}{1.09}
\setsymbol{HFI:beam:ellipticity:Mars:545GHz}{1.25}
\setsymbol{HFI:beam:ellipticity:Mars:857GHz}{1.03}

\setsymbol{HFI:CMB:relative:calibration:100GHz}{$\lsim 1\%$}
\setsymbol{HFI:CMB:relative:calibration:143GHz}{$\lsim 1\%$}
\setsymbol{HFI:CMB:relative:calibration:217GHz}{$\lsim 1\%$}
\setsymbol{HFI:CMB:relative:calibration:353GHz}{$\lsim 1\%$}
\setsymbol{HFI:CMB:relative:calibration:545GHz}{}
\setsymbol{HFI:CMB:relative:calibration:857GHz}{}

\setsymbol{HFI:CMB:absolute:calibration:100GHz}{$\lsim 2\%$}
\setsymbol{HFI:CMB:absolute:calibration:143GHz}{$\lsim 2\%$}
\setsymbol{HFI:CMB:absolute:calibration:217GHz}{$\lsim 2\%$}
\setsymbol{HFI:CMB:absolute:calibration:353GHz}{$\lsim 2\%$}
\setsymbol{HFI:CMB:absolute:calibration:545GHz}{}
\setsymbol{HFI:CMB:absolute:calibration:857GHz}{}

\setsymbol{HFI:FIRAS:gain:calibration:accuracy:statistical:100GHz}{}
\setsymbol{HFI:FIRAS:gain:calibration:accuracy:statistical:143GHz}{}
\setsymbol{HFI:FIRAS:gain:calibration:accuracy:statistical:217GHz}{}
\setsymbol{HFI:FIRAS:gain:calibration:accuracy:statistical:353GHz}{2.5\%}
\setsymbol{HFI:FIRAS:gain:calibration:accuracy:statistical:545GHz}{1\%}
\setsymbol{HFI:FIRAS:gain:calibration:accuracy:statistical:857GHz}{0.5\%}

\setsymbol{HFI:FIRAS:gain:calibration:accuracy:systematic:100GHz}{}
\setsymbol{HFI:FIRAS:gain:calibration:accuracy:systematic:143GHz}{}
\setsymbol{HFI:FIRAS:gain:calibration:accuracy:systematic:217GHz}{}
\setsymbol{HFI:FIRAS:gain:calibration:accuracy:systematic:353GHz}{}
\setsymbol{HFI:FIRAS:gain:calibration:accuracy:systematic:545GHz}{7\%}
\setsymbol{HFI:FIRAS:gain:calibration:accuracy:systematic:857GHz}{7\%}

\setsymbol{HFI:FIRAS:zero:point:accuracy:100GHz:units}{0.8\MJysr}
\setsymbol{HFI:FIRAS:zero:point:accuracy:143GHz:units}{}
\setsymbol{HFI:FIRAS:zero:point:accuracy:217GHz:units}{}
\setsymbol{HFI:FIRAS:zero:point:accuracy:353GHz:units}{1.4\MJysr}
\setsymbol{HFI:FIRAS:zero:point:accuracy:545GHz:units}{2.2\MJysr}
\setsymbol{HFI:FIRAS:zero:point:accuracy:857GHz:units}{1.7\MJysr}

\setsymbol{HFI:FIRAS:zero:point:accuracy:100GHz}{0.8}
\setsymbol{HFI:FIRAS:zero:point:accuracy:143GHz}{}
\setsymbol{HFI:FIRAS:zero:point:accuracy:217GHz}{}
\setsymbol{HFI:FIRAS:zero:point:accuracy:353GHz}{1.4}
\setsymbol{HFI:FIRAS:zero:point:accuracy:545GHz}{2.2}
\setsymbol{HFI:FIRAS:zero:point:accuracy:857GHz}{1.7}

\setsymbol{HFI:unit:conversion:100GHz:units}{0.2415\MJysrmK}
\setsymbol{HFI:unit:conversion:143GHz:units}{0.3694\MJysrmK}
\setsymbol{HFI:unit:conversion:217GHz:units}{0.4811\MJysrmK}
\setsymbol{HFI:unit:conversion:353GHz:units}{0.2883\MJysrmK}
\setsymbol{HFI:unit:conversion:545GHz:units}{0.05826\MJysrmK}
\setsymbol{HFI:unit:conversion:857GHz:units}{0.002238\MJysrmK}

\setsymbol{HFI:unit:conversion:100GHz}{0.2415}
\setsymbol{HFI:unit:conversion:143GHz}{0.3694}
\setsymbol{HFI:unit:conversion:217GHz}{0.4811}
\setsymbol{HFI:unit:conversion:353GHz}{0.2883}
\setsymbol{HFI:unit:conversion:545GHz}{0.05826}
\setsymbol{HFI:unit:conversion:857GHz}{0.002238}

\setsymbol{HFI:colour:correction:alpha=-2:V1.01:100GHz}{0.9893}
\setsymbol{HFI:colour:correction:alpha=-2:V1.01:143GHz}{0.9759}
\setsymbol{HFI:colour:correction:alpha=-2:V1.01:217GHz}{1.0007}
\setsymbol{HFI:colour:correction:alpha=-2:V1.01:353GHz}{1.0028}
\setsymbol{HFI:colour:correction:alpha=-2:V1.01:545GHz}{1.0019}
\setsymbol{HFI:colour:correction:alpha=-2:V1.01:857GHz}{0.9889}

\setsymbol{HFI:colour:correction:alpha=0:V1.01:100GHz}{1.0008}
\setsymbol{HFI:colour:correction:alpha=0:V1.01:143GHz}{1.0148}
\setsymbol{HFI:colour:correction:alpha=0:V1.01:217GHz}{0.9909}
\setsymbol{HFI:colour:correction:alpha=0:V1.01:353GHz}{0.9888}
\setsymbol{HFI:colour:correction:alpha=0:V1.01:545GHz}{0.9878}
\setsymbol{HFI:colour:correction:alpha=0:V1.01:857GHz}{1.0014}

\providecommand{\sorthelp}[1]{}

\newcommand{\StokesI}{I}                    
\newcommand{\StokesQ}{Q}                    
\newcommand{\StokesU}{U}                    
\newcommand{\polfrac}{p}                    
\newcommand{\polang}{ \psi}                  
\newcommand{\sigpolfrac}{\sigma_{\polfrac}}   
\newcommand{\sigpolang}{\sigma_{\polang}}     
\newcommand{\polangsky}{\gamma}             
\newcommand{\DeltaAng}{\mathcal{S}}          
\newcommand{\pmax}{p_\mathrm{max}}		

\newcommand{\DeltaAngName}{angle dispersion function}
\newcommand{\DeltaAngNameMaj}{Angle dispersion function}
\newcommand{\IntrinsicpName}{intrinsic polarization fraction}

\newcommand{\planck}{\Planck}  

\newcommand{\healpix}{{\tt HEALPix}}

\newcommand{\viewangle}{\alpha}

\begin{document}

\title{\Planck~intermediate results. XX. Comparison of polarized thermal emission from Galactic dust with simulations of MHD turbulence}
\titlerunning{Comparison of polarized thermal emission from Galactic dust with simulations of MHD turbulence}
\author{\small
Planck Collaboration:
P.~A.~R.~Ade\inst{71}
\and
N.~Aghanim\inst{50}
\and
D.~Alina\inst{76, 9}
\and
M.~I.~R.~Alves\inst{50}
\and
G.~Aniano\inst{50}
\and
C.~Armitage-Caplan\inst{74}
\and
M.~Arnaud\inst{61}
\and
D.~Arzoumanian\inst{50}
\and
M.~Ashdown\inst{58, 5}
\and
F.~Atrio-Barandela\inst{16}
% ...
}
\institute{\small
APC, AstroParticule et Cosmologie, Universit\'{e} Paris Diderot, CNRS/IN2P3, CEA/lrfu, Observatoire de Paris, Sorbonne Paris Cit\'{e}, 10, rue Alice Domon et L\'{e}onie Duquet, 75205 Paris Cedex 13, France\\
\and
African Institute for Mathematical Sciences, 6-8 Melrose Road, Muizenberg, Cape Town, South Africa\\
\and
Agenzia Spaziale Italiana Science Data Center, Via del Politecnico snc, 00133, Roma, Italy\\
\and
Agenzia Spaziale Italiana, Viale Liegi 26, Roma, Italy\\
\and
Astrophysics Group, Cavendish Laboratory, University of Cambridge, J J Thomson Avenue, Cambridge CB3 0HE, U.K.\\
\and
Astrophysics \& Cosmology Research Unit, School of Mathematics, Statistics \& Computer Science, University of KwaZulu-Natal, Westville Campus, Private Bag X54001, Durban 4000, South Africa\\
\and
Atacama Large Millimeter/submillimeter Array, ALMA Santiago Central Offices, Alonso de Cordova 3107, Vitacura, Casilla 763 0355, Santiago, Chile\\
\and
CITA, University of Toronto, 60 St. George St., Toronto, ON M5S 3H8, Canada\\
\and
CNRS, IRAP, 9 Av. colonel Roche, BP 44346, F-31028 Toulouse cedex 4, France\\
\and
California Institute of Technology, Pasadena, California, U.S.A.\\
\and
Centro de Estudios de F\'{i}sica del Cosmos de Arag\'{o}n (CEFCA), Plaza San Juan, 1, planta 2, E-44001, Teruel, Spain\\
\and
Computational Cosmology Center, Lawrence Berkeley National Laboratory, Berkeley, California, U.S.A.\\
\and
DSM/Irfu/SPP, CEA-Saclay, F-91191 Gif-sur-Yvette Cedex, France\\
\and
DTU Space, National Space Institute, Technical University of Denmark, Elektrovej 327, DK-2800 Kgs. Lyngby, Denmark\\
\and
D\'{e}partement de Physique Th\'{e}orique, Universit\'{e} de Gen\`{e}ve, 24, Quai E. Ansermet,1211 Gen\`{e}ve 4, Switzerland\\
\and
Departamento de F\'{\i}sica Fundamental, Facultad de Ciencias, Universidad de Salamanca, 37008 Salamanca, Spain\\
\and
Departamento de F\'{\i}sica, Universidad de Oviedo, Avda. Calvo Sotelo s/n, Oviedo, Spain\\
\and
Department of Astronomy and Astrophysics, University of Toronto, 50 Saint George Street, Toronto, Ontario, Canada\\
\and
Department of Astrophysics/IMAPP, Radboud University Nijmegen, P.O. Box 9010, 6500 GL Nijmegen, The Netherlands\\
\and
Department of Physics \& Astronomy, University of British Columbia, 6224 Agricultural Road, Vancouver, British Columbia, Canada\\
\and
Department of Physics and Astronomy, Dana and David Dornsife College of Letter, Arts and Sciences, University of Southern California, Los Angeles, CA 90089, U.S.A.\\
\and
Department of Physics and Astronomy, University College London, London WC1E 6BT, U.K.\\
\and
Department of Physics, Florida State University, Keen Physics Building, 77 Chieftan Way, Tallahassee, Florida, U.S.A.\\
\and
Department of Physics, Gustaf H\"{a}llstr\"{o}min katu 2a, University of Helsinki, Helsinki, Finland\\
\and
Department of Physics, Princeton University, Princeton, New Jersey, U.S.A.\\
\and
Department of Physics, University of California, Santa Barbara, California, U.S.A.\\
\and
Department of Physics, University of Illinois at Urbana-Champaign, 1110 West Green Street, Urbana, Illinois, U.S.A.\\
\and
Dipartimento di Fisica e Astronomia G. Galilei, Universit\`{a} degli Studi di Padova, via Marzolo 8, 35131 Padova, Italy\\
\and
Dipartimento di Fisica e Scienze della Terra, Universit\`{a} di Ferrara, Via Saragat 1, 44122 Ferrara, Italy\\
\and
Dipartimento di Fisica, Universit\`{a} La Sapienza, P. le A. Moro 2, Roma, Italy\\
\and
Dipartimento di Fisica, Universit\`{a} degli Studi di Milano, Via Celoria, 16, Milano, Italy\\
\and
Dipartimento di Fisica, Universit\`{a} degli Studi di Trieste, via A. Valerio 2, Trieste, Italy\\
\and
Dipartimento di Fisica, Universit\`{a} di Roma Tor Vergata, Via della Ricerca Scientifica, 1, Roma, Italy\\
\and
Discovery Center, Niels Bohr Institute, Blegdamsvej 17, Copenhagen, Denmark\\
\and
European Southern Observatory, ESO Vitacura, Alonso de Cordova 3107, Vitacura, Casilla 19001, Santiago, Chile\\
\and
European Space Agency, ESAC, Planck Science Office, Camino bajo del Castillo, s/n, Urbanizaci\'{o}n Villafranca del Castillo, Villanueva de la Ca\~{n}ada, Madrid, Spain\\
\and
European Space Agency, ESTEC, Keplerlaan 1, 2201 AZ Noordwijk, The Netherlands\\
\and
Helsinki Institute of Physics, Gustaf H\"{a}llstr\"{o}min katu 2, University of Helsinki, Helsinki, Finland\\
\and
INAF - Osservatorio Astrofisico di Catania, Via S. Sofia 78, Catania, Italy\\
\and
INAF - Osservatorio Astronomico di Padova, Vicolo dell'Osservatorio 5, Padova, Italy\\
\and
INAF - Osservatorio Astronomico di Roma, via di Frascati 33, Monte Porzio Catone, Italy\\
\and
INAF - Osservatorio Astronomico di Trieste, Via G.B. Tiepolo 11, Trieste, Italy\\
\and
INAF/IASF Bologna, Via Gobetti 101, Bologna, Italy\\
\and
INAF/IASF Milano, Via E. Bassini 15, Milano, Italy\\
\and
INFN, Sezione di Bologna, Via Irnerio 46, I-40126, Bologna, Italy\\
\and
INFN, Sezione di Roma 1, Universit\`{a} di Roma Sapienza, Piazzale Aldo Moro 2, 00185, Roma, Italy\\
\and
INFN/National Institute for Nuclear Physics, Via Valerio 2, I-34127 Trieste, Italy\\
\and
Imperial College London, Astrophysics group, Blackett Laboratory, Prince Consort Road, London, SW7 2AZ, U.K.\\
\and
Infrared Processing and Analysis Center, California Institute of Technology, Pasadena, CA 91125, U.S.A.\\
\and
Institut d'Astrophysique Spatiale, CNRS (UMR8617) Universit\'{e} Paris-Sud 11, B\^{a}timent 121, Orsay, France\\
\and
Institut d'Astrophysique de Paris, CNRS (UMR7095), 98 bis Boulevard Arago, F-75014, Paris, France\\
\and
Institute for Space Sciences, Bucharest-Magurale, Romania\\
\and
Institute of Astronomy, University of Cambridge, Madingley Road, Cambridge CB3 0HA, U.K.\\
\and
Institute of Theoretical Astrophysics, University of Oslo, Blindern, Oslo, Norway\\
\and
Instituto de F\'{\i}sica de Cantabria (CSIC-Universidad de Cantabria), Avda. de los Castros s/n, Santander, Spain\\
\and
Jet Propulsion Laboratory, California Institute of Technology, 4800 Oak Grove Drive, Pasadena, California, U.S.A.\\
\and
Jodrell Bank Centre for Astrophysics, Alan Turing Building, School of Physics and Astronomy, The University of Manchester, Oxford Road, Manchester, M13 9PL, U.K.\\
\and
Kavli Institute for Cosmology Cambridge, Madingley Road, Cambridge, CB3 0HA, U.K.\\
\and
LAL, Universit\'{e} Paris-Sud, CNRS/IN2P3, Orsay, France\\
\and
LERMA, CNRS, Observatoire de Paris, 61 Avenue de l'Observatoire, Paris, France\\
\and
Laboratoire AIM, IRFU/Service d'Astrophysique - CEA/DSM - CNRS - Universit\'{e} Paris Diderot, B\^{a}t. 709, CEA-Saclay, F-91191 Gif-sur-Yvette Cedex, France\\
\and
Laboratoire Traitement et Communication de l'Information, CNRS (UMR 5141) and T\'{e}l\'{e}com ParisTech, 46 rue Barrault F-75634 Paris Cedex 13, France\\
\and
Laboratoire de Physique Subatomique et de Cosmologie, Universit\'{e} Joseph Fourier Grenoble I, CNRS/IN2P3, Institut National Polytechnique de Grenoble, 53 rue des Martyrs, 38026 Grenoble cedex, France\\
\and
Laboratoire de Physique Th\'{e}orique, Universit\'{e} Paris-Sud 11 \& CNRS, B\^{a}timent 210, 91405 Orsay, France\\
\and
Lawrence Berkeley National Laboratory, Berkeley, California, U.S.A.\\
\and
Max-Planck-Institut f\"{u}r Astrophysik, Karl-Schwarzschild-Str. 1, 85741 Garching, Germany\\
\and
National University of Ireland, Department of Experimental Physics, Maynooth, Co. Kildare, Ireland\\
\and
Niels Bohr Institute, Blegdamsvej 17, Copenhagen, Denmark\\
\and
Observational Cosmology, Mail Stop 367-17, California Institute of Technology, Pasadena, CA, 91125, U.S.A.\\
\and
SISSA, Astrophysics Sector, via Bonomea 265, 34136, Trieste, Italy\\
\and
School of Physics and Astronomy, Cardiff University, Queens Buildings, The Parade, Cardiff, CF24 3AA, U.K.\\
\and
Space Sciences Laboratory, University of California, Berkeley, California, U.S.A.\\
\and
Special Astrophysical Observatory, Russian Academy of Sciences, Nizhnij Arkhyz, Zelenchukskiy region, Karachai-Cherkessian Republic, 369167, Russia\\
\and
Sub-Department of Astrophysics, University of Oxford, Keble Road, Oxford OX1 3RH, U.K.\\
\and
UPMC Univ Paris 06, UMR7095, 98 bis Boulevard Arago, F-75014, Paris, France\\
\and
Universit\'{e} de Toulouse, UPS-OMP, IRAP, F-31028 Toulouse cedex 4, France\\
% ...
}

\authorrunning{\Planck~Collaboration}
\date{Received ...; accepted ...}

  \abstract{ Polarized emission observed by \Planck~HFI at 353\,GHz
towards a sample of nearby fields is presented, focusing on
the statistics of polarization fractions $\polfrac$ and angles $\polang$. The polarization fractions and column densities in
these nearby fields are representative of the range of values obtained
over the whole sky. We find that: (i) the largest polarization fractions are reached in the most diffuse
fields; (ii) the maximum polarization
fraction $\pmax$ decreases with column density $N_\mathrm{H}$ in the more opaque
fields with $N_\mathrm{H} > 10^{21}\,\mathrm{cm}^{-2}$; and (iii) the polarization fraction along a given line of sight is correlated
with the local spatial coherence of the polarization angle. These observations are compared to
polarized emission maps computed in simulations of anisotropic
magnetohydrodynamical (MHD) turbulence in which we assume a
uniform \IntrinsicpName~of the dust grains.  We find that an estimate of this parameter may be recovered from the maximum polarization fraction
$\pmax$ in diffuse regions where the magnetic field is ordered on large
scales and perpendicular to the line of sight. This emphasizes the impact of anisotropies of the magnetic field on the emerging polarization signal. The decrease of the polarization fraction with column density in
nearby molecular clouds is well reproduced in the simulations,
indicating that it is essentially due to the turbulent
structure of the magnetic field: an accumulation of variously
polarized structures along the line of sight leads to such an
anti-correlation. 
In the simulations, polarization fractions are also found to anti-correlate with the \DeltaAngName~$\DeltaAng$. However, the dispersion of
the polarization angle for a given polarization fraction is found to be
larger in the simulations than in the observations, suggesting a shortcoming in the physical content of these numerical models. In summary, we find that the
turbulent structure of the magnetic field is able to reproduce the
main statistical properties of the dust polarization as observed in a
variety of nearby clouds, dense cores excluded, and that the large-scale field orientation
with respect to the line of sight plays a major role in the quantitative analysis of these
statistical properties.
 }

   \keywords{ISM: general, dust, magnetic fields, clouds -- Infrared: ISM -- Submillimetre: ISM -- Methods: observational, numerical, statistical}

   \maketitle

\begin{figure*}
\centerline{\includegraphics[width=18cm]{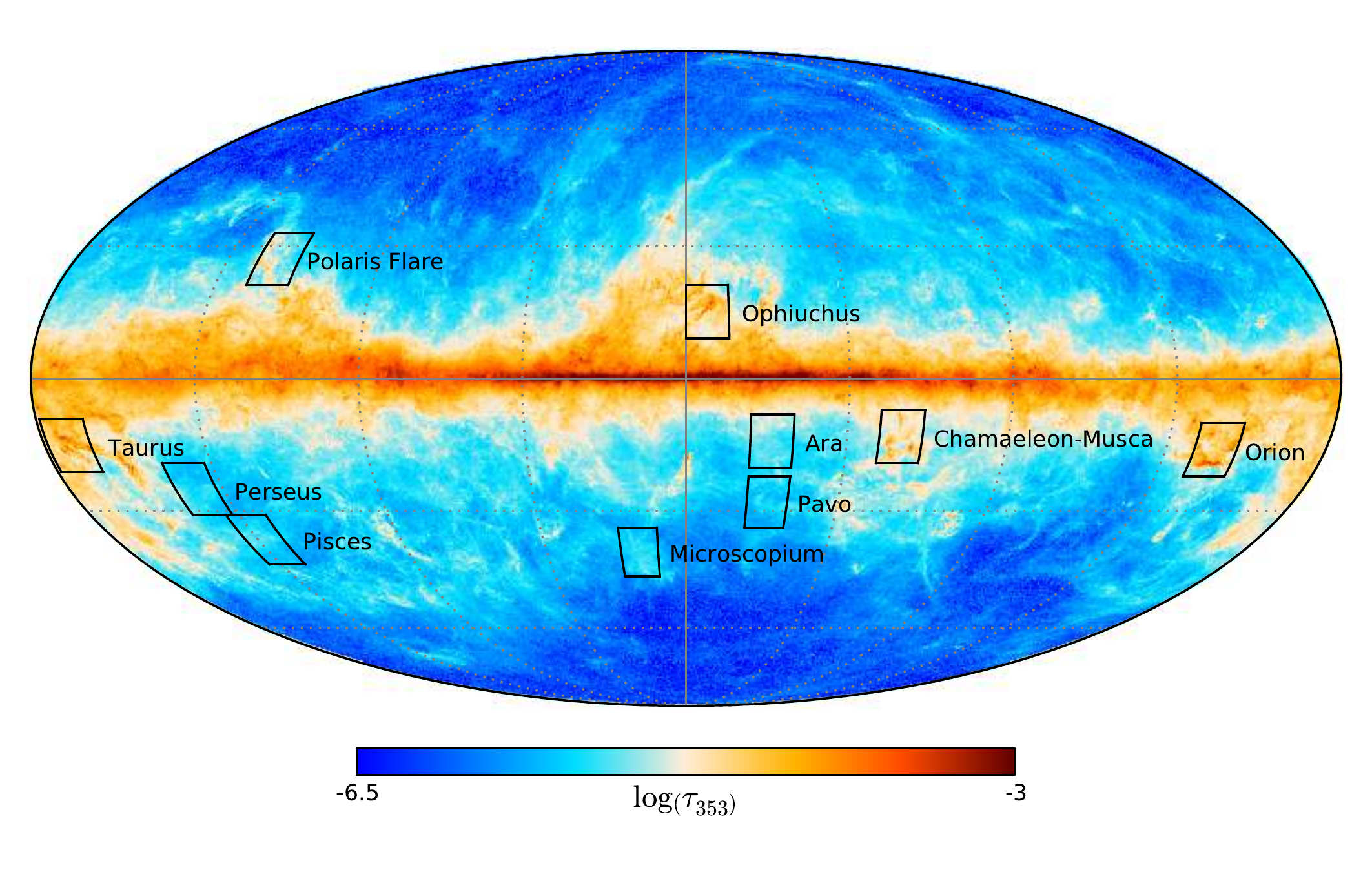}}
\caption{Locations of the selected nearby fields. The background map represents optical depth $\tau_{353}$ at 353\,GHz on a logarithmic scale, at 5\arcm~resolution~\citep{planck2013-p06b}. The map uses a Mollweide projection in Galactic coordinates, with $(l,b)=(0\deg{,}0\deg)$ at the centre.}
\label{fig:regions}
\end{figure*}

\begin{table*}[tmb]                 
\begingroup
\newdimen\tblskip \tblskip=5pt
\caption{Locations and properties of the selected fields: Galactic longitudes $l$ and latitudes $b$ of the centre of the $12^{\circ}\times 12^{\circ}$ fields; estimates of distances, masses and ages, where available; average and maximum column densities at 15\arcm~resolution; fraction $f_{22}$ of the pixels with $N_\mathrm{H}>10^{22}\,\mathrm{cm}^{-2}$; and  fraction $f_{21}$ of the pixels with $N_\mathrm{H}<10^{21}\,\mathrm{cm}^{-2}$. These fields are the same as several of those listed in Table 1 of~\cite{planck2014-XIX}.}
\label{table-fields}                            
\nointerlineskip
\vskip -3mm
\footnotesize
\setbox\tablebox=\vbox{
   \newdimen\digitwidth 
   \setbox0=\hbox{\rm 0} 
   \digitwidth=\wd0 
   \catcode`*=\active 
   \def*{\kern\digitwidth}
   \newdimen\signwidth 
   \setbox0=\hbox{+} 
   \signwidth=\wd0 
   \catcode`!=\active 
   \def!{\kern\signwidth}
\halign{\hbox to 1.15in{#\leaderfil}\tabskip 2.2em&
\hfil#&
\hfil#&
\hfil#&
\hfil#&
\hfil#&
\hfil#&
\hfil#&
\hfil#&
\hfil#\tabskip 0pt\cr
\noalign{\doubleline}
\omit&\hfil$l$\hfil&\hfil$b$\hfil&\hfil{Distance}$^a$\hfil&\hfil{Mass}$^b$\hfil&\hfil{Age}$^c$\hfil&\hfil$\left<N_\mathrm{H}\right>$\hfil&\hfil$\mathrm{max}\left(N_\mathrm{H}\right)$\hfil&\hfil$f_{22}$\hfil&\hfil$f_{21}$\hfil\cr
\omit\hfil Field\hfil&\hfil[\deg]\hfil&\hfil[\deg]\hfil&[pc]\hfil&\hfil[M$_\odot]\hfil$&\hfil[Myr]\hfil&\hfil[$10^{21}$~cm$^{-2}$]\hfil&\hfil[$10^{21}$~cm$^{-2}$]\hfil&\hfil{[\%]}\hfil&\hfil{[\%]}\hfil\cr
\noalign{\vskip 4pt\hrule\vskip 6pt}
Polaris Flare&120&$27$&$\hfil130$--$140\hfil$&---\hfil&---\hfil&\hfil1.1\hfil&\hfil\phantom{0}5.0\hfil&\hfil0\phantom{.0}\hfil&\hfil58\phantom{.0}\hfil\cr
Taurus&173&$-15$&140\hfil&$2\times10^4$\hfil&20\hfil&\hfil4.1\hfil&\hfil26\phantom{.0}\hfil&\hfil4.2\hfil&\hfil\phantom{0}0.8\hfil\cr
Orion&211&$-16$&414\hfil&\hfil$3\times10^5$\hfil&$>$12\hfil&\hfil4.0\hfil&\hfil40\phantom{.0}\hfil&\hfil5.4\hfil&\hfil\phantom{0}7.3\hfil\cr
Chamaeleon-Musca&300&$-13$&\hfil$160$--$180$\hfil&\hfil$5\times10^3$\hfil&$>$2\hfil&\hfil2.0\hfil&\hfil21\phantom{.0}\hfil&\hfil0.5\hfil&\hfil\phantom{0}7.5\hfil\cr
Ophiuchus&354&$15$&$\hfil120$--$140\hfil$&\hfil$3\times10^4$\hfil&$>2$--$5$\hfil&\hfil3.1\hfil&\hfil62\phantom{.0}\hfil&\hfil2.2\hfil&\hfil\phantom{0}3.8\hfil\cr
\noalign{\vskip 3pt\hrule\vskip 4pt}
Microscopium&15&$-40$&---\hfil&---\hfil&---\hfil&\hfil0.4\hfil&\hfil\phantom{0}1.1\hfil&\hfil0\phantom{.0}\hfil&\hfil99\phantom{.0}\hfil\cr
Pisces&133&$-37$&---\hfil&---\hfil&---\hfil&\hfil0.4\hfil&\hfil\phantom{0}1.9\hfil&\hfil0\phantom{.0}\hfil&\hfil99\phantom{.0}\hfil\cr
Perseus&143&$-25$&---\hfil&---\hfil&---\hfil&\hfil0.4\hfil&\hfil\phantom{0}1.5\hfil&\hfil0\phantom{.0}\hfil&\hfil99\phantom{.0}\hfil\cr
Ara&336&$-14$&---\hfil&---\hfil&---\hfil&\hfil0.8\hfil&\hfil\phantom{0}2.1\hfil&\hfil0\phantom{.0}\hfil&\hfil75\phantom{.0}\hfil\cr
Pavo&336&$-28$&---\hfil&---\hfil&---\hfil&\hfil0.4\hfil&\hfil\phantom{0}1.4\hfil&\hfil0\phantom{.0}\hfil&\hfil99\phantom{.0}\hfil\cr
\noalign{\vskip 3pt\hrule\vskip 4pt}}}
\endPlancktablewide                 
\tablenote a Estimates  of distances are from \cite{elias_78} for Taurus, \cite{zagury_et_al_99} for Polaris Flare, \cite{dezeeuw_et_al_99} for Ophiuchus, \cite{whittet_et_al_97} for Chamaeleon-Musca, and \cite{draine_11} for Orion.\par
\tablenote b Estimates of masses are from \cite{ungerechts_thaddeus_87} for Taurus, \cite{loren_89} for Ophiuchus, \cite{luhman_08} for Chamaeleon-Musca, and \cite{draine_11} for Orion.\par
\tablenote c Estimates of ages are from \cite{palla_stahler_02} for Taurus, \cite{wilking_et_al_08} for Ophiuchus, {\cite{luhman_08}} for Chamaeleon-Musca, and \cite{bally_08} for Orion.\par
\endgroup
\end{table*}  

\section{Introduction}
\label{sec-intro}

\Planck\footnote{\Planck\ (\url{http://www.esa.int/Planck}) is a project of the European Space Agency (ESA) with instruments provided by two scientific consortia funded by ESA member states (in particular the lead countries France and Italy), with contributions from NASA (USA) and telescope reflectors provided by a collaboration between ESA and a scientific consortium led and funded by Denmark.}~(\citealt{tauber2010a}, \citealt{planck2011-1.1}) is the third generation space-mission aimed at mapping the anisotropies of the cosmic microwave background (CMB). With its unprecedented sensitivity and large spectral coverage (nine channels from 30\,GHz to 857\,GHz) it has provided exquisite maps of that relic radiation~\citep{planck2013-p01}. With its polarimetric capabilities up to 353\,GHz, \Planck~will also provide clues on the physics of the early Universe, by measuring the CMB polarization. However, dominant foreground emission is also partially polarized, masking the primordial signal. In the range of the High Frequency Instrument~\citep[HFI,][]{lamarre2010}, from 100\,GHz to 857\,GHz, the main contribution to the observed radiation, besides point sources, is thermal emission from dust grains. 

The angular momenta of aspherical and spinning grains tend to align with the local magnetic field, although the details of how this alignment proceeds are still the subject of study : see for instance \cite{andersson_12} for a review on observational constraints regarding grain alignment with respect to current dust models. Submillimetre thermal dust emission is therefore polarized and represents a powerful tool to study interstellar magnetic fields and dust properties. Ideally, we would like to know where in interstellar clouds, and with what efficiency the dust emission and extinction is polarized. This would allow us to use polarization data to infer the spatial structure of the magnetic field. There is an extensive literature on this topic based on observations of starlight polarization, which have been interpreted from two different viewpoints, i.e., grain alignement and magnetic field structure, without achieving a clear understanding of the respective roles of these processes in accounting for variations of polarization across the sky. A number of papers (e.g., \citealt{pereyra_magalhaes_07}, \citealt{alves_et_al_08}, \citealt{marchwinski_et_al_12}) use the data to infer the magnetic field strength using the Chandreskar-Fermi method~\citep{chandrasekhar_fermi_53}. Other papers focus on the observed decrease of polarization fraction $\polfrac$ with $N_\mathrm{H}$ to interpret the data as a decrease of the dust alignment efficiency in dense clouds (\citealt{lazarian_et_al_97}, \citealt{whittet_et_al_08}, \citealt{chapman_et_al_11}).

Magnetohydrodynamical (MHD) simulations provide a theoretical framework to consider both aspects in the interpretation of polarization datasets. \cite{ostriker_et_al_01} were among the first to present simulated polarization maps from MHD simulations, for comparison with data and to study the field structure beyond the simple Chandrasekhar-Fermi method. \cite{falceta-goncalves_et_al_08} used a similar technique to study the effect of the Alfv\'enic Mach number, while \cite{pelkonen_09} added to this approach the modelling of the alignment process by radiative torques~\citep{hoang_lazarian_08}.

\Planck~has mapped the polarized dust emission with great sensitivity and resolution~\citep{planck2014-XIX}, allowing us to characterize spatial variations of dust polarization and compare data with MHD simulations with unprecedented statistics. This paper is the second in a series of four dealing with a first presentation of the \Planck~polarized thermal emission from Galactic dust. The other three are the following:~\cite{planck2014-XIX} describes the polarized dust emission at 353\,GHz as seen by \Planck~over the whole sky and shows in particular that the maximum polarization fraction $\pmax$ at a given total gas column density $N_\mathrm{H}$ decreases as $N_\mathrm{H}$ increases, and that there is an anti-correlation between polarization fractions $p$ and {\DeltaAngName}s $\DeltaAng$, an effect which has also been seen with starlight polarization data \citep{hatano_et_al_13}. \cite{planck2014-XXI} compares polarized thermal emission from dust at 353\,GHz to polarization in extinction in the visible towards a sample of stars. Finally, \cite{planck2014-XXII} discusses the variation of polarized thermal emission from dust with frequency, from 70 to 353\,GHz. Both \cite{planck2014-XXI} and \cite{planck2014-XXII} aim at providing constraints for models of interstellar dust.

In this paper, we use \Planck~polarization data at 353\,GHz to present statistics of polarization fractions and angles in nearby interstellar clouds seen outside the Galactic plane. We then compare the \Planck~results with simulated observations of polarized thermal dust emission at 353\,GHz built from a three-dimensional MHD simulation of the formation of a molecular cloud within colliding flows~\citep{hennebelle_08}. 

In these simulated observations, we work under the assumption that the optical properties and the \IntrinsicpName~of dust grains are constant. At this stage we do not aim at testing models of grain alignment. In this picture, it is expected that the polarization fraction should be maximal when the magnetic field is in the plane of the sky and should, in this case, yield valuable information on the \IntrinsicpName. That is why we first focus on the decrease of the maximum value of $\polfrac$, rather than its mean or median values, with increasing column density. We then consider the correlation between polarization fractions and local measures of the dispersion in polarization angles, as it is expected that larger angular dispersions should lower the observed polarization fraction. 

The paper is organized as follow. Section~\ref{sec:planck-data} describes the \Planck~data used and the statistics drawn from them in the selected regions. Section~\ref{sec:simus} presents simulated polarized emission observations based on an MHD simulation of interstellar turbulence and compares their statistical properties with those found towards similar fields in the \Planck~data. Conclusions are given in Sect.~\ref{sec:conclusions}. Appendix~\ref{extra-figures} presents supplementary figures, and Appendix~\ref{sec:Stokes-LD85} details the derivation of the equations yielding the Stokes parameters for dust emission.

\section{\Planck~observations of polarized dust emission}
\label{sec:planck-data}
\subsection{\Planck~all-sky data post-processing}
\label{subsec:dataproc}
The data processing of \Planck~HFI is presented in \cite{planck2013-p03}, \cite{planck2013-p03c}, \cite{planck2013-p03f}, \cite{planck2013-p03d}, and \cite{planck2013-p03e}. The specifics of the data processing in terms of polarization are given in~\cite{planck2014-XIX}. We use the same \Planck~data set as that presented in \cite{planck2014-XIX}, i.e., full 5-survey HFI mission data for Stokes $\StokesI$, $\StokesQ$, and $\StokesU$ at 353\,GHz (which is the \Planck~channel offering the best signal-to-noise ratio for dust polarization) from the ``DR3'' internal data release. Bandpass mismatch between individual elements of a pair of polarization sensitive bolometers (PSBs) is corrected using in-flight measurements for the dust emission but not for the negligible CO $J$=3$\rightarrow$2 emission~\citep{planck2013-p03d}. From the total intensity map we subtract the offset $\StokesI_\mathrm{offset}=0.0887\,\mathrm{MJy\,sr^{-1}}$ to set the Galactic zero level at 353\,GHz~\citep{planck2013-p06b}. Note that this value includes the cosmic infrared background (CIB) monopole and is slightly different from the one given in~\cite{planck2013-p06b}, as the maps are not the same (full mission vs. nominal mission). We do not correct for zodiacal light emission, nor for the residual dipole identified by~\cite{planck2013-p06b} at 353\,GHz. CMB and CIB fluctuations are ignored, since the regions selected in this study are outside the CMB-CIB mask described in \cite{planck2014-XIX}, so the polarized emission there is dominated by the dust.

The Planck polarization and intensity data that we use in this analysis have been generated in exactly the same manner as the data publicly released in March 2013 and described in~\cite{planck2013-p01} and associated papers. Note, however, that the publicly available data include only temperature maps based on the first two surveys. \cite{planck2013-p11} shows the very good consistency of cosmological models derived from intensity only with polarization data at small scales (high CMB multipoles). However, as detailed in \citet{planck2013-p03} (see their Fig.~27), the 2013 polarization data are known to be affected by systematic effects at low multipoles which were not yet fully corrected, and thus these data were not used for cosmology\footnote{The full mission maps for intensity as well as for polarization will be made publicly available in the fall of 2014.}. We have been careful to check that the Galactic science results in this paper are robust with respect to these systematics\footnote{The error-bars we quote include uncertainties associated with residual systematics as estimated by repeating the analysis on different subsets of the data. We have also checked our data analysis on the latest version of the maps available to the consortium to check that the results we find are consistent within the error-bars quoted in this paper.}.

We focus in this paper on the polarization fractions $\polfrac$ and the polarization angles $\polang$ derived from the Stokes $I$, $Q$, and $U$ maps obtained by \Planck~at 353\,GHz and at an angular resolution of 15\arcm. In the absence of noise, $\polfrac$ and $\polang$ are defined by

\begin{equation}
\label{eq:polfrac}
\polfrac=\frac{\sqrt{Q^2+U^2}}{I},
\end{equation}
and
\begin{equation}
\label{eq:polang}
\polang=\frac{1}{2}\mathrm{atan}\left(U,Q\right).
\end{equation}
Note that $\polang$ is here defined in the \healpix\footnote{\url{http://healpix.jpl.nasa.gov} See in particular the latest version of the \healpix~primer, available at \url{http://healpix.jpl.nasa.gov/pdf/intro.pdf}.}~convention~\citep{gorski_et_al_05}, which means that angles are counted positively clockwise from the north-south direction. Working in that convention instead of the IAU one, which is anti-clockwise~\citep{planck2014-XXI}, has however no impact on the results presented here. Additionally, since we work on ratios of Stokes parameters, no colour correction is necessary.

When (possibly correlated) noise affects the Stokes parameters, the polarization fraction computed directly using Eq.~\ref{eq:polfrac} is biased. We call this one the ``na\"ive'' estimator of $\polfrac$, but various methods have been devised to correct for the bias~\citep{pma1}, and their respective efficiencies are compared in~\cite{pma2}. Among them is the modified asymptotic (MAS) estimator introduced by~\cite{plaszczynski_et_al_14}, which is computed from the naive estimator and the noise covariance matrix pertaining to $Q$ and $U$. Another estimator of the polarization fraction and angle is the Bayesian estimator described in~\cite{pma1} and~\cite{planck2014-XIX}, which has the advantage of taking into account the full noise covariance matrix in $I$, $Q$ and $U$, and also taking into account the uncertainty on the zero-level offset for $I$. In the rest of this paper, except where noted, the maps of polarization fraction $\polfrac$ and polarization angle $\polang$ at 353\,GHz refer to these Bayesian estimators. The Bayesian method also provides maps of the polarization fraction and angle uncertainties, $\sigpolfrac$ and $\sigpolang$. 

For the total hydrogen column density map $N_\mathrm{H}$, we use a conversion from the optical depth at 353\,GHz, $\tau_{353}$, derived from~\cite{planck2013-p06b}: for $N_\mathrm{H}\gtrsim2\times10^{21}\,\mathrm{cm}^{-2}$, the dust opacity is approximately constant, with $\sigma_{353}=\tau_{353}/N_\mathrm{H}\simeq 1.2\times10^{-26}\,\mathrm{cm}^{2}$. We are aware that this conversion is crude, with possible variations in dust opacity of the order of $20\%$ to $25\%$, but our findings do not critically depend on that calibration. 

All of the maps used in this study have a \healpix~resolution $N_\mathrm{side}=1024$.

\subsection{Overview of the statistics of polarized emission in various fields}
\label{sec:overview-statistics}

\begin{figure}[htbp]
\centerline{\includegraphics[width=8.8cm,trim=60 0 70 0,clip=true]{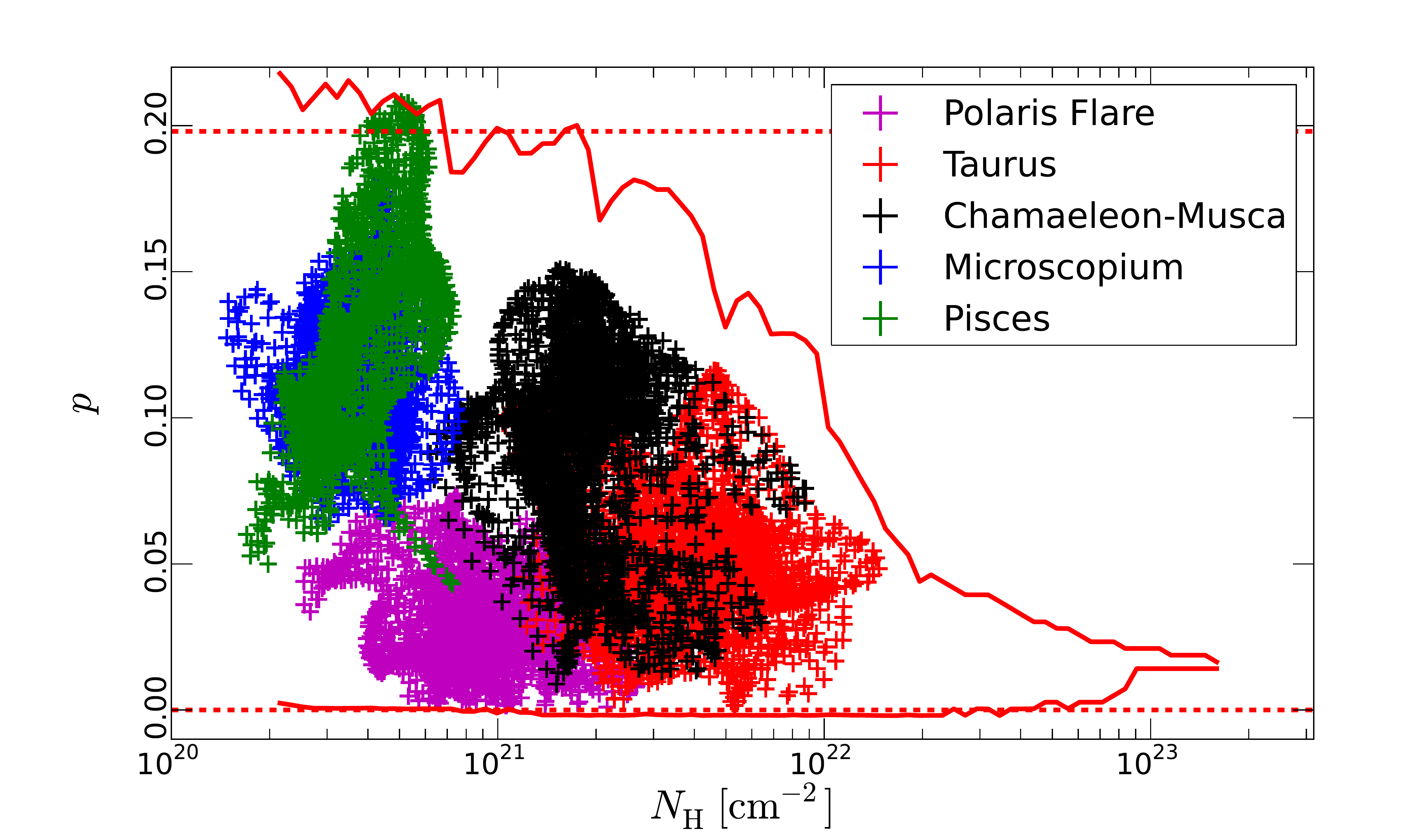}}
\centerline{\includegraphics[width=8.8cm,trim=60 0 70 0,clip=true]{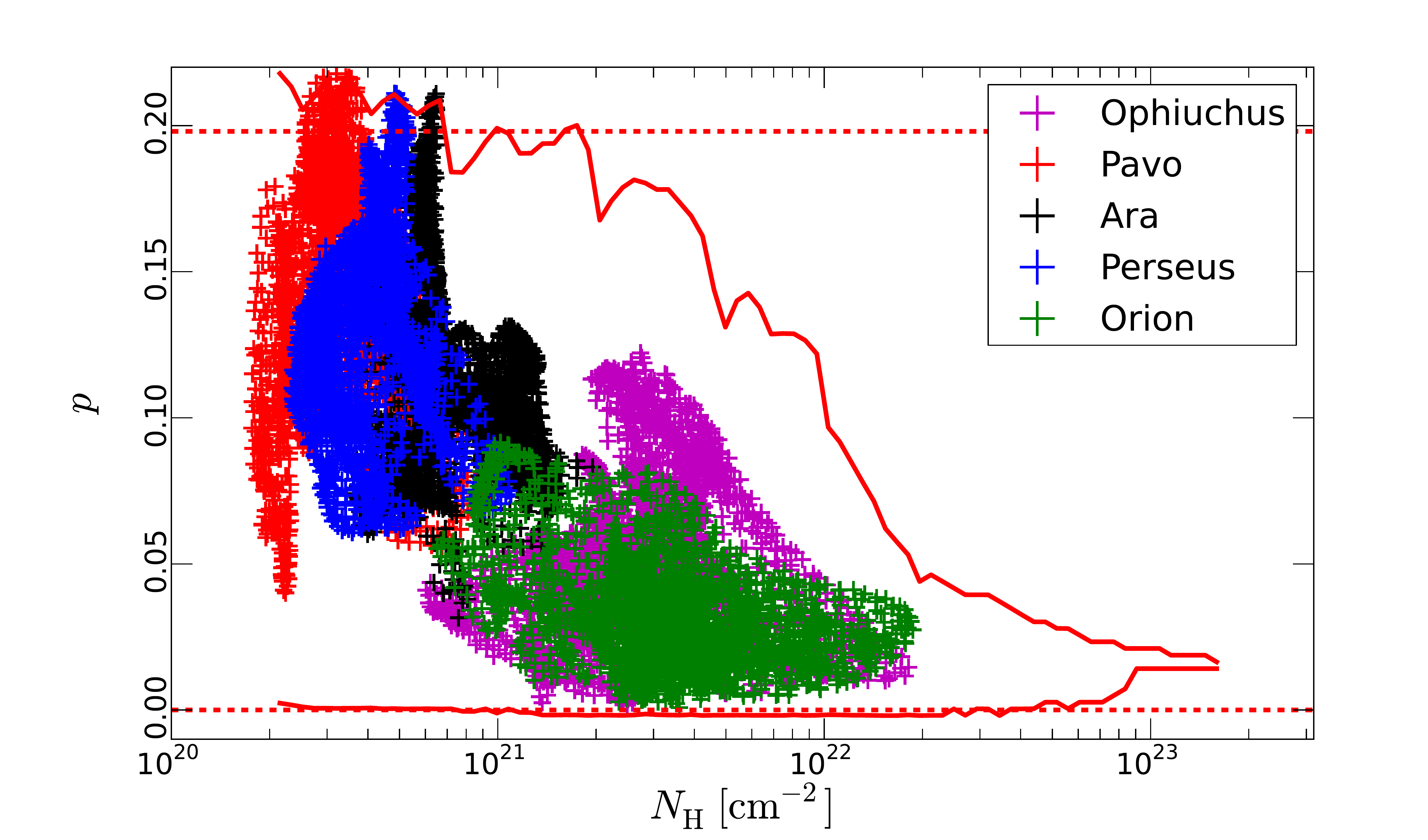}}
\caption{Two-dimensional distribution functions of polarization fraction $\polfrac$ and column density $N_\mathrm{H}$ in the fields highlighted in Fig.~\ref{fig:regions}. \emph{Top}: Polaris Flare (magenta), Taurus (red), Chamaeleon-Musca (black), Microscopium (blue), and Pisces (green). \emph{Bottom}: Ophiuchus (magenta), Pavo (red), Ara (black), Perseus (blue), and Orion (green). On both panels, the solid red lines show the upper and lower envelopes (see text) of the large-scale distribution of $\polfrac$ and $N_\mathrm{H}$, clipped below $N_\mathrm{H}=2\times10^{20}\,\mathrm{cm}^{-2}$, while the dashed red lines correspond to $\polfrac=0$ and the maximum value $\polfrac=0.198$ (i.e., 19.8\%) quoted in~\cite{planck2014-XIX}.}
\label{fig:PI_vs_NH_all}
\end{figure}

We have selected ten regions, each $12^{\circ}\times 12^{\circ}$ in size, that are highlighted in Fig.~\ref{fig:regions} and whose locations are given in Table~\ref{table-fields}. These are the same as some of the individual regions mentioned in~\cite{planck2014-XIX}. All of these fields are outside the Galactic plane and probe nearby interstellar material, but they exhibit very different physical conditions, from the diffuse, turbulent ISM with little to no star-forming activity (Polaris Flare), to self-gravitating, star-forming clouds (Orion). They also differ in terms of polarized emission. Some diffuse regions have high polarization fractions (e.g., Pavo), while some have low polarization fractions (e.g., Polaris Flare). This variety of conditions in terms of polarization fraction and gas content is emphasized in Fig.~\ref{fig:PI_vs_NH_all}, which shows the distribution of $\polfrac$ and $N_\mathrm{H}$ in these regions, compared with the large-scale distribution shown in~\cite{planck2014-XIX}. The latter is represented by its upper and lower envelopes, computed from the 0.01\% and 99.99\% percentiles of the $\polfrac$ distribution within each bin in column density. All the envelopes of two-dimensional distribution functions shown in this paper are computed in this fashion. Note that to facilitate the comparison with~\cite{planck2014-XIX}, Fig.~\ref{fig:PI_vs_NH_all} uses maps at $1\deg$ resolution. In the rest of the paper, as already stated, we use 15\arcm~resolution maps.

It appears that for column densities between a few times $10^{20}~\mathrm{cm}^{-2}$ and a few times $10^{22}~\mathrm{cm}^{-2}$, the selected fields probe most of the range of polarization fractions observed over the whole sky in this range of column densities. The diffuse Polaris Flare field shows low polarization, while high polarization fractions are reached at similar column densities in the Chamaeleon-Musca complex, which, being closer to the Galactic plane, is threaded by the large-scale Galactic magnetic field. Another notable feature of Fig.~\ref{fig:PI_vs_NH_all} is the fact that in regions with the largest column densities (Taurus, Orion, and Ophiuchus) the maximum polarization fraction decreases with increasing $N_\mathrm{H}$, and that the slopes are comparable to the large-scale trend. 

In the following, we perform statistical analyses of the polarization data in these nearby fields by simply selecting \healpix~pixels whose centres fall within the region of interest, directly from the large-scale maps. Only pixels for which $\polfrac/\sigpolfrac>3$ are retained. This threshold is a reasonable value above which the polarization signal-to-noise ratio is properly estimated \citep{pma2}. Note that some of the fields in Table~\ref{table-fields} are quite diffuse (e.g., Pavo), so that the dynamic range in column densities is too small to exhibit a significant relationship between $\polfrac_\mathrm{max}$ and $N_\mathrm{H}$. These diffuse fields are therefore discarded in the later analysis.

\begin{figure*}[htbp]
\centerline{\includegraphics[width=9cm,trim=0 50 0 0,clip=true]{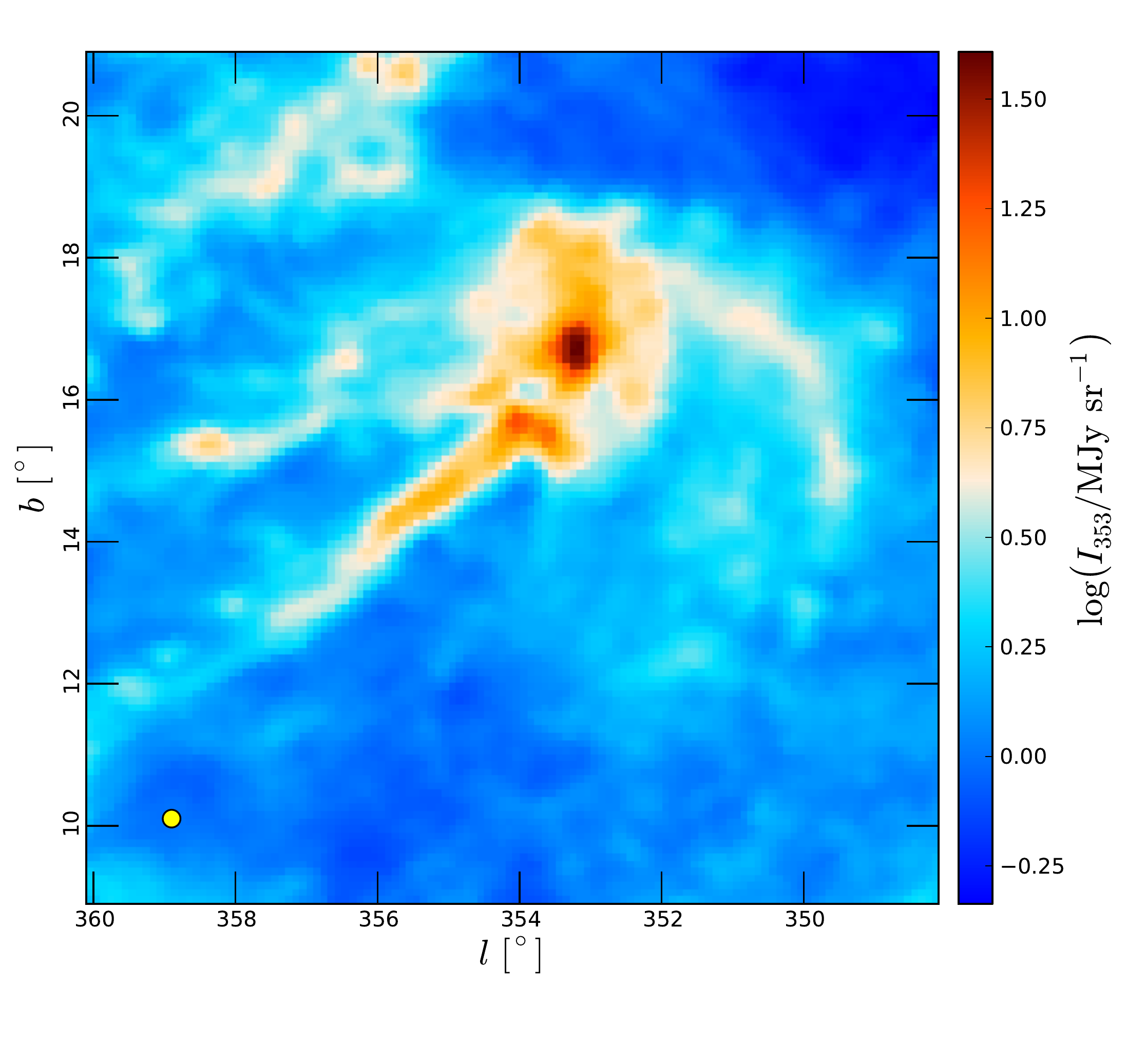}\includegraphics[width=9cm,trim=0 50 0 0,clip=true]{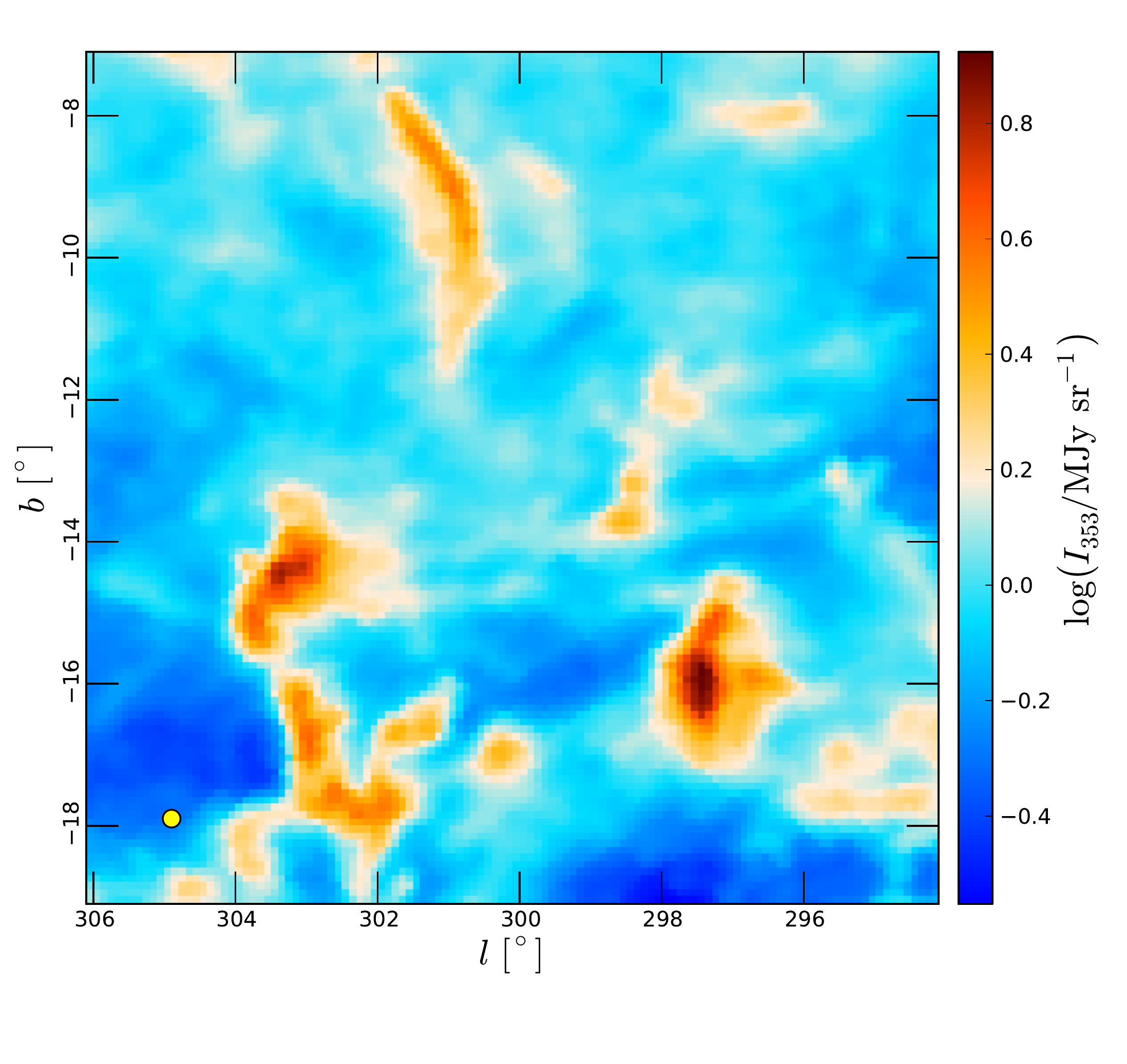}}
\centerline{\includegraphics[width=9cm,trim=0 50 0 20,clip=true]{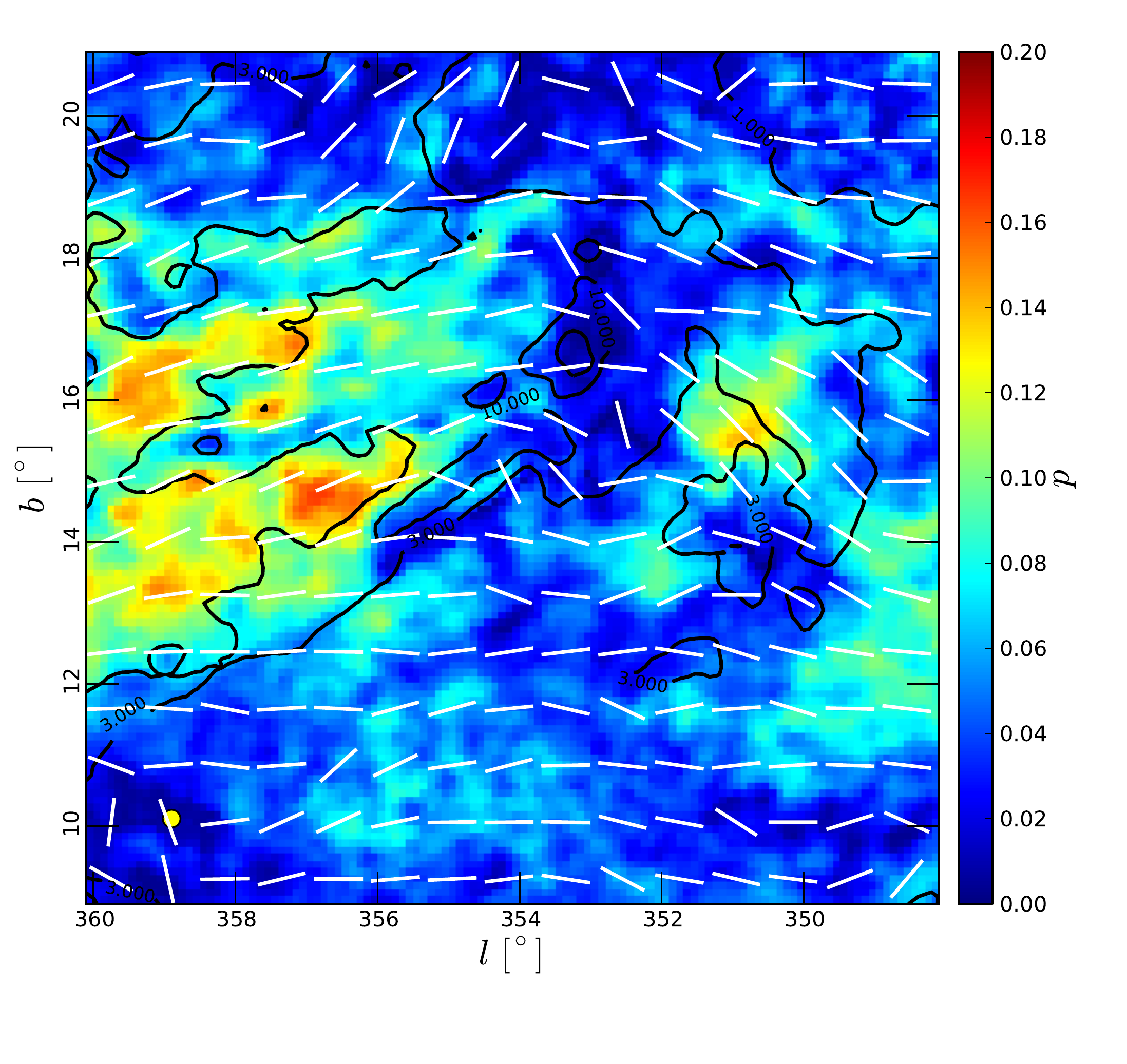}\includegraphics[width=9cm,trim=0 50 0 20,clip=true]{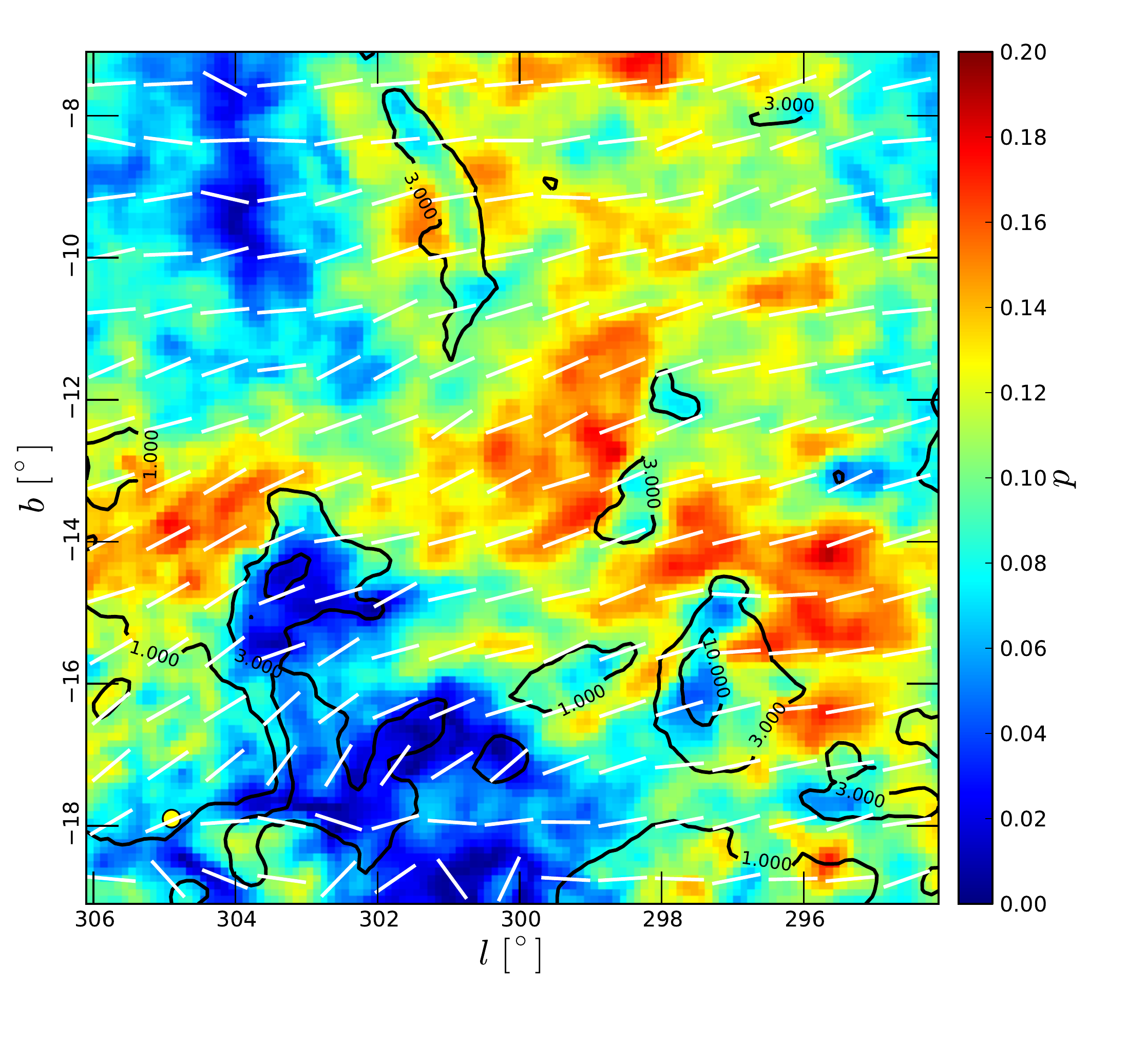}}
\centerline{\includegraphics[width=9cm,trim=0 50 0 20,clip=true]{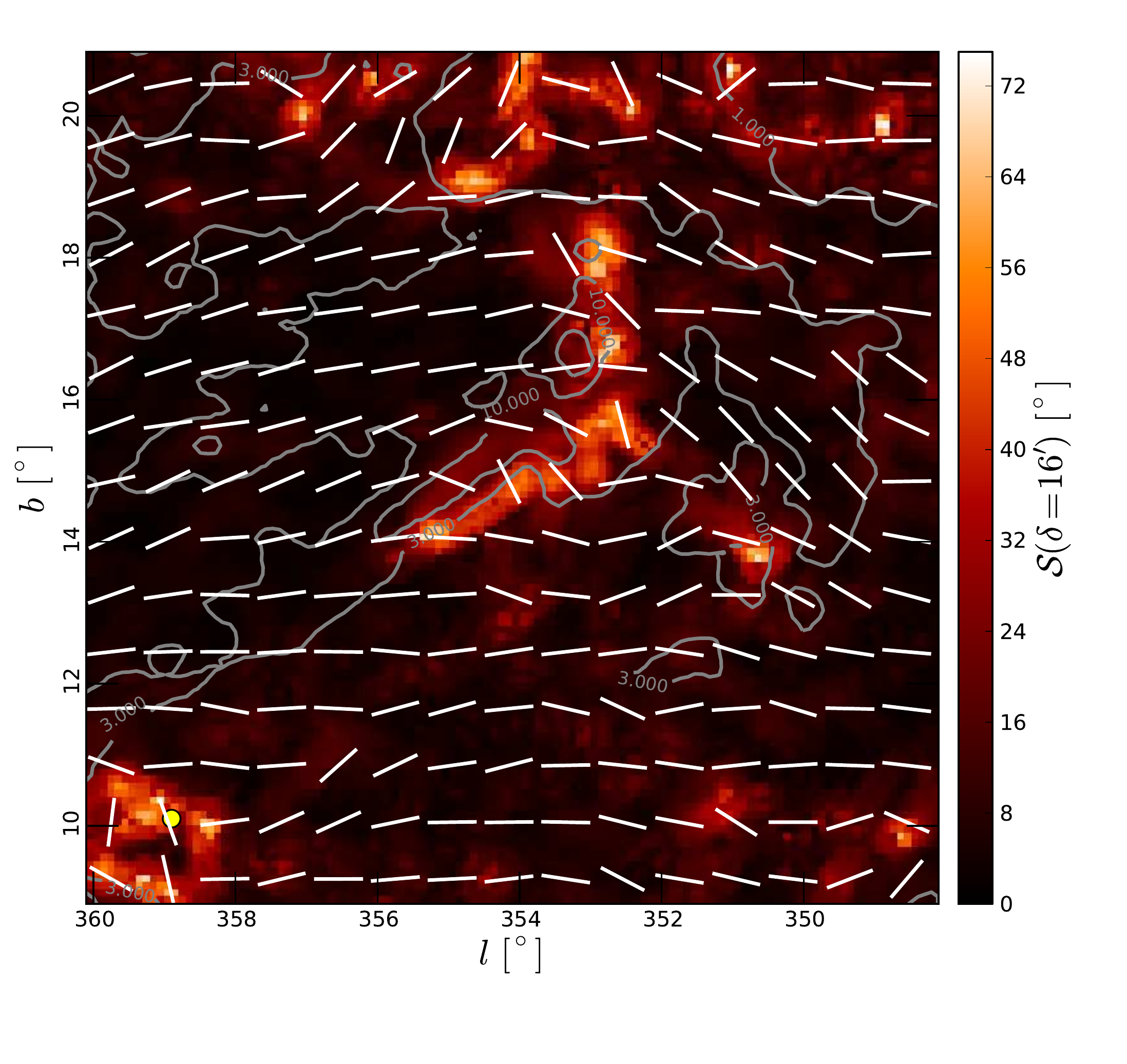}\includegraphics[width=9cm,trim=0 50 0 20,clip=true]{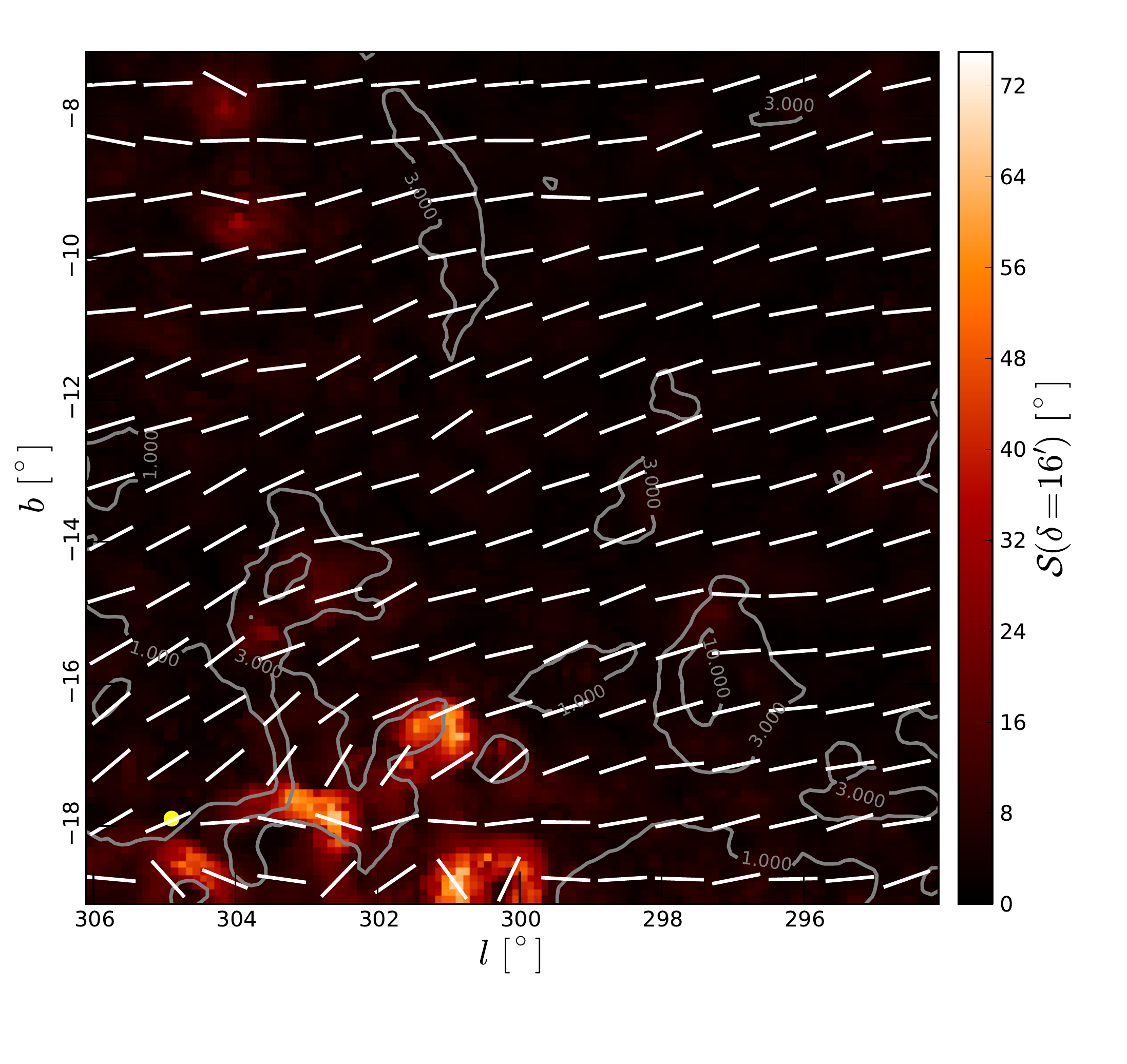}}
\caption{Maps of the Ophiuchus and Chamaeleon-Musca fields. \emph{Left}: Ophiuchus field. \emph{Right}: Chamaeleon-Musca field. \emph{Top}: Total intensity at 353\,GHz. \emph{Middle}: Polarization fraction $\polfrac$, column density $N_\mathrm{H}$ (contours in units of $10^{21}\,\mathrm{cm}^{-2}$), and magnetic orientation (bars, see text). \emph{Bottom}: \DeltaAngNameMaj~$\DeltaAng$ with lag $\delta=16\arcm$ (see Sect.~\ref{sec:PI-vs-H}) with contours and bars identical to the middle row. In all maps, the 15\arcm~beam is shown in the lower-left corner.}
\label{fig:PI-B-NH_Ophiuchus-Chamaeleon}
\end{figure*}

We also build local maps of polarized emission using gnomic projections of the \healpix~maps. These are shown in the middle row panels of Fig.~\ref{fig:PI-B-NH_Ophiuchus-Chamaeleon} for the Ophiuchus and Chamaeleon-Musca fields. Similar figures for all other fields are given in Appendix~\ref{extra-figures}. On all these maps, which share the same scale, we show the polarization fractions $\polfrac$ at 353\,GHz (colour scale) overlaid with contours of the total gas density and bars of constant length giving the orientation of the apparent projection of the magnetic field on the plane of the sky. These are built by rotating the 353\,GHz polarization bars by $90\deg$ so as to recover the average magnetic field orientation in the plane of the sky. In the rest of the paper, we will refer to the rotated polarization bars as the magnetic orientation bars. Note that although they are plotted once every few pixels only, to improve visibility, each of these bars represents the orientation at the given pixel. In other words, beyond the 15$\arcmin$ smoothing performed on the Stokes maps, no further averaging is done to plot the orientation bars on Fig.~\ref{fig:PI-B-NH_Ophiuchus-Chamaeleon} and similar plots. 

The large-scale structure of the Galactic magnetic field appears clearly (see e.g., the top part of the Chamaeleon-Musca field, Fig.~\ref{fig:PI-B-NH_Ophiuchus-Chamaeleon}). There is also a strong correlation between the coherence of the polarization orientation and the level of polarization fraction, in the sense that more ordered regions have higher polarization fractions. This feature, which is already seen at $1\deg$ resolution in~\cite{planck2014-XIX}, is discussed later on in Sect. \ref{sec:PI-vs-H}. 

A final qualitative aspect of these maps is that regions with higher column densities tend to be less polarized than their surroundings. An example of this effect can be seen in the Chamaeleon-Musca field (Fig.~\ref{fig:PI-B-NH_Ophiuchus-Chamaeleon}, center right panel) near $(l,b)=(301\deg{,}-9\deg)$, where $\polfrac \simeq 10\%$, while it is surrounded by more diffuse material with $\polfrac \simeq 15\%$. A future paper~\citep{planck2014-pip113} will discuss in more detail the structure of the polarized thermal emission with respect to the morphology of the clouds themselves.

\begin{table*}[tmb]
\begingroup
\newdimen\tblskip \tblskip=5pt
\caption{Polarization statistics in the selected fields: absolute maximum polarization fraction at 15\arcm~resolution; linear fit parameters $m$ and $c$ to the decrease of $\polfrac_\mathrm{max}$ with $\log\left(N_\mathrm{H}/\mathrm{cm^{-2}}\right)$, with fitting range indicated; and linear fit parameters of the $\log(\polfrac)$ vs. $\log\left(\DeltaAng\right)$ correlation. See text for the derivation of the listed uncertainties. The figures given here are for a signal-to-noise threshold $\polfrac/\sigpolfrac>3$.}
\label{table-fields-properties}
\nointerlineskip
\vskip -3mm
\footnotesize
\setbox\tablebox=\vbox{
   \newdimen\digitwidth 
   \setbox0=\hbox{\rm 0} 
   \digitwidth=\wd0 
   \catcode`*=\active 
   \def*{\kern\digitwidth}
   \newdimen\signwidth 
   \setbox0=\hbox{+} 
   \signwidth=\wd0 
   \catcode`!=\active 
   \def!{\kern\signwidth}
\halign{\hbox to 1.15in{#\leaderfil}\tabskip 2.2em&
\hfil#\hfil&
\hfil#\hfil&
\hfil#\hfil&
\hfil#\hfil&
\hfil#\hfil&
\hfil#\hfil\cr
\noalign{\doubleline}
\omit\hfil Field\hfil&\hfil$\polfrac_\mathrm{max}$\hfil&\multispan2\hfil$\polfrac_\mathrm{max}=m\log\left(N_\mathrm{H}/\mathrm{cm^{-2}}\right)+c$\hfil&\hfil$N_\mathrm{H}$ range\hfil&\multispan2\hfil$\log\left(\DeltaAng\right)=m'\log(\polfrac)+c'$\hfil\cr
\omit\hfil&\omit&\hfil$m$\hfil&\hfil$c$\hfil&\hfil[$10^{21}~\mathrm{cm}^{-2}$]\hfil&\hfil$m'$\hfil&\hfil$c'$\hfil\cr
\noalign{\vskip 4pt\hrule\vskip 6pt}
Polaris Flare&$0.134\pm0.015$&$-0.114\pm0.014$&$2.5\pm0.3$&$\phantom{.0}1$--$4\phantom{.0}$&$-0.56\pm0.08$&$\phantom{-}0.25\pm0.17$\cr
Taurus&$0.149\pm0.011$&$-0.140\pm0.004$&$3.2\pm0.1$&$\phantom{.0}5$--$25\phantom{.}$&$-0.87\pm0.09$&$-0.31\pm0.11$\cr
Orion&$0.129\pm0.014$&$-0.068\pm0.003$&$1.6\pm0.1$&$\phantom{.0}3$--$40\phantom{.}$&$-0.87\pm0.11$&$-0.25\pm0.13$\cr
Chamaeleon-Musca&$0.190\pm0.008$&$-0.134\pm0.003$&$3.0\pm0.1$&$\phantom{.0}3$--$20\phantom{.}$&$-0.94\pm0.03$&$-0.39\pm0.02$\cr
Ophiuchus&$0.166\pm0.006$&$-0.129\pm0.004$&$2.9\pm0.1$&$\phantom{.0}3$--$40\phantom{.}$&$-0.92\pm0.05$&$-0.30\pm0.04$\cr
\noalign{\vskip 3pt\hrule\vskip 4pt}
Microscopium&$0.24\phantom{0}\pm0.05\phantom{0}$&---&---&---&$-0.41\pm0.07$&$\phantom{-}0.38\pm0.07$\cr
Pisces&$0.30\phantom{0}\pm0.11\phantom{0}$&---&---&---&$-0.67\pm0.13$&$\phantom{-}0.21\pm0.12$\cr
Perseus&$0.33\phantom{0}\pm0.09\phantom{0}$&---&---&---&$-0.46\pm0.09$&$\phantom{-}0.37\pm0.06$\cr
Ara&$0.27\phantom{0}\pm0.03\phantom{0}$&---&---&---&$-0.48\pm0.07$&$\phantom{-}0.15\pm0.06$\cr
Pavo&$0.48\phantom{0}\pm0.18\phantom{0}$&---&---&---&$-0.27\pm0.05$&$\phantom{-}0.57\pm0.03$\cr
\noalign{\vskip 3pt\hrule\vskip 4pt}}}
\endPlancktable
\endgroup
\end{table*}   

\subsection{Maximum polarization fraction}

We give in Table~\ref{table-fields-properties} the maximum polarization fractions $\polfrac_\mathrm{max}$ in all the selected fields. Note that for the most diffuse fields Microscopium, Pisces, Perseus, Ara, and Pavo, the quoted values should be taken with caution, since most pixels in these regions have $N_\mathrm{H}\leqslant 10^{21}\,\mathrm{cm}^{-2}$, which corresponds roughly to $I_{353}\leqslant 0.5\,\mathrm{MJy\,sr^{-1}}$, and therefore the effect of the (uncertain) zero-level offset on the polarization fraction $\polfrac$ may not be negligible. 

In the less diffuse fields, the values of $\polfrac_\mathrm{max}$ are noticeably larger than those found in the same fields at 1$\deg$ resolution in~\cite{planck2014-XIX}\footnote{See their Table 1, which also lists extrema, mean and median values for $\polfrac$, as well as median values for $\polang$.}, which shows the strong effect of spatial resolution on polarization measurements. The uncertainties $\sigma_{\polfrac_\mathrm{max}}$ on the maximum polarization fractions, listed in Table~\ref{table-fields-properties}, are derived from the various sources of uncertainty involved. 

First, the noise properties on the Stokes parameters $I$, $Q$, and $U$ in each pixel are described in the data by the noise covariance matrices, which are input in the Bayesian method of~\cite{pma1} and~\cite{planck2014-XIX}, and lead to a map of the uncertainty $\sigpolfrac$ on the polarization fraction. This includes the $0.0068\,\mathrm{MJy\,sr^{-1}}$ uncertainty on the zero-level offset. We then compute the difference $\sigma_{\polfrac_\mathrm{max},p}$ between the maximum polarization fractions found in the maps of $\polfrac-\sigpolfrac$ and $\polfrac+\sigpolfrac$.

Second, there is a part of the uncertainty related to the method used to debias the data~\citep{pma1}. We have computed the standard deviation $\sigma_{\polfrac_\mathrm{max},\mathrm{d}}$ of the maximum polarization fractions obtained in each field when using the ``na\"ive'' $\sqrt{Q^2+U^2}/I$, modified asymptotic~(MAS, \citealt{plaszczynski_et_al_14}) and Bayesian estimators of $\polfrac$. 

Third, we have computed the standard deviation $\sigma_{\polfrac_\mathrm{max},\mathrm{s}}$ of the maximum polarization fractions obtained in each field when considering subsets of the data, namely half-ring maps (one half of each stable pointing period) and detector set maps (one half of the detectors). 

The final uncertainty quoted in Table~\ref{table-fields-properties} is then given by the quadratic sum
\begin{equation}
\sigma^2_{\polfrac_\mathrm{max}}=\sigma^2_{\polfrac_\mathrm{max},p}+\sigma^2_{\polfrac_\mathrm{max},\mathrm{d}}+\sigma^2_{\polfrac_\mathrm{max},\mathrm{s}}/2
\end{equation}
It should be noted that the last contribution is usually the dominant one in the selected fields, and that the uncertainty related to the debiasing method is much smaller than the other two.

\subsection{Polarization fraction vs. column density}

We show the distributions of $\polfrac$ and $N_\mathrm{H}$ for the Ophiuchus and Chamaeleon-Musca fields in Fig.~\ref{fig:PI-vs-NH} and for  all other fields in Appendix~\ref{extra-figures}. The decrease in maximum polarization fraction $\pmax$ at higher column densities is apparent for all fields, above a given threshold in $N_\mathrm{H}$ that depends on the field and is of the order of $10^{21}$ to $3\times10^{21}\,\mathrm{cm}^{-2}$, corresponding to visual extinctions $A_\mathrm{V}\simeq 0.6$ to $1.7$, for the fields that are not too diffuse (Polaris Flare, Taurus, Orion, Chamaeleon-Musca, and Ophiuchus). Below this threshold, the polarization fraction may be related to the background more than to the clouds themselves. To quantify the decrease in maximum polarization fraction $\pmax$ with increasing $N_\mathrm{H}$, we consider the upper envelope of the distribution of $\polfrac$ and $N_\mathrm{H}$, computed as described in Sect.~\ref{sec:overview-statistics}, and fit this curve with a function $\pmax=m\log{\left(N_\mathrm{H}/\mathrm{cm^{-2}}\right)}+c$, restricted to a range of column densities that depends on the field considered (see Table~\ref{table-fields-properties}). Note that we perform this fit for the above five fields only, for which there is a large enough dynamic range in column density.

Results of these fits are shown as solid black lines on each panel of Fig.~\ref{fig:PI-vs-NH}, and values of the slopes $m$ and intercepts $c$ are listed in Table~\ref{table-fields-properties}. Uncertainties on these parameters are derived in the same way as for the maximum polarization fractions $\polfrac_\mathrm{max}$ in the previous section. The slopes $m$ range between $-0.068$ for Orion and $-0.140$ for Taurus, and regions exhibiting stronger column density peaks (e.g., Orion) tend to have shallower slopes than more diffuse molecular clouds (e.g., Polaris Flare).

As mentioned before, the pixels selected for plotting Fig.~\ref{fig:PI-vs-NH} and performing the fits are those for which the polarization signal-to-noise ratio is $\polfrac/\sigpolfrac>3$. We have checked that modifying this threshold does not change our results, as can be seen in Fig.~\ref{fig:PI-vs-NH-SNR10}, which shows the same as the top panel of Fig.~\ref{fig:PI-vs-NH} but with a signal-to-noise ratio threshold $\polfrac/\sigpolfrac>10$. The effect of that stricter selection is to remove points below the original lower envelope, but leaves the upper envelope unchanged. Consequently, both the absolute maximum polarization fraction $\polfrac_\mathrm{max}$ and the slope of the decrease of $\polfrac_\mathrm{max}$ at the high end of column densities are quite robust.

\begin{figure}[htbp]
\centerline{\includegraphics[width=8.8cm,trim=120 0 60 0,clip=true]{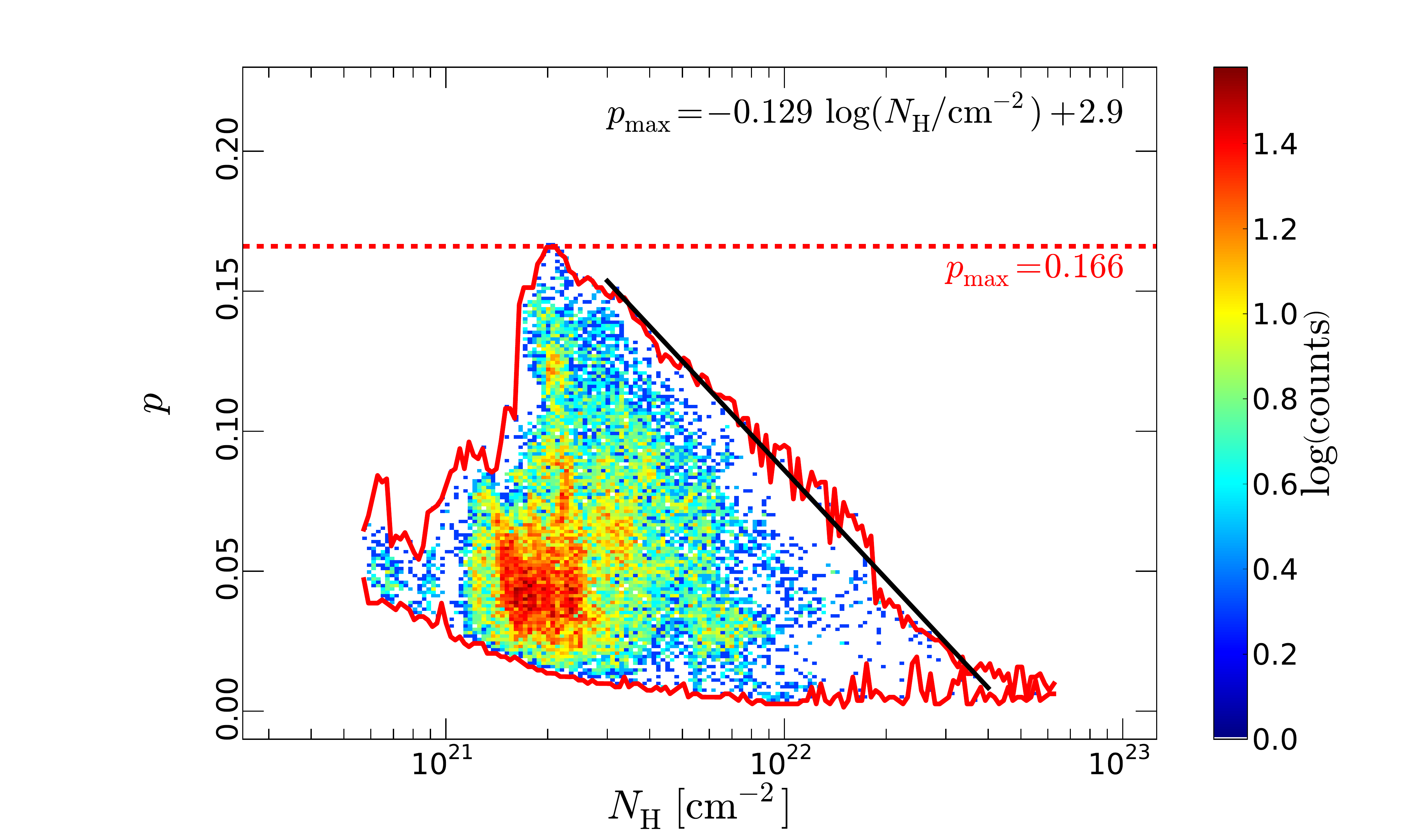}}
\centerline{\includegraphics[width=8.8cm,trim=120 0 60 0,clip=true]{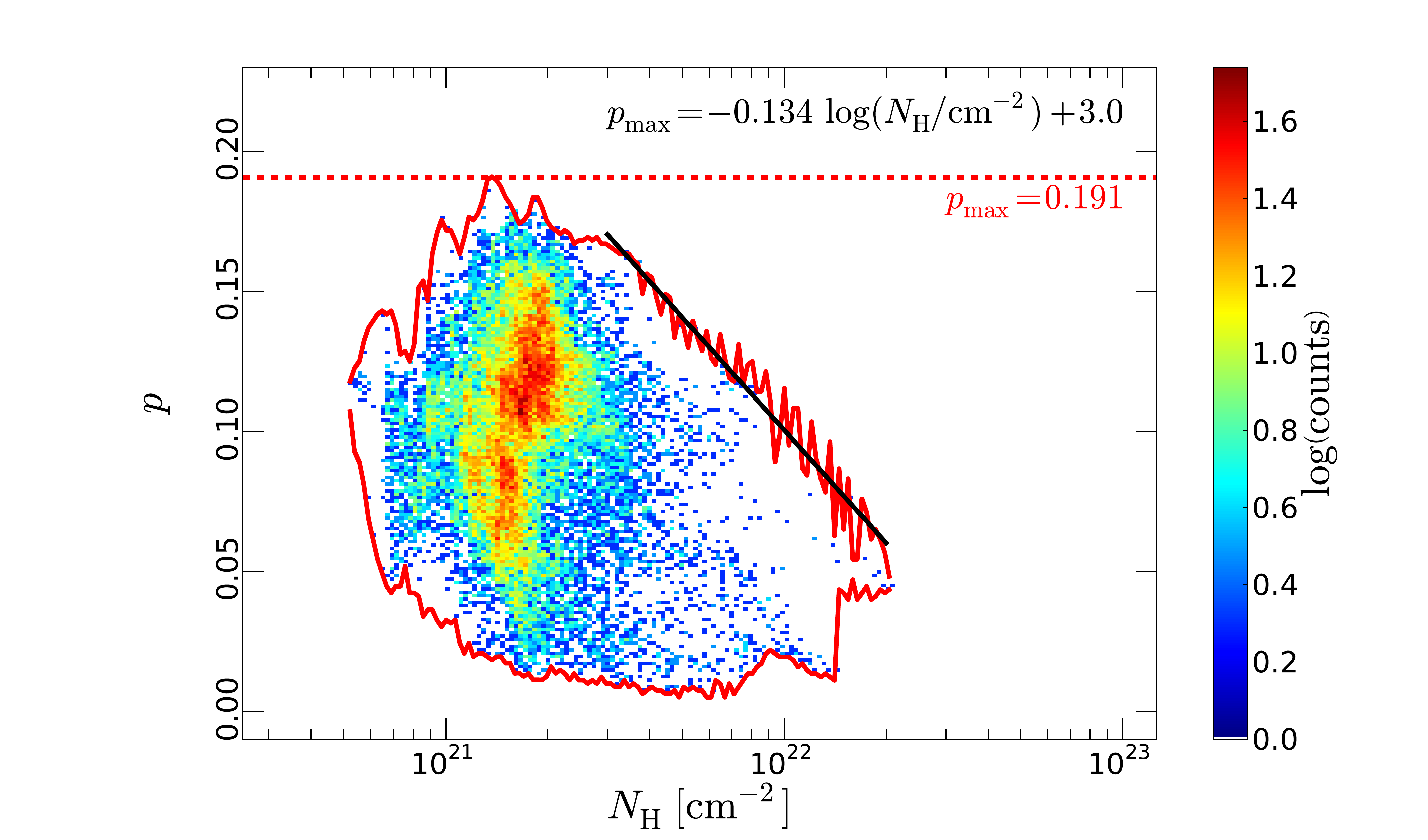}}
\caption{Two-dimensional distribution function of polarization fraction $\polfrac$ and column density $N_\mathrm{H}$. \emph{Top}: Ophiuchus field. \emph{Bottom}: Chamaeleon-Musca field. The distribution functions are presented in logarithmic colour scale and include only points for which $p/\sigma_p>3$. The dashed red lines correspond to the absolute maximum polarization fractions $\polfrac_\mathrm{max}$ and the solid red curves show the upper and lower envelopes of $\polfrac$ as functions of $N_\mathrm{H}$. The solid black line is a linear fit $\polfrac_\mathrm{max}=m\log{\left(N_\mathrm{H}/\mathrm{cm^{-2}}\right)}+c$ to the decrease of the maximum polarization fraction with column density at the high end of $N_\mathrm{H}$ (see Table~\ref{table-fields-properties} for the fitting ranges and fit parameters).}
\label{fig:PI-vs-NH}
\end{figure}

\begin{figure}[htbp]
\centerline{\includegraphics[width=8.8cm,trim=120 0 60 0,clip=true]{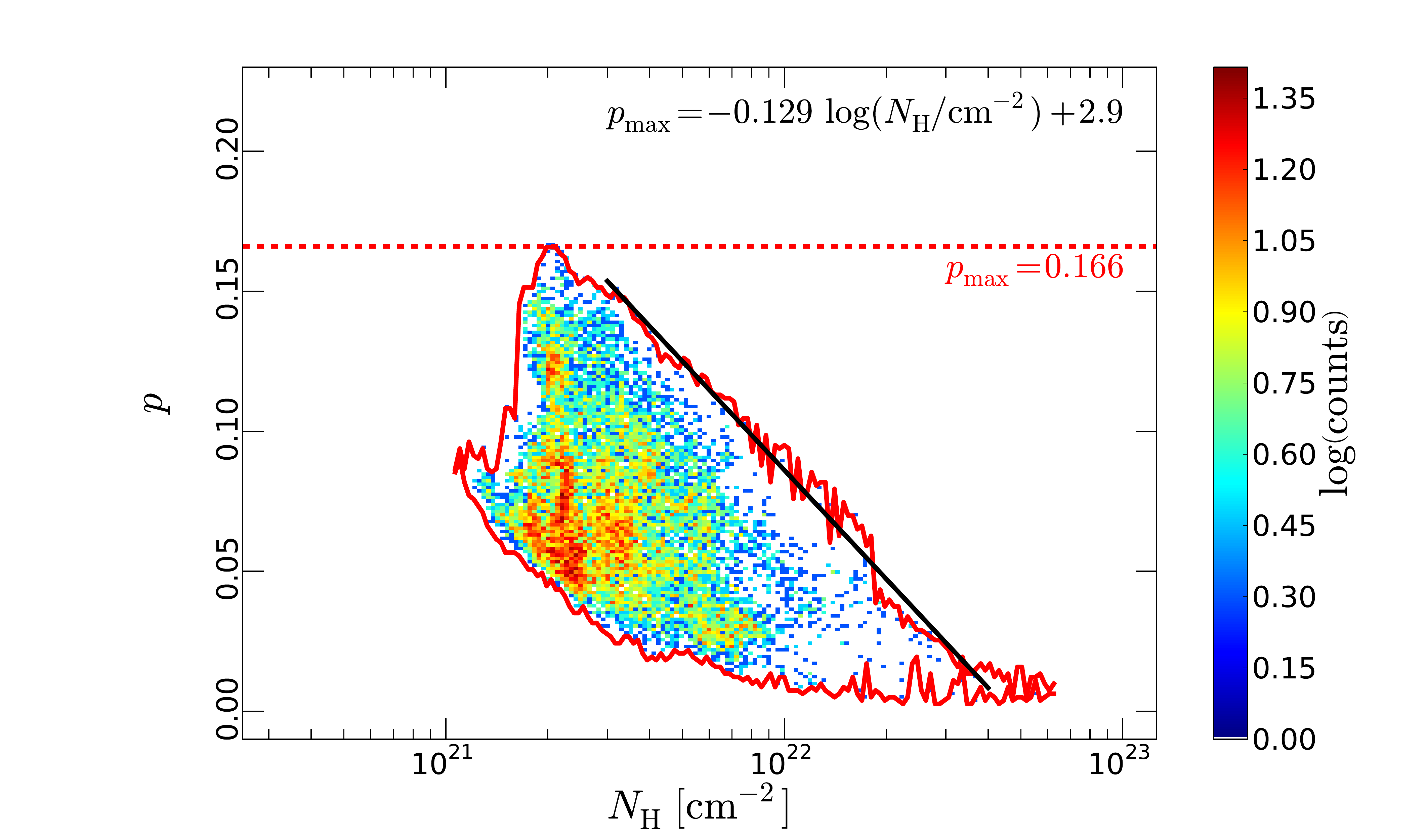}}
\caption{Same as the top panel of Fig.~\ref{fig:PI-vs-NH}, but using only pixels for which $\polfrac/\sigpolfrac>10$.}
\label{fig:PI-vs-NH-SNR10}
\end{figure}

\subsection{Polarization angle coherence vs. polarization fraction}
\label{sec:PI-vs-H}

We show in the bottom row panels of Fig.~\ref{fig:PI-B-NH_Ophiuchus-Chamaeleon} the maps of the {\DeltaAngName}s $\DeltaAng$ for the Ophiuchus and Chamaeleon-Musca fields. Similar maps for all other fields are shown in Appendix~\ref{extra-figures}. We recall that this function, defined in~\cite{planck2014-XIX}, is
\begin{equation}
\DeltaAng(\boldsymbol{r},\delta)=\sqrt{\frac{1}{N}\sum_{i=1}^N\left[\polang\left(\boldsymbol{r}\right)-\polang\left(\boldsymbol{r}+\boldsymbol{\delta}_i\right)\right]^2},
\end{equation}
where the sum extends over pixels whose distances from the central pixel $\boldsymbol{r}$ are between $\delta/2$ and $3\delta/2$. Here they are computed at a lag $\delta=16\arcm$, comparable to the size of the beam's FWHM. One can readily see filamentary structures that correspond to regions where the polarization angle is less ordered or where it changes abruptly. These filaments are already noted at $1\deg$ resolution in~\cite{planck2014-XIX} over several degrees. These regions of large angular dispersions correspond to regions of low polarization fraction, as can be seen for instance by comparing the middle and bottom row panels of Fig.~\ref{fig:PI-B-NH_Ophiuchus-Chamaeleon}. 

\begin{figure}[htbp]
\centerline{\includegraphics[width=9cm,trim=0 50 0 0,clip=true]{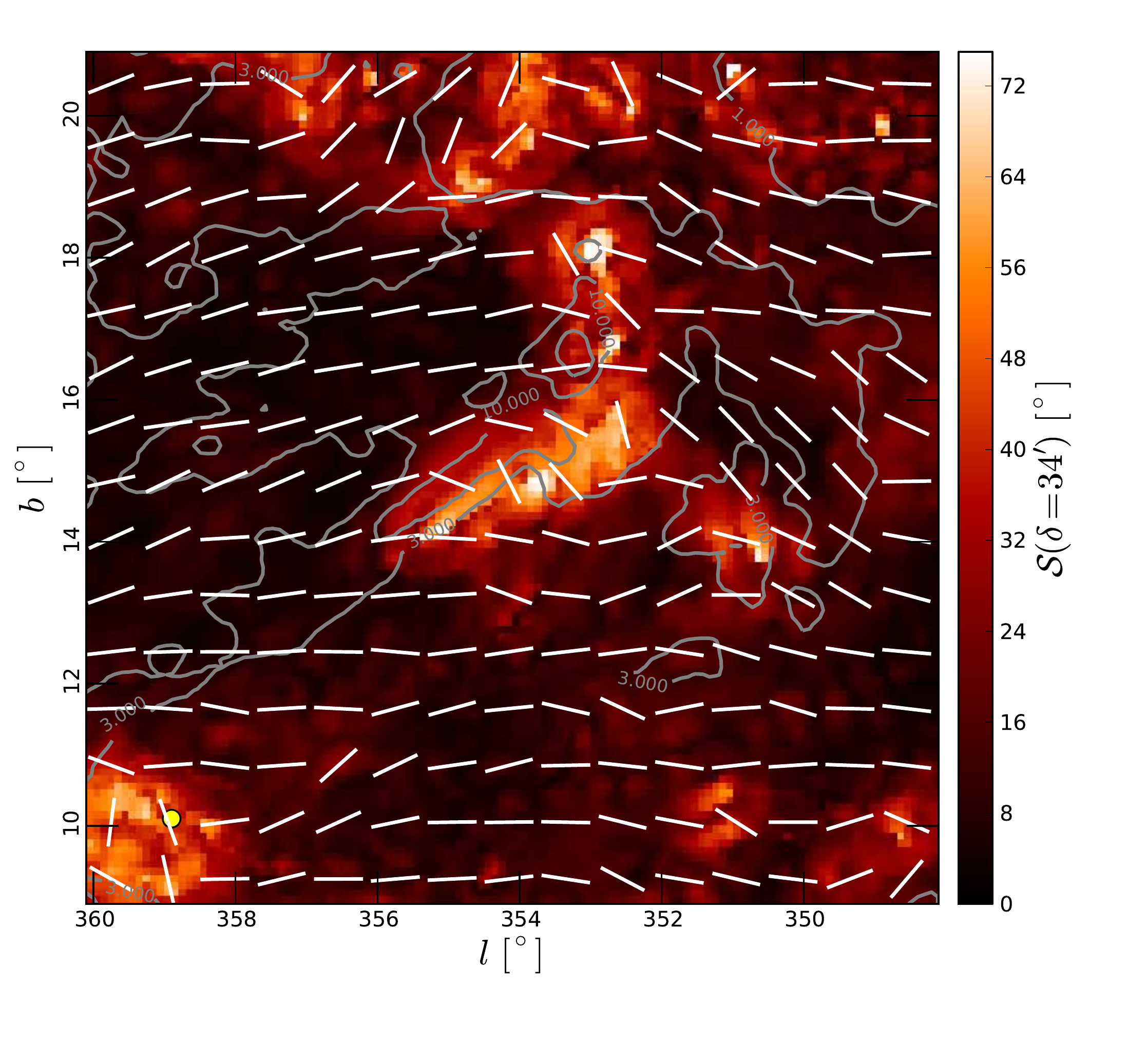}}
\caption{Map of $\DeltaAng$ for the Ophiuchus field computed at $\delta=34\arcm$. Contours are the same as in the map at $\delta=16\arcm$ (bottom left panel of Fig.~\ref{fig:PI-B-NH_Ophiuchus-Chamaeleon}).}
\label{fig:dpsi-Ophiuchus-varying-lag}
\end{figure}

When increasing the value of the lag $\delta$, we obtain maps of $\DeltaAng$ such as that presented in Fig.~\ref{fig:dpsi-Ophiuchus-varying-lag} for the Ophiuchus field at $\delta=34\arcm$ (approximately twice the FWHM). It appears that the overall value of $\DeltaAng$ increases with lag, as already noted in~\cite{hildebrand_et_al_09} and~\cite{planck2014-XIX}. However, since $\DeltaAng$ has an upper limit of $90\deg$, this means that the anti-correlation with $\polfrac$ (see below) will flatten out at large lags. Note however that a completely random sample yields $\DeltaAng=\pi/\sqrt{12}\simeq52\deg$~\citep{planck2014-XIX}. Values larger than this are few, but they do exist, as can be seen on the maps of $\DeltaAng$ in Figs.~\ref{fig:PI-B-NH_Ophiuchus-Chamaeleon} and~\ref{fig:dpsi-Ophiuchus-varying-lag}. They may be linked to sharp boundaries between two well-ordered regions: for instance, the \DeltaAngName~at the interface between two half-planes with orthogonal magnetic orientations is $\DeltaAng=\pi/\sqrt{8}\simeq64\deg$.

To confirm the visual impression that the spatial coherence of the polarization angle is anti-correlated with the polarization fraction, we show the distribution function of these two quantities for the Ophiuchus and Chamaeleon-Musca fields in Figs.~\ref{fig:PI-dpsi-data} and~\ref{fig:PI-dpsi-data-Chamaeleon}, respectively, and for all other fields in Appendix~\ref{extra-figures}.

The large-scale anti-correlation seen in~\cite{planck2014-XIX} at 1$\deg$ resolution and $\delta=30\arcm$ is also present when using a lag close to the beam size. With $\delta=1\pdeg07$, we find it to be $\log{\left(\DeltaAng\right)}=-0.75\log{\polfrac}-0.06$, where $\DeltaAng$ is measured in degrees. Since in this case the ratio $\delta/\mathrm{FWHM}$ is the same as for our higher resolution maps ($\mathrm{FWHM}=15\arcm$ and $\delta=16\arcm$), we compare the anti-correlations found in the selected fields to this law. Note that the slope $-0.75$ is similar to the value $-0.834$ quoted in~\cite{planck2014-XIX}, but the intercept is larger ($-0.06$ vs. $-0.504$). This points to a global increase of $\DeltaAng$ at larger $\delta/\mathrm{FWHM}$ values, which we interpret as a decorrelation of polarization angles at larger lags.

The distributions of $\polfrac$ and $\DeltaAng$ in the various fields considered show an anti-correlation very similar to the large-scale trend, with slopes and intercepts of the fits through the data points that are very close to the large-scale fit values. When increasing the lag at the same resolution, however, $\DeltaAng$ increases and the anti-correlation with $\polfrac$ flattens out, as can be seen in Fig.~\ref{fig:PI-dpsi-data-Ophiuchus-R34}. The linear fits $\log{\left(\DeltaAng\right)}=m'\log{\polfrac}+c'$ for the individual fields are listed in Table~\ref{table-fields-properties}. The uncertainties on the parameters $m'$ and $c'$ are the quadratic sums of uncertainties obtained in three ways: (i) by performing the linear regression using the three estimators of $\polfrac$, i.e., the ``na\"ive'', MAS and Bayesian ones; (ii) by using half-ring maps and detector set maps; (iii) via a Monte-Carlo simulation using the maps of polarization fraction uncertainty $\sigpolfrac$ and \DeltaAngName~uncertainty $\sigma_{\DeltaAng}$~\citep{planck2014-XIX}.

\begin{figure}[htbp]
\centerline{\includegraphics[width=8.8cm,trim=120 0 60 0,clip=true]{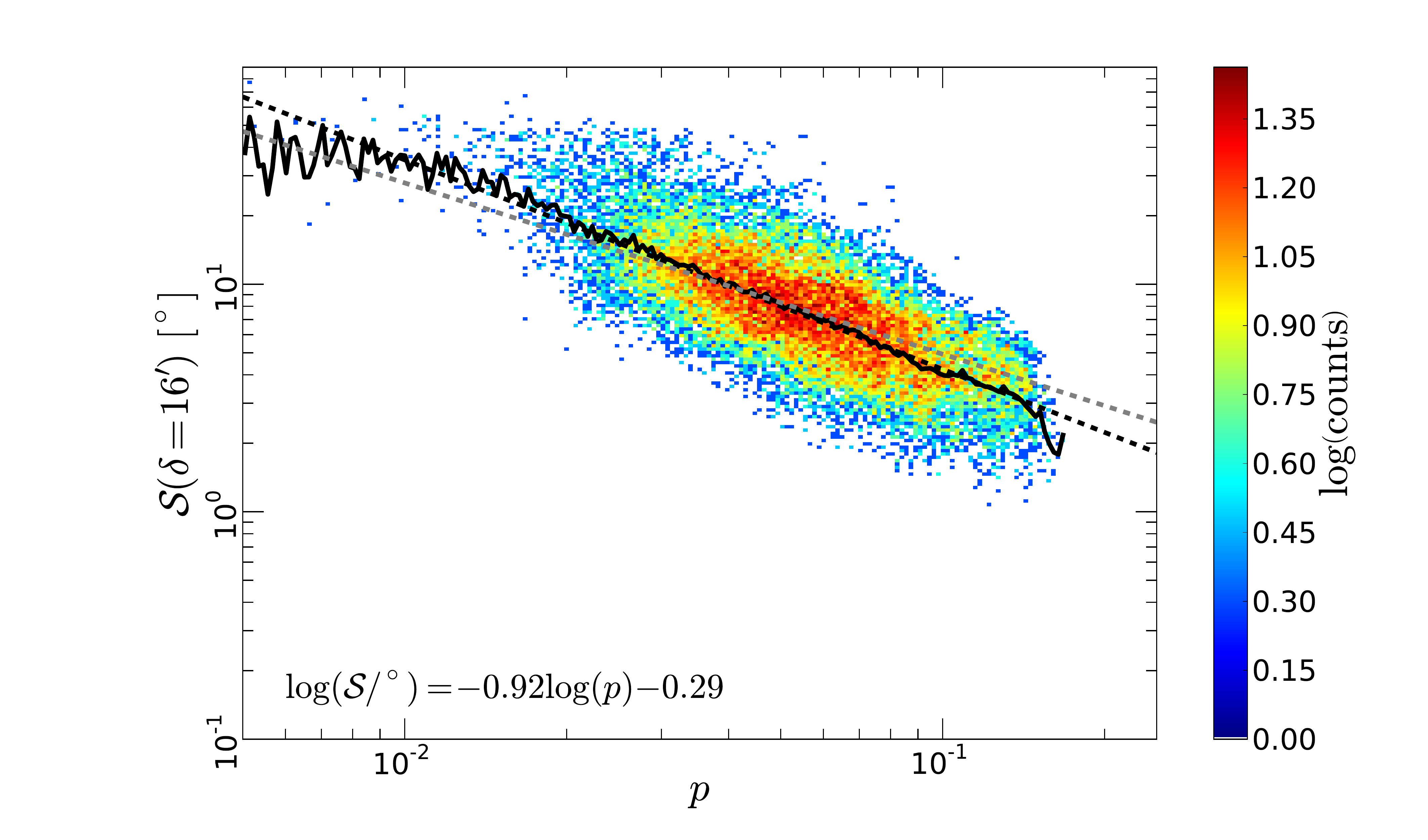}}
\caption{Two-dimensional distribution function of $\DeltaAng$ and polarization fraction $\polfrac$ for the Ophiuchus field. The {\DeltaAngName} $\DeltaAng$ is computed at a lag $\delta=16\arcm$. Only pixels for which $\polfrac/\sigpolfrac>3$ are retained. The dashed grey line is the large-scale fit (with $\mathrm{FWHM}=1\deg$ and $\delta=1\pdeg07$) $\log{\left(\DeltaAng\right)}=-0.75\log\left({\polfrac}\right)-0.06$, the solid black line shows the mean $\DeltaAng$ for each bin in $\polfrac$ (the bin size is $\Delta\log(\polfrac)=0.008$) and the dashed black line is a linear fit of that curve in log-log space, restricted to bins in $\polfrac$ which contain at least 1\% of the total number of points (so about 150 points per bin).}
\label{fig:PI-dpsi-data}
\end{figure}

\begin{figure}[htbp]
\centerline{\includegraphics[width=8.8cm,trim=120 0 60 0,clip=true]{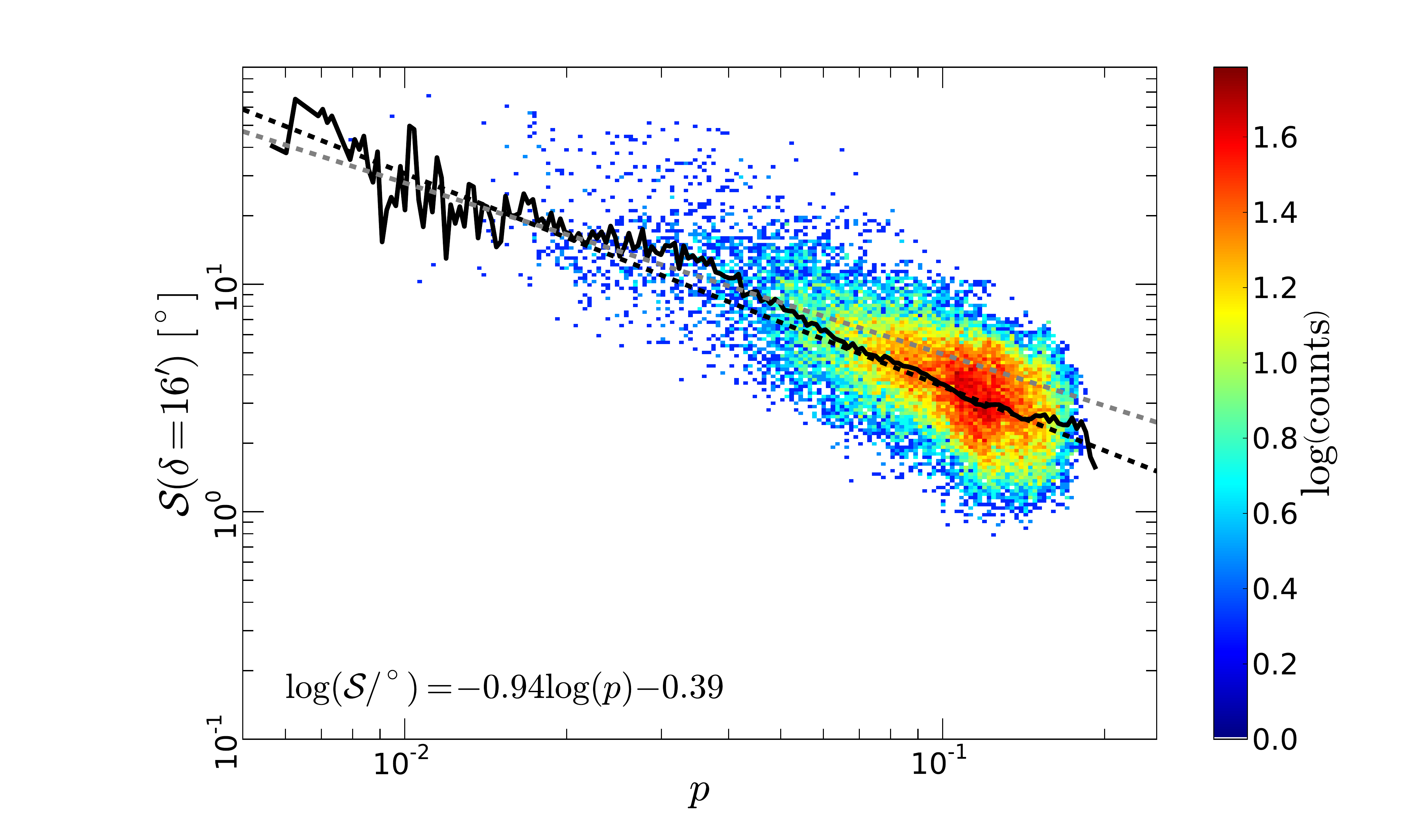}}
\caption{Same as Fig.~\ref{fig:PI-dpsi-data}, but for the Chamaeleon-Musca field.}
\label{fig:PI-dpsi-data-Chamaeleon}
\end{figure}

\begin{figure}[htbp]
\centerline{\includegraphics[width=8.8cm,trim=120 0 60 0,clip=true]{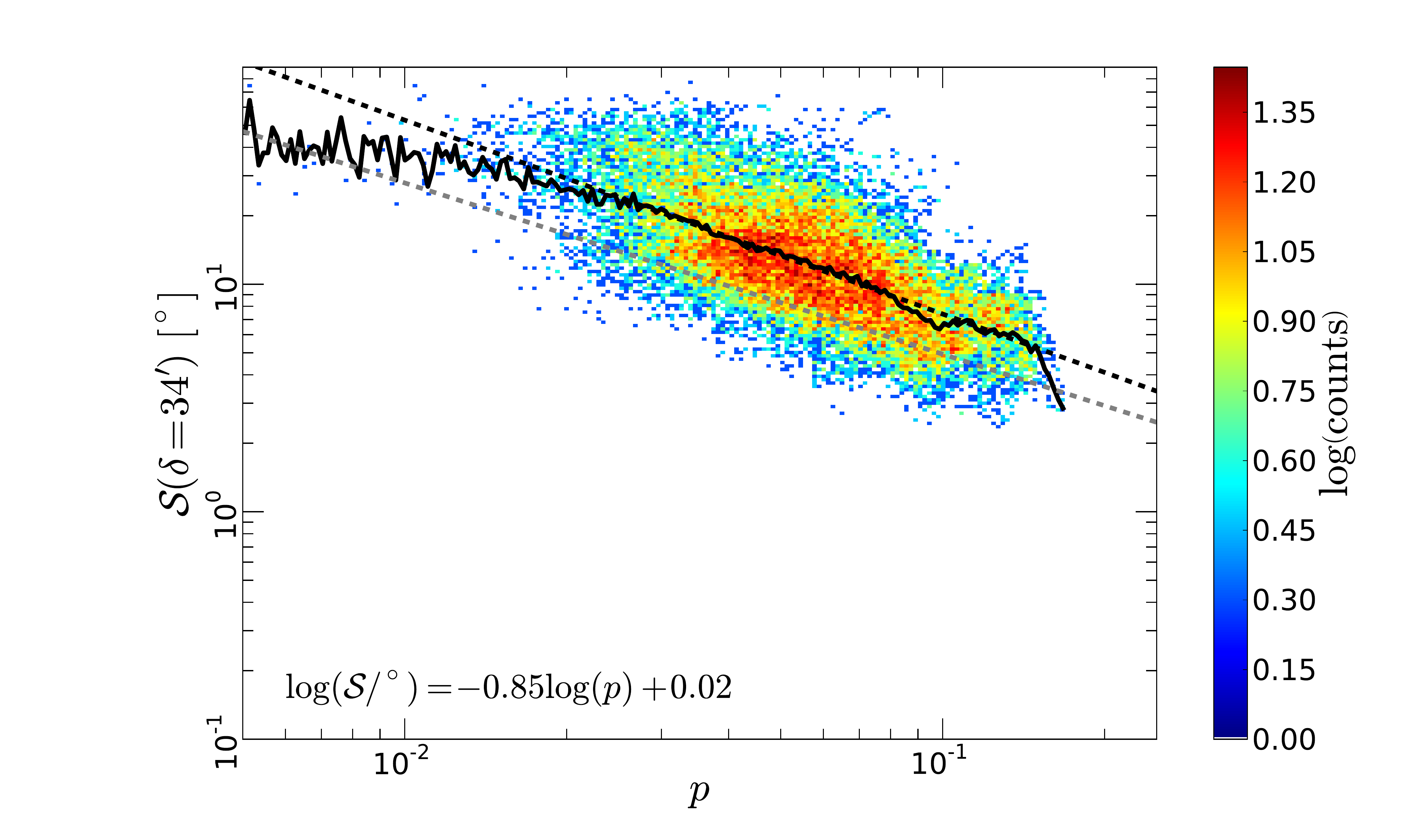}}
\caption{Same as Fig.~\ref{fig:PI-dpsi-data}, but for a lag $\delta=34\arcm$.}
\label{fig:PI-dpsi-data-Ophiuchus-R34}
\end{figure}

\section{Simulations of polarized emission}
\label{sec:simus}
\subsection{Simulations of MHD turbulence}

We aim to compare the observed polarization statistics in the selected fields to predictions built on the results 
of a numerical simulation of MHD turbulence. This simulation is described in detail in \cite{hennebelle_08}\footnote{It was performed with the {\tt RAMSES} code (\citealt{teyssier_02}, \citealt{fromang_06}), whose adaptive mesh refinement capabilities allow for a locally high spatial sampling. It is freely available via the STARFORMAT project, \url{http://starformat.obspm.fr/}. To be precise, it is the {\it Fiducial} run under the tab {\it Colliding flow simulation}.}. It follows the formation of clumps of dense and cold gas (cold neutral medium, CNM) out of magnetized warm 
neutral atomic gas (warm neutral medium, WNM) in an open box of $50\,\mathrm{pc}$ on each side, without reaching the stage when cold cores of column density larger than 
$2 \times 10^{22}\,\mathrm{cm}^{-2}$ form. 

The simulation cube initially contains a uniform distribution of WNM with density $n_\mathrm{H}=1\,\textrm{cm}^{-3}$ and temperature $T=8000\,\textrm{K}$, and two converging flows of that same gas are injected from opposing faces along the $x$ axis with a velocity $\Delta V_x\simeq 40\,\mathrm{km\,s}^{-1}$ relative to each other. Spatial modulations of the velocity are imposed on the incoming flows, with amplitudes relative to the mean flow of about unity and a periodicity of about 10\,pc. Periodic boundary conditions are applied on the remaining four faces. The total mass contained in the cube continuously increases with time. The magnetic field's initial direction is along that of the incoming flows, and its intensity is about 5\,$\mu\textrm{G}$, consistent with observational values at these densities \citep{crutcher_et_al_10}. There is therefore a large-scale anisotropic component of the magnetic field throughout the simulation, as well as a turbulent component linked to the velocity perturbations imposed on the converging flows. 

These flows collide near the midplane, where the combined effects of cooling and self-gravity eventually lead to the formation of dense ($n_\mathrm{H}> 100\,\textrm{cm}^{-3}$) clumps of cold gas ($T$ of the order of $10$--$50\,\textrm{K}$) \citep{hennebelle_07}. To follow that condensation, the grid is adaptively refined, with an effective (maximum) resolution of 0.05\,pc. 

In this paper, we select a cubic subset ($18\,\mathrm{pc}\,\times18\,\mathrm{pc}\,\times18\,\mathrm{pc}$) of the density and magnetic field in the simulation snapshot timed at $t=10.9\,\textrm{Myr}$, which corresponds to an evolved state of the simulation, given the crossing time $t_\mathrm{c}\simeq 2.4\,\mathrm{Myr}$. The structures present in the simulation are due to the collision of the incoming flows and not to a pure gravitational collapse, since the initial free-fall time is $t_\mathrm{ff}\simeq 44\,\mathrm{Myr}$. However, some of the densest structures ($n_\mathrm{H}>10^4\,\mathrm{cm}^{-3}$) may have had time to collapse. 

\begin{figure}[htbp]
\centerline{\includegraphics[width=8.8cm,trim=40 0 70 0,clip=true]{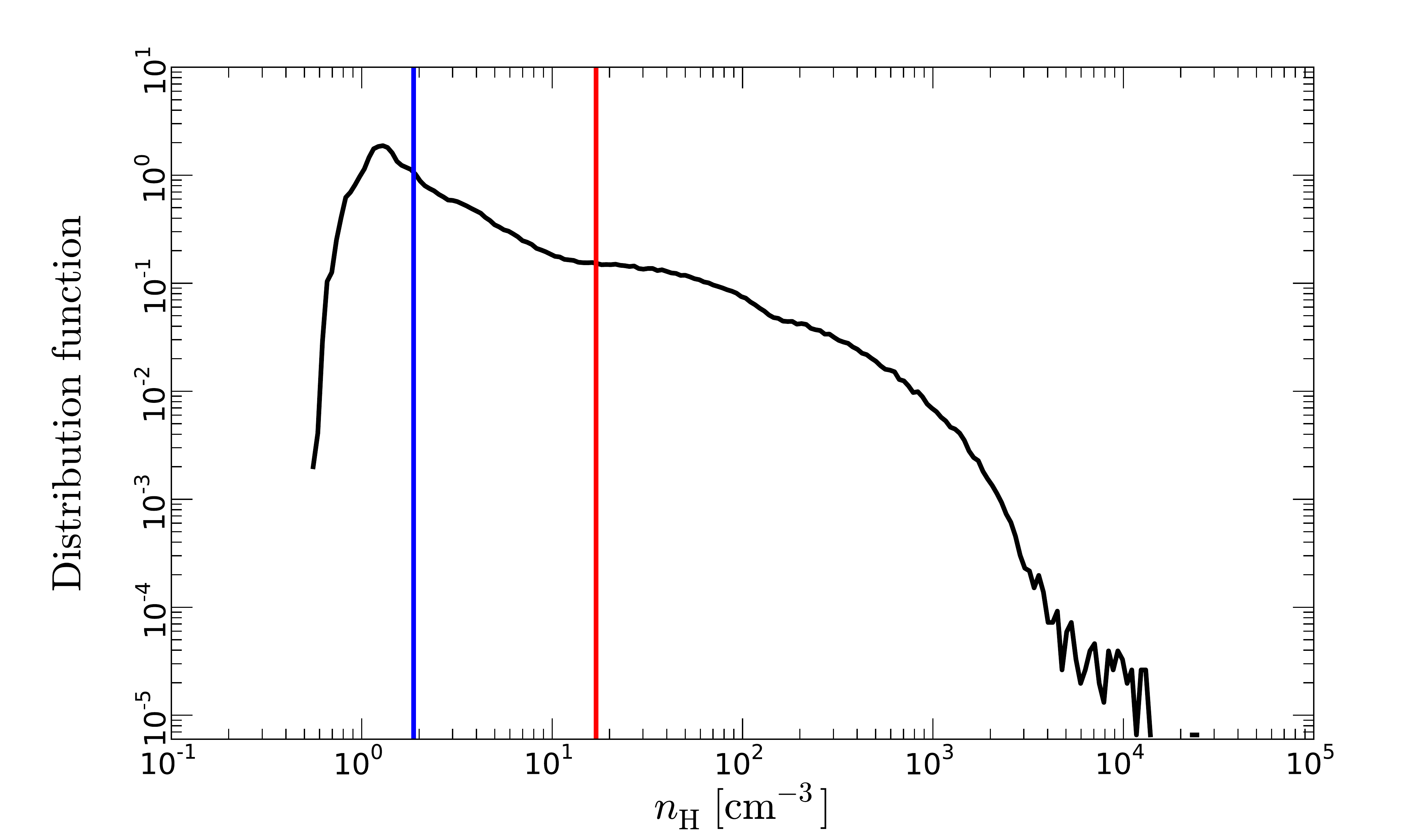}}
\caption{Distribution function of the total gas density $n_\mathrm{H}$ in the selected subset of the simulation, with cell sizes $0.1\,\mathrm{pc}\times0.1\,\mathrm{pc}\times0.1\,\mathrm{pc}$. The solid red line shows the mean value $\left<n_\mathrm{H}\right>=17~\mathrm{cm}^{-3}$ and the solid blue line the median value $n_\mathrm{H}^{\mathrm{med}}=2~\mathrm{cm}^{-3}$.}
\label{fig:n_stats}
\end{figure}

\begin{figure}[htbp]
\centerline{\includegraphics[width=8.8cm,trim=40 0 70 0,clip=true]{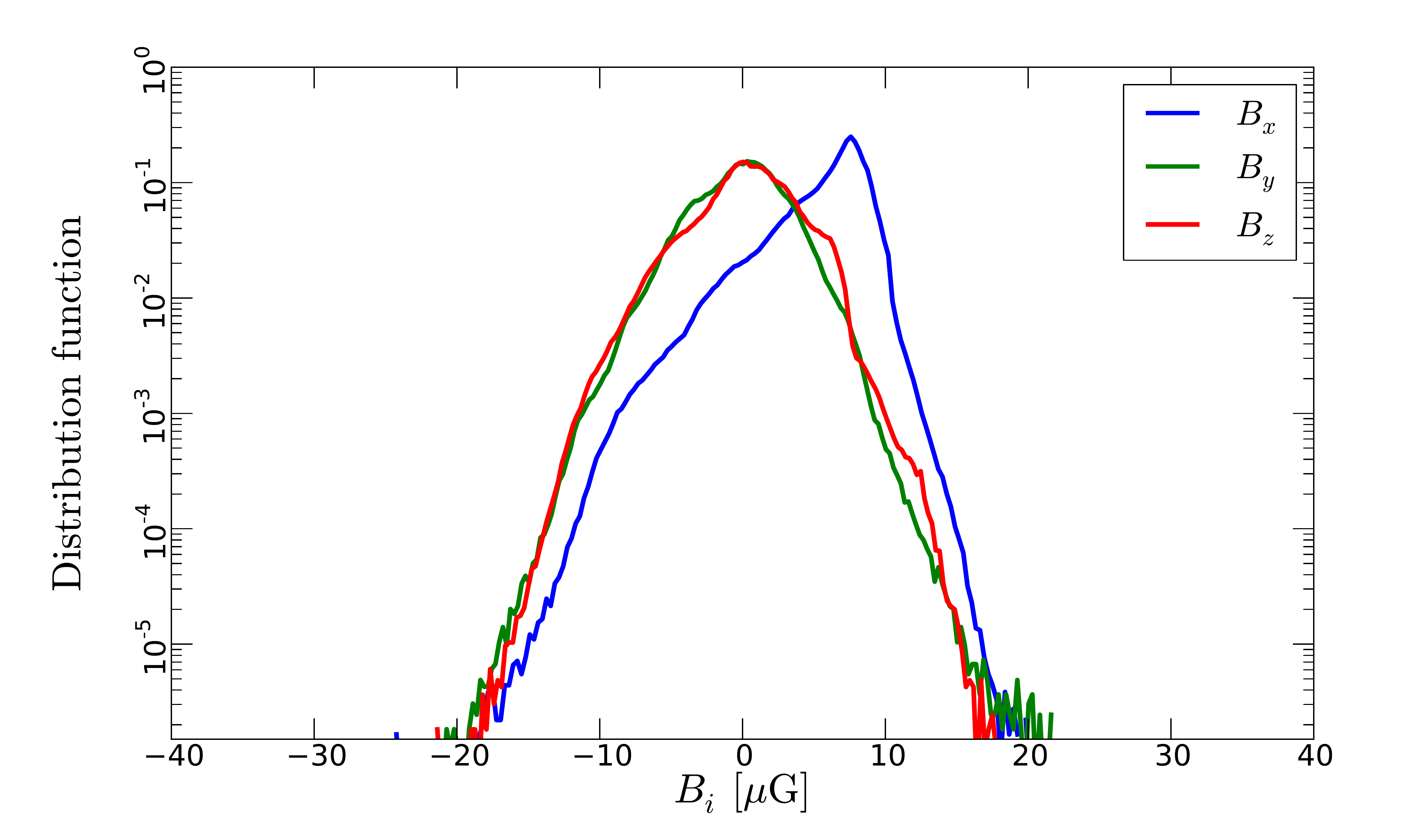}}
\caption{Distribution functions of the components of the magnetic field, $B_x$ (blue), $B_y$ (green), and $B_z$ (red), in the selected subset of the simulation, with cell sizes $0.1\,\mathrm{pc}\times0.1\,\mathrm{pc}\times0.1\,\mathrm{pc}$.}
\label{fig:b_stats}
\end{figure}

The chosen subset is located near the midplane, so that the influence of boundary conditions is minimal. It contains approximately $3200\,\Msolar$ of gas; its physical properties are listed in Table~\ref{table-clump}, and the distribution functions of total gas density $n_\mathrm{H}$ and magnetic field components $B_x$, $B_y$, $B_z$ are shown in Figs.~\ref{fig:n_stats} and~\ref{fig:b_stats}, respectively. The standard deviations are very similar for all three magnetic field components, but only the $x$ component has a significant mean value, which shows that the mean magnetic field within the cube is approximatively aligned with the $x$ axis, that is with the incoming flows.

We would like to stress here that the MHD simulations we use for comparison
with the \Planck~polarization data do not faithfully reproduce the 
 whole range of densities and column densities spanned by the cloud sample of Table~\ref{table-fields}, i.e., from
  diffuse molecular clouds (Polaris Flare) to massive star-forming clouds (Orion).
However, as shown in Table~\ref{table-fields}, only a few percent of the pixels
(at most 5.4\% in Orion) have column densities larger than $10^{22}\,\mathrm{cm}^{-2}$ in these 
fields, the regions of star formation filling only a small fraction of the area in each field. 
The MHD simulations with their broad range of densities (Fig.~\ref{fig:n_stats}) and column densities 
reaching\footnote{This value is computed over the whole range of viewing angles $\viewangle$.} $N_\mathrm{H}=1.6\times10^{22}\,\mathrm{cm}^{-2}$ are therefore
representative of the dynamics of the bulk of the gas. 
Together with their anisotropy, due to the large-scale magnetic field pervading the cube,
these simulations are particularly well suited to analyse 
the polarization properties of nearby molecular clouds immersed in their low density
and large-scale environment.

To compute simulated polarization fractions $\polfrac$, the local gas density $n_\mathrm{H}$ and magnetic field components $B_x$, $B_y$, $B_z$ are extracted from the simulation and interpolated on a regular grid at the next-to-highest spatial resolution available, so that pixel sizes are approximately $\Delta x=0.1\,\mathrm{pc}$. These cubes are used in the following section to build simulated polarized emission maps. However, they are first rotated around the $y$ axis, as sketched out in Fig.~\ref{fig:cuberotation}, to explore the full range of possible angles between the mean magnetic field and the line of sight, and therefore to test the effects of the large-scale magnetic field's anisotropy. The viewing angle $\alpha$ introduced in Fig.~\ref{fig:cuberotation} is such that the mean magnetic field is approximately in the plane of the sky for $\viewangle=0\deg$, and along the line of sight for $\viewangle=90\deg$.

\begin{figure}[htbp]
\centerline{\includegraphics[width=8.8cm,trim=40 0 70 0,clip=true]{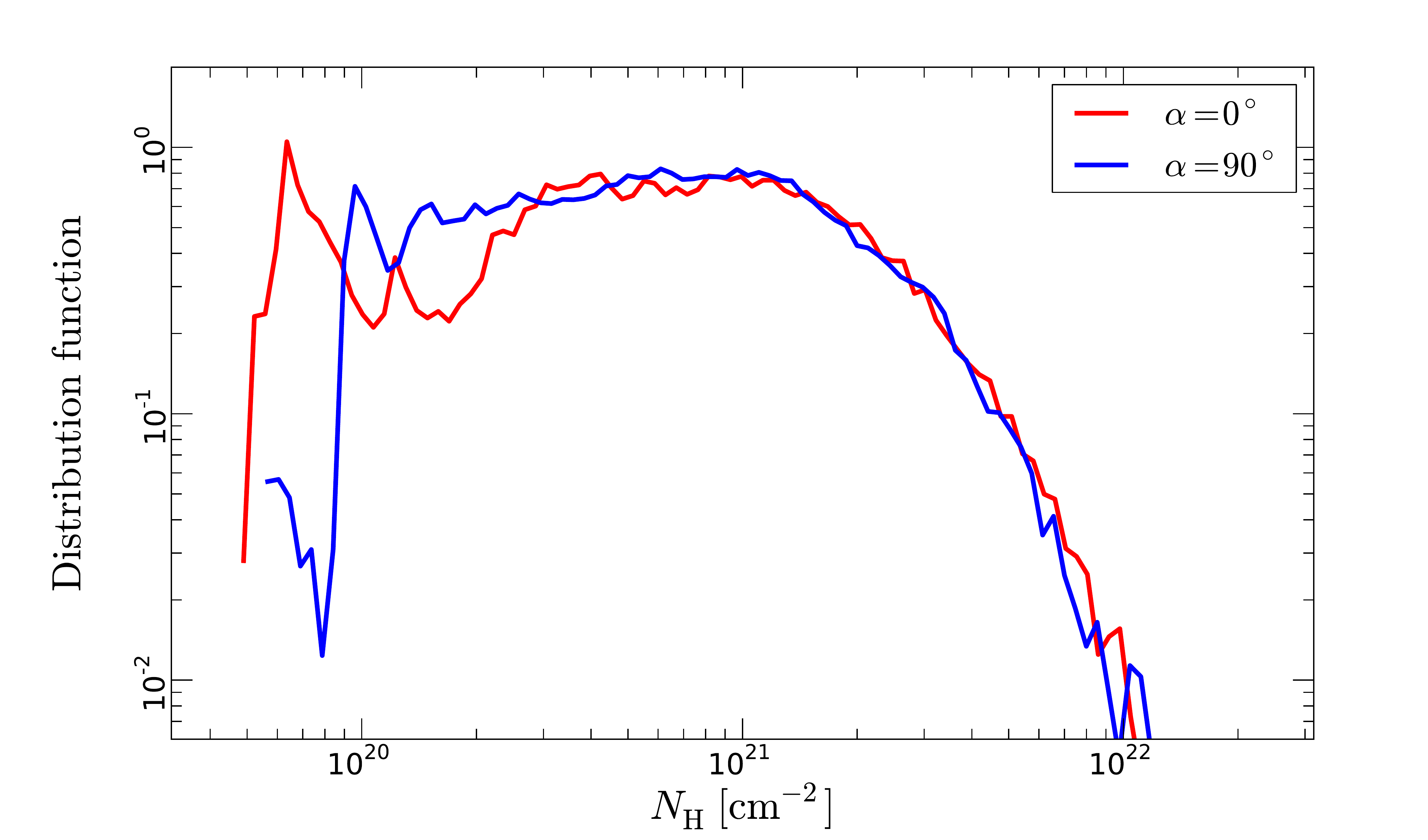}}
\caption{Distribution functions of the total gas column density $N_\mathrm{H}$ in the selected subset of the simulation, using viewing angles $\viewangle=0\deg$ (red) and $\viewangle=90\deg$ (blue). These distribution functions are computed after convolution with the 15\arcm~beam.}
\label{fig:NH_stats}
\end{figure}

\begin{table}[htbp]                 
\begingroup
\newdimen\tblskip \tblskip=5pt
\caption{Physical properties of the subset of the simulation. These values correspond to $\viewangle=0\deg$ (see text and Fig.~\ref{fig:cuberotation}).}                          
\label{table-clump}                            
\nointerlineskip
\vskip -3mm
\footnotesize
\setbox\tablebox=\vbox{
   \newdimen\digitwidth 
   \setbox0=\hbox{\rm 0} 
   \digitwidth=\wd0 
   \catcode`*=\active 
   \def*{\kern\digitwidth}
   \newdimen\signwidth 
   \setbox0=\hbox{+} 
   \signwidth=\wd0 
   \catcode`!=\active 
   \def!{\kern\signwidth}
\halign{\hbox to 1.15in{#\leaderfil}\tabskip 2.2em&
\hfil#\hfil&
\hfil#\hfil&
\hfil#\hfil&
\hfil#\hfil\tabskip 0pt\cr
\noalign{\doubleline\vskip 2pt}
\omit\hfil $F$\hfil&$\left<F\right>$&$\mathrm{min}(F)$&$\mathrm{max}(F)$&$\sigma(F)$\cr
\noalign{\vskip 4pt\hrule\vskip 6pt}
$N_\mathrm{H}$ $[10^{21}~\mathrm{cm}^{-2}]$&$1.0$&$0.05$&$13.4$&1.0\cr
$n_\mathrm{H}$ $[\mathrm{cm}^{-3}]$&$16.4$&$0.5$&$4.1\times10^{4}$&92\cr
$B_x$ $[\mu\mathrm{G}]$&$\phantom{-}5.8$&$-32.5$&$25.8$&3.2\cr
$B_y$ $[\mu\mathrm{G}]$&$-0.1$&$-26.1$&$26.5$&3.0\cr
$B_z$ $[\mu\mathrm{G}]$&$\phantom{-}0.3$&$-22.3$&$30.6$&3.3\cr
\noalign{\vskip 3pt\hrule\vskip 4pt}}}
\endPlancktable                    
\endgroup
\end{table}  

\begin{figure}[htbp]
\centerline{\includegraphics[width=9cm]{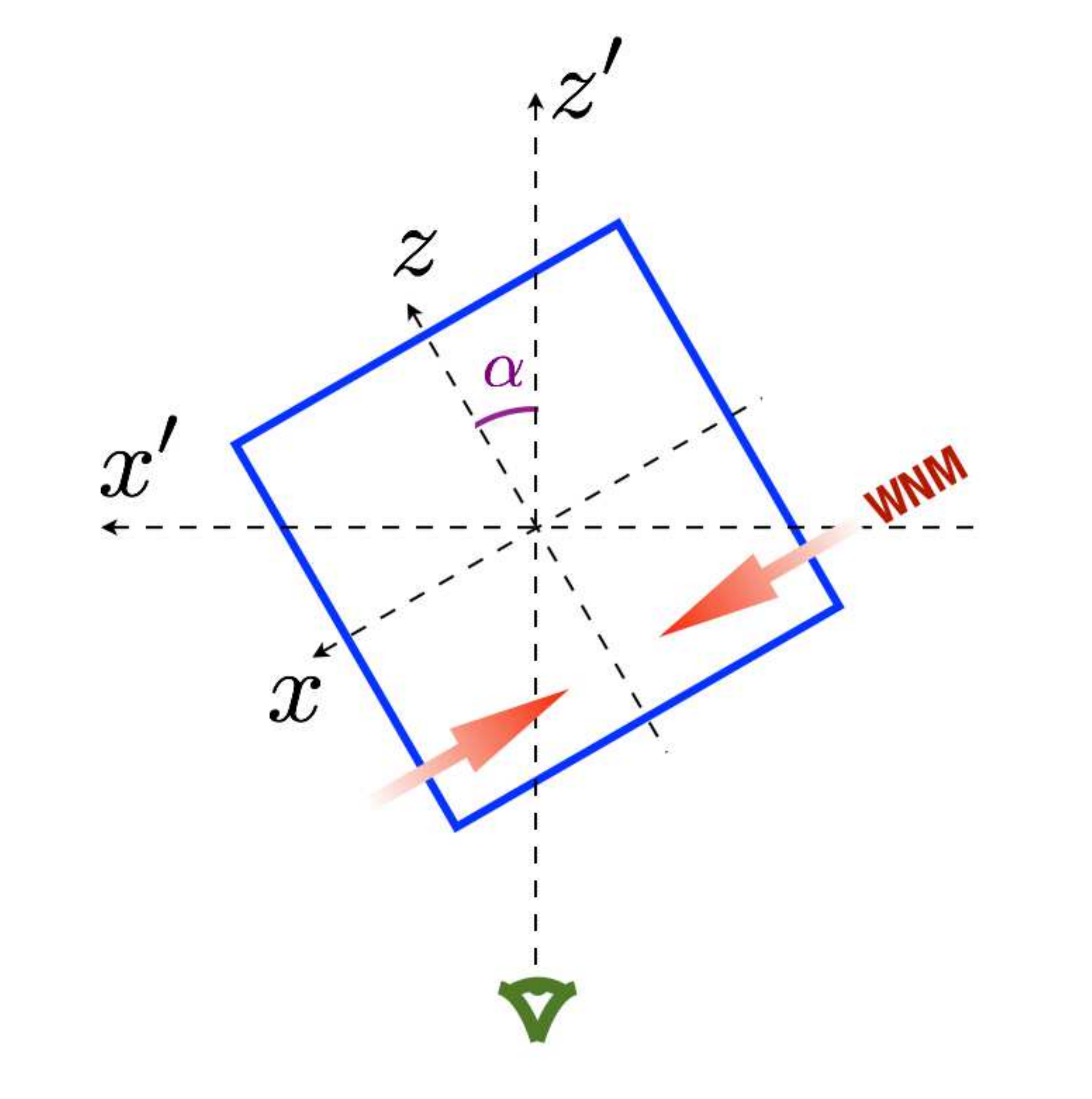}}
\caption{Sketch of the rotation of the simulation subset.}
\label{fig:cuberotation}
\end{figure}

\subsection{Simulated Planck observations}
\begin{figure}[htbp]
\centerline{\includegraphics[width=9.5cm]{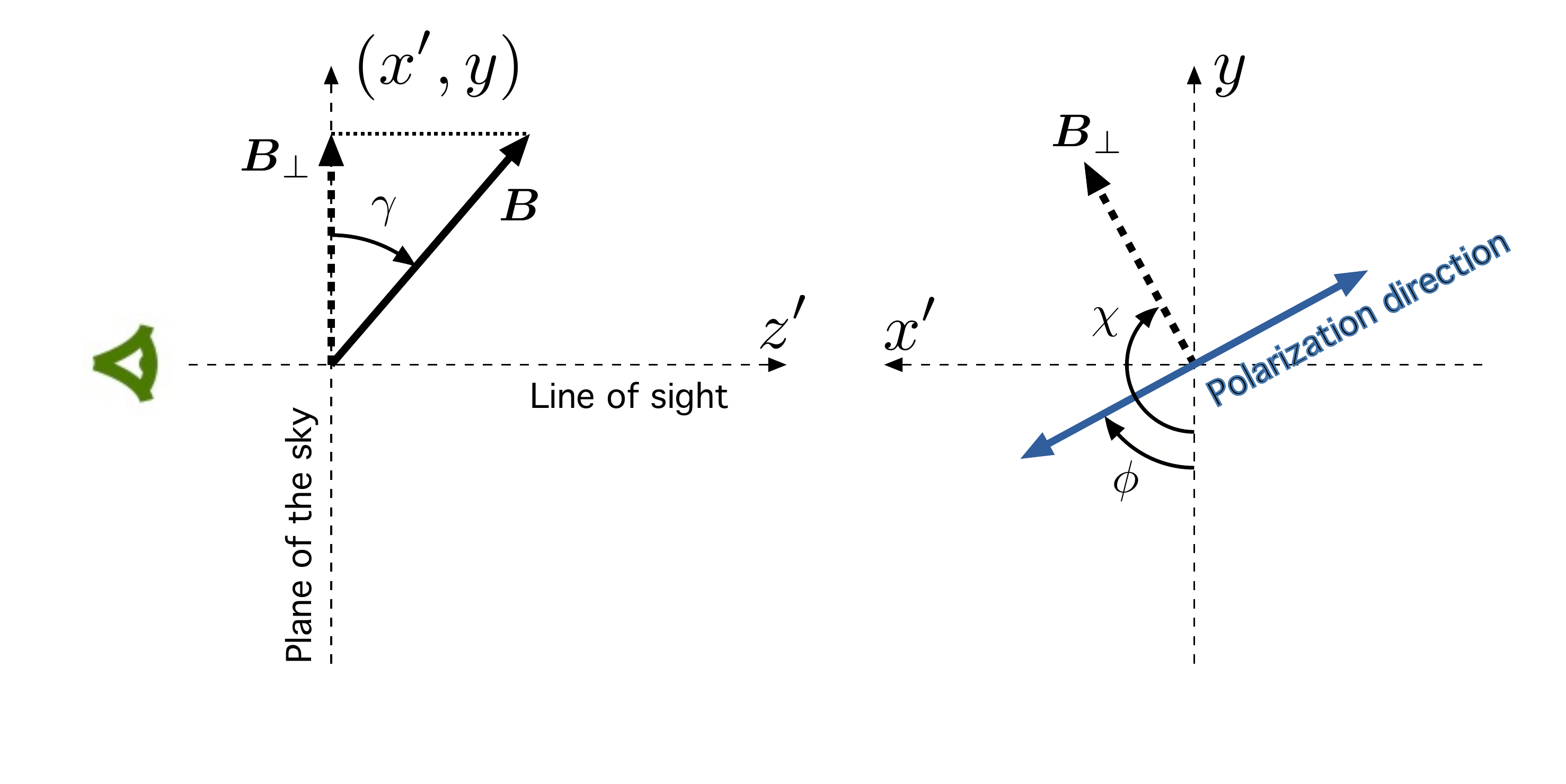}}
\caption{Definition of angles. Here the line of sight is along the $z'$ axis (see Fig.~\ref{fig:cuberotation}), $\polangsky$ is the angle the magnetic field $\vec{B}$ makes with the plane of the sky, $\phi$ is the local polarization angle, and $\chi$ is the position angle of the plane of the sky projection $\vec{B}_\perp$, both in the \healpix~convention, so counted positively clockwise from the north-south direction, while the IAU convention is anti-clockwise~\citep{planck2014-XIX}.}
\label{fig:anglesdefinition}
\end{figure}

We build simulated Stokes $I$, $Q$, and $U$ maps by integrating along the line of sight ($z'$ in Fig.~\ref{fig:cuberotation}) through the rotated simulation cube, following the method in \cite{wardle_90}, \cite{fiege_00}, \cite{pelkonen_09}, and \cite{padovani_et_al_12}. Because of a number of inconsistencies in the literature, we give the correct derivation in Appendix~\ref{sec:Stokes-LD85}, drawing on the works of \cite{lee_85} and \cite{wardle_90}. This results in:
\begin{equation}
\label{eq:simulated_I}
\StokesI=\int S_\nu\,e^{-\tau_\nu}\left[1-p_0\left(\cos^2\polangsky-\frac{2}{3}\right)\right]\mathrm{d}\tau_\nu;
\end{equation}
\begin{equation}
\label{eq:simulated_Q}
\StokesQ=\int p_0\,S_\nu\,e^{-\tau_\nu}\cos\left(2\phi\right)\cos^2\polangsky\,\mathrm{d}\tau_\nu;
\end{equation}
\begin{equation}
\label{eq:simulated_U}
\StokesU=\int p_0\,S_\nu\,e^{-\tau_\nu}\sin\left(2\phi\right)\cos^2\polangsky\,\mathrm{d}\tau_\nu.
\end{equation}
Here $p_0$ is a polarization fraction parameter related to the \IntrinsicpName~(see Eq.~\ref{eq:pmax} and Appendix~\ref{sec:Stokes-LD85}), $\polangsky$ is the angle that the local magnetic field makes with the plane of the sky, and $\phi$ is the local polarization angle in the \healpix~convention. This angle differs by $90^\circ$ from the angle $\chi$ of the plane of the sky projection of the magnetic field, as defined in Fig.~\ref{fig:anglesdefinition}, and should not be confused with the actual polarization angle $\polang$. These angles are equal ($\phi=\polang$) only for a uniform magnetic field along the line of sight.

Note that the corrective term in Eq.~\ref{eq:simulated_I} is incorrectly written in \cite{fiege_00}, \cite{goncalves_05}, \cite{pelkonen_09}, and \cite{padovani_et_al_12}, with $p_0/2$ instead of $p_0$.

The hypotheses made here, besides the absence of background radiation, are that $p_0=0.2$ is uniform, that the source function $S_\nu=B_\nu(T_\mathrm{d})$ is that of a blackbody with an assumed uniform dust temperature $T_\mathrm{d}=18~\mathrm{K}$, and that since we are working at 353\,GHz the optical depth is simply given by $\mathrm{d}\tau_\nu=\sigma_{353}\,n_\mathrm{H}\,\mathrm{d}z'$. We use the value $\sigma_{353}=1.2\times10^{-26}\,\mathrm{cm}^{2}$ (see Sect.~\ref{subsec:dataproc}), and $n_\mathrm{H}$ is the total gas density in the simulation. Given the maximum gas column density in the simulation subset computed over all possible viewing angles $\alpha$, $N_{\mathrm{H},\mathrm{max}}=1.6\times10^{22}\,\mathrm{cm}^{-2}$, the maximum optical depth at 353\,GHz using this conversion factor is $\tau_\mathrm{max}=1.9\times10^{-4}$, so we may safely neglect optical depth effects and take $e^{-\tau_\nu}=1$ in the $I$, $Q$, and $U$ integrals. We are aware~\citep{planck2013-p06b} that the opacity actually varies with $N_\mathrm{H}$, but the variation is at most a factor of 3 from the value assumed here, so the optical depth is in any case much lower than unity. Moreover, the choice of the conversion factor has no impact on the simulated maps of polarization fractions and angles, provided that a constant value is assumed along each line of sight. 

We note that the dense cores that exist in our simulated cube are only weakly shielded from the ambient UV radiation field. Indeed, the mean column density through the cube is about $10^{21}~\mathrm{cm}^{-2}$ (corresponding to $A_\mathrm{V}\simeq 0.6$), which is comparable to the values in the simulation of~\cite{pelkonen_09}, but over a much larger volume (18\,pc box compared to less than 1\,pc); the bulk of the gas is therefore more fragmented and radiation penetrates more easily~\citep{levrier_et_al_12}. That is why we take a uniform parameter $p_0$.

The maps of Stokes parameters are placed at a distance of $D=100~\mathrm{pc}$ and convolved with a circular 15\arcm~FWHM Gaussian beam (corresponding to a physical size $0.44\,\mathrm{pc}$). The resulting field of view is a little less than 10$\deg$ across, which is comparable to the selected \Planck~fields, and small enough that separate smoothing of Stokes $I$, $Q$, and $U$ is not an issue~\citep[see Appendix A of][]{planck2014-XIX}. Maps of polarization fractions and angles are then built from these convolved Stokes parameter maps using Eqs.~\ref{eq:polfrac}--\ref{eq:polang} for consistency with the \planck~data. Let us stress that $\polang$ is defined in the \healpix~convention, which means that it is counted positively clockwise from the north-south direction, and not in the IAU convention (anti-clockwise).

\begin{figure*}[htbp]
\centerline{\includegraphics[width=9cm,trim=0 50 0 0,clip=true]{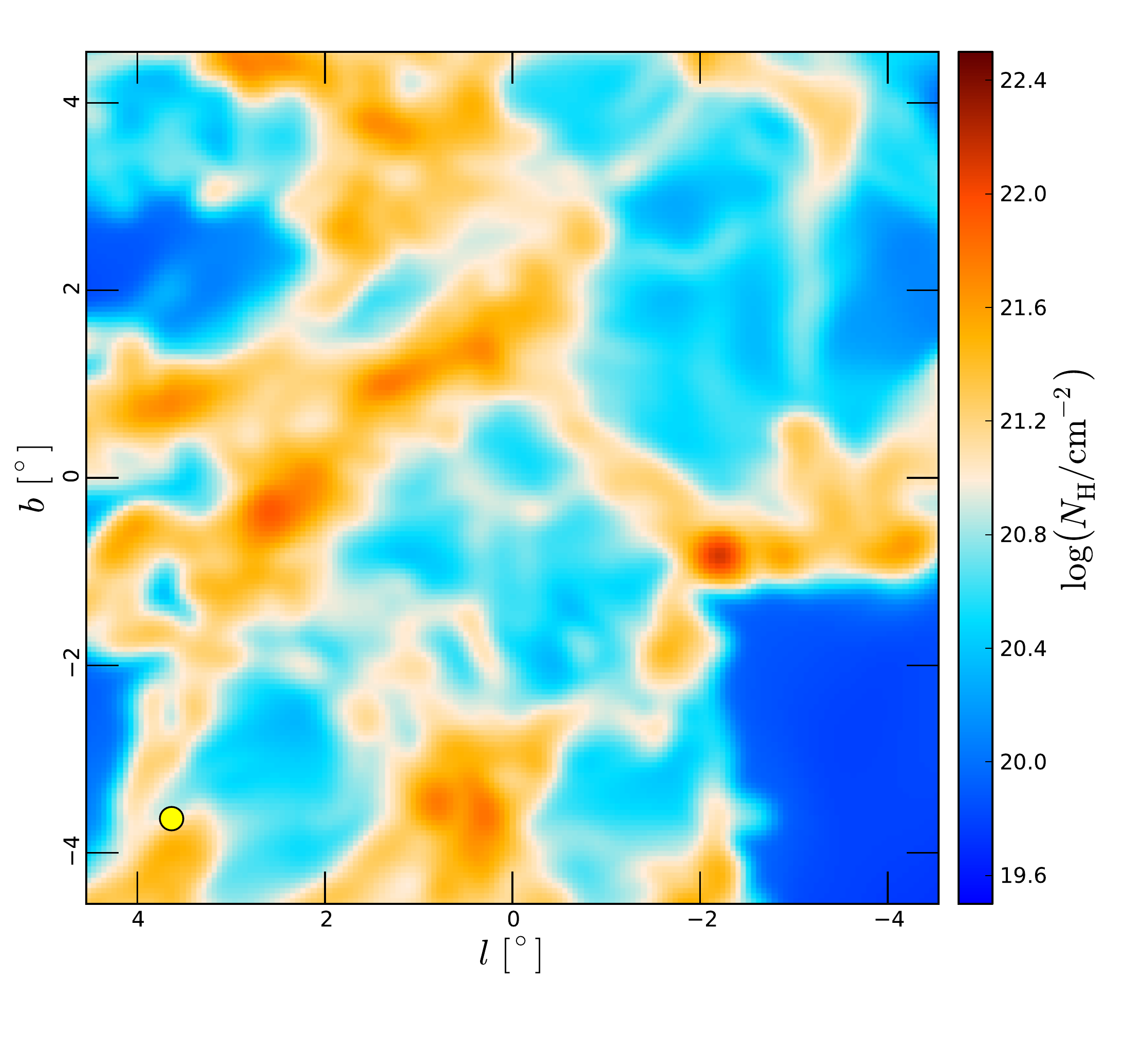}\includegraphics[width=9cm,trim=0 50 0 0,clip=true]{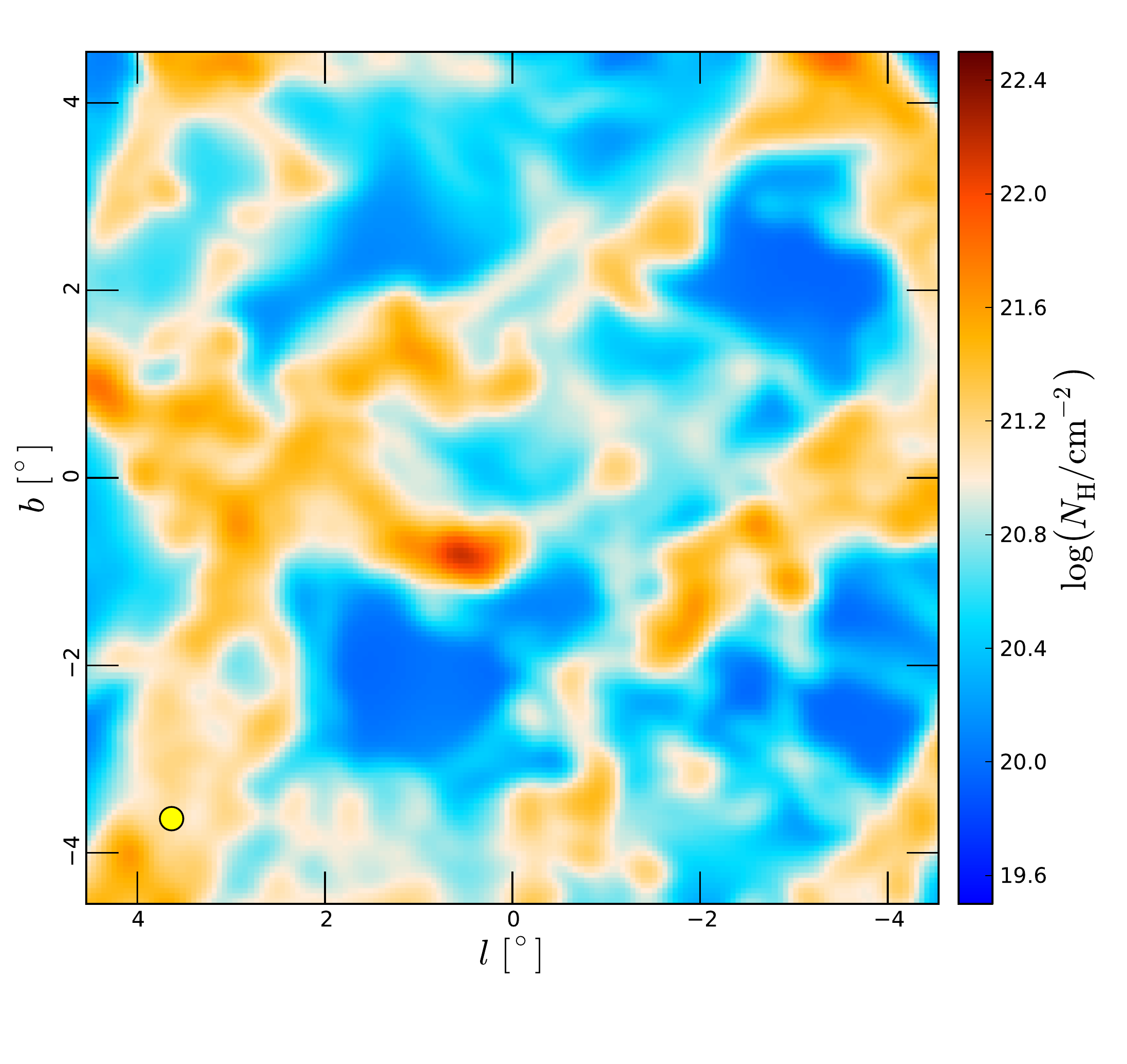}}
\centerline{\includegraphics[width=9cm,trim=0 50 0 20,clip=true]{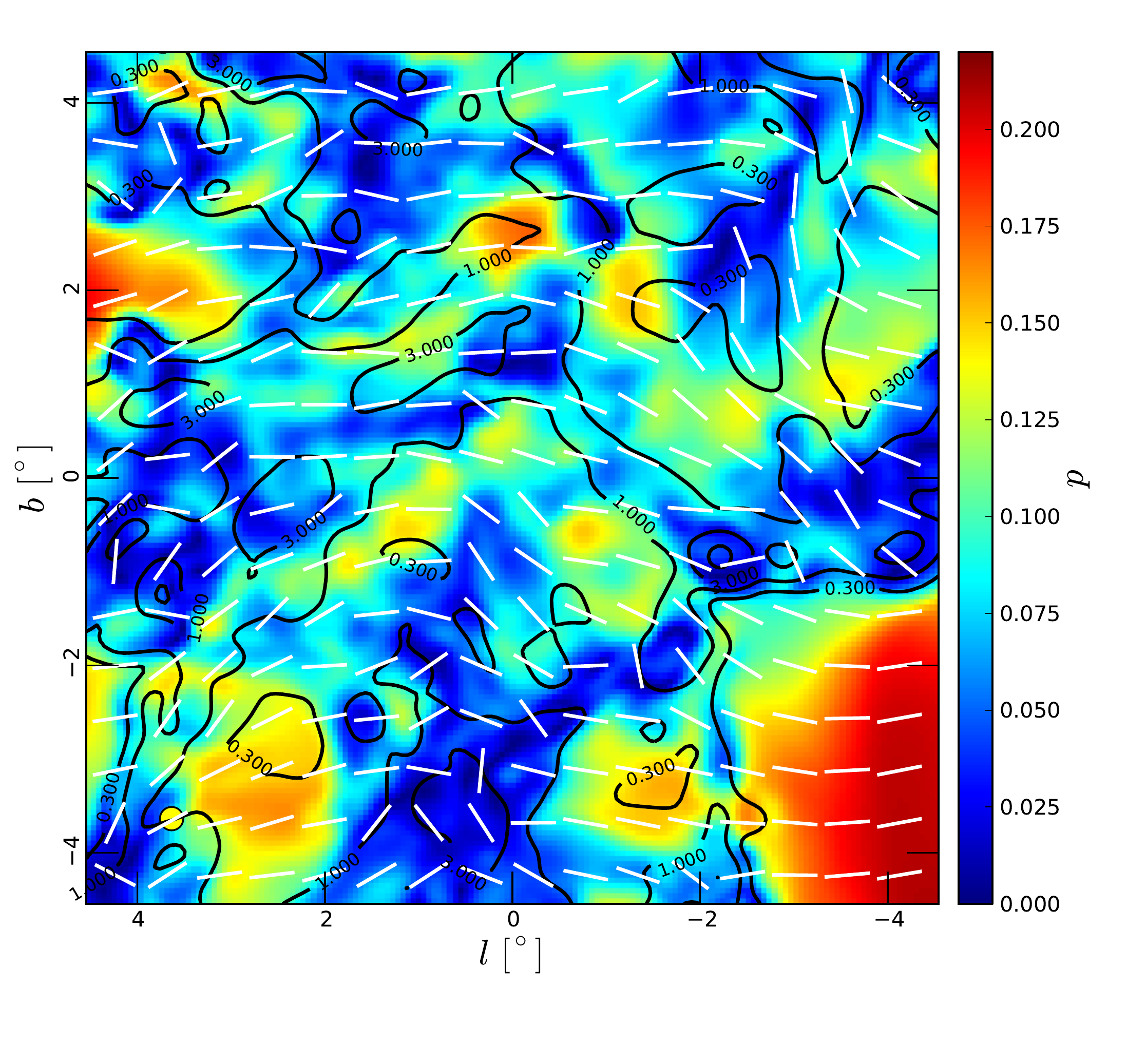}\includegraphics[width=9cm,trim=0 50 0 20,clip=true]{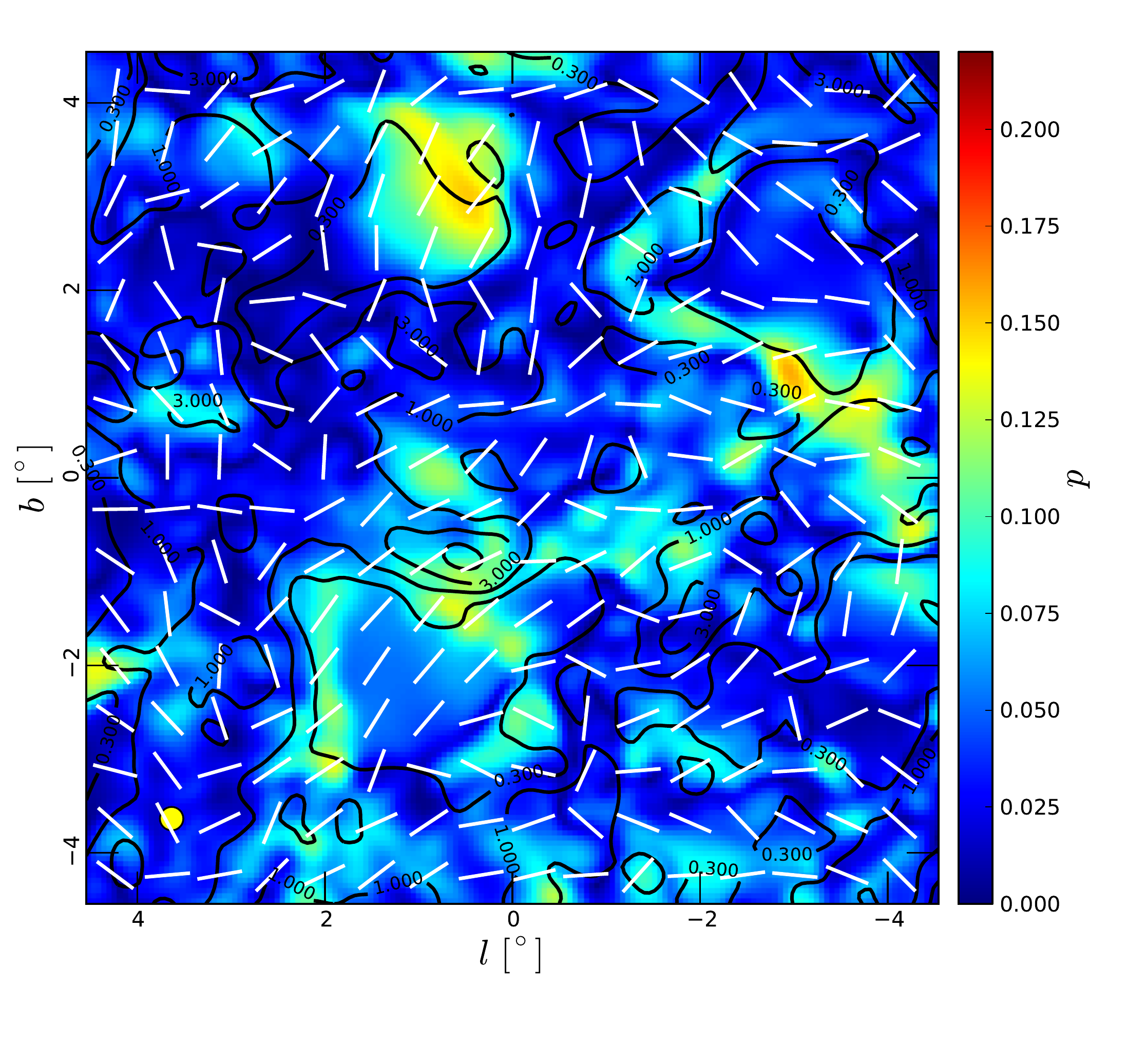}}
\centerline{\includegraphics[width=9cm,trim=0 50 0 20,clip=true]{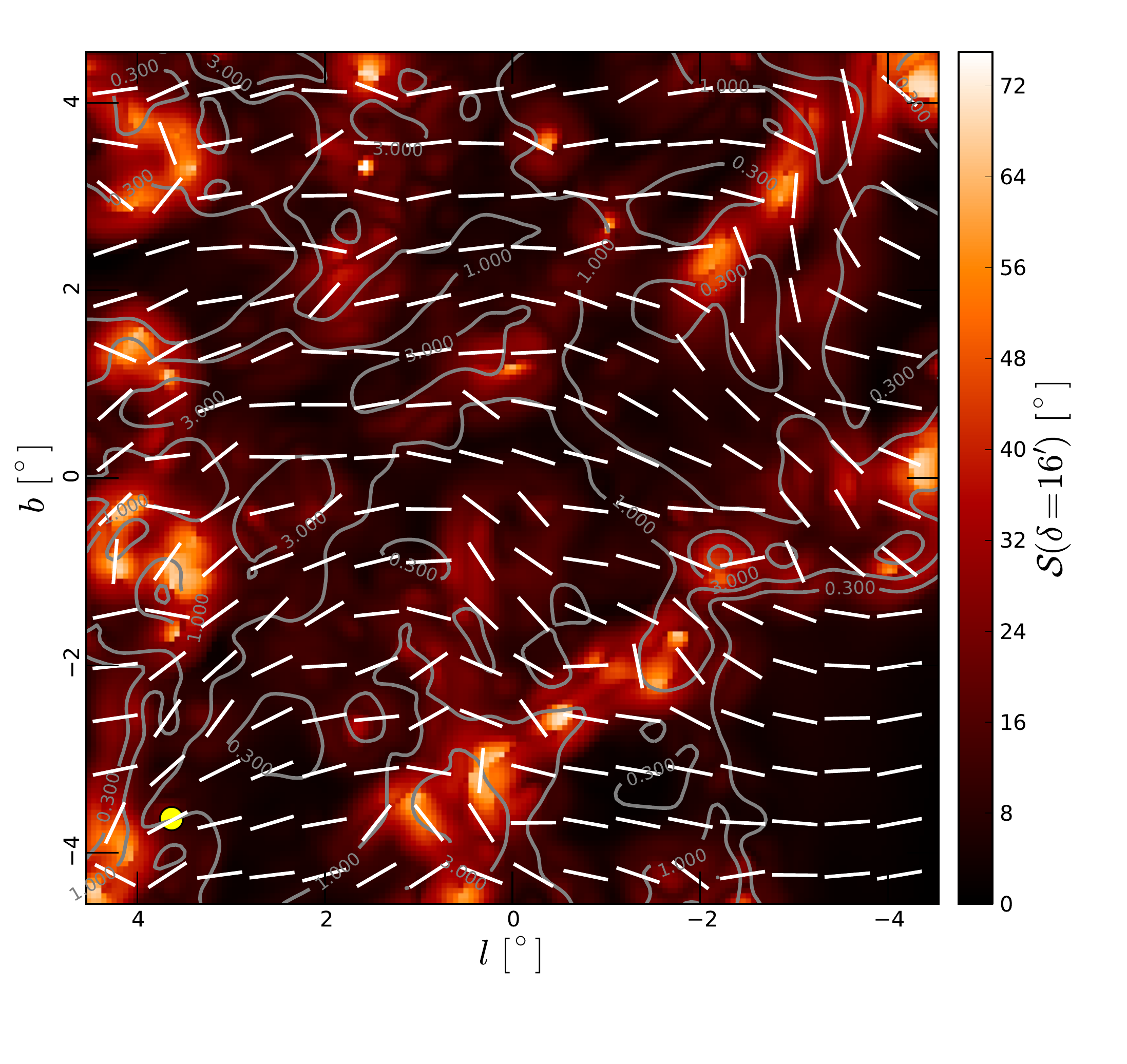}\includegraphics[width=9cm,trim=0 50 0 20,clip=true]{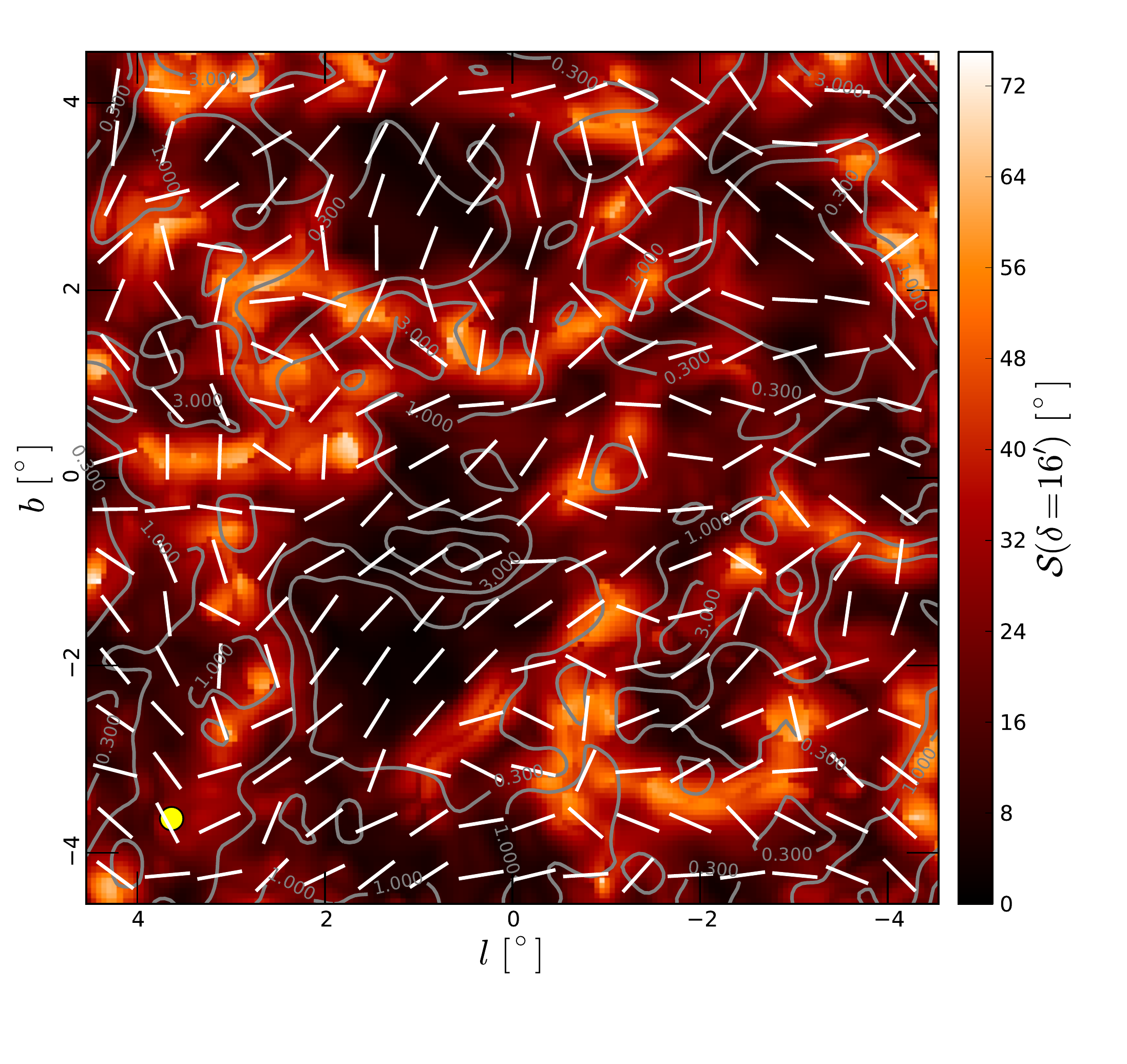}}
\caption{Simulated \Planck~maps. \emph{Top}: Total gas column density. \emph{Middle}: Polarization fraction and angle, with contours being the column density at values indicated in units of $10^{21}~\mathrm{cm}^{-2}$, and the bars indicate magnetic orientation. \emph{Bottom}: \DeltaAngNameMaj~at lag $\delta=16\arcm$, with the same contours and bars as in the middle row. \emph{Left}: viewing angle $\viewangle=0\deg$. \emph{Right}: viewing angle $\viewangle=90\deg$. In each row, the same colour scale is used. In the lower left corner of each plot (yellow circle) is the 15\arcm~FWHM beam. }
\label{fig:PI-B-NH}
\end{figure*}

Figure~\ref{fig:PI-B-NH} (middle row) shows the maps of polarization fraction $p$ and magnetic orientation in these simulated observations, when integrating along the mean magnetic field ($\alpha=90\deg$), and perpendicular to it ($\alpha=0\deg$). The large-scale component of the magnetic field is clearly visible in several regions, for instance in the lower right corner of the $\alpha=0\deg$ case: it leads to long-range coherence in the polarization angle, which correlates with the highest polarization fractions and lowest column densities. Conversely, when integrating along the direction of the large-scale field ($\alpha=90\deg$, right column), $p$ is on average much lower, and no such long-range ordering of $\chi$ is visible, although some local correlations are present. These effects are expected from the vectorial nature of the polarization: with the magnetic field more or less aligned with the line of sight, only its transverse fluctuations lead to a signal in polarization, and these fluctuations are isotropic in the plane of the sky, so they cancel out in the integration (along the line of sight and also through beam dilution). This correlation between $p$ and spatial coherence of the polarization angle is discussed later on (Sect. \ref{sec:PI-vs-H-sims}).

\begin{figure}[htbp]
\centerline{\includegraphics[width=8.8cm,trim=60 0 70 0,clip=true]{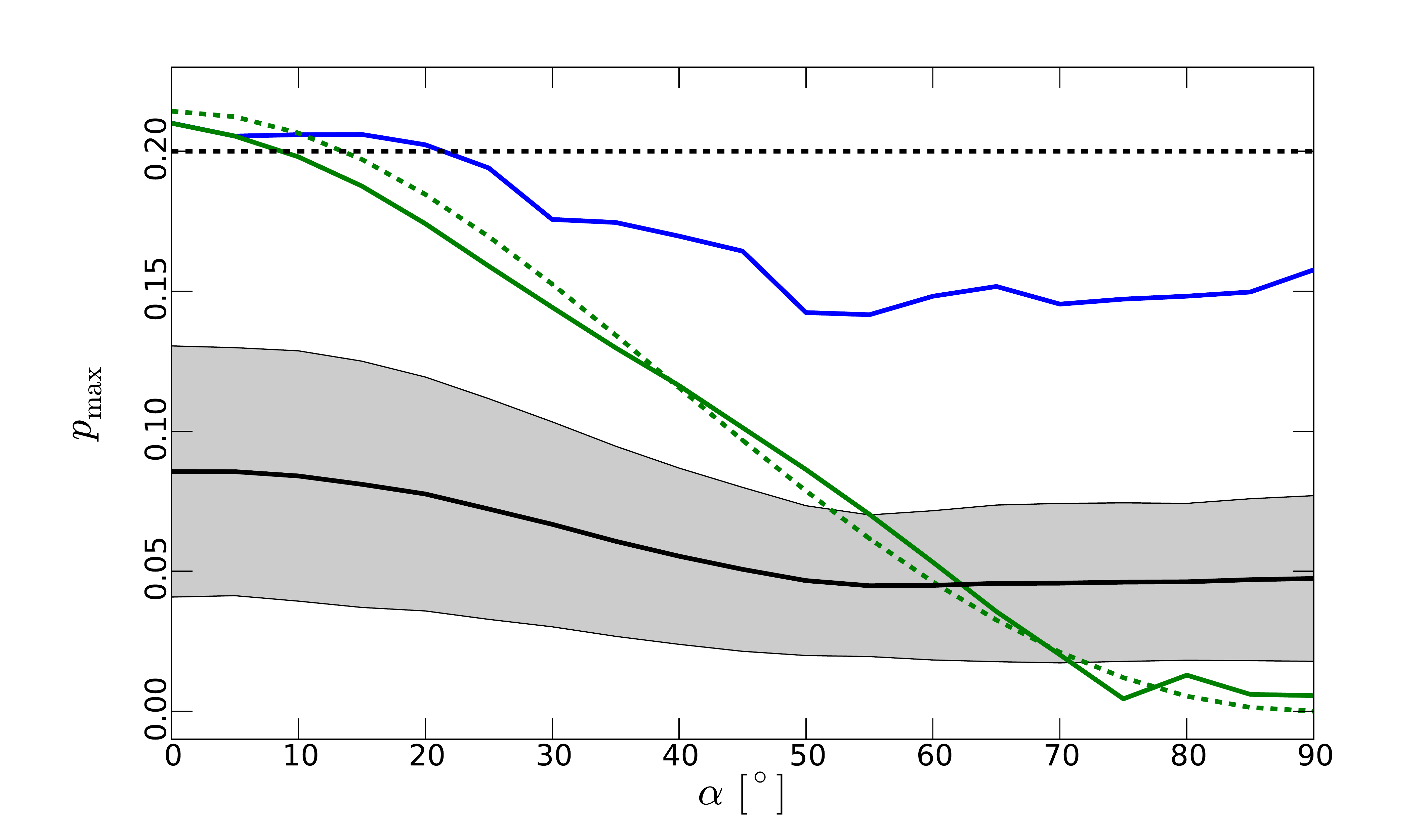}}
\caption{Statistics of polarization fractions in the simulated \Planck~observations as a function of viewing angle $\viewangle$ (see Fig.~\ref{fig:cuberotation}). The solid blue line shows $\pmax$, the solid black line shows the mean $p$, and the solid green line shows the value of $p$ for the most diffuse lines of sight in the map. The dashed black line marks the polarization fraction parameter $p_0$ and the dashed green line gives the theoretical polarization fraction in the case where the density and magnetic field are homogeneous and the latter makes an angle $\viewangle$ with the plane of the sky (see text). The grey region shows the $\pm1\sigma$ spread around the mean $p$.}
\label{fig:pmax-sims}
\end{figure}

Statistics of simulated maps of the polarization fraction (maximum, mean and standard deviation) are shown as a function of the viewing angle $\viewangle$ in Fig.~\ref{fig:pmax-sims}. We find the maximum polarization fraction to be $\pmax\simeq 0.14$--$0.21$ (depending on the viewing angle $\viewangle$). On some lines of sight, in the most tenuous parts of the map integrated perpendicularly to the large-scale $\vec{B}$ (e.g., in the lower right corner of the map in the $\viewangle=0\deg$ case), $\pmax$ almost reaches the theoretical maximum value possible, which is the \IntrinsicpName,
\begin{equation}
\label{eq:pmax}
\polfrac_\mathrm{i}=\frac{p_0}{\displaystyle 1-\frac{p_0}{3}},
\end{equation}
obtained when the medium is homogeneous and the magnetic field is uniform and parallel to the plane of the sky ($\polangsky=0\deg$). Fig.~\ref{fig:pmax-sims} emphasizes the importance of the magnetic field geometry on the measured $\polfrac_\mathrm{max}$, as that value varies by about $40\%$ over the range of viewing angles.

\subsection{Polarization fraction vs. column density}

\begin{figure}[htbp]
\centerline{\includegraphics[width=8.8cm,trim=120 0 60 0,clip=true]{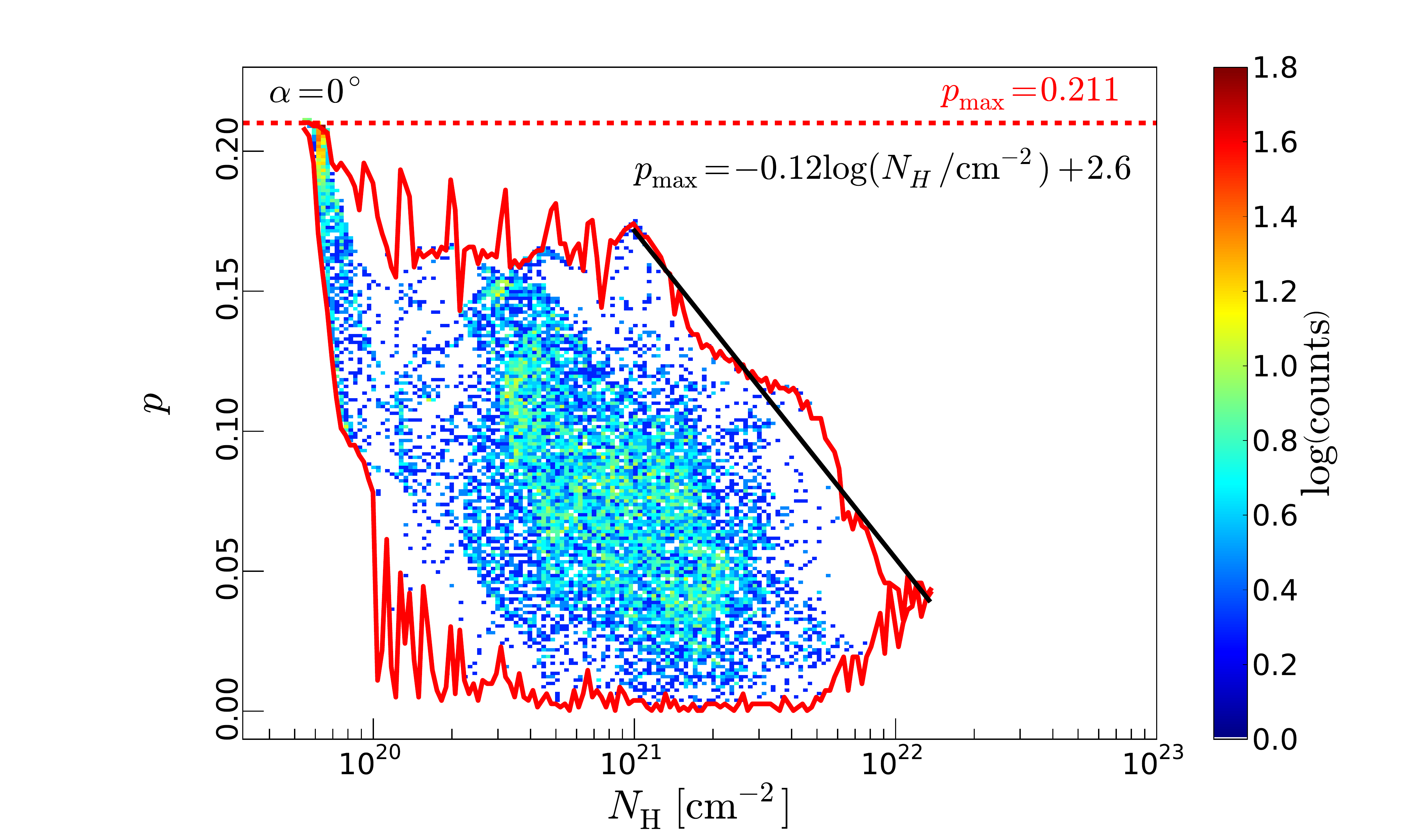}}
\centerline{\includegraphics[width=8.8cm,trim=120 0 60 0,clip=true]{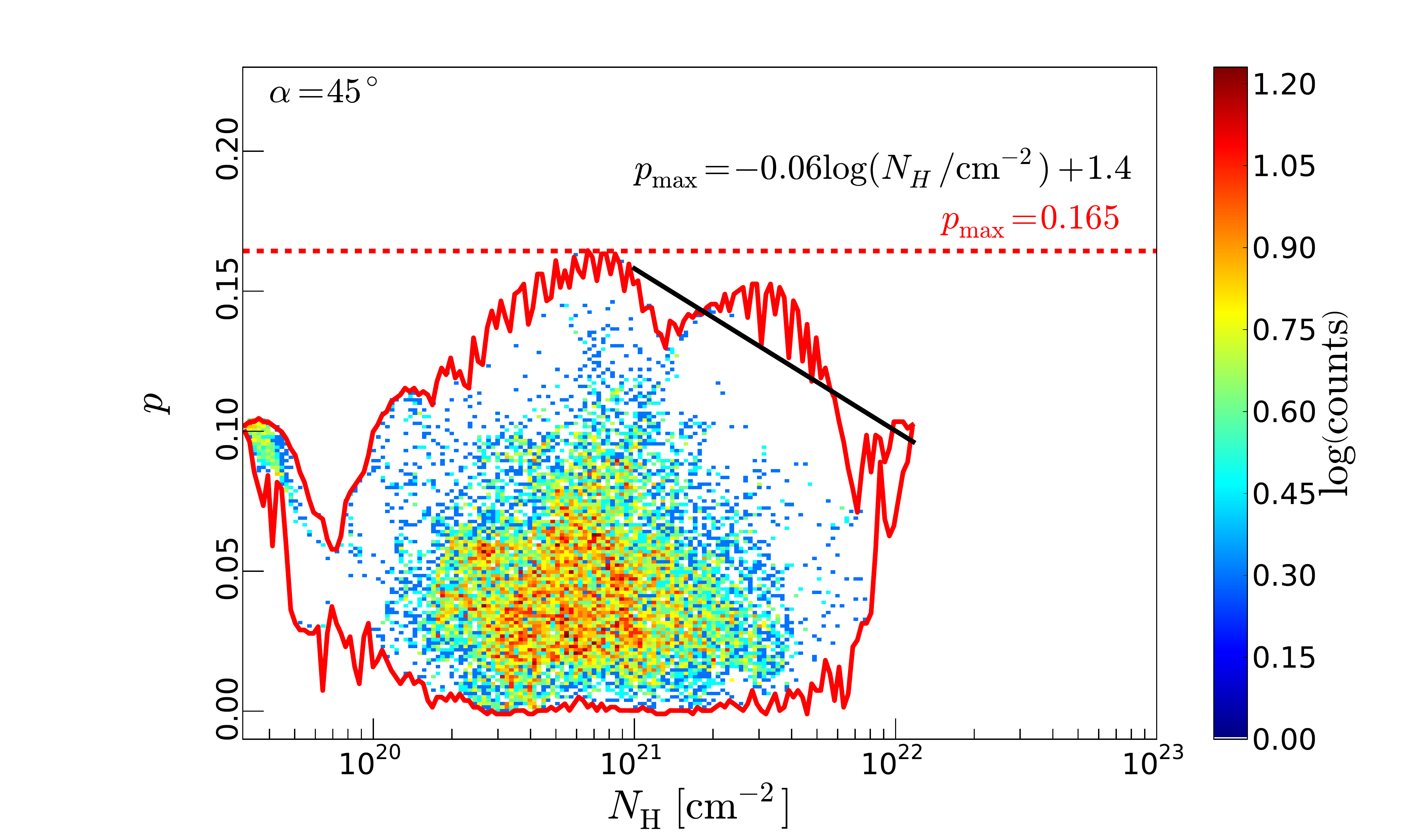}}
\centerline{\includegraphics[width=8.8cm,trim=120 0 60 0,clip=true]{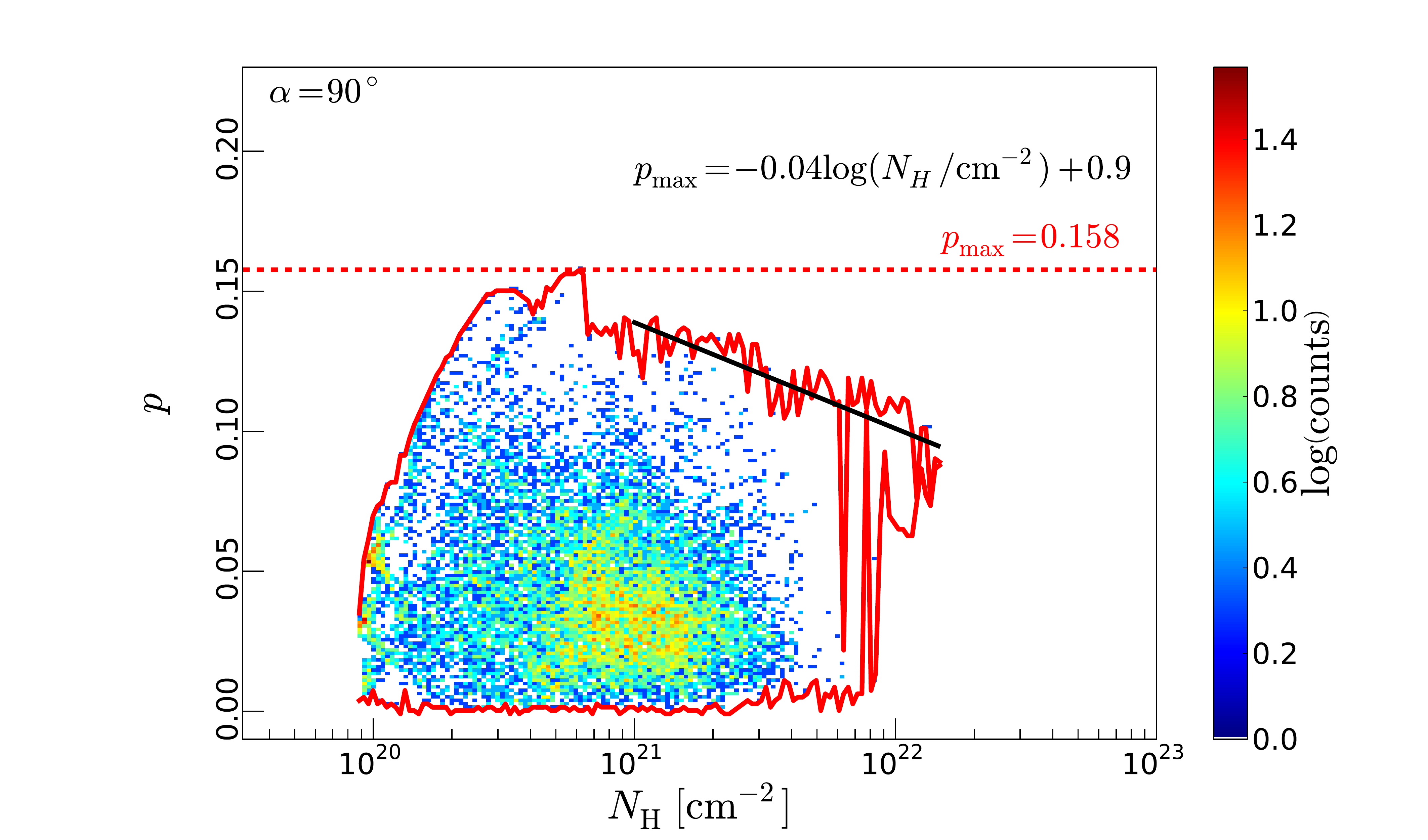}}
\caption{Two-dimensional distribution functions of polarization fractions and logarithmic column densities in the simulated \Planck~observations. \emph{Top}: viewing angle $\viewangle=0\deg$. \emph{Middle}: viewing angle $\viewangle=45\deg$. \emph{Bottom}: viewing angle $\viewangle=90\deg$. The dashed red horizontal lines and the solid red and black lines are the same as in Fig.~\ref{fig:PI-vs-NH}. The fits to the upper envelopes are performed for $N_\mathrm{H}>10^{21}~\mathrm{cm}^{-2}$.}
\label{fig:PI-I-sims}
\end{figure}

We show in Fig.~\ref{fig:PI-I-sims} the joint distribution function of polarization fractions $\polfrac$ and total gas column densities $N_\mathrm{H}$ in the simulated observations when integrating along both directions used in Fig.~\ref{fig:PI-B-NH}, and in the intermediate case $\viewangle=45\deg$. The most striking feature of the plots in Fig.~\ref{fig:PI-I-sims} is the different behaviour at low column densities $N_\mathrm{H}<10^{20}\,\mathrm{cm}^{-2}$. Along these lines of sight, the density is essentially uniform, with $n_\mathrm{H}$ of about $2\,\mathrm{cm}^{-3}$, so the computed polarization is entirely due to magnetic field geometry; when we integrate with $\viewangle=0\deg$ the mean magnetic field is almost in the plane of the sky, $\polangsky\simeq 0\deg$, and polarized emission is at its highest, while when we integrate with $\viewangle=90\deg$, then the ordered field is almost along the line of sight, so $\polangsky\simeq 90\deg$ and no polarized emission appears. In fact, for each value of $\viewangle$, polarization fractions observed towards the most diffuse lines of sight are well reproduced by the formula for a homogeneous medium, easily derived from Eqs.~\ref{eq:simulated_I}--\ref{eq:simulated_U},
\begin{equation}
\polfrac=\frac{p_0\cos^2\viewangle}{\displaystyle 1-p_0\left(\cos^2\viewangle-\frac{2}{3}\right)}
\end{equation}
as can be seen in Fig.~\ref{fig:pmax-sims}. We may therefore only derive the polarization fraction parameter $p_0$ from the maximum observed value $\pmax$ if the angle between the magnetic field and the plane of the sky is known, which is a strong assumption. 

The second striking feature of Fig.~\ref{fig:PI-I-sims} is the decrease of the maximum polarization fraction with increasing column density, as observed in the data. The same linear fit yields slopes $\Delta\pmax/\Delta\log\left(N_\mathrm{H}/\mathrm{cm}^{-2}\right)$ that span values from $-0.025$ (for $\viewangle=80\deg$) to $-0.15$ (for $\viewangle=-15\deg$), the latter being comparable to those found in the data for the selected fields.

\begin{figure}[htbp]
\centerline{\includegraphics[width=8.8cm,trim=120 0 60 0, clip=true]{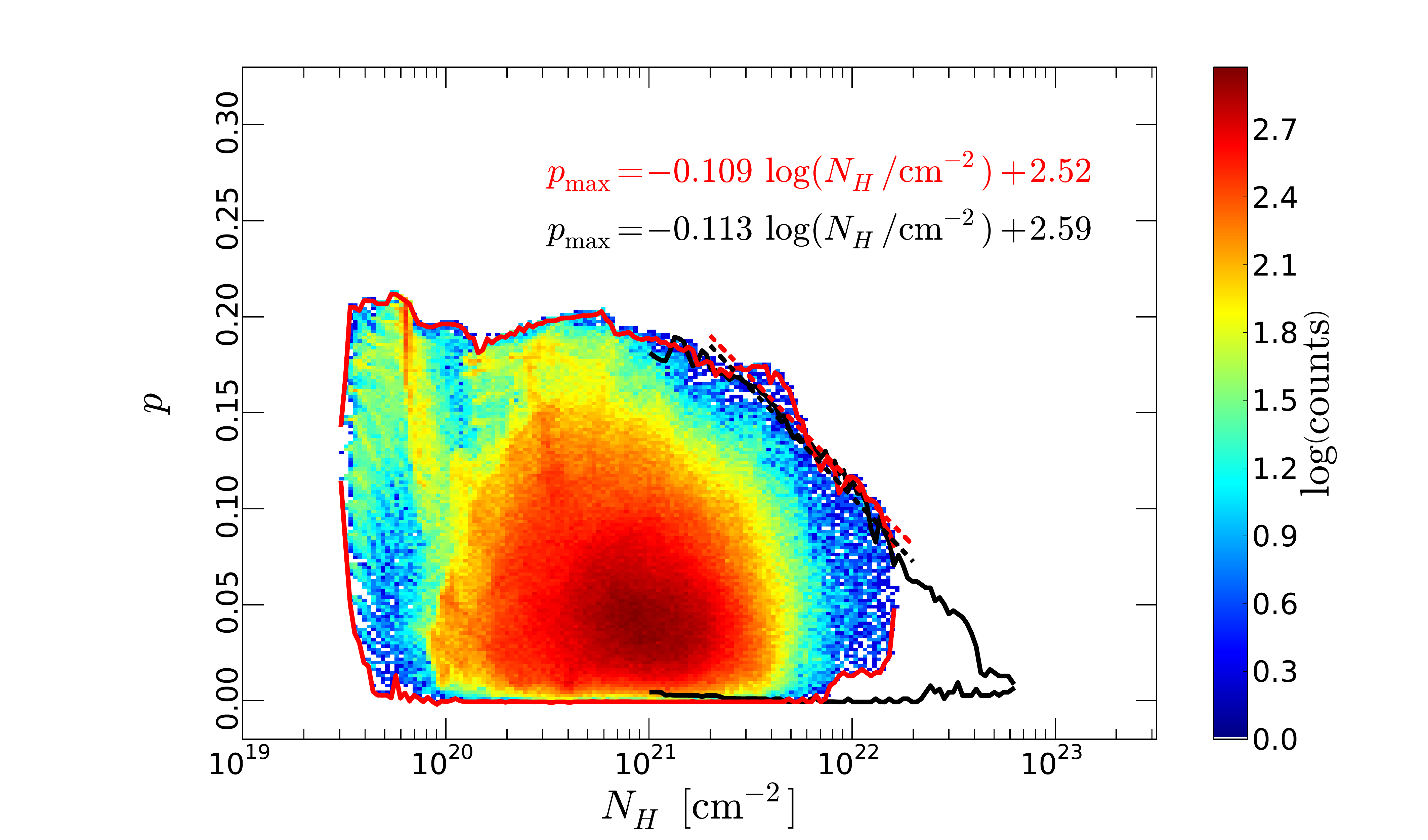}}
\caption{Comparison between the distributions of the polarization fractions $\polfrac$ and logarithmic column densities in the simulations (colour scale, all viewing angles combined, with upper and lower envelopes in solid red lines) and those of the observations in the selected fields (solid black lines). Note that the latter are restricted to $N_\mathrm{H}>10^{21}\,\mathrm{cm}^{-2}$. Dashed lines are linear fits of the form $\polfrac_\mathrm{max}=m\log\left(N_\mathrm{H}/\mathrm{cm}^{-2}\right)+c$ on the distributions' upper envelopes, restricted to a common range of column densities $2\times 10^{21}~\mathrm{cm}^{-2} < N_\mathrm{H} < 2\times 10^{22}~\mathrm{cm}^{-2}$.}
\label{fig:PI-NH-allobs-allsims}
\end{figure}

For a global comparison between simulations and observations, we show in Fig.~\ref{fig:PI-NH-allobs-allsims} the distribution of $\polfrac$ and $N_\mathrm{H}$ for all the simulated fields, with their upper and lower envelopes, together with the envelope for the selected sky fields. Linear fits to the distributions' upper envelopes are performed, restricted to a common range of column densities $2\times 10^{21}~\mathrm{cm}^{-2} < N_\mathrm{H} < 2\times 10^{22}~\mathrm{cm}^{-2}$. They yield similar values in terms of both slopes ($m=-0.109$ for simulations, compared to $m=-0.113$ for the selected fields) and intercepts ($c=2.52$ for simulations, compared to $c=2.59$ for the selected fields). Note that the ``ripple'' pattern in the density plot at low $N_\mathrm{H}$ is due to the sampling in viewing angles $\viewangle$, and is a signature of the decrease of $\polfrac$ with viewing angle for the most diffuse lines of sight, as already noted in Fig.~\ref{fig:pmax-sims}.

\subsection{Polarization angle coherence vs. polarization fraction}
\label{sec:PI-vs-H-sims}

\begin{figure}[htbp]
\centerline{\includegraphics[width=8.8cm,trim=120 0 60 0,clip=true]{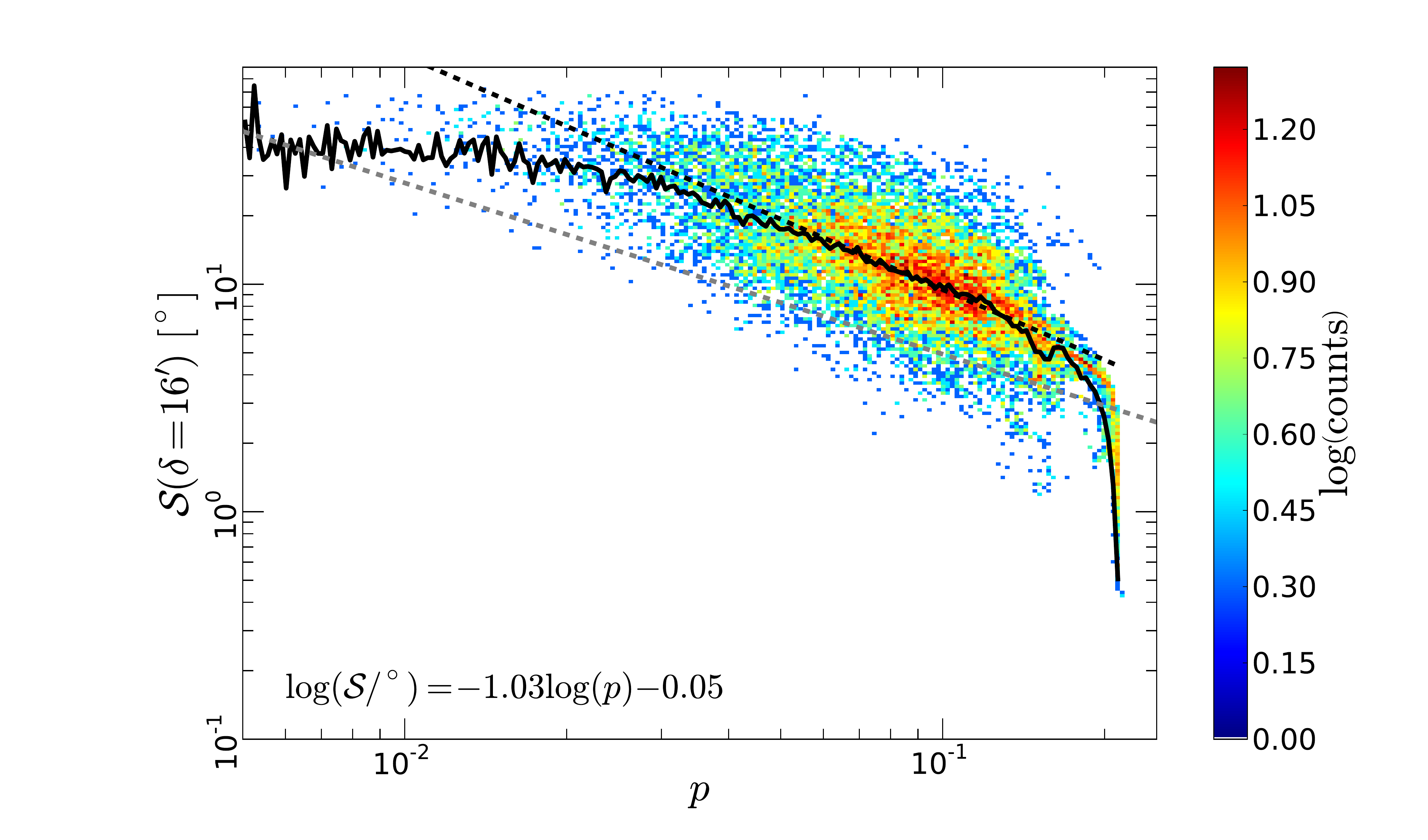}}
\caption{Two-dimensional distribution function of $\log{(\polfrac)}$ and $\log{\left(\DeltaAng\right)}$ in the simulated observations for $\delta=16\arcm$ and $\viewangle=0\deg$. The solid black curve represents the evolution of the mean $\log\left(\DeltaAng\right)$ per bin of $\log(\polfrac)$. A linear fit $\log{\left(\DeltaAng\right)}=m'\log{(\polfrac)}+c'$ is performed, restricted to bins in $\log(\polfrac)$ that contain at least 1\% of the total number of points. This fit is shown as the dashed black line. The dashed grey line is the large-scale fit presented in Sect.~\ref{sec:PI-vs-H}.}
\label{fig:PI-deltapsi-sims}
\end{figure}

\begin{figure}[htbp]
\centerline{\includegraphics[width=8.8cm,trim=50 0 70 0,clip=true]{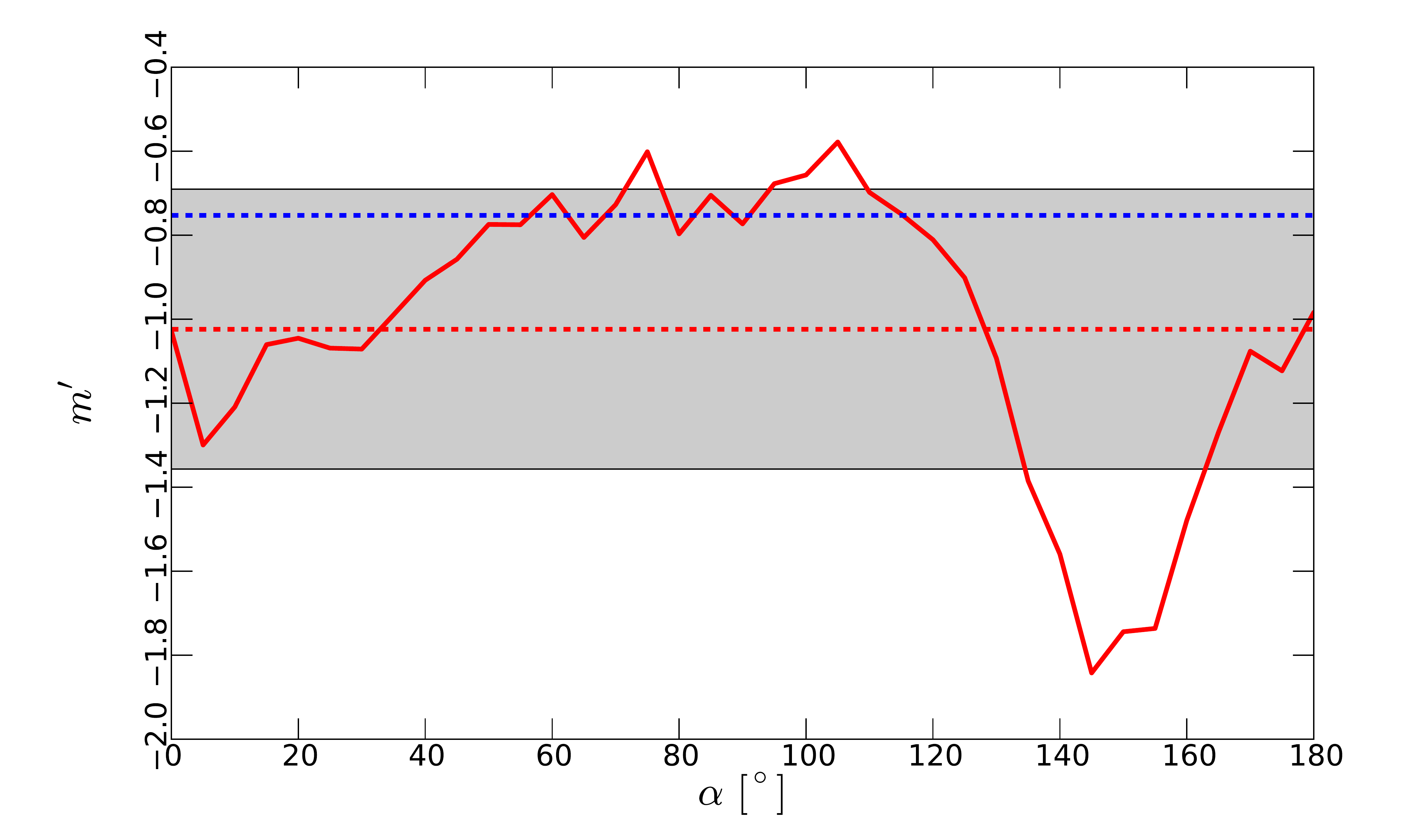}}
\centerline{\includegraphics[width=8.8cm,trim=50 0 70 0,clip=true]{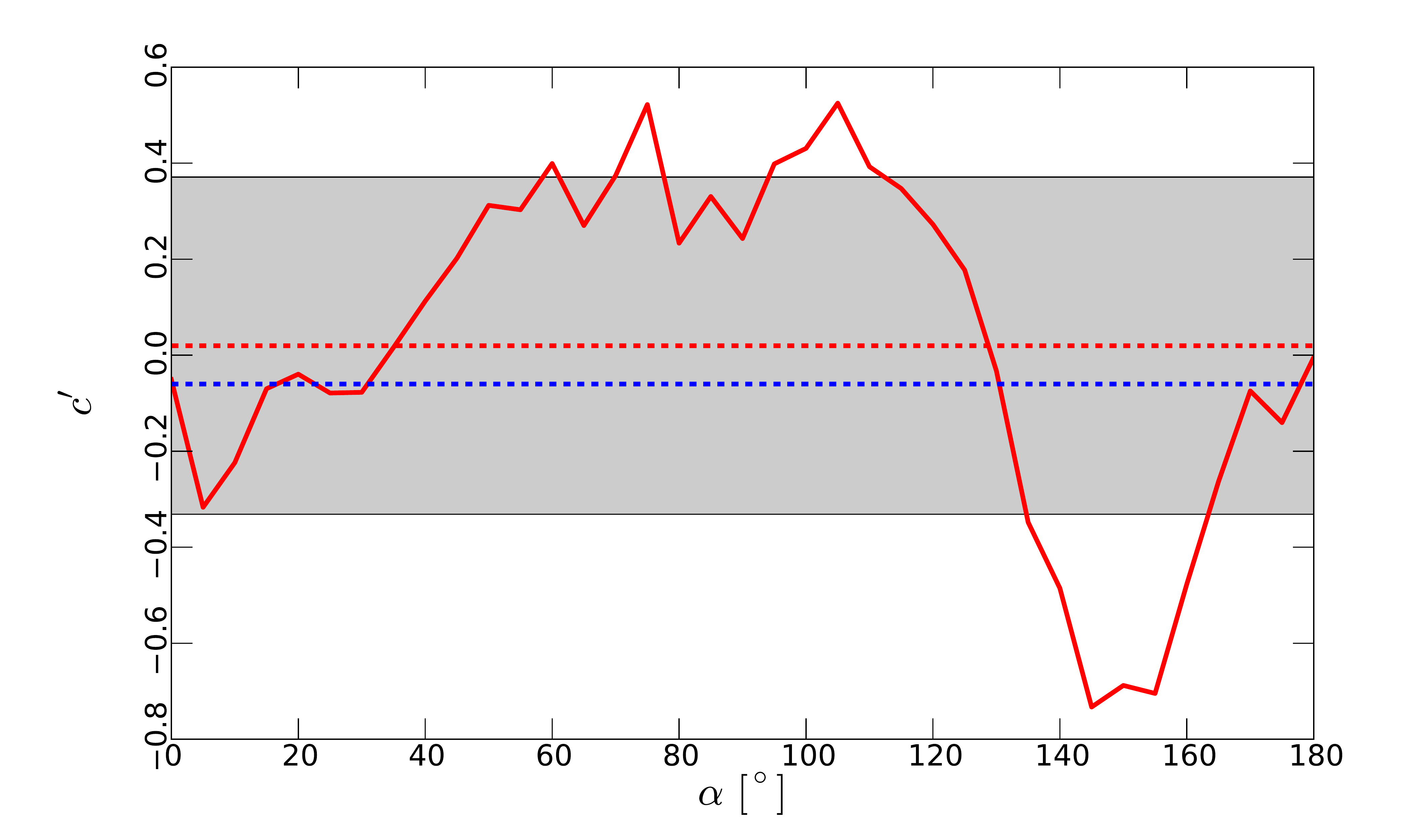}}
\caption{Slopes $m'$ {\it (top)} and intercepts $c'$ {\it (bottom)} of the linear fits $\log\left({\DeltaAng}\right)=m'\log{(\polfrac)}+c'$ to the distribution of $\log{(\polfrac)}$ and $\log\left({\DeltaAng}\right)$ in the simulated observations, as a function of viewing angle $\viewangle$. The lag is $\delta=16\arcm$. The dashed blue lines indicate the values for the large-scale fit presented in Sect.~\ref{sec:PI-vs-H}, the dashed red lines represent the average slope and intercept over the range of $\viewangle$, and the grey areas indicate $\pm 1\sigma$ around the mean, with the standard deviation $\sigma$ computed statistically over all angles.}
\label{fig:slopes-intercepts}
\end{figure}

The \DeltaAngName~$\DeltaAng$ is computed from the simulated $\polang$ maps, using a lag $\delta=16\arcm$, as we did for the data. We first note that the mean \DeltaAngName~is larger when the large-scale magnetic field is oriented along the line of sight, with $\langle\DeltaAng\rangle\simeq12\deg$ for $\viewangle=0\deg$ and $\langle\DeltaAng\rangle\simeq20\deg$ for $\viewangle=90\deg$, a result that is consistent with the findings of~\cite{falceta-goncalves_et_al_08}. Maps of $\DeltaAng$ (for the $\viewangle=0\deg$ and $\viewangle=90\deg$ cases) can be seen in the lower row panels of Fig.~\ref{fig:PI-B-NH}, exhibiting filamentary patterns similar to those found in observations. These filaments of high $\DeltaAng$ also correspond to regions where the polarization angle rotates on small scales, and are correlated with regions of low polarization fraction $\polfrac$ (compare with the middle row panels of Fig.~\ref{fig:PI-B-NH}). This anti-correlation is clearly seen in distribution functions of $\log({\polfrac})$ and $\log{\left(\DeltaAng\right)}$, as shown in Fig.~\ref{fig:PI-deltapsi-sims} for the $\viewangle=0\deg$ case. A linear fit $\log{\left(\DeltaAng\right)}=m'\log({\polfrac})+c'$ to the mean $\log\left({\DeltaAng}\right)$ per bin of $\log(\polfrac)$ is performed, restricted to bins which contain at least 1\% of the total number of points and limited to $\polfrac<p_0$ to avoid the most diffuse lines of sight. The slope and intercept of the anti-correlation observed in the data are fairly well reproduced ($m'=-1.0\pm0.3$ and $c'=0.02\pm0.34$ over the range of $\viewangle$, compared to $m'=-0.75$ and $c'=-0.06$ in observations) with steeper slopes for viewing angles $\viewangle\simeq0\deg$ and shallower slopes for viewing angles $\viewangle\simeq90\deg$ (see Fig.~\ref{fig:slopes-intercepts}). However, since the slopes in simulations are generally steeper than what is observed, but with very similar intercepts at $\polfrac=1$, the \DeltaAngName~$\DeltaAng$ in simulations is globally higher than in observations for a given polarization fraction.

This result suggests that, in the simulations, the \DeltaAngName~is too large for a given polarization fraction, i.e., that the magnetic field is too tangled. Since the physical processes one can think of to reduce the field's tangling (e.g., larger field intensity with respect to turbulence or partial ion-neutral decoupling)  would also affect $\polfrac$, we propose that this difference comes from the lack of power in the low frequency modes of the simulated turbulence, as illustrated by the fact that the power spectra of the velocity and magnetic field components flatten out at small wavenumber $k$. In reality, molecular clouds are organized in a self-similar structure over a broad range of scales and that is therefore not properly reproduced in the simulations we used. In short, the large-scale fluctuations of the magnetic field are closer to random in simulations than in reality.

\subsection{Statistics on the magnetic field fluctuations in the simulations}
\label{sec:geometry}

\begin{figure}[htbp]
\centerline{\includegraphics[width=8.8cm,trim=120 0 60 0,clip=true]{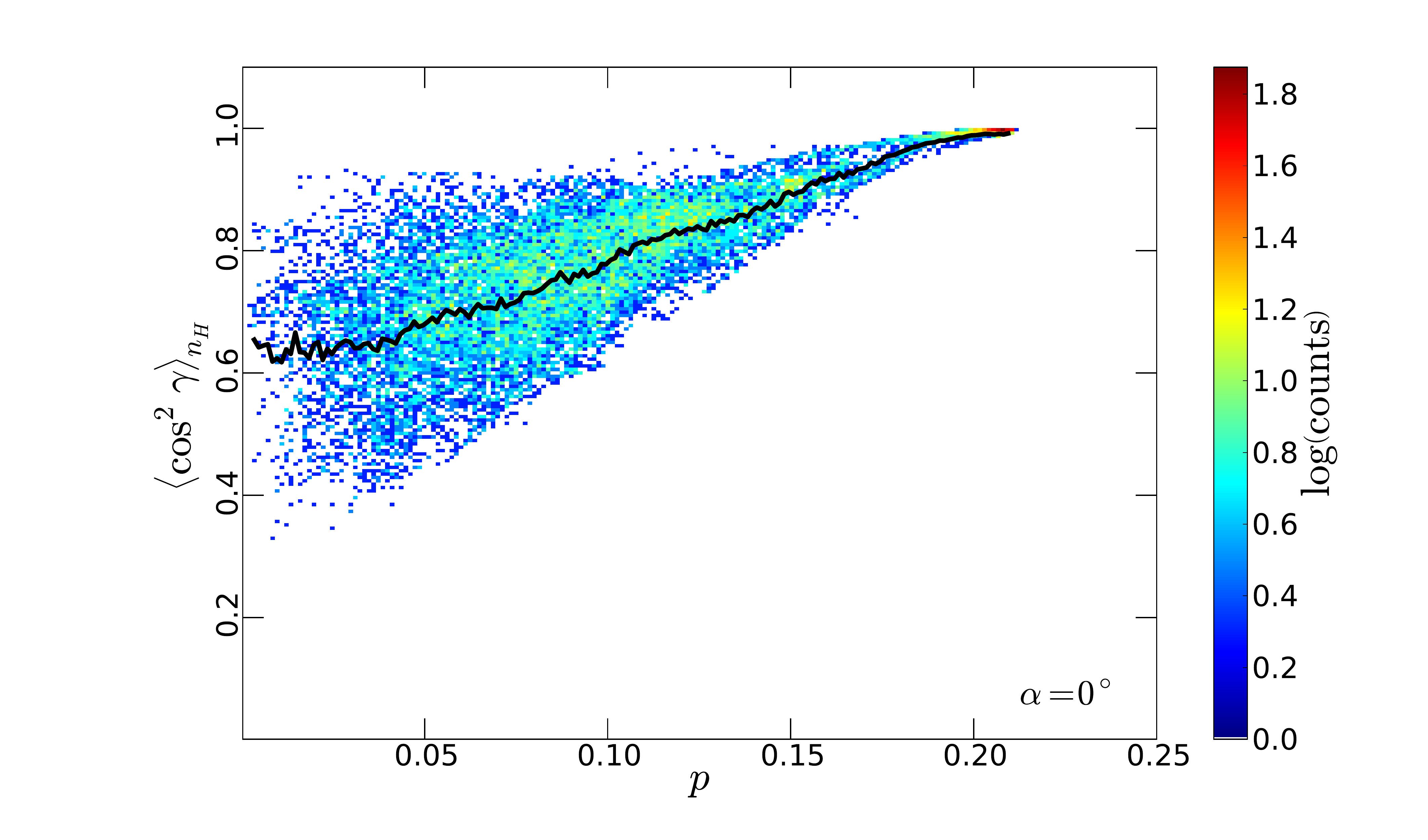}}
\centerline{\includegraphics[width=8.8cm,trim=120 0 60 0,clip=true]{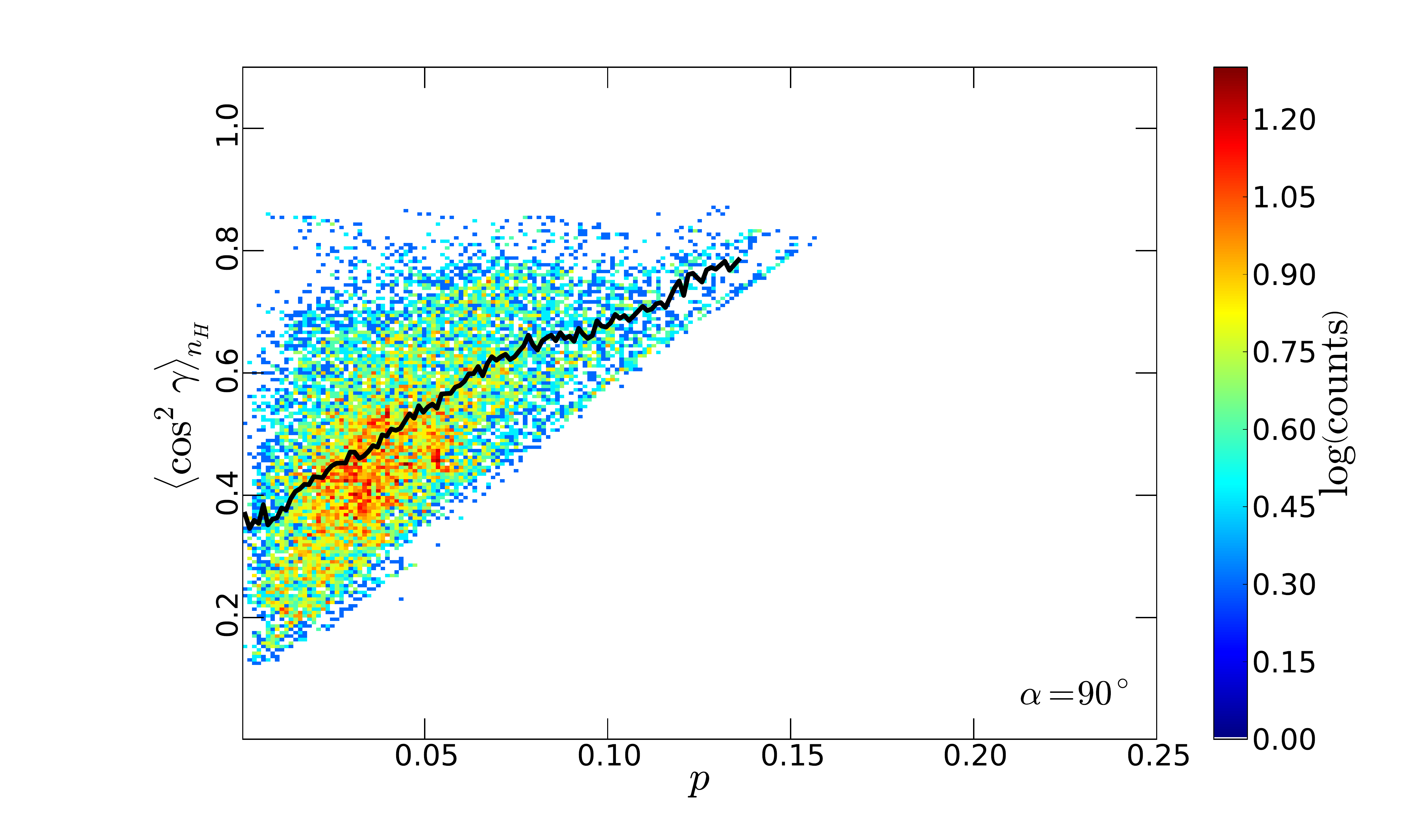}}
\caption{Distribution functions of polarization fraction $\polfrac$ and density-weighted mean of $\cos^2\polangsky$ ($\polangsky$ is the angle of the magnetic field with respect to the plane of the sky, see Fig.~\ref{fig:anglesdefinition}) along the line of sight $z'$ in the simulation cube. \emph{Top}: viewing angle $\viewangle=0\deg$. \emph{Bottom}: viewing angle $\viewangle=90\deg$. The solid black lines show the mean values per bin of $\polfrac$.}
\label{fig:geometry-AB}
\end{figure}

\begin{figure}[htbp]
\centerline{\includegraphics[width=8.8cm,trim=120 0 60 0,clip=true]{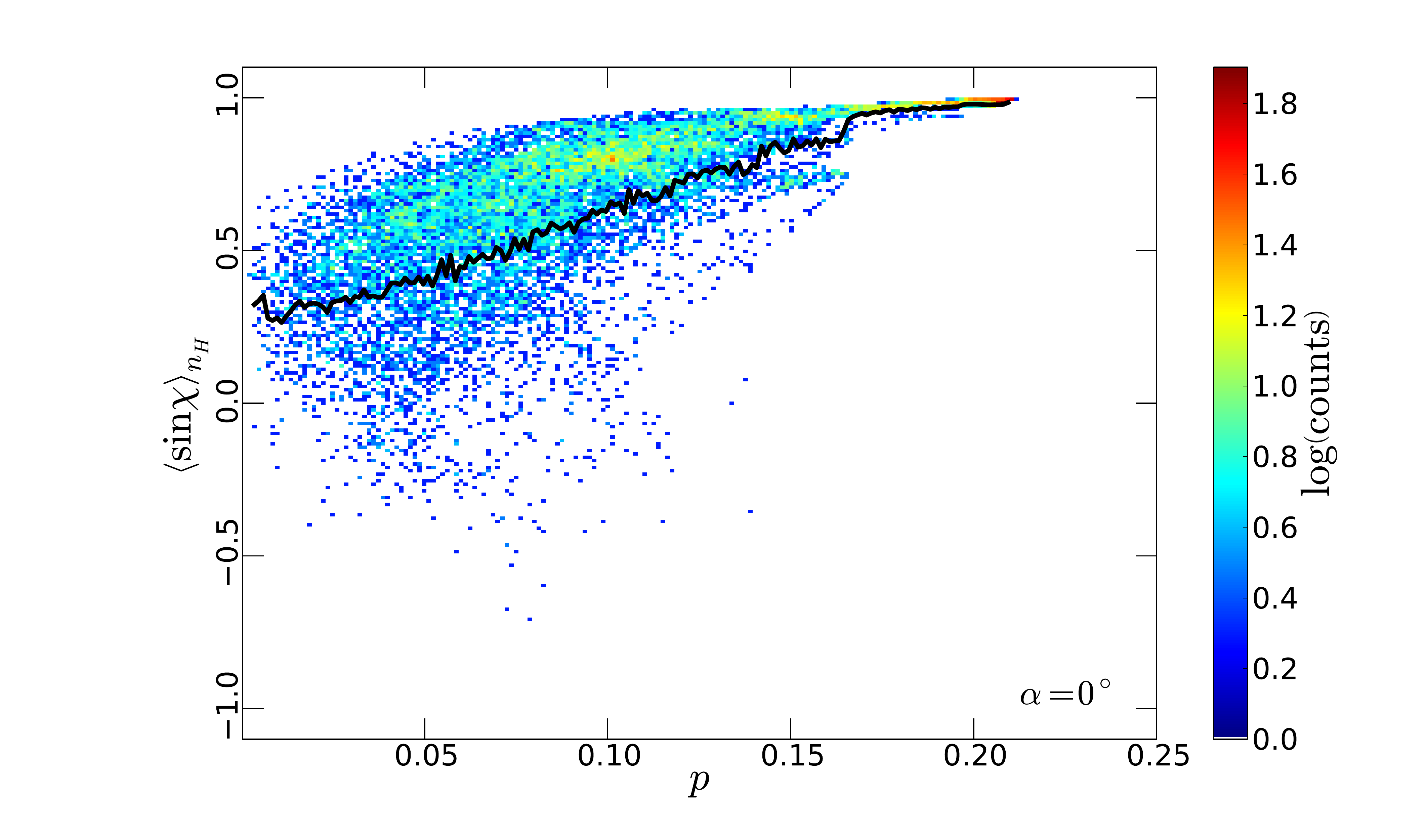}}
\centerline{\includegraphics[width=8.8cm,trim=120 0 60 0,clip=true]{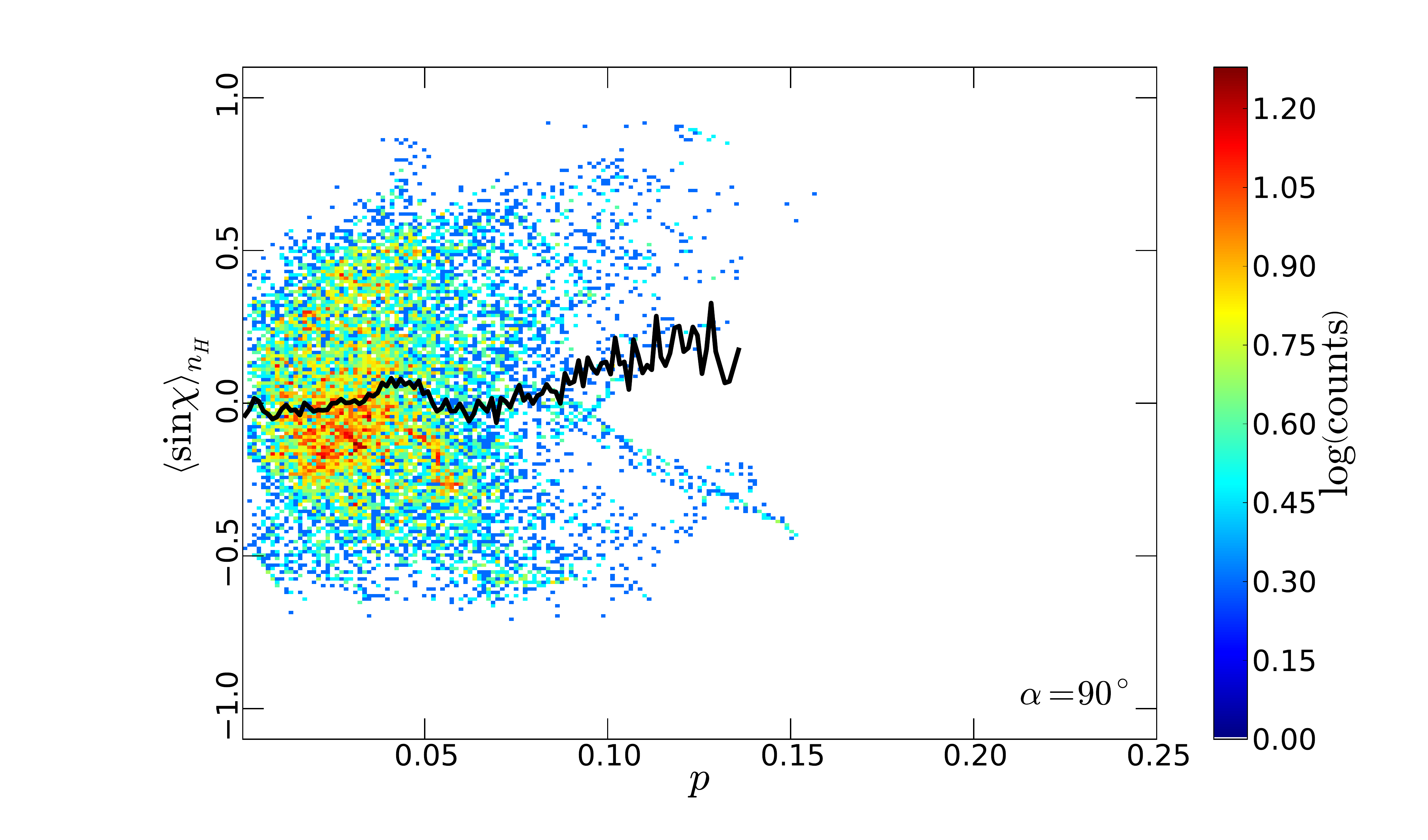}}
\caption{Distribution functions of polarization fraction $\polfrac$ and density-weighted mean of $\sin\chi$ ($\chi$ is the position angle of the projection of the magnetic field in the plane of the sky, see Fig.~\ref{fig:anglesdefinition}) along the line of sight $z'$ in the simulation cube. \emph{Top}: viewing angle $\viewangle=0\deg$. \emph{Bottom}: viewing angle $\viewangle=90\deg$. The solid black lines show the mean values per bin of $\polfrac$.}
\label{fig:geometry-ABprime}
\end{figure}

\begin{figure}[htbp]
\centerline{\includegraphics[width=8.8cm,trim=120 0 60 0,clip=true]{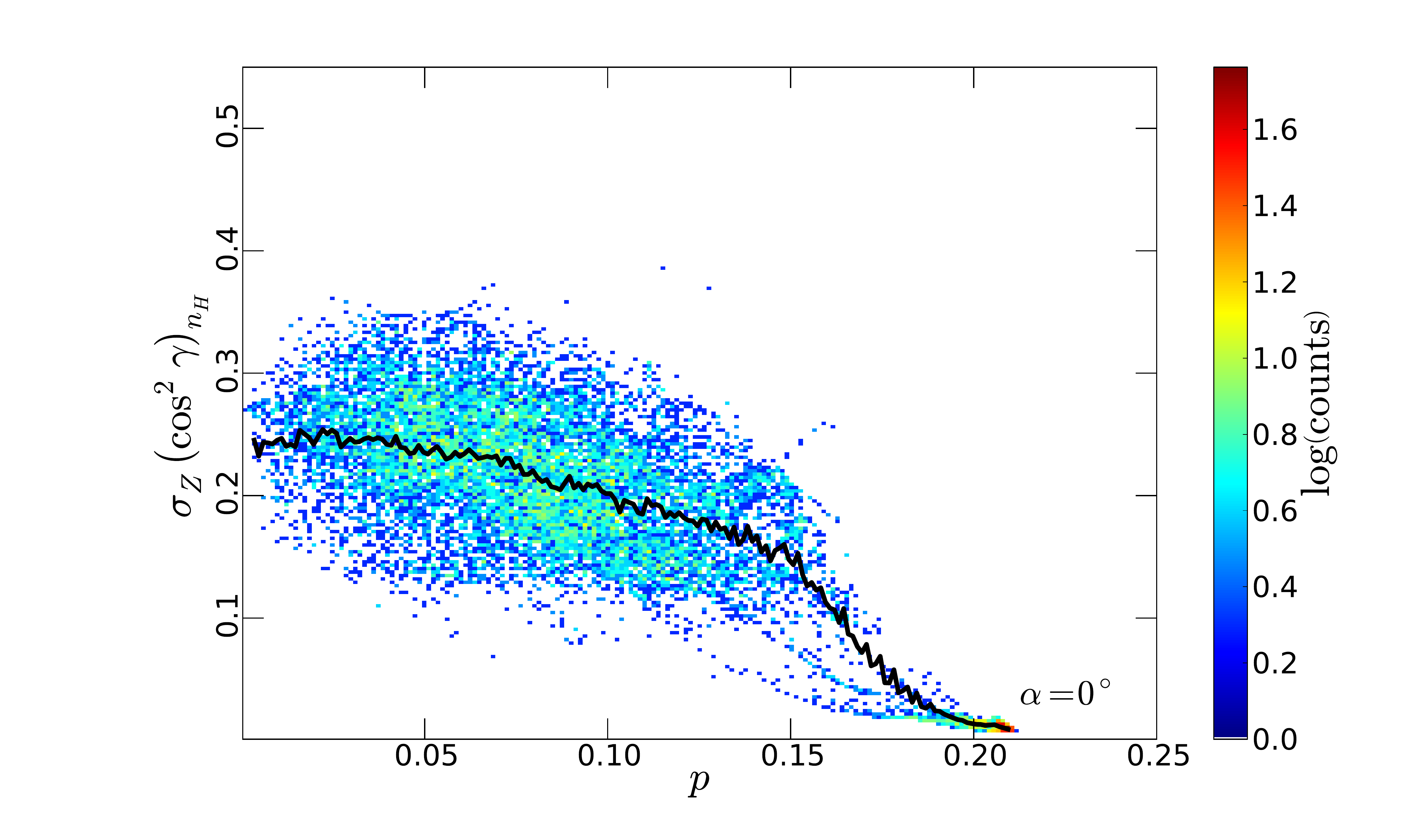}}
\centerline{\includegraphics[width=8.8cm,trim=120 0 60 0,clip=true]{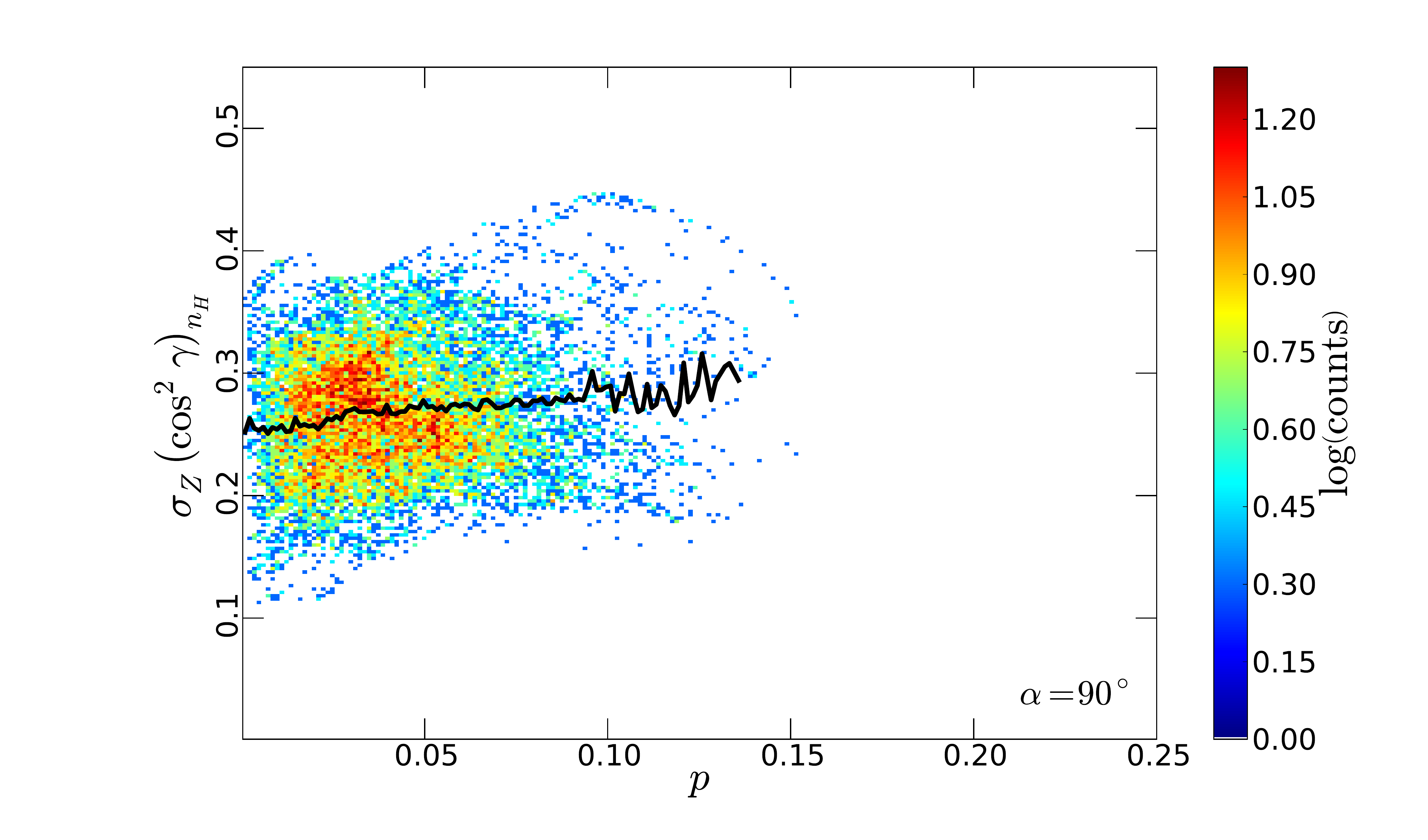}}
\caption{Distribution functions of polarization fraction $\polfrac$ and density-weighted standard deviation of $\cos^2\polangsky$ along the line of sight $z'$ in the simulation cube. \emph{Top}: viewing angle $\viewangle=0\deg$. \emph{Bottom}: viewing angle $\viewangle=90\deg$. The solid black lines show the mean values per bin of $\polfrac$.}
\label{fig:geometry-A1}
\end{figure}

\begin{figure}[htbp]
\centerline{\includegraphics[width=8.8cm,trim=120 0 60 0,clip=true]{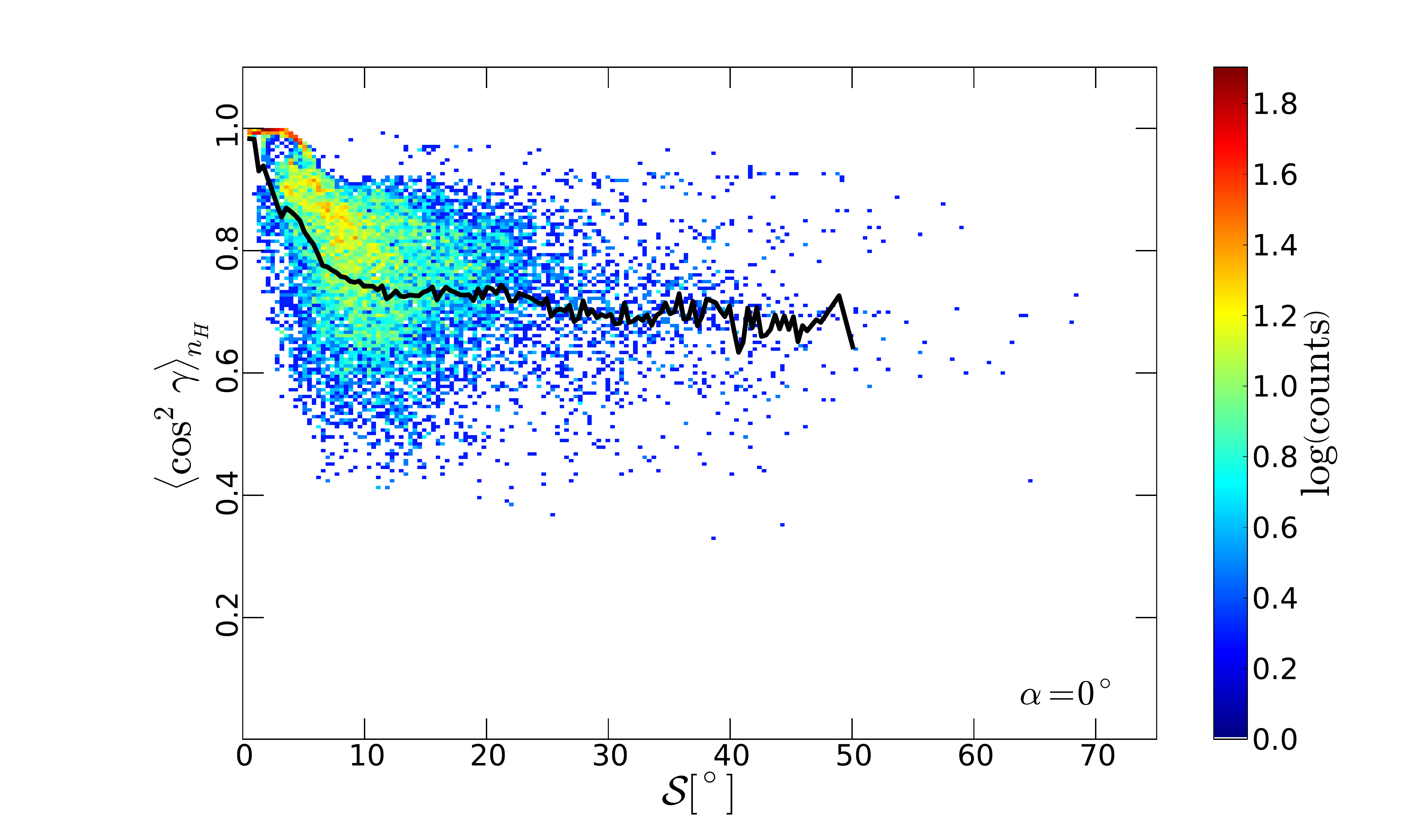}}
\centerline{\includegraphics[width=8.8cm,trim=120 0 60 0,clip=true]{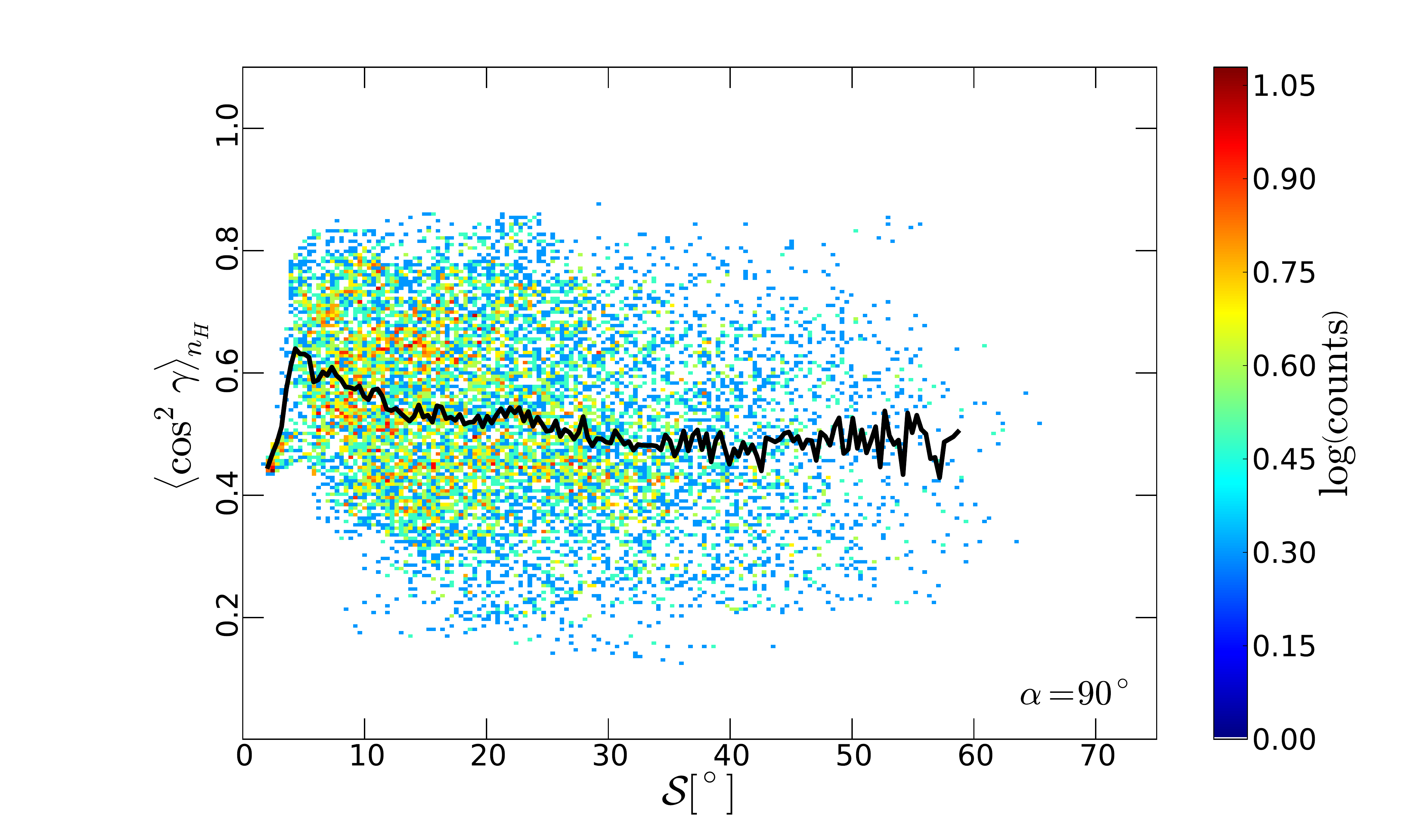}}
\caption{Distribution functions of \DeltaAngName~$\DeltaAng\left(\delta=16\arcm\right)$ and density-weighted mean of $\cos^2\polangsky$ in the simulation cube. \emph{Top}: viewing angle $\viewangle=0\deg$. \emph{Bottom}: viewing angle $\viewangle=90\deg$. The solid black lines show the mean values per bin of $\DeltaAng$.}
\label{fig:geometry-Aprime}
\end{figure}

We investigate here the possible causes of the variations in the polarization
fraction $\polfrac$ and the dispersion of the polarization angle $\DeltaAng$
in the simulations, i.e., what are the respective roles of the field
tangling and the orientation of the large-scale field in the
variations of $\polfrac$ and $\DeltaAng$. To quantify these roles, we
compute the average and dispersion along the line of sight of both
$\cos^2\polangsky$ and $\sin\chi$ (see Fig.~\ref{fig:anglesdefinition} for the definition of angles). These quantities are computed 
for different viewing angles. In the following, we write the magnetic field as $\boldsymbol{B}=\boldsymbol{B}_0+\Delta\boldsymbol{B}$, where $\boldsymbol{B}_0$ is the large-scale ordered field and $\Delta\boldsymbol{B}$ is the fluctuating part of $\boldsymbol{B}$.

The role of the average values of the angles $\polangsky$ and $\chi$ along the line of sight is
illustrated in Figs.~\ref{fig:geometry-AB} and~\ref{fig:geometry-ABprime}.  First, the role of the
large-scale field $\boldsymbol{B}_0$ is clear: the largest values of $\polfrac$ are obtained
when $\boldsymbol{B}_0$ is viewed in the plane of the sky ($\alpha\simeq0\deg$). The largest $\polfrac$
values are obtained when the average of $\cos^2\polangsky$ along the line of sight stays close to unity.  In that
case, the field perturbations are such that they keep the field close to the plane of the sky, on average, hence the large $\polfrac$. The
same effect is visible in the top panel of Fig.~\ref{fig:geometry-ABprime} where the largest
polarization fractions are obtained for average values of $\chi$
close to $90^\circ$.

However, even in this configuration ($\alpha=0\deg$), small values of $\polfrac$
are obtained.  The fraction of low $\polfrac$ values is clearly larger when the
large-scale field is viewed along the line of sight ($\alpha=90\deg$). The
remarkable feature visible in Fig.~\ref{fig:geometry-AB} (bottom panel) is the proportionality of
$\polfrac_\mathrm{max}$ with the average of $\cos^2\polangsky$: the smaller this average,
the closer $\polangsky$ is to $90^\circ$, therefore the closer the field is
aligned with the line of sight, and the smaller the resulting value of
$\polfrac_\mathrm{max}$. One also sees in Fig.~\ref{fig:geometry-AB} that $\left<\cos^2\polangsky\right>$ reaches much
smaller values when $\boldsymbol{B}_0$ is along the line of sight (bottom panel), producing lower values of $\polfrac$ than in the
case where $\boldsymbol{B}_0$ is in the plane of the sky (top panel).

We note, interestingly, that the same effect is not visible in Fig.~\ref{fig:geometry-ABprime}, which displays the 
line of sight average of $\sin\chi$ versus $\polfrac$: there is no such upper value of $\polfrac$ that would 
scale with the average of $\sin\chi$ because this fluctuation 
of the field direction is measured in the plane of the sky and does not affect the maximal 
polarization fraction that can be obtained. Instead, when $\boldsymbol{B}_0$ is along the line of sight for instance, the scatter of $\langle\sin\chi\rangle$ along the 
line of sight is the largest and the resulting values of $\polfrac$ are low.
  
Figure~\ref{fig:geometry-A1} also illustrates the effect of the field tangling:
the larger the dispersion of $\cos^2\polangsky$ along the line of sight (and the
larger the scatter of this dispersion), the smaller $\polfrac$ is. Obviously,
when the line of sight is dominated by the large-scale field, the scatter is the
lowest.
  
Figure~\ref{fig:geometry-Aprime} shows the joint distribution of the average of $\cos^2\polangsky$ and
$\DeltaAng$, where one recognizes the role of the large-scale field when $\alpha=0\deg$:
the lowest values of $\DeltaAng$ are obtained when $\polangsky$ stays close to 0\deg, meaning that the field 
is more or less in the plane of the sky. 
Clearly, the largest values of $\DeltaAng$ are obtained when the influence of the large-scale field is minimized ($\alpha=90\deg$, bottom panel).


\section{Conclusions}
\label{sec:conclusions}

To summarize, the maximum polarization fraction $\polfrac_\mathrm{max}$ observed towards the sample of nearby fields selected in this study is reached in the most diffuse fields. The large-scale decrease of $\polfrac_\mathrm{max}$ with increasing $N_\mathrm{H}$ is seen in the individual fields considered here, as soon as $N_\mathrm{H}>10^{21}\,\mathrm{cm}^{-2}$. This trend is fairly well reproduced by numerical simulations of anisotropic MHD turbulence, even assuming uniform dust temperatures and grain alignment efficiencies in the gas weakly shielded from the UV radiation. The polarization of thermal dust emission observed by \Planck~towards these regions is essentially related to the geometry of the magnetic field and in particular to its orientation at large scales with respect to the line of sight. We do not discuss the evolution of polarization fractions at large column densities $N_\mathrm{H}>3\times 10^{22}\,\mathrm{cm}^{-2}$, for which the MHD simulation considered is not suitable. It is clear, however, that additional processes must be at work to achieve the change of slope in the $\polfrac_\mathrm{max}$ vs. $\log{\left(N_\mathrm{H}/\mathrm{cm^{-2}}\right)}$ relation observed towards the most opaque lines of sight. This change is probably related to variations in the properties of dust alignment, as pointed out by \cite{soler_et_al_13}. We also find that polarization fractions observed by \planck~towards these nearby regions correlate well with the local coherence of the polarization angle, which is measured using the \DeltaAngName~$\DeltaAng$. This correlation is also found in simulations, with slopes that are very close to observational values. In simulations, however, values of $\DeltaAng$ for a given polarization fraction are globally too high compared to observations, which points to a possible limitation of the specific MHD simulation used. 

\begin{acknowledgements}
 The development of \Planck\ has been supported by: ESA; CNES and CNRS/INSU-IN2P3-INP (France); 
  ASI, CNR, and INAF (Italy); NASA and DoE (USA); STFC and UKSA (UK); 
  CSIC, MICINN, JA and RES (Spain); Tekes, AoF and CSC (Finland); 
  DLR and MPG (Germany); CSA (Canada); DTU Space (Denmark); SER/SSO (Switzerland); 
  RCN (Norway); SFI (Ireland); FCT/MCTES (Portugal); and PRACE (EU). 
  A description of the \Planck\ Collaboration and a list of its members, 
  including the technical or scientific activities in which they have been involved, 
  can be found at \url{http://www.sciops.esa.int/index.php?project=planck&page=Planck_Collaboration}. Some of the results in this paper have been derived using the \healpix~package. The authors would like to thank Charles Beichman for his careful reading of the manuscript and useful comments. The research leading to these results has received funding from the European Research Council under the European Union's Seventh Framework Programme (FP7/2007-2013) / ERC grant agreement n$^\circ$ 267934.
\end{acknowledgements}
\bibliographystyle{aa}

\appendix
\section{Extra figures}
\label{extra-figures}

\begin{figure}[htbp]
\centerline{\includegraphics[width=9cm,trim=0 50 0 0,clip=true]{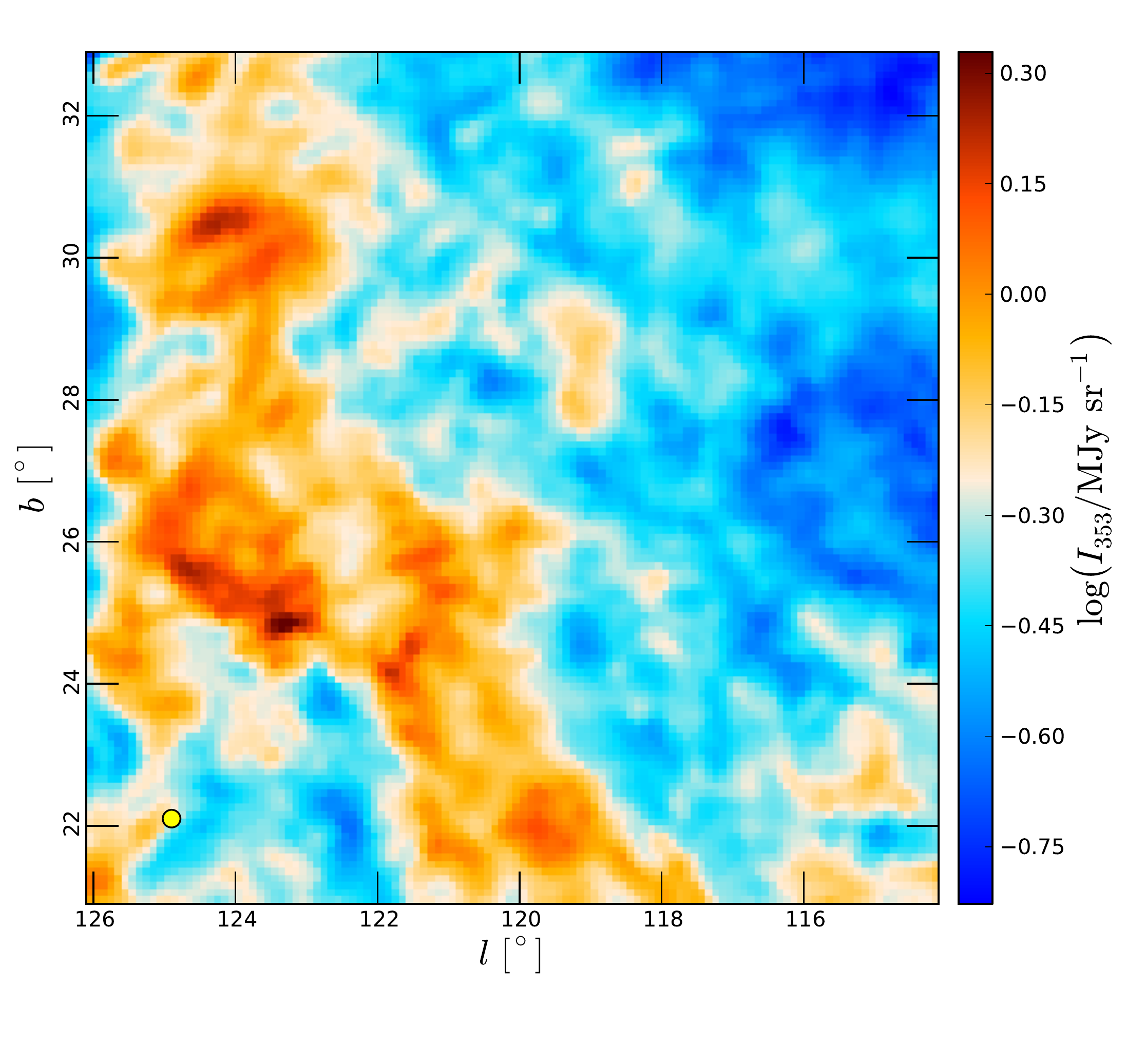}}
\centerline{\includegraphics[width=9cm,trim=0 50 0 20,clip=true]{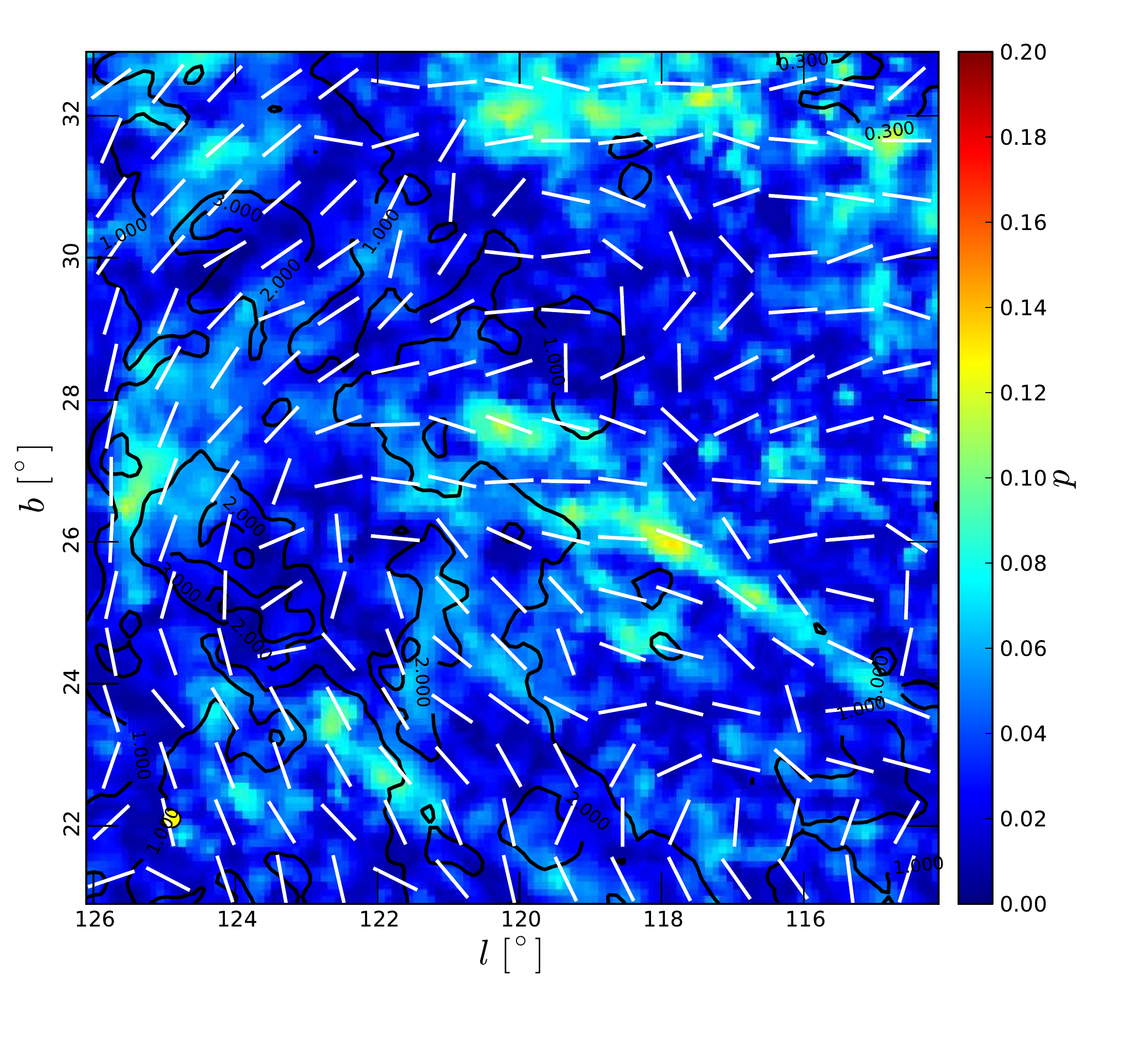}}
\centerline{\includegraphics[width=9cm,trim=0 50 0 20,clip=true]{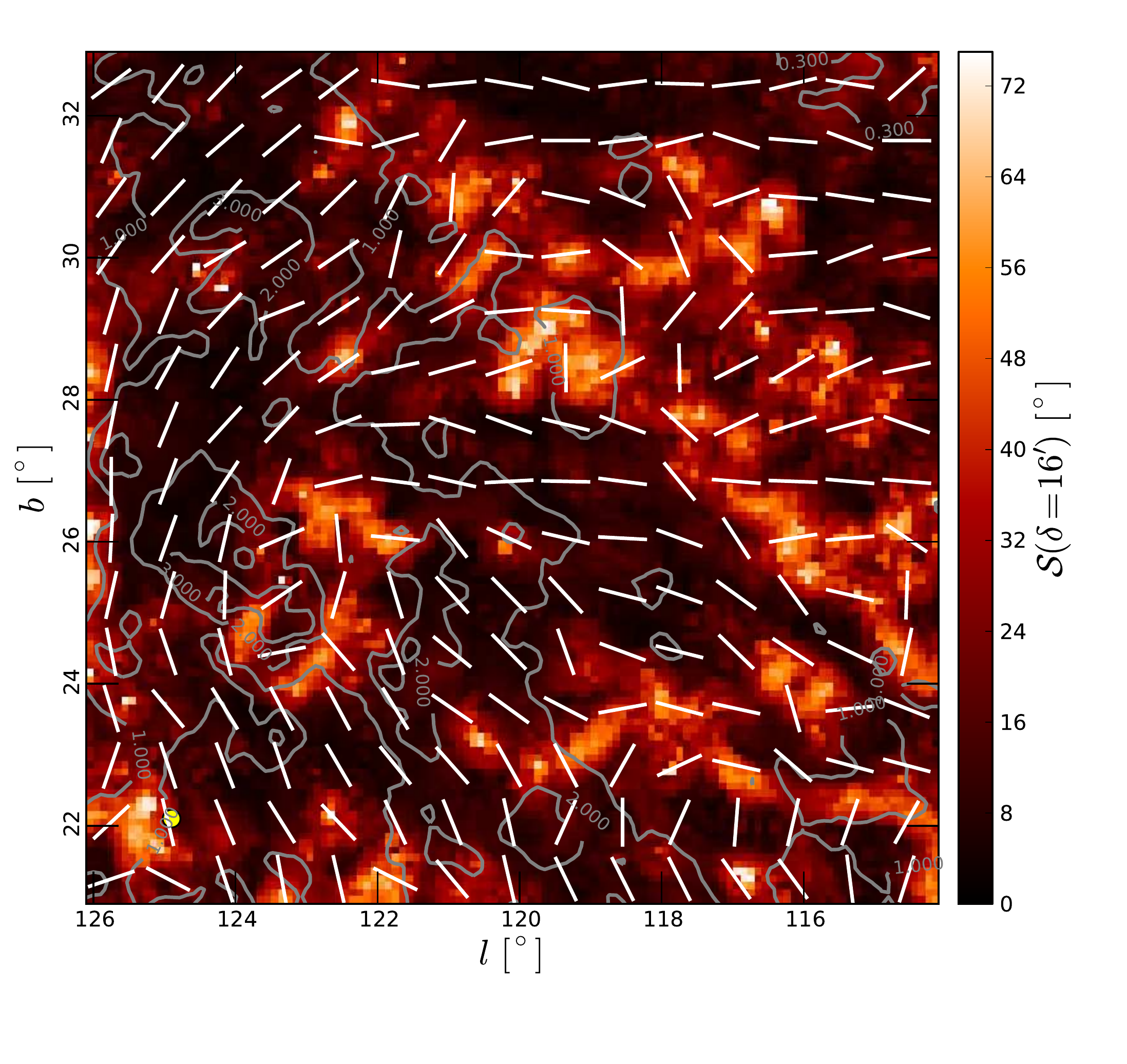}}
\caption{Same as Fig.~\ref{fig:PI-B-NH_Ophiuchus-Chamaeleon}, but for the Polaris Flare field. \emph{Top}: Total intensity at 353\,GHz. \emph{Middle}: Polarization fraction $\polfrac$, column density $N_\mathrm{H}$ (contours in units of $10^{21}\,\mathrm{cm}^{-2}$), and magnetic orientation (bars). \emph{Bottom}: \DeltaAngNameMaj~$\DeltaAng$ with lag $\delta=16\arcm$ (see Sect.~\ref{sec:PI-vs-H}) with contours and bars identical to the middle row. Note that contours values are different from those of Fig.~\ref{fig:PI-B-NH_Ophiuchus-Chamaeleon}.}
\label{fig:PI-B-NH_Polaris}
\end{figure}

\begin{figure}[htbp]
\centerline{\includegraphics[width=9cm,trim=0 50 0 0,clip=true]{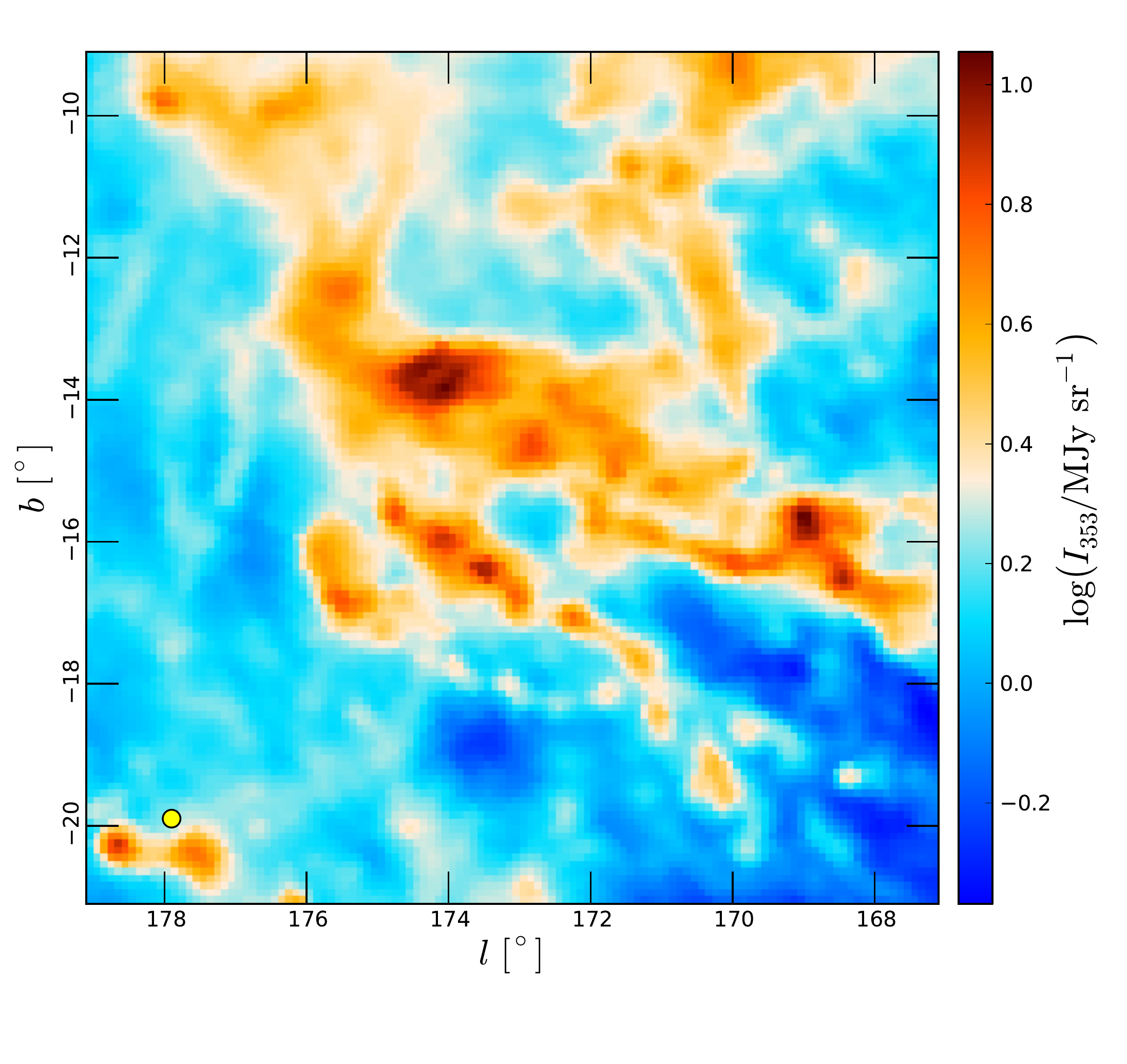}}
\centerline{\includegraphics[width=9cm,trim=0 50 0 20,clip=true]{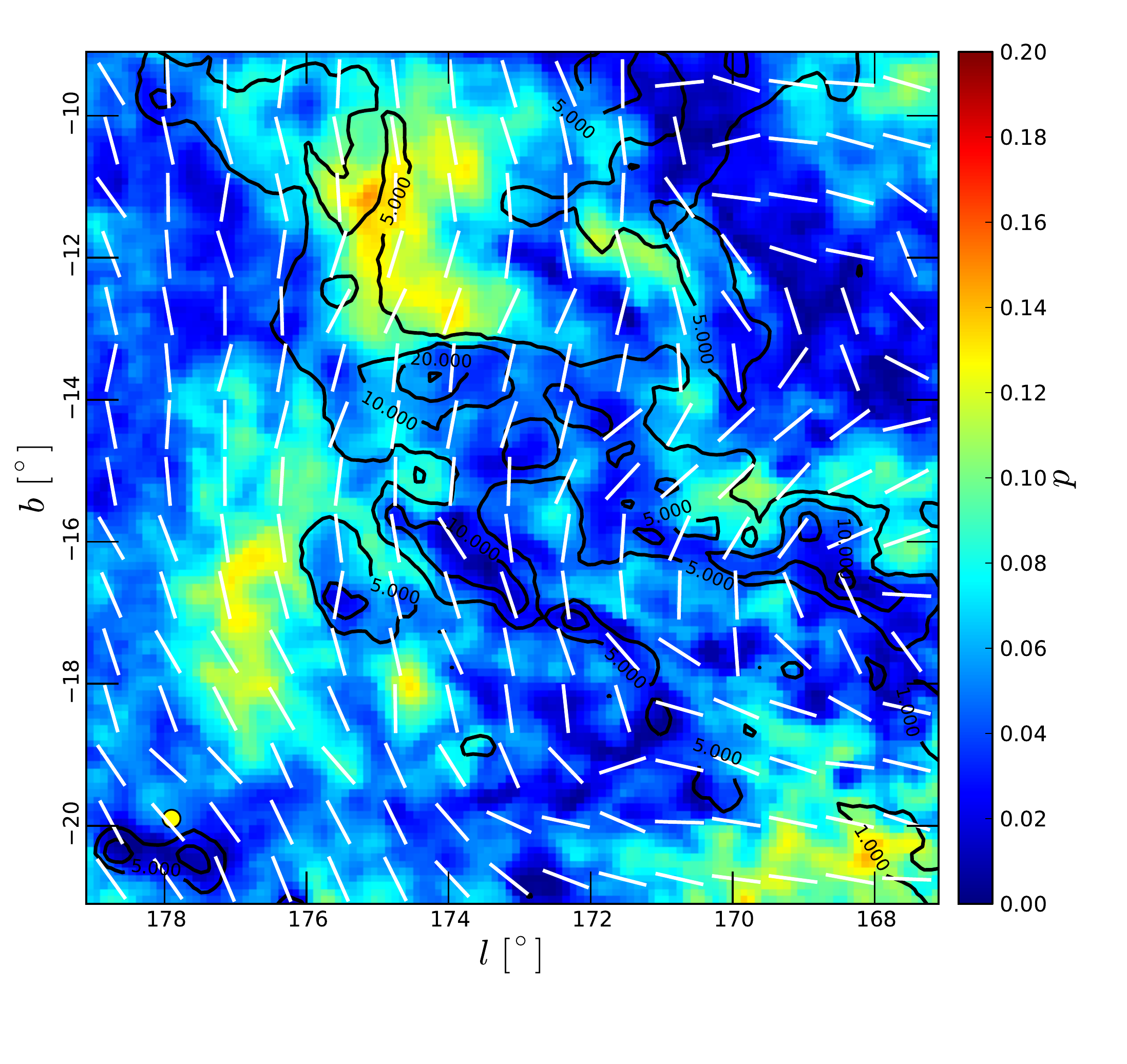}}
\centerline{\includegraphics[width=9cm,trim=0 50 0 20,clip=true]{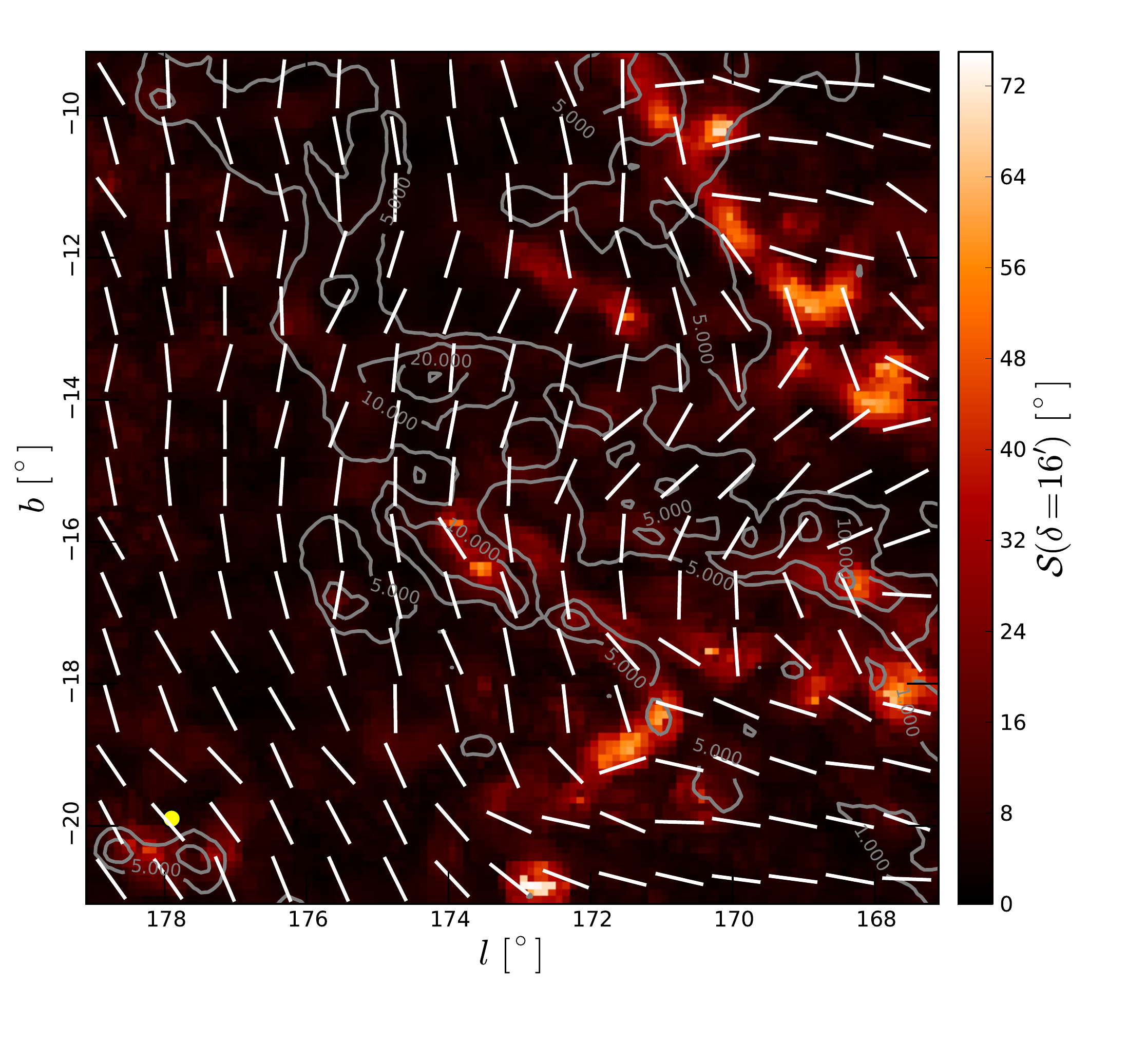}}
\caption{Same as Fig.~\ref{fig:PI-B-NH_Ophiuchus-Chamaeleon}, but for the Taurus field.}
\label{fig:PI-B-NH_Taurus}
\end{figure}

\begin{figure}[htbp]
\centerline{\includegraphics[width=9cm,trim=0 50 0 0,clip=true]{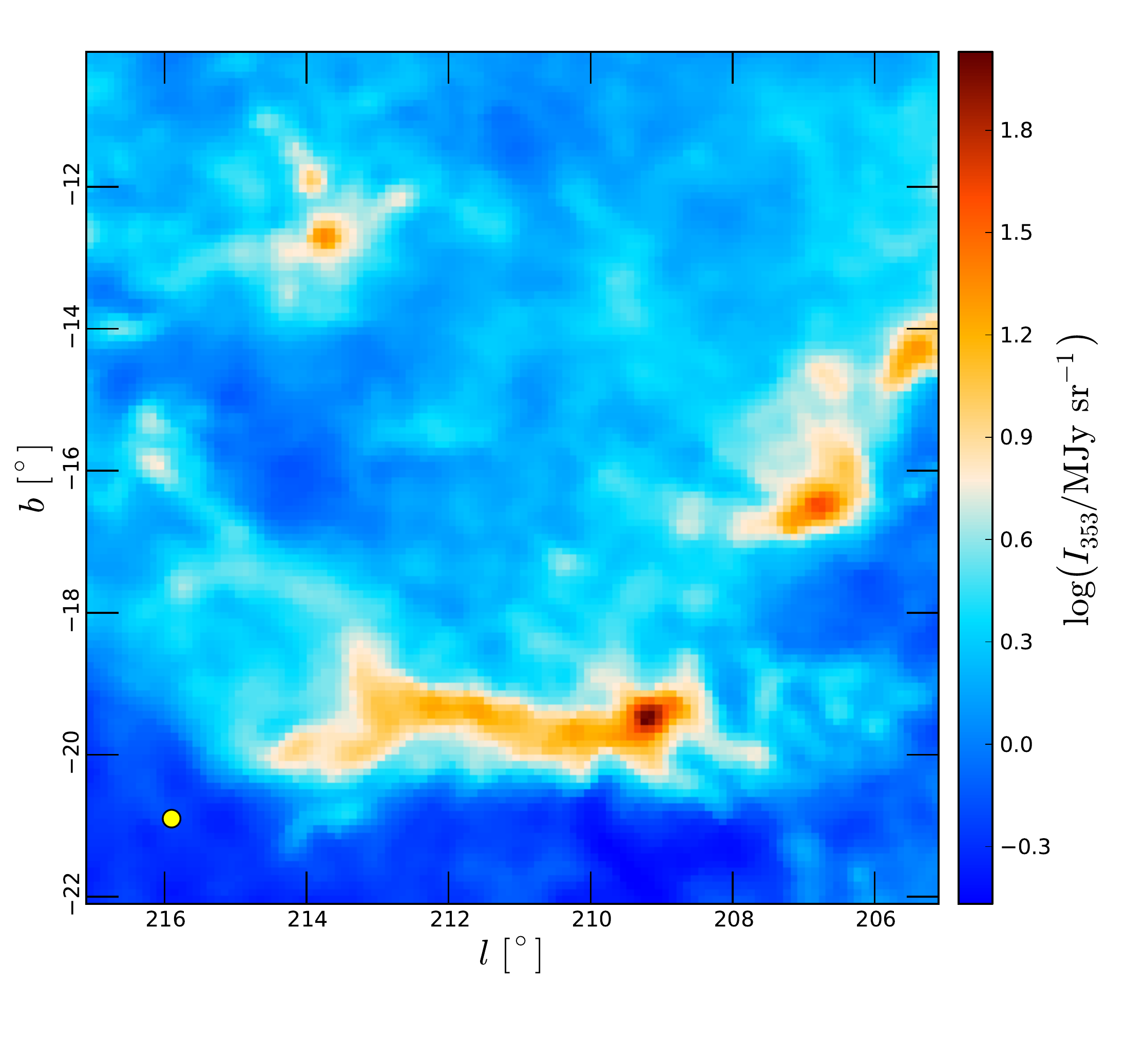}}
\centerline{\includegraphics[width=9cm,trim=0 50 0 20,clip=true]{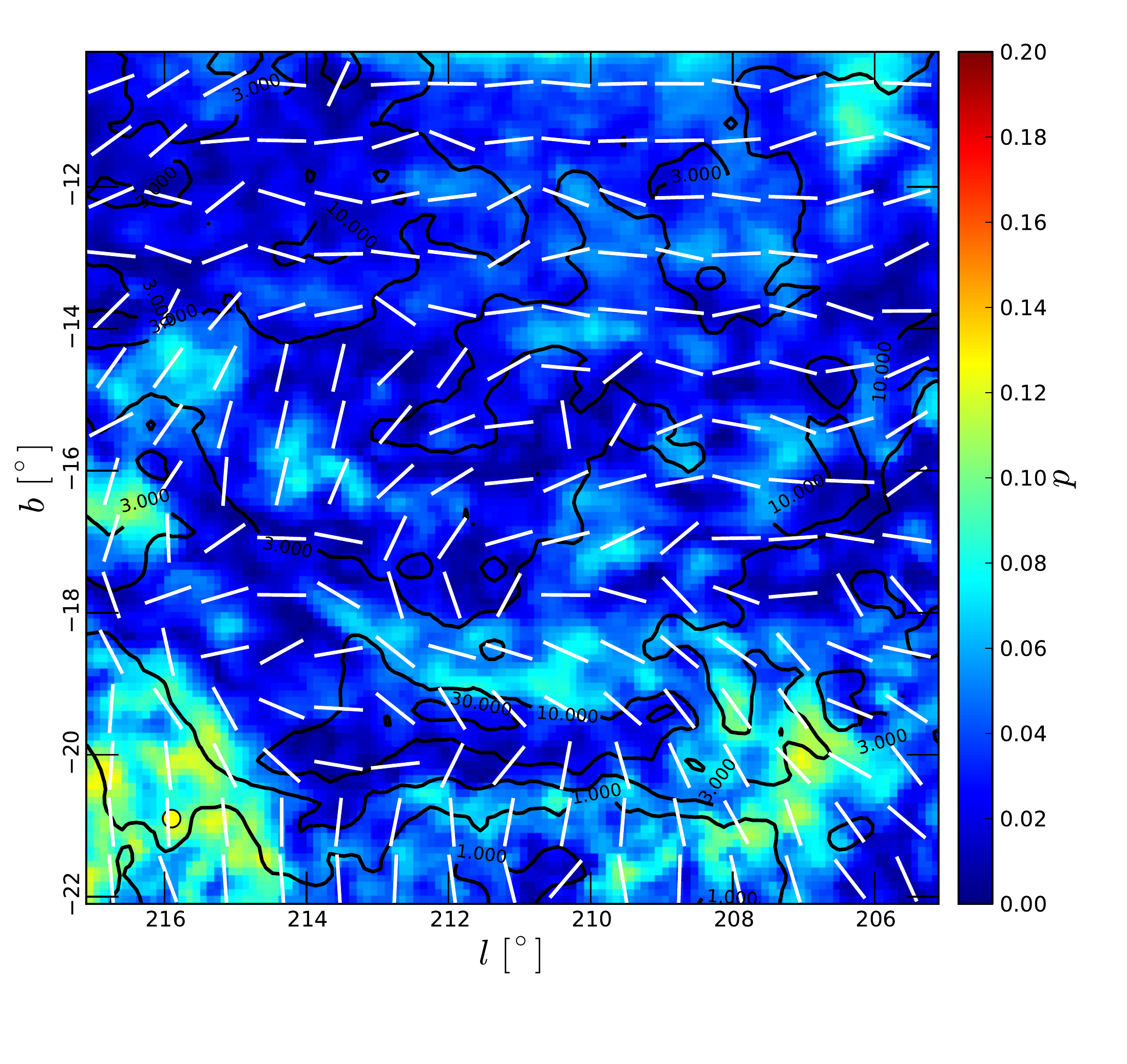}}
\centerline{\includegraphics[width=9cm,trim=0 50 0 20,clip=true]{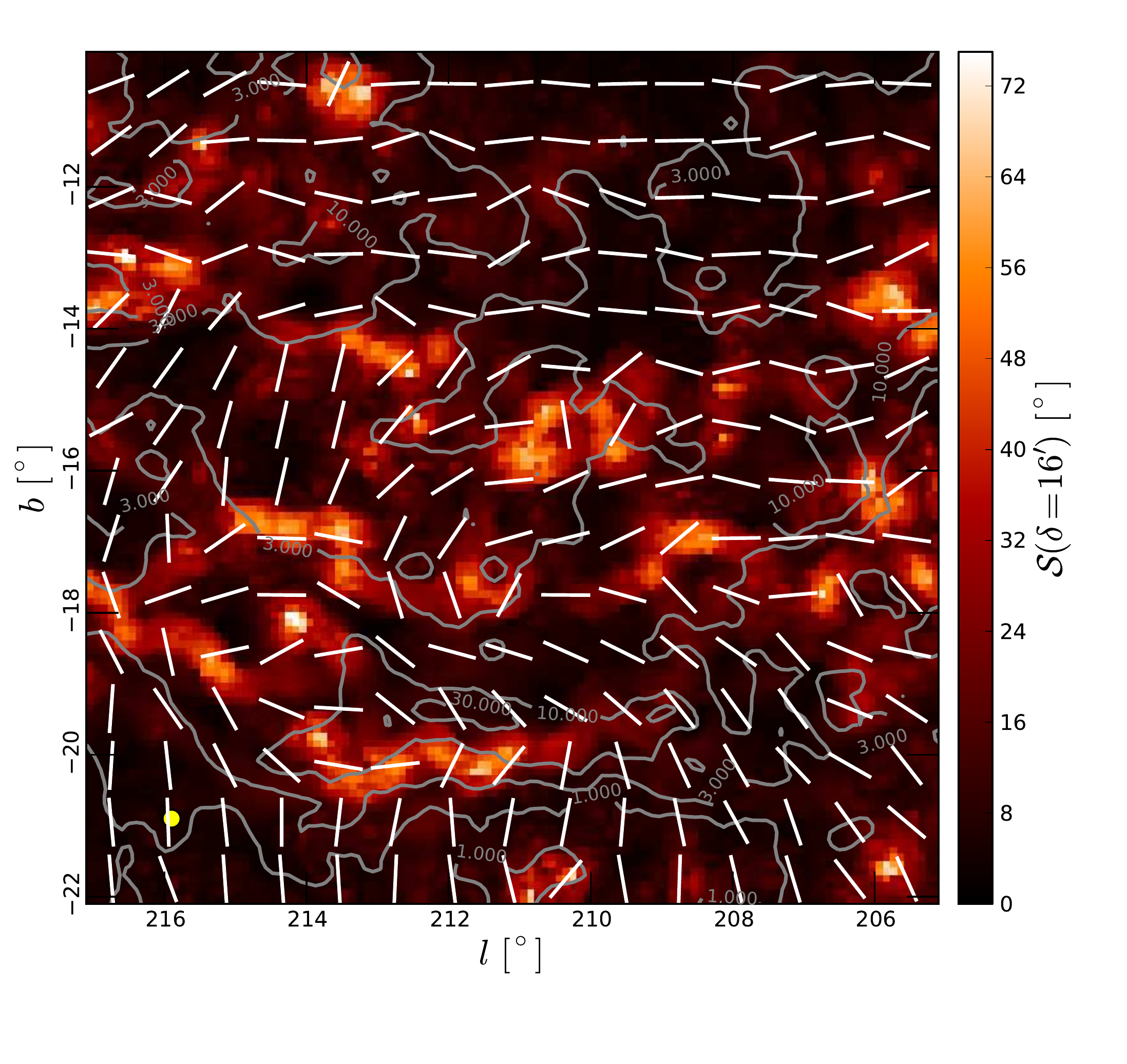}}
\caption{Same as Fig.~\ref{fig:PI-B-NH_Ophiuchus-Chamaeleon}, but for the Orion field.}
\label{fig:PI-B-NH_Orion}
\end{figure}

\begin{figure}[htbp]
\centerline{\includegraphics[width=9cm,trim=0 50 0 0,clip=true]{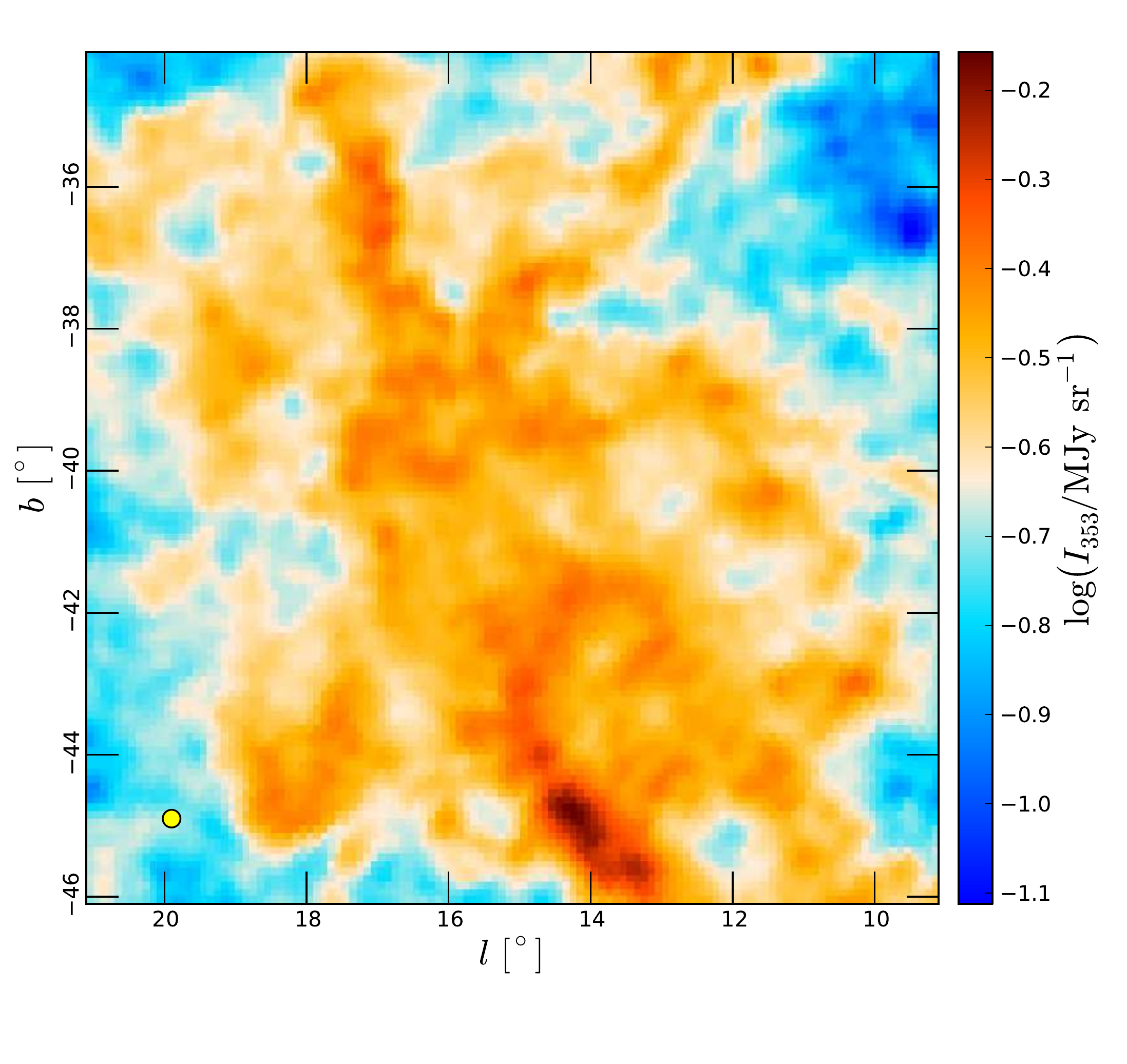}}
\centerline{\includegraphics[width=9cm,trim=0 50 0 20,clip=true]{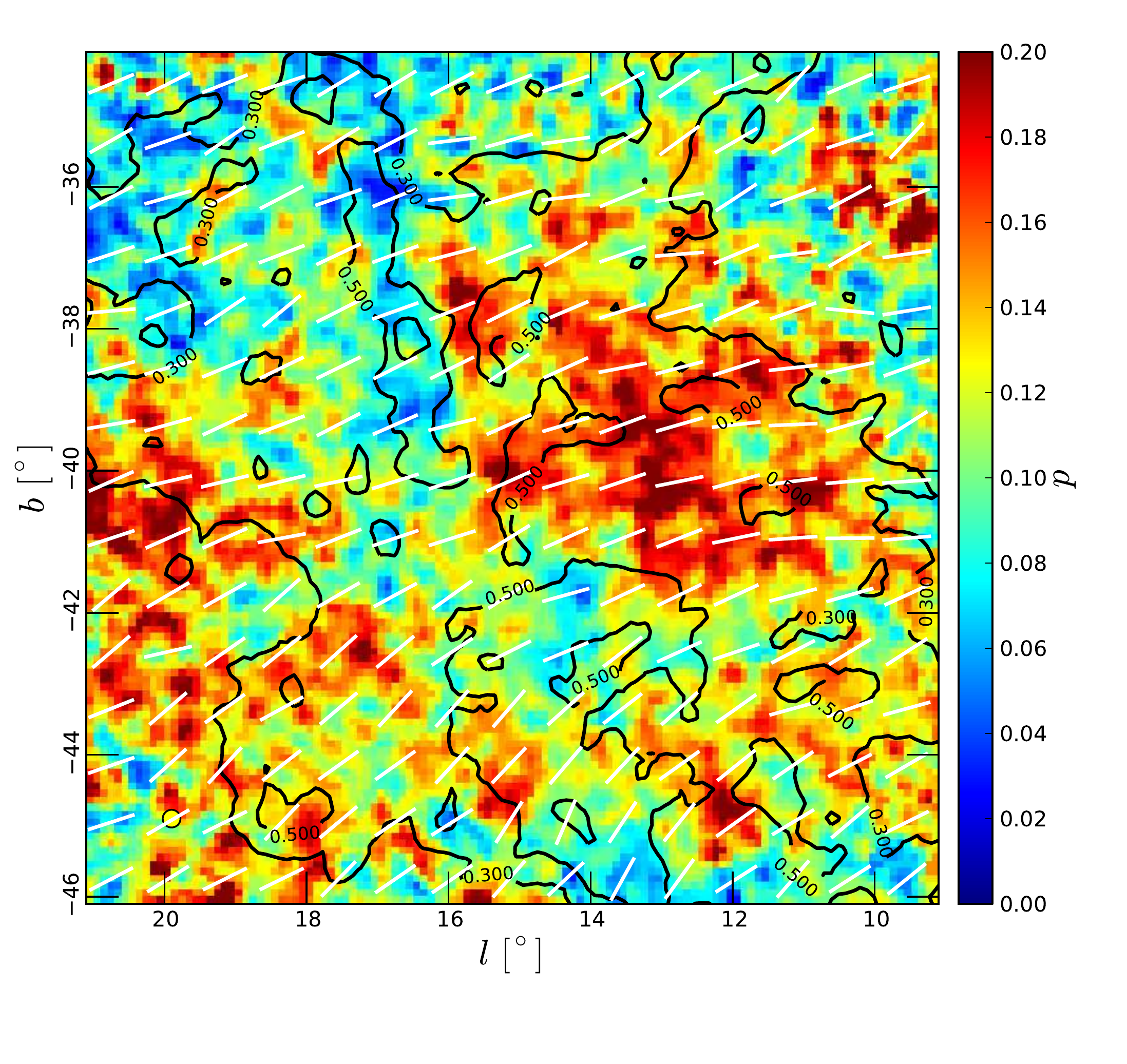}}
\centerline{\includegraphics[width=9cm,trim=0 50 0 20,clip=true]{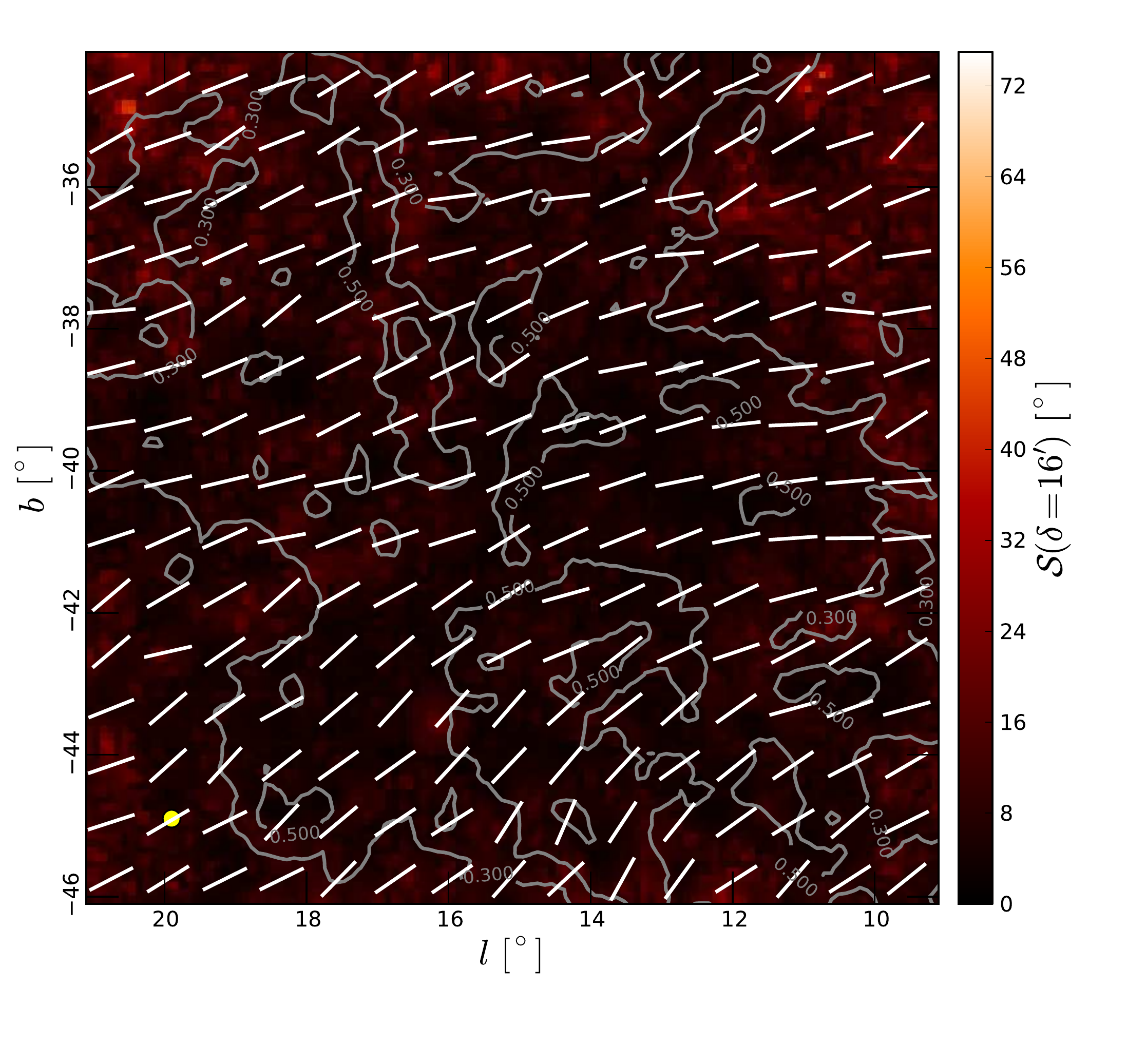}}
\caption{Same as Fig.~\ref{fig:PI-B-NH_Ophiuchus-Chamaeleon}, but for the Microscopium field.}
\label{fig:PI-B-NH_Microscopium}
\end{figure}

\begin{figure}[htbp]
\centerline{\includegraphics[width=9cm,trim=0 50 0 0,clip=true]{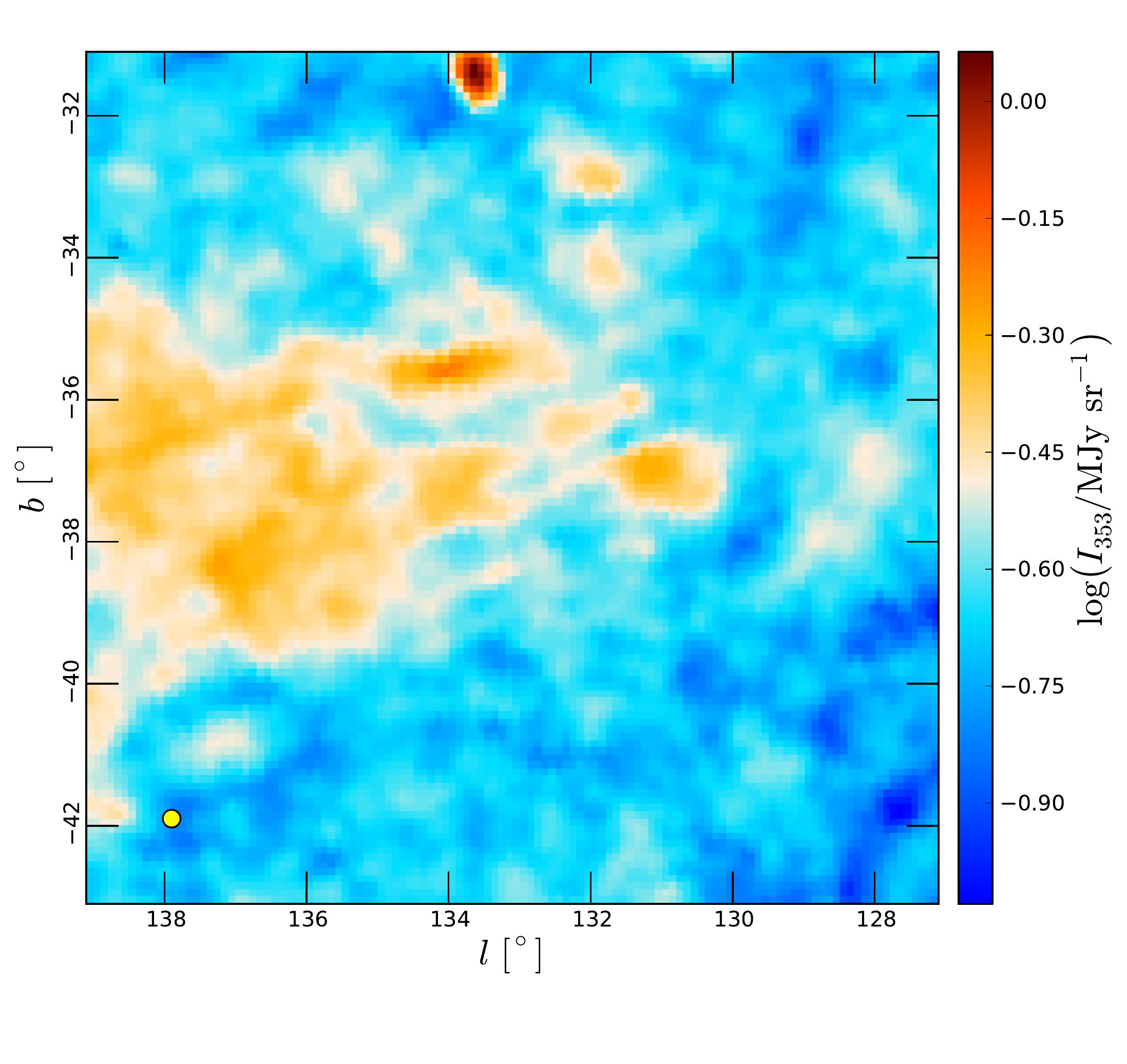}}
\centerline{\includegraphics[width=9cm,trim=0 50 0 20,clip=true]{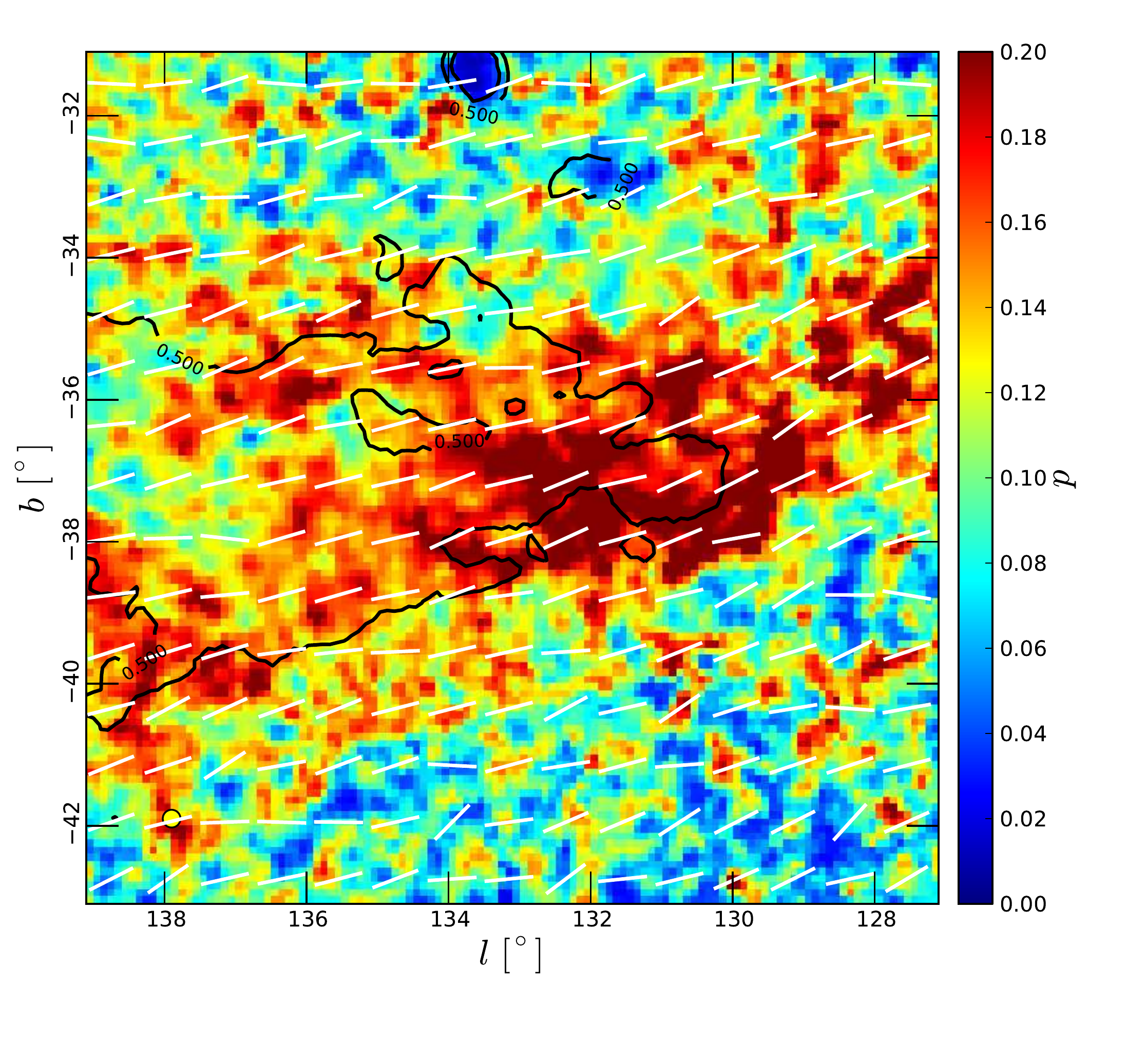}}
\centerline{\includegraphics[width=9cm,trim=0 50 0 20,clip=true]{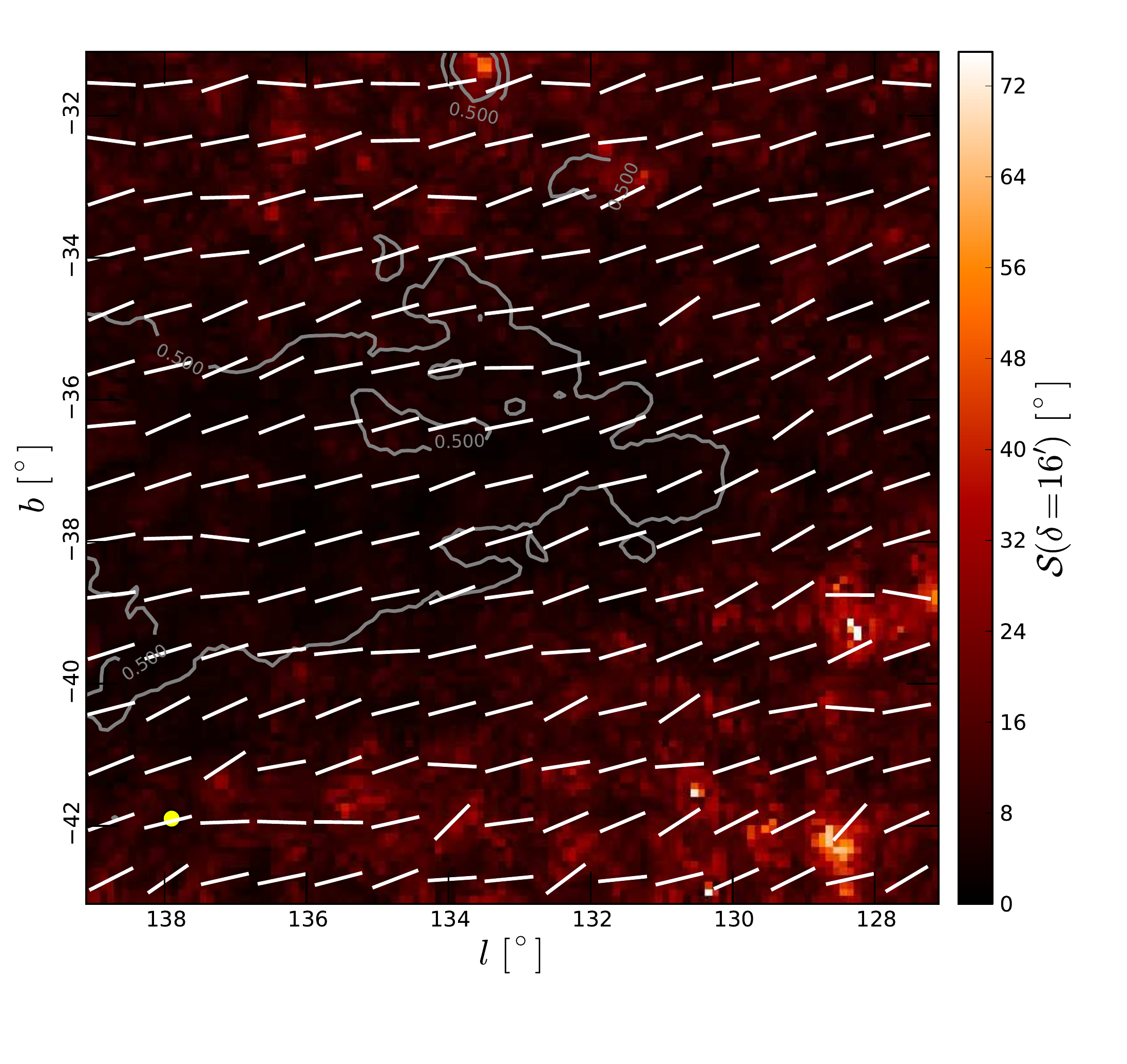}}
\caption{Same as Fig.~\ref{fig:PI-B-NH_Ophiuchus-Chamaeleon}, but for the Pisces field.}
\label{fig:PI-B-NH_Pisces}
\end{figure}

\begin{figure}[htbp]
\centerline{\includegraphics[width=9cm,trim=0 50 0 0,clip=true]{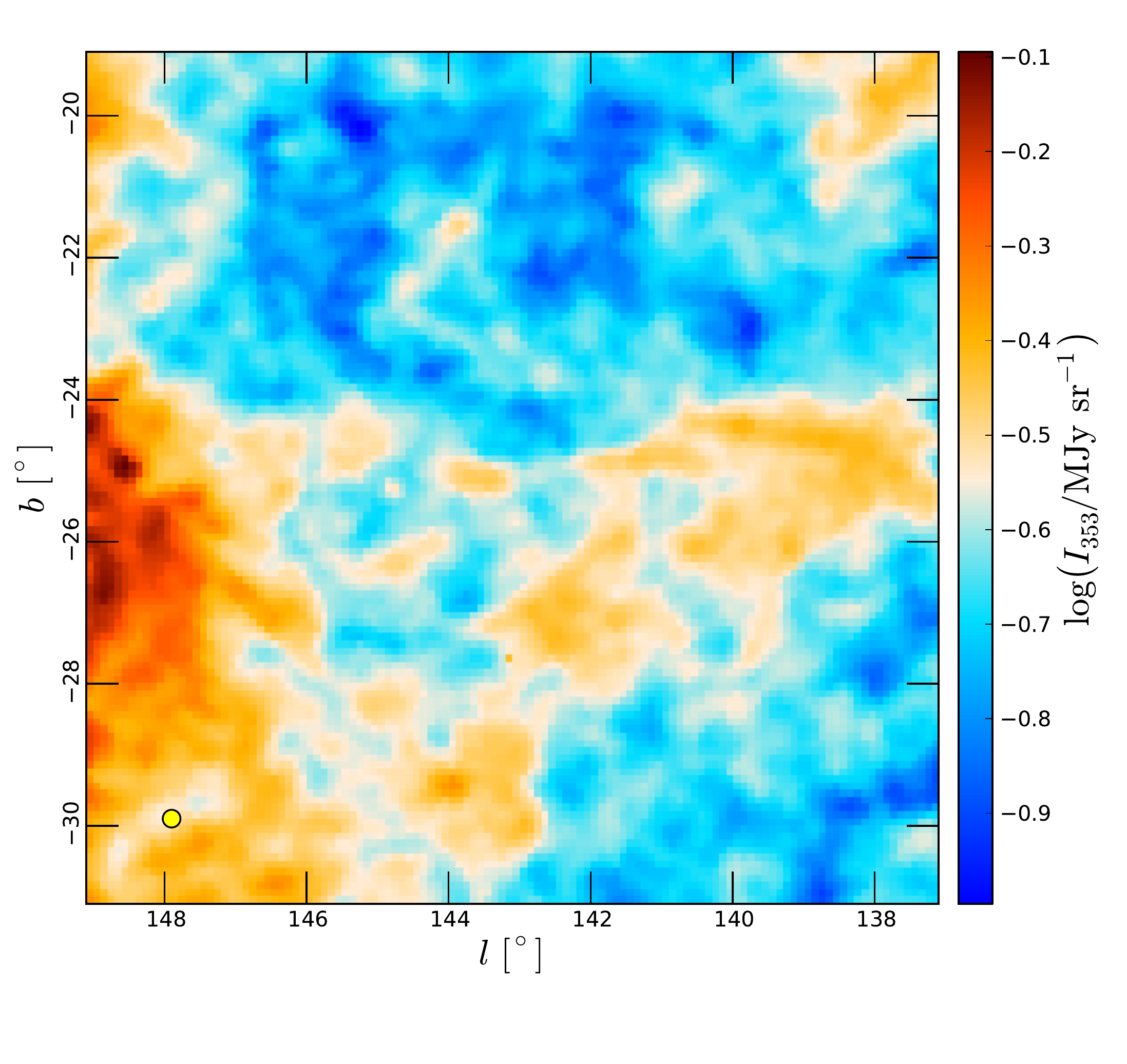}}
\centerline{\includegraphics[width=9cm,trim=0 50 0 20,clip=true]{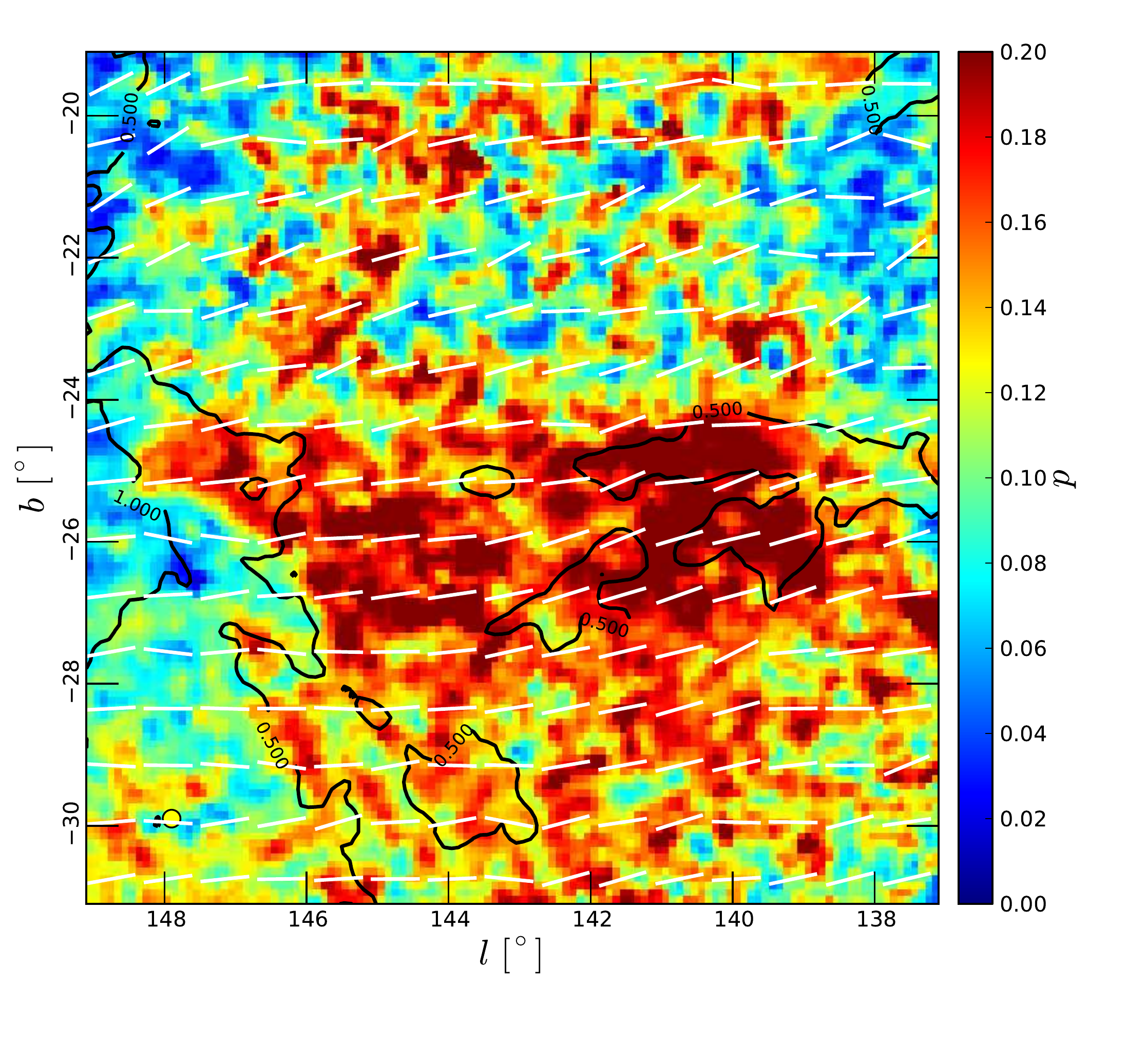}}
\centerline{\includegraphics[width=9cm,trim=0 50 0 20,clip=true]{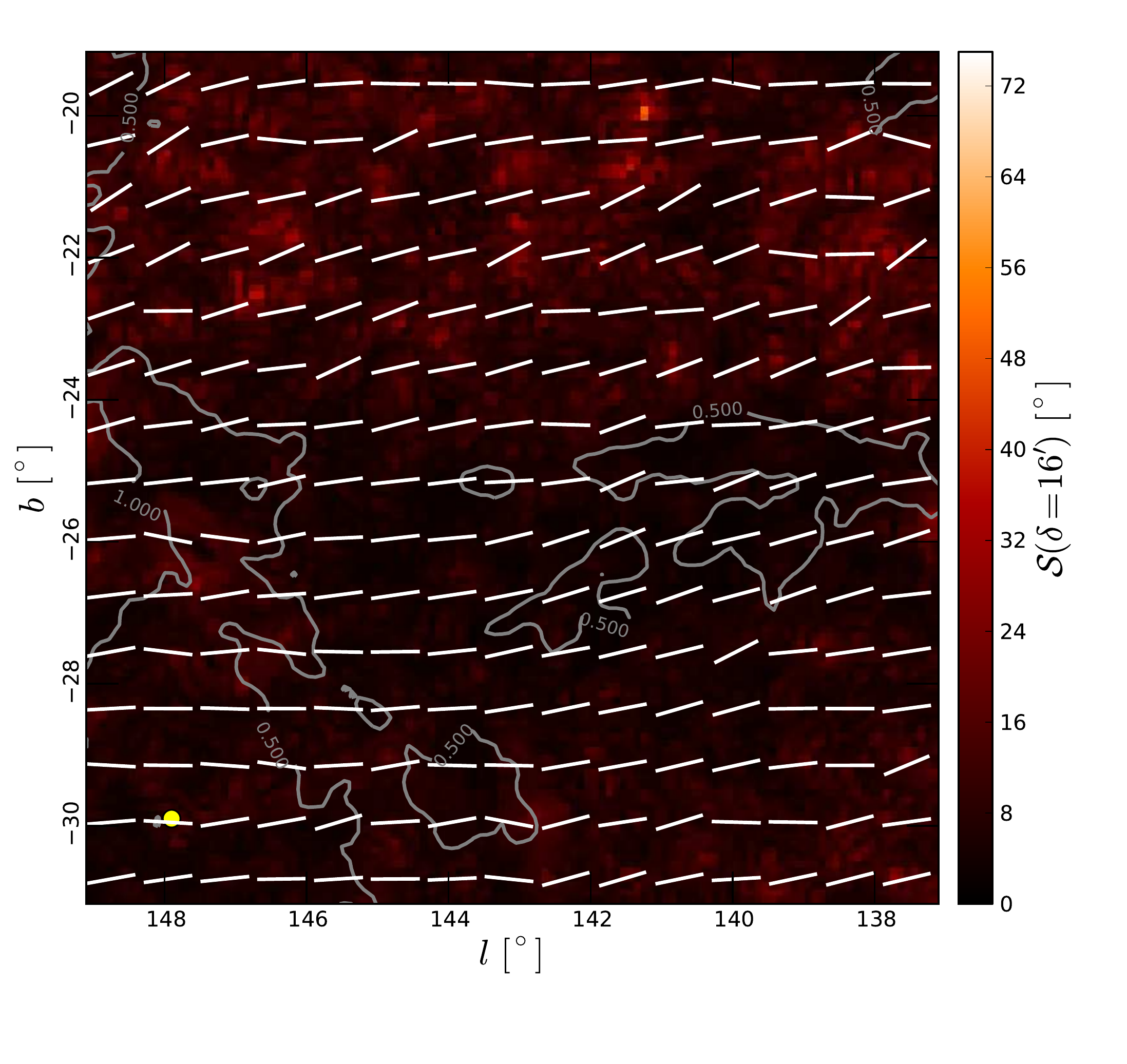}}
\caption{Same as Fig.~\ref{fig:PI-B-NH_Ophiuchus-Chamaeleon}, but for the Perseus field.}
\label{fig:PI-B-NH_Perseus}
\end{figure}

\begin{figure}[htbp]
\centerline{\includegraphics[width=9cm,trim=0 50 0 0,clip=true]{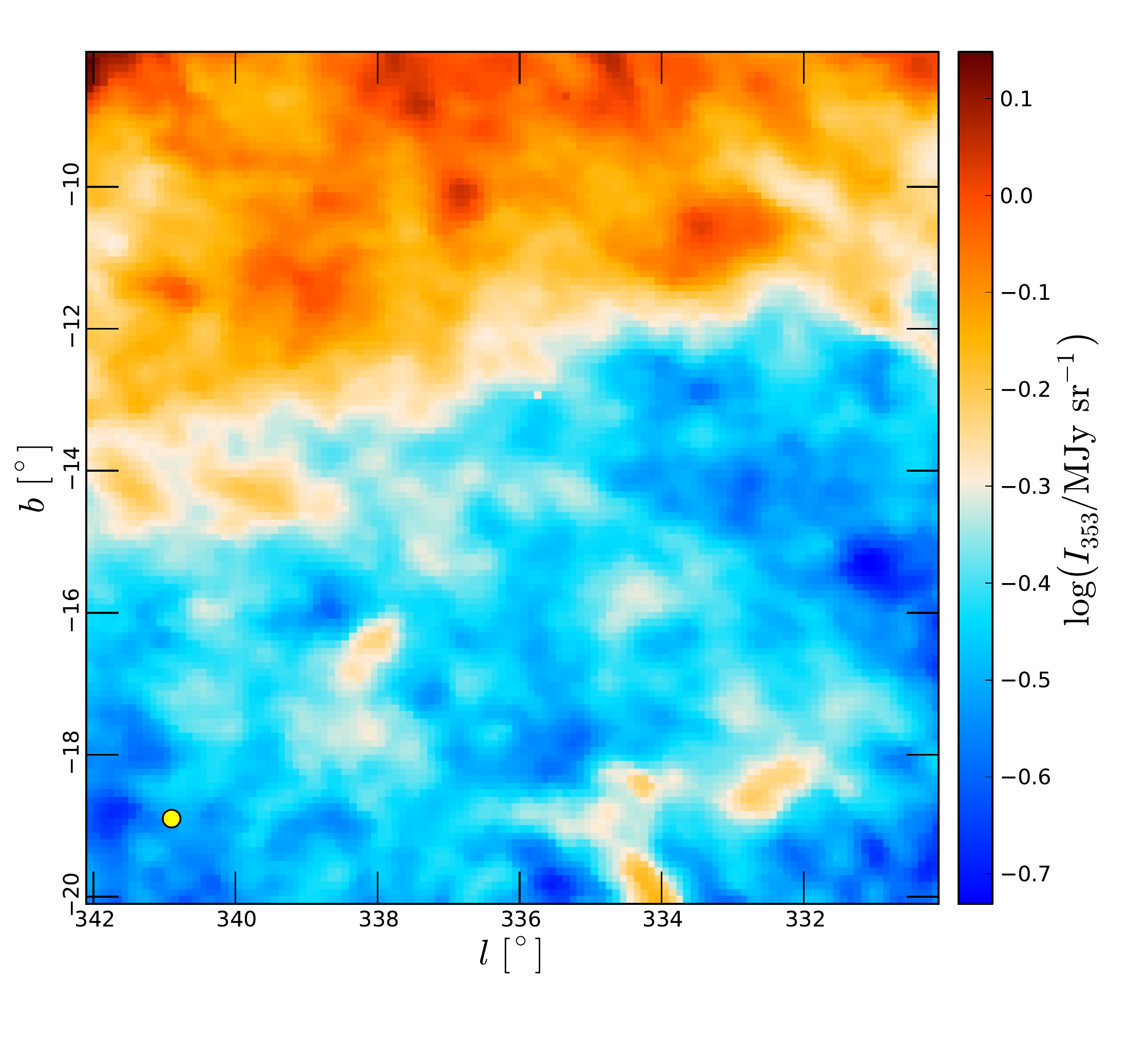}}
\centerline{\includegraphics[width=9cm,trim=0 50 0 20,clip=true]{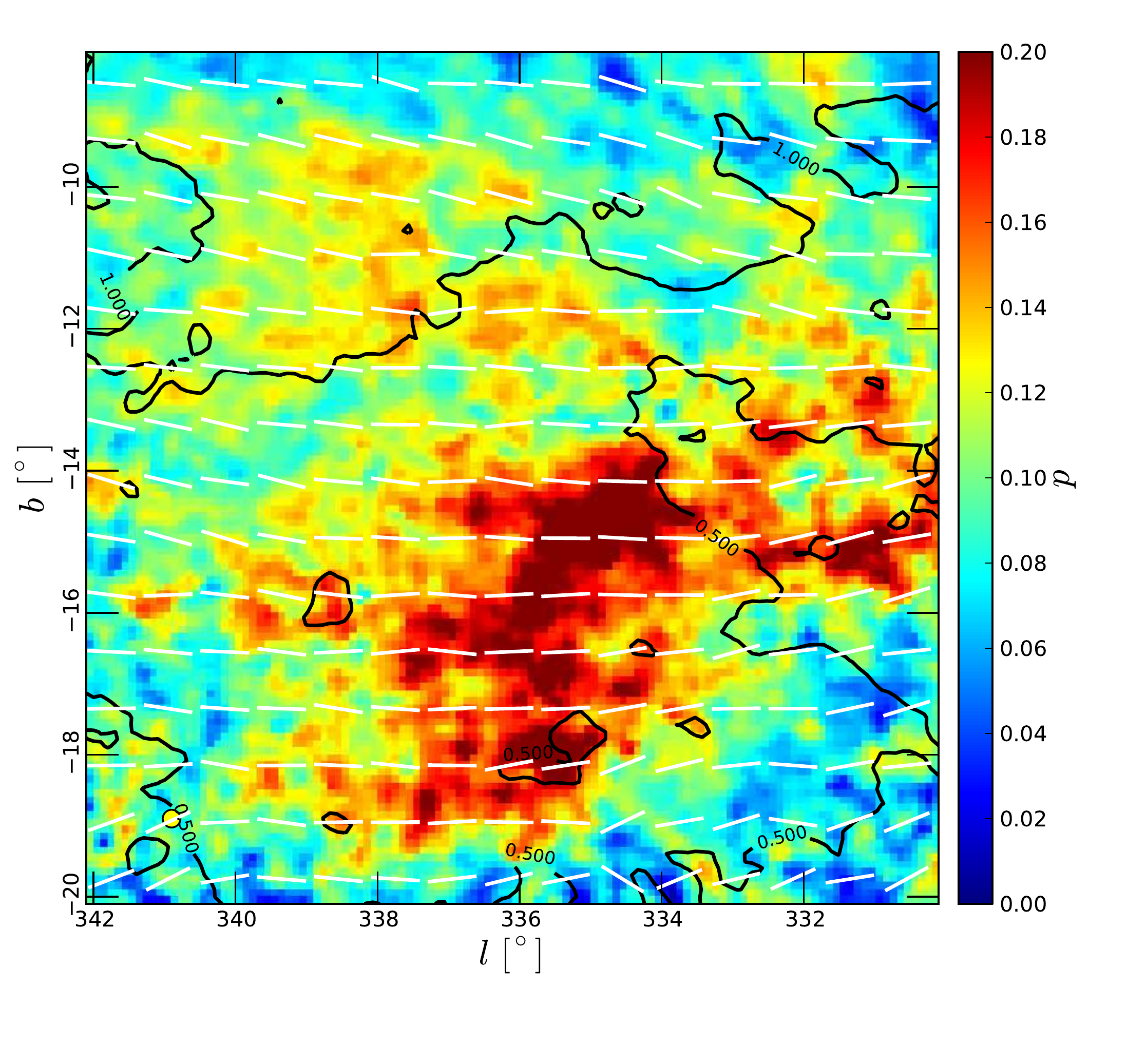}}
\centerline{\includegraphics[width=9cm,trim=0 50 0 20,clip=true]{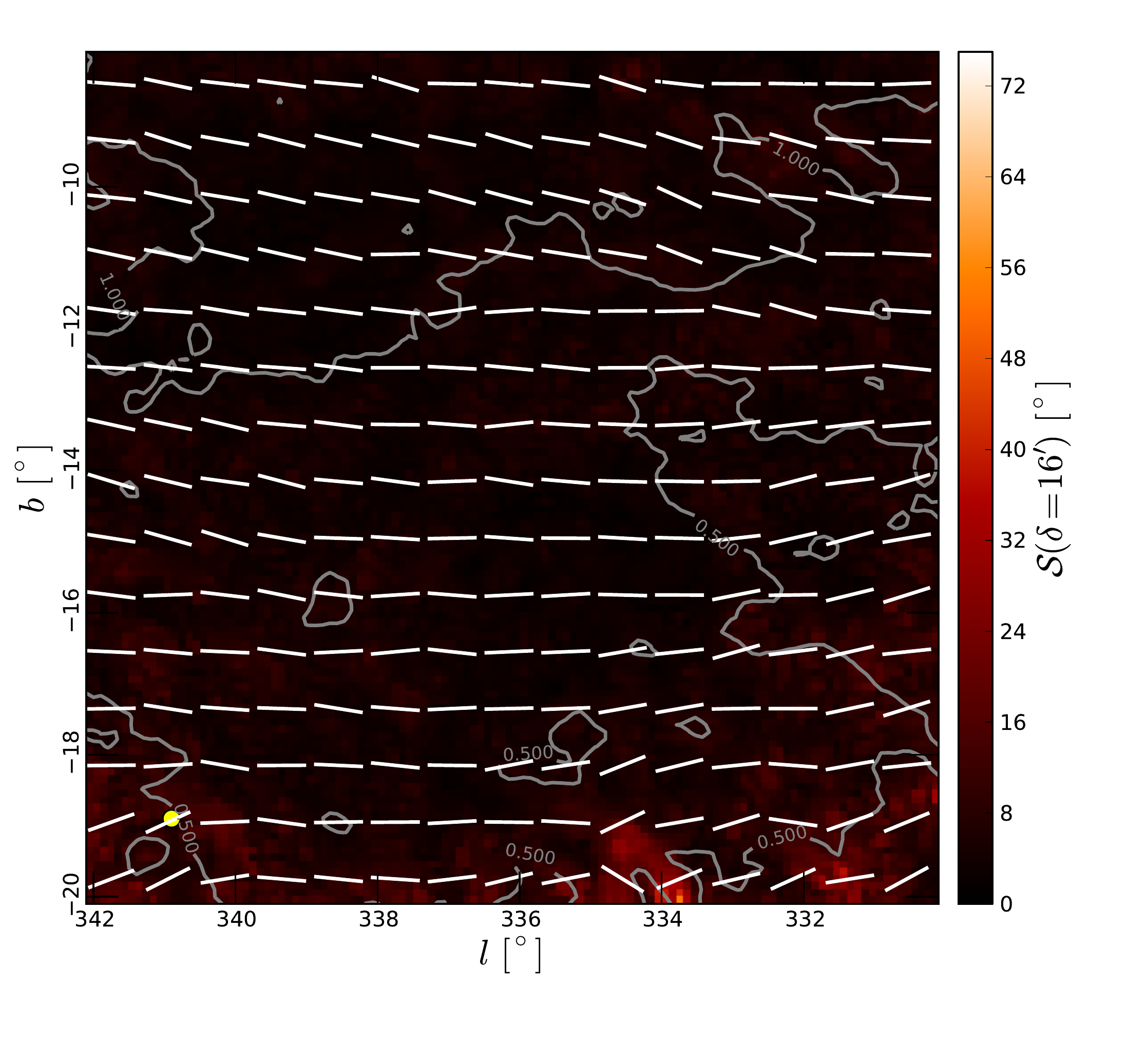}}
\caption{Same as Fig.~\ref{fig:PI-B-NH_Ophiuchus-Chamaeleon}, but for the Ara field.}
\label{fig:PI-B-NH_Ara}
\end{figure}

\begin{figure}[htbp]
\centerline{\includegraphics[width=9cm,trim=0 50 0 0,clip=true]{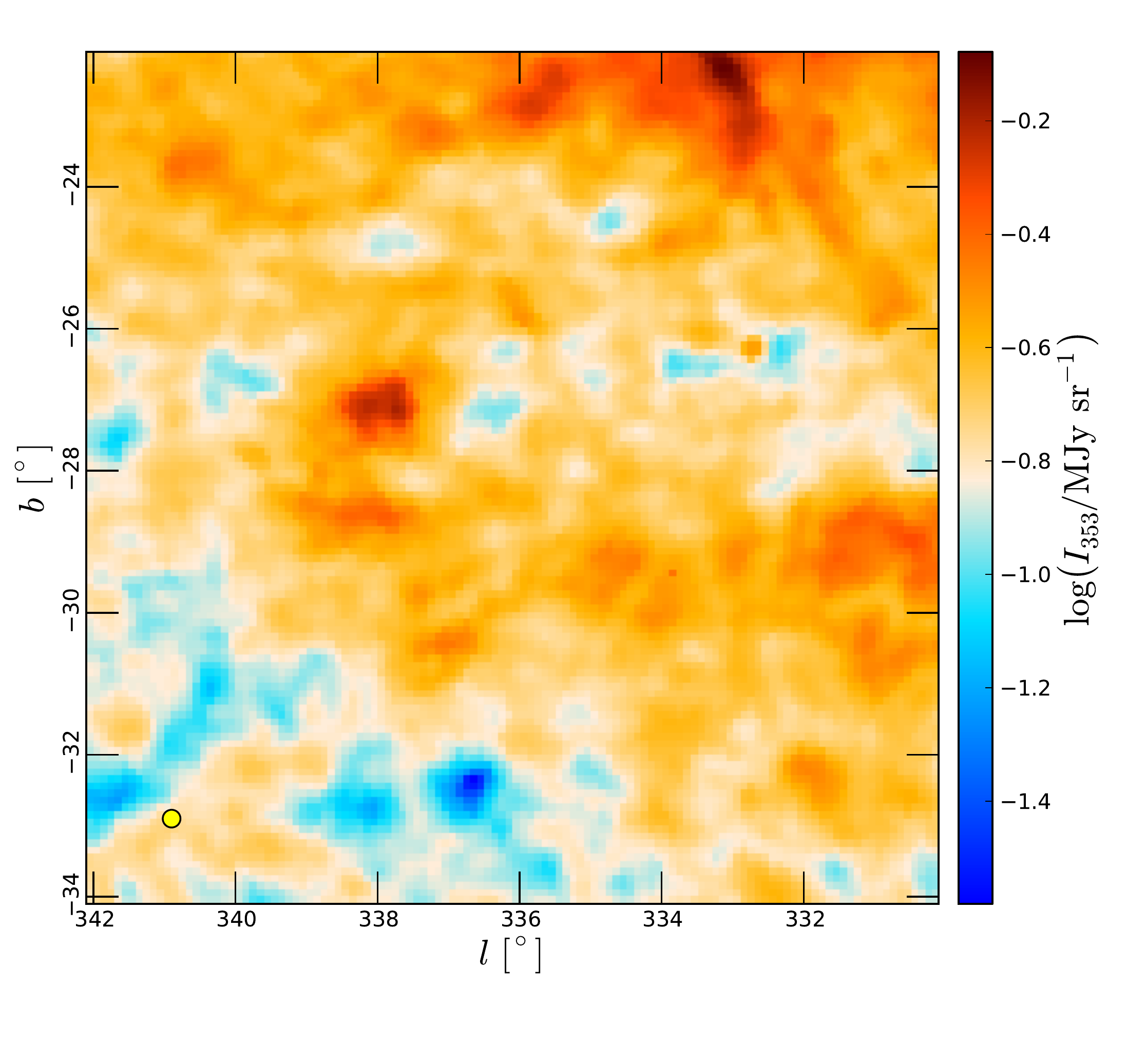}}
\centerline{\includegraphics[width=9cm,trim=0 50 0 20,clip=true]{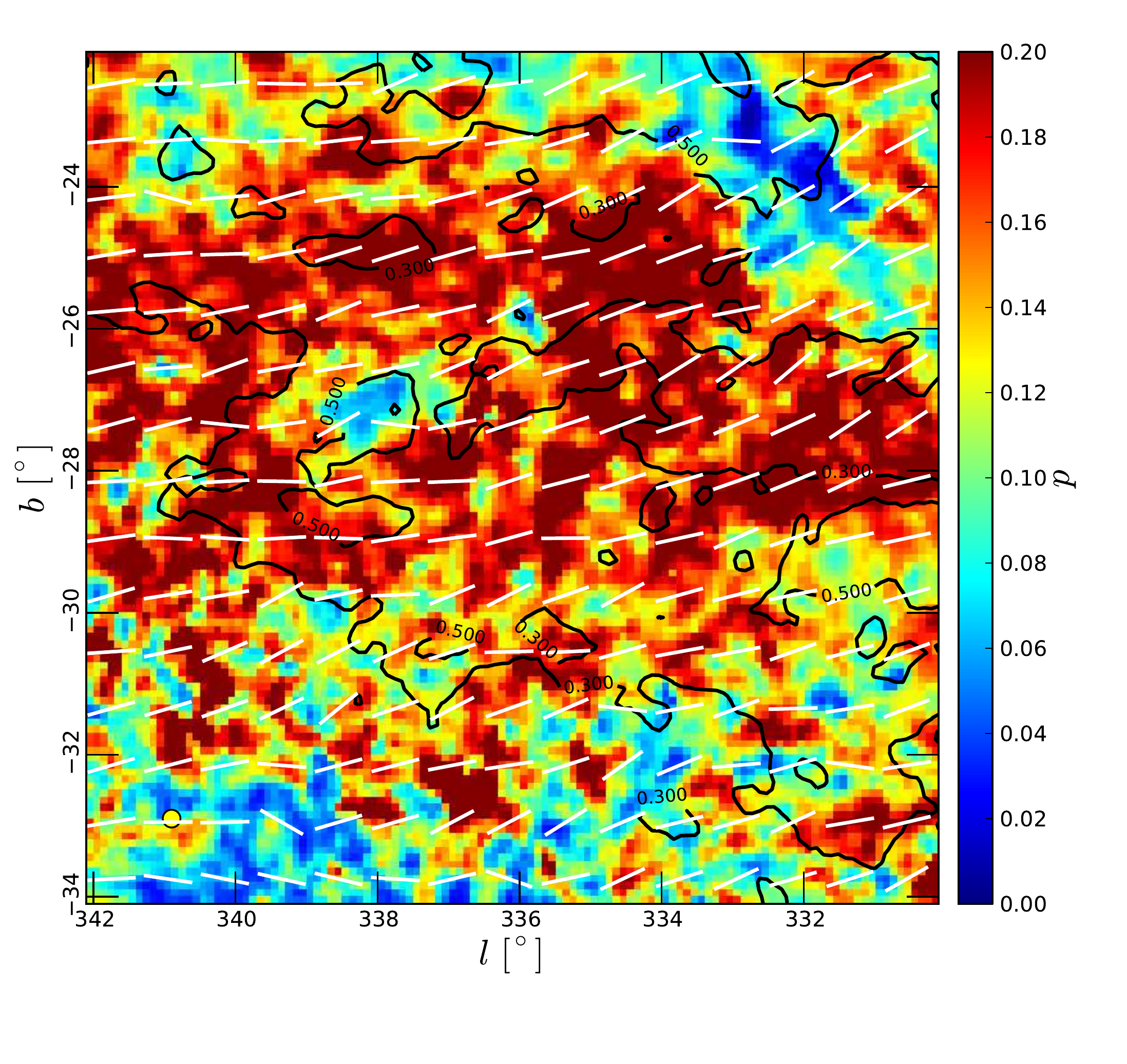}}
\centerline{\includegraphics[width=9cm,trim=0 50 0 20,clip=true]{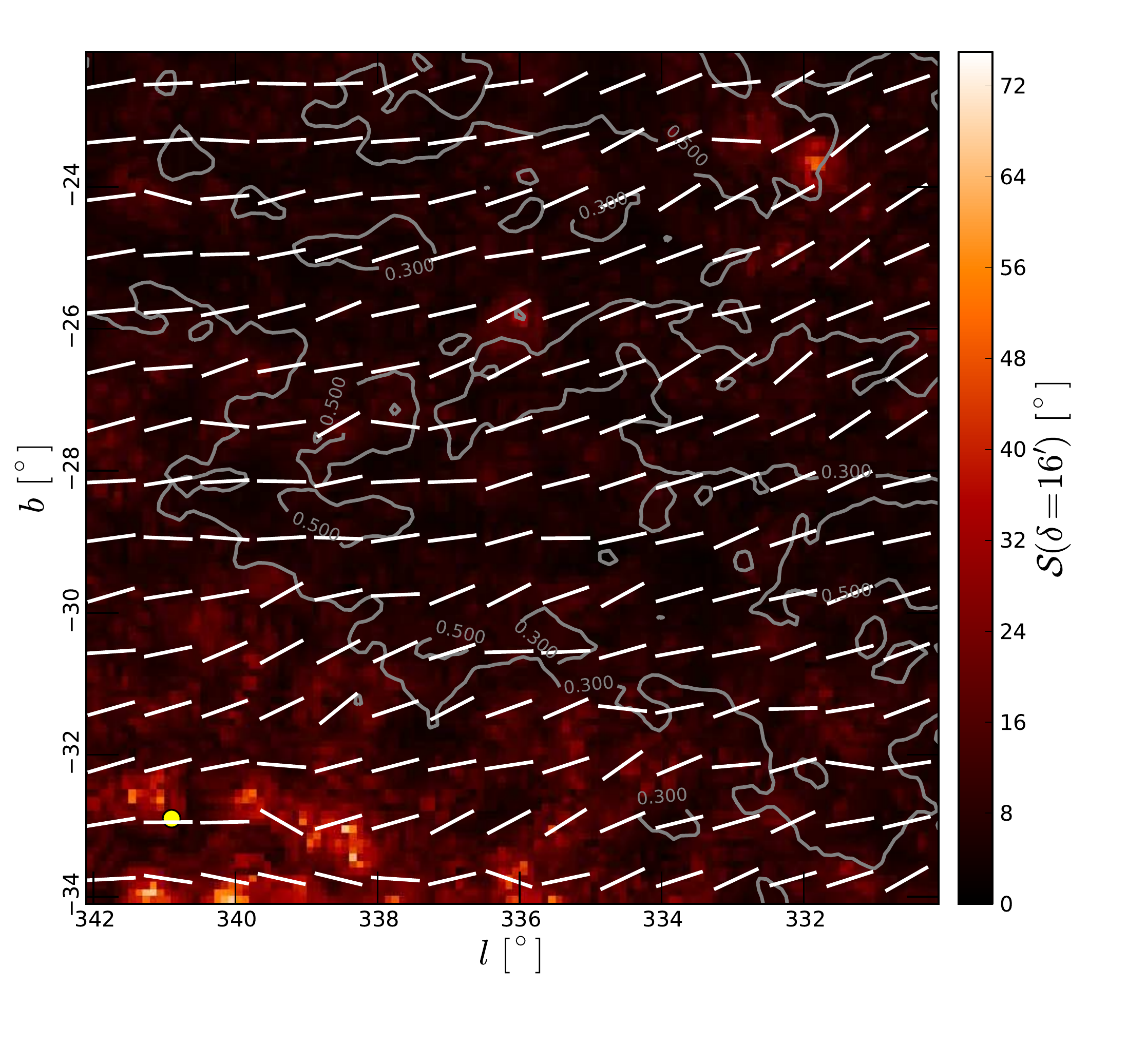}}
\caption{Same as Fig.~\ref{fig:PI-B-NH_Ophiuchus-Chamaeleon}, but for the Pavo field.}
\label{fig:PI-B-NH_Pavo}
\end{figure}

\begin{figure}[htbp]
\centerline{\includegraphics[width=8.8cm,trim=120 0 60 0,clip=true]{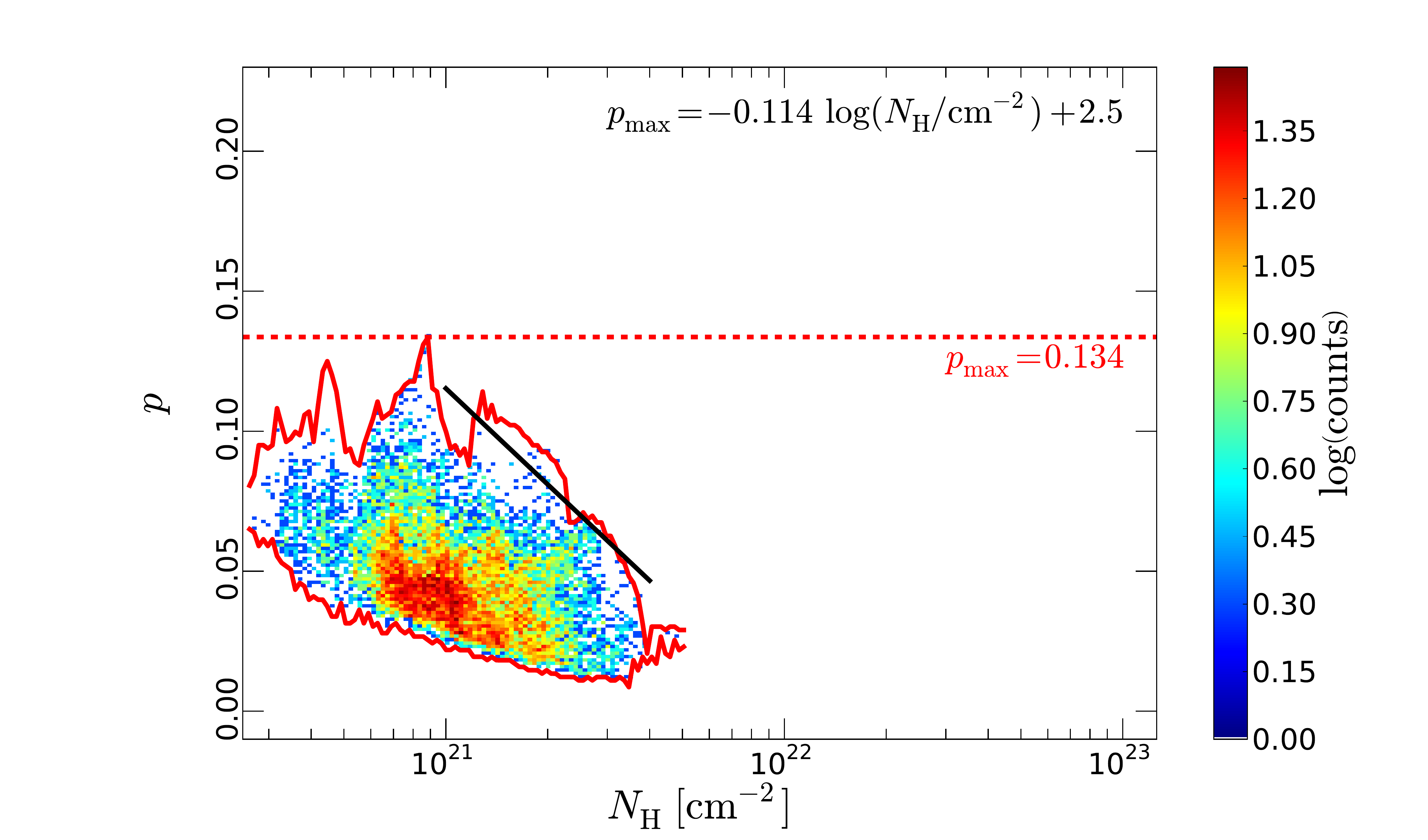}}
\caption{Same as Fig.~\ref{fig:PI-vs-NH}, but for the Polaris Flare field. Two-dimensional distribution function of polarization fraction $\polfrac$ and column density $N_\mathrm{H}$. The distribution function is presented in logarithmic colour scale and includes only points for which $p/\sigma_p>3$. The dashed red line corresponds to the absolute maximum polarization fraction $\polfrac_\mathrm{max}$ and the solid red curves show the upper and lower envelopes of $\polfrac$ as functions of $N_\mathrm{H}$. The solid black line is a linear fit $\polfrac_\mathrm{max}=m\log{\left(N_\mathrm{H}/\mathrm{cm^{-2}}\right)}+c$ to the decrease of the maximum polarization fraction with column density at the high end of $N_\mathrm{H}$ (see Table~\ref{table-fields-properties} for the fitting ranges and fit parameters).}
\label{fig:PI-vs-NH-Polaris}
\end{figure}

\begin{figure}[htbp]
\centerline{\includegraphics[width=8.8cm,trim=120 0 60 0,clip=true]{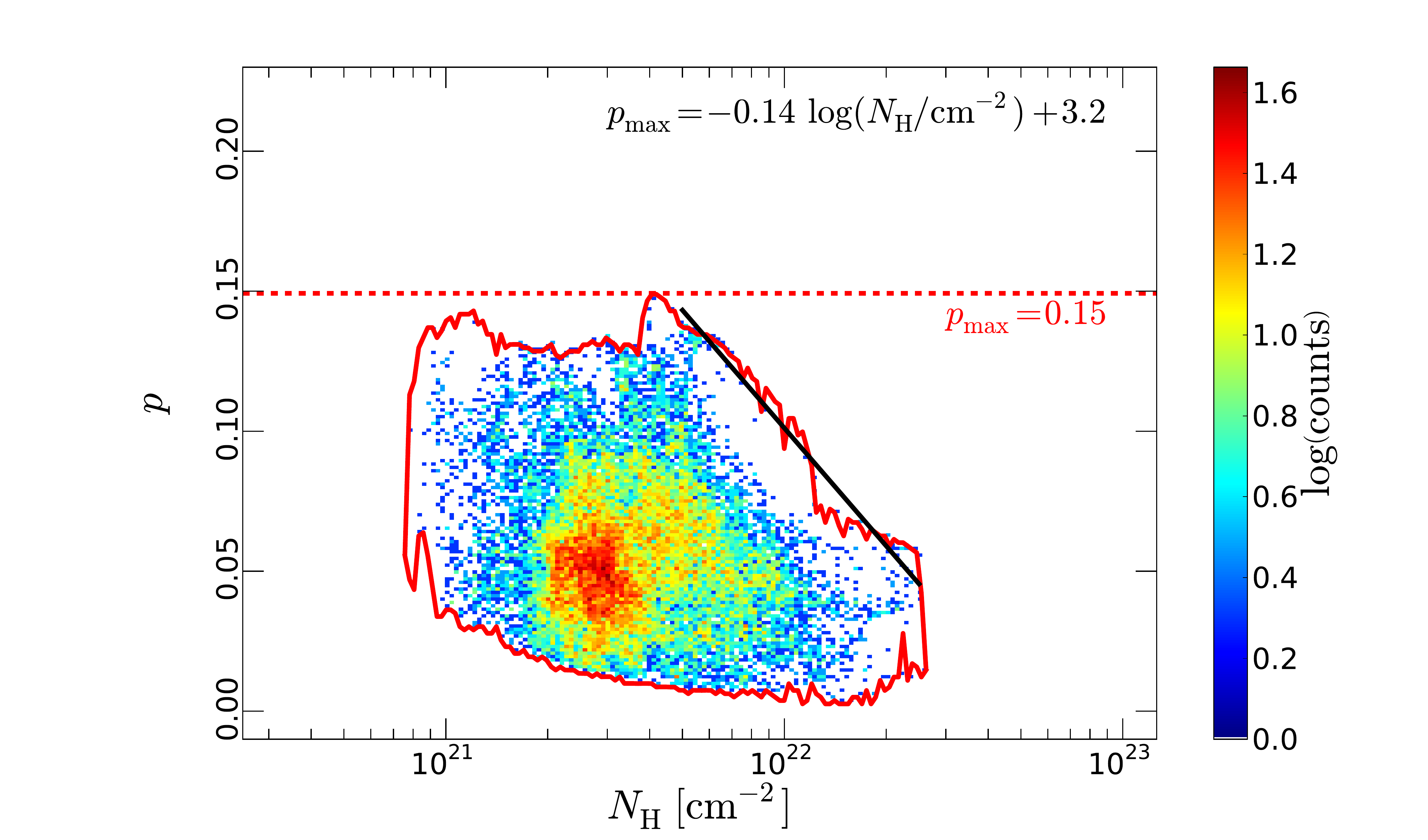}}
\caption{Same as Fig.~\ref{fig:PI-vs-NH}, but for the Taurus field.}
\label{fig:PI-vs-NH-Taurus}
\end{figure}

\begin{figure}[htbp]
\centerline{\includegraphics[width=8.8cm,trim=120 0 60 0,clip=true]{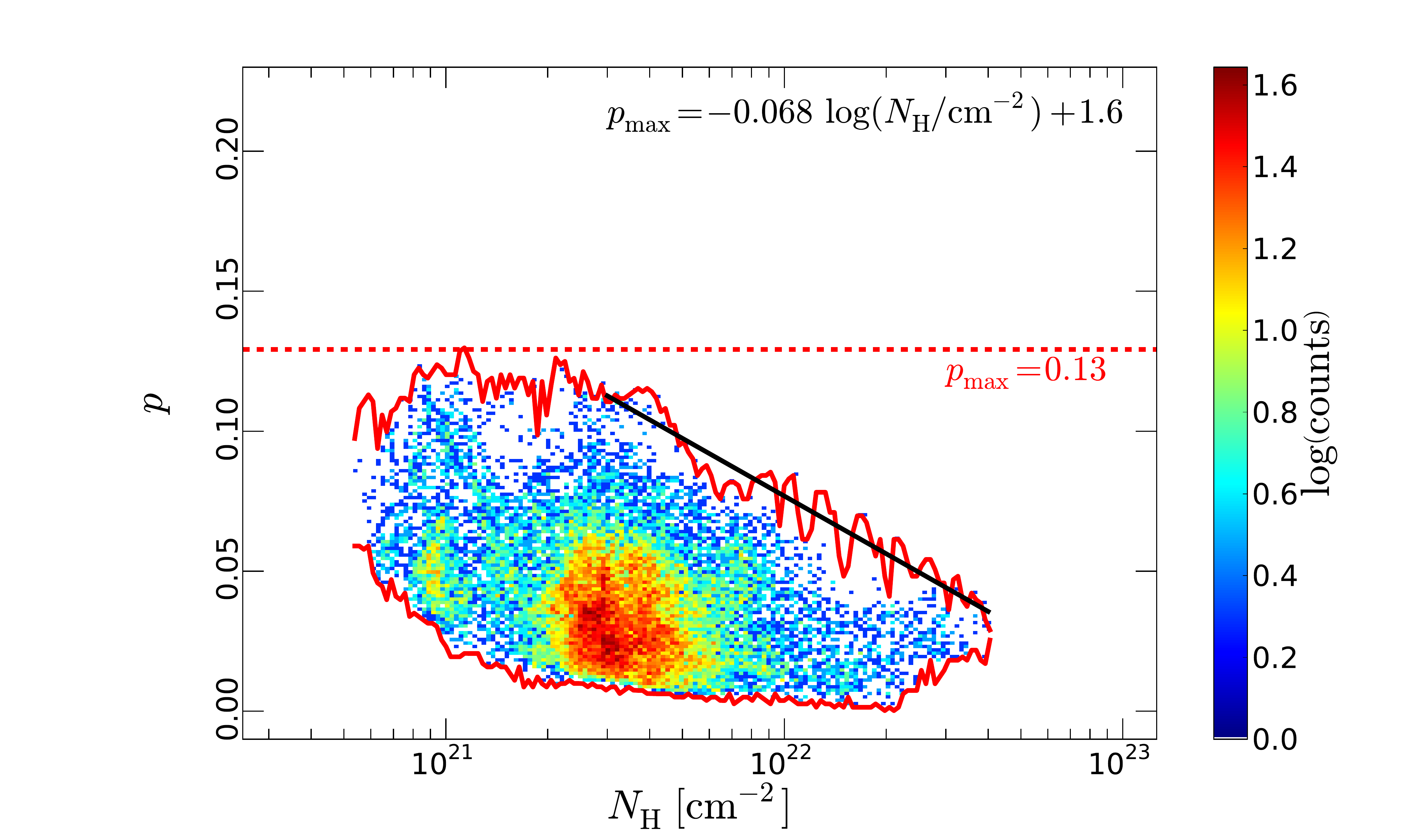}}
\caption{Same as Fig.~\ref{fig:PI-vs-NH}, but for the Orion field.}
\label{fig:PI-vs-NH-Orion}
\end{figure}

\begin{figure}[htbp]
\centerline{\includegraphics[width=8.8cm,trim=120 0 60 0,clip=true]{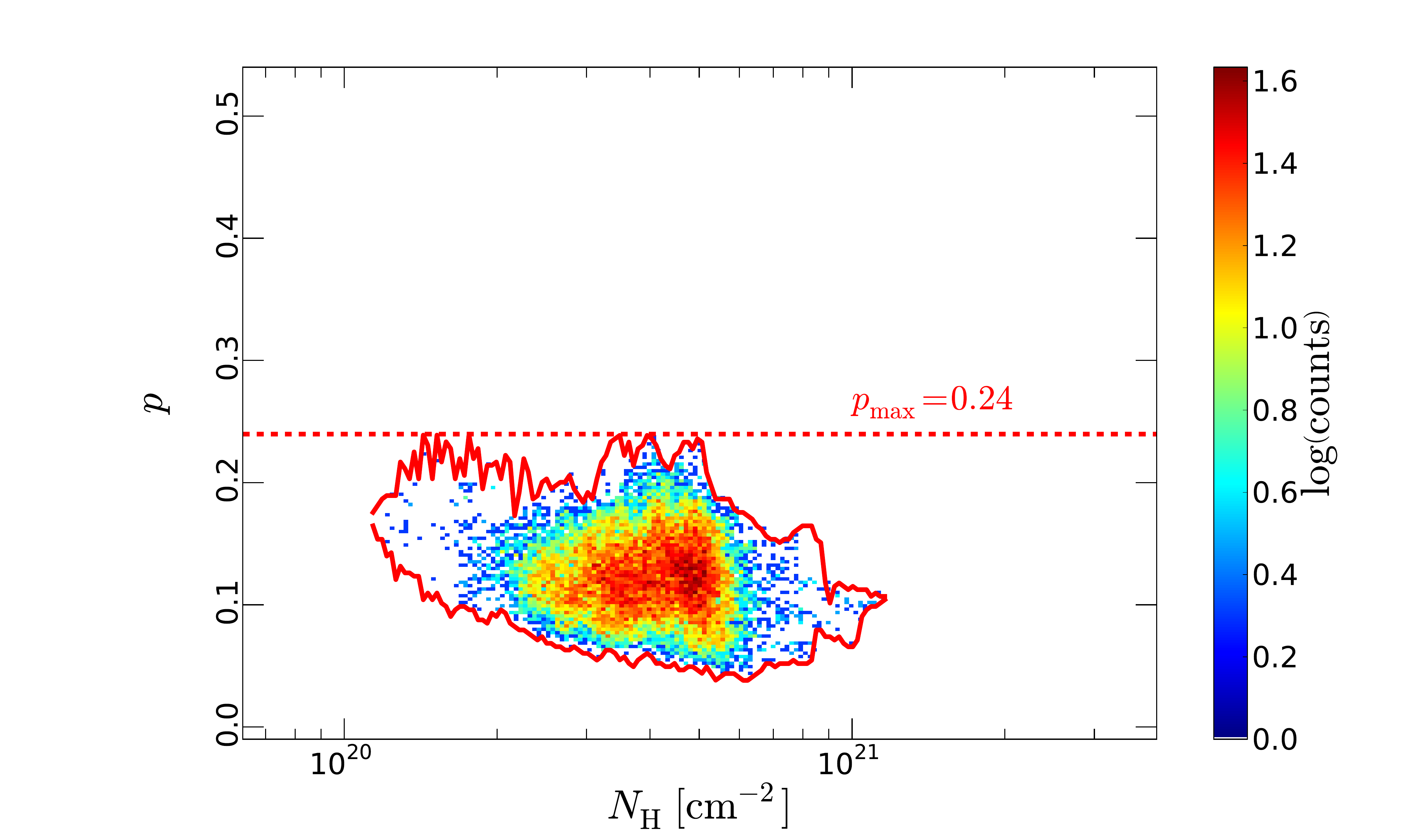}}
\caption{Same as Fig.~\ref{fig:PI-vs-NH}, but for the Microscopium field. Note that the ranges in $N_\mathrm{H}$ and $\polfrac$ are different from Fig.~\ref{fig:PI-vs-NH}, and that no fit is performed.}
\label{fig:PI-vs-NH-Microscopium}
\end{figure}

\begin{figure}[htbp]
\centerline{\includegraphics[width=8.8cm,trim=120 0 60 0,clip=true]{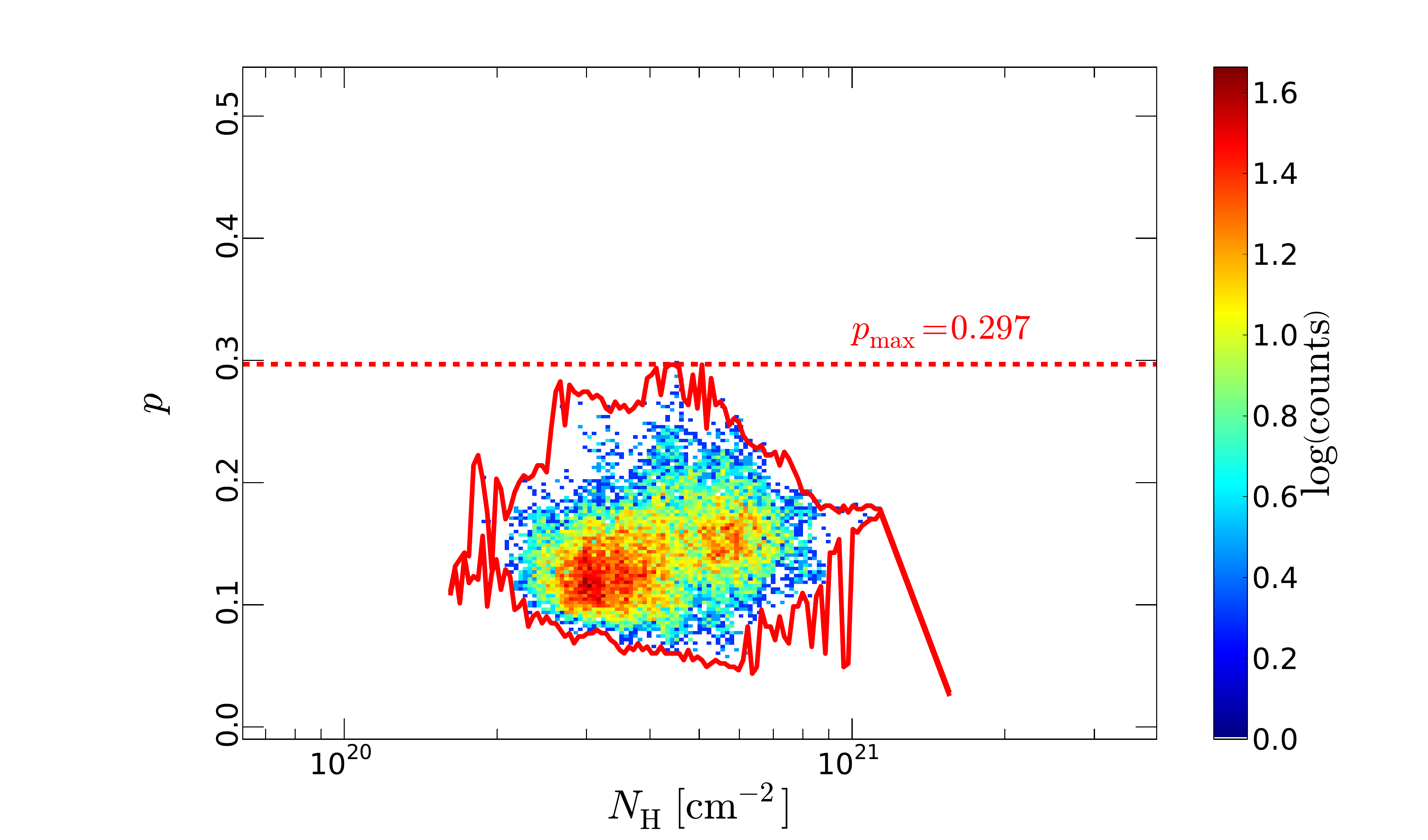}}
\caption{Same as Fig.~\ref{fig:PI-vs-NH}, but for the Pisces field. Note that the ranges in $N_\mathrm{H}$ and $\polfrac$ are different from Fig.~\ref{fig:PI-vs-NH}, and that no fit is performed.}
\label{fig:PI-vs-NH-Pisces}
\end{figure}

\begin{figure}[htbp]
\centerline{\includegraphics[width=8.8cm,trim=120 0 60 0,clip=true]{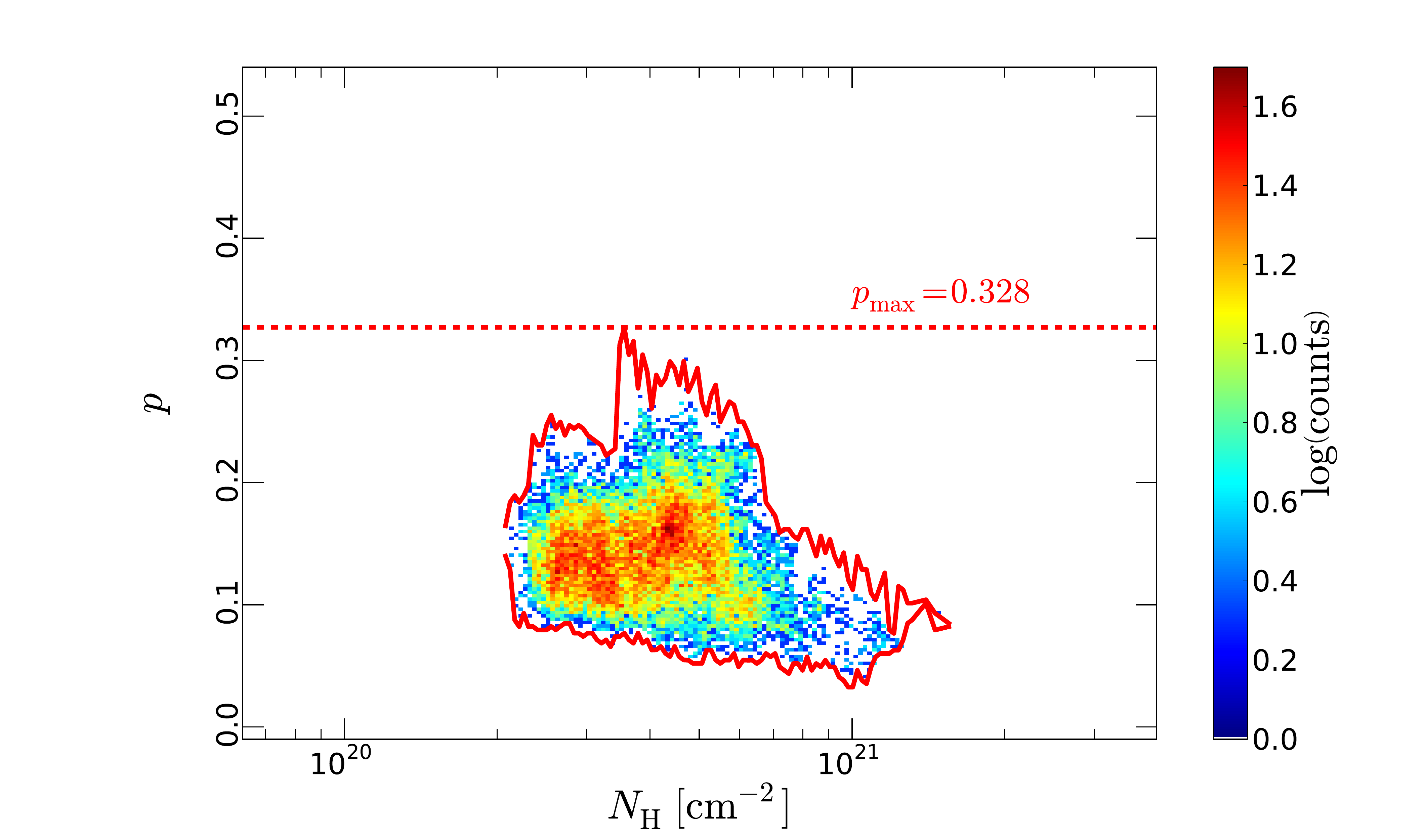}}
\caption{Same as Fig.~\ref{fig:PI-vs-NH}, but for the Perseus field. Note that the ranges in $N_\mathrm{H}$ and $\polfrac$ are different from Fig.~\ref{fig:PI-vs-NH}, and that no fit is performed.}
\label{fig:PI-vs-NH-Perseus}
\end{figure}

\begin{figure}[htbp]
\centerline{\includegraphics[width=8.8cm,trim=120 0 60 0,clip=true]{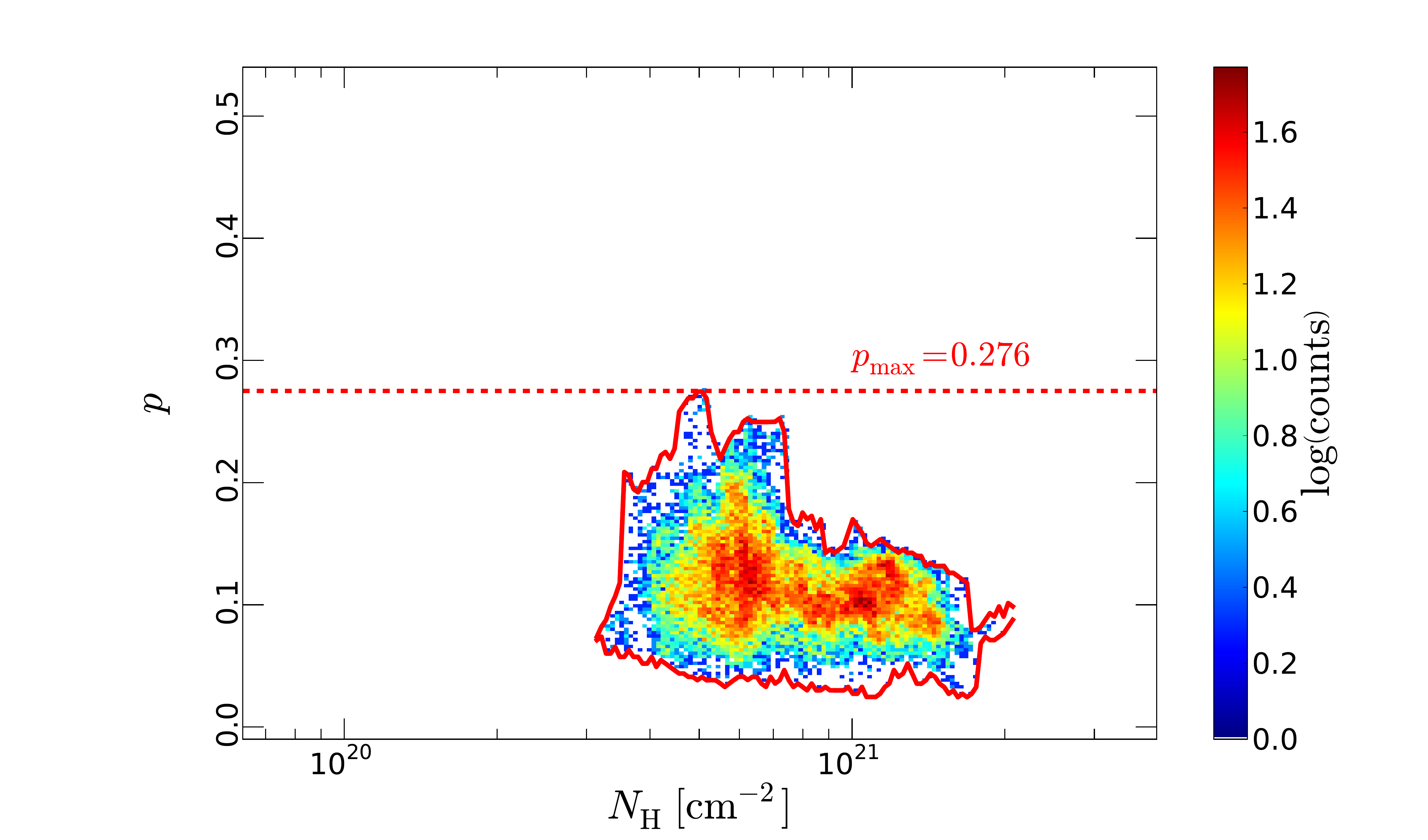}}
\caption{Same as Fig.~\ref{fig:PI-vs-NH}, but for the Ara field. Note that the ranges in $N_\mathrm{H}$ and $\polfrac$ are different from Fig.~\ref{fig:PI-vs-NH}, and that no fit is performed.}
\label{fig:PI-vs-NH-Ara}
\end{figure}

\begin{figure}[htbp]
\centerline{\includegraphics[width=8.8cm,trim=120 0 60 0,clip=true]{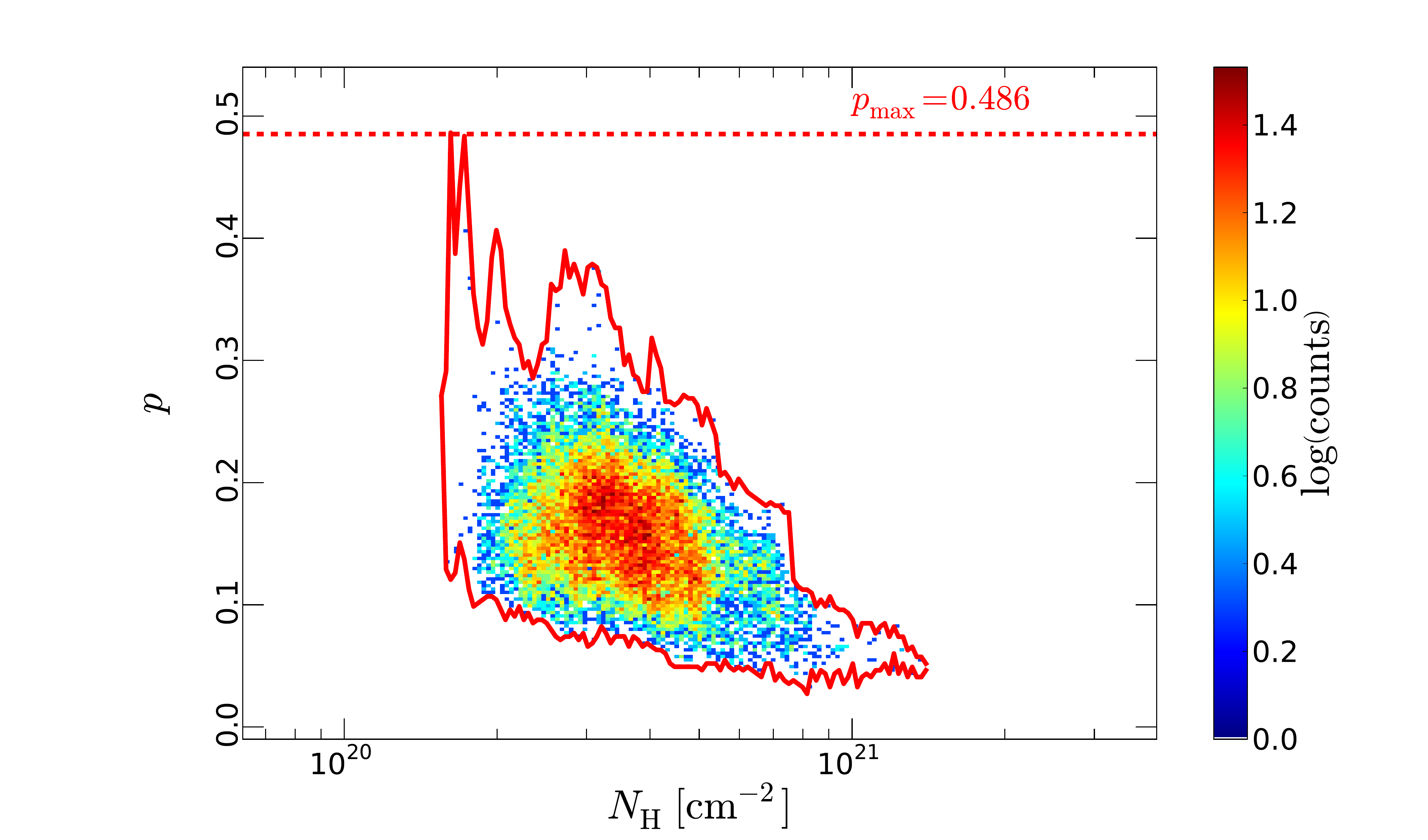}}
\caption{Same as Fig.~\ref{fig:PI-vs-NH}, but for the Pavo field. Note that the ranges in $N_\mathrm{H}$ and $\polfrac$ are different from Fig.~\ref{fig:PI-vs-NH}, and that no fit is performed.}
\label{fig:PI-vs-NH-Pavo}
\end{figure}

\begin{figure}[htbp]
\centerline{\includegraphics[width=8.8cm,trim=120 0 60 0,clip=true]{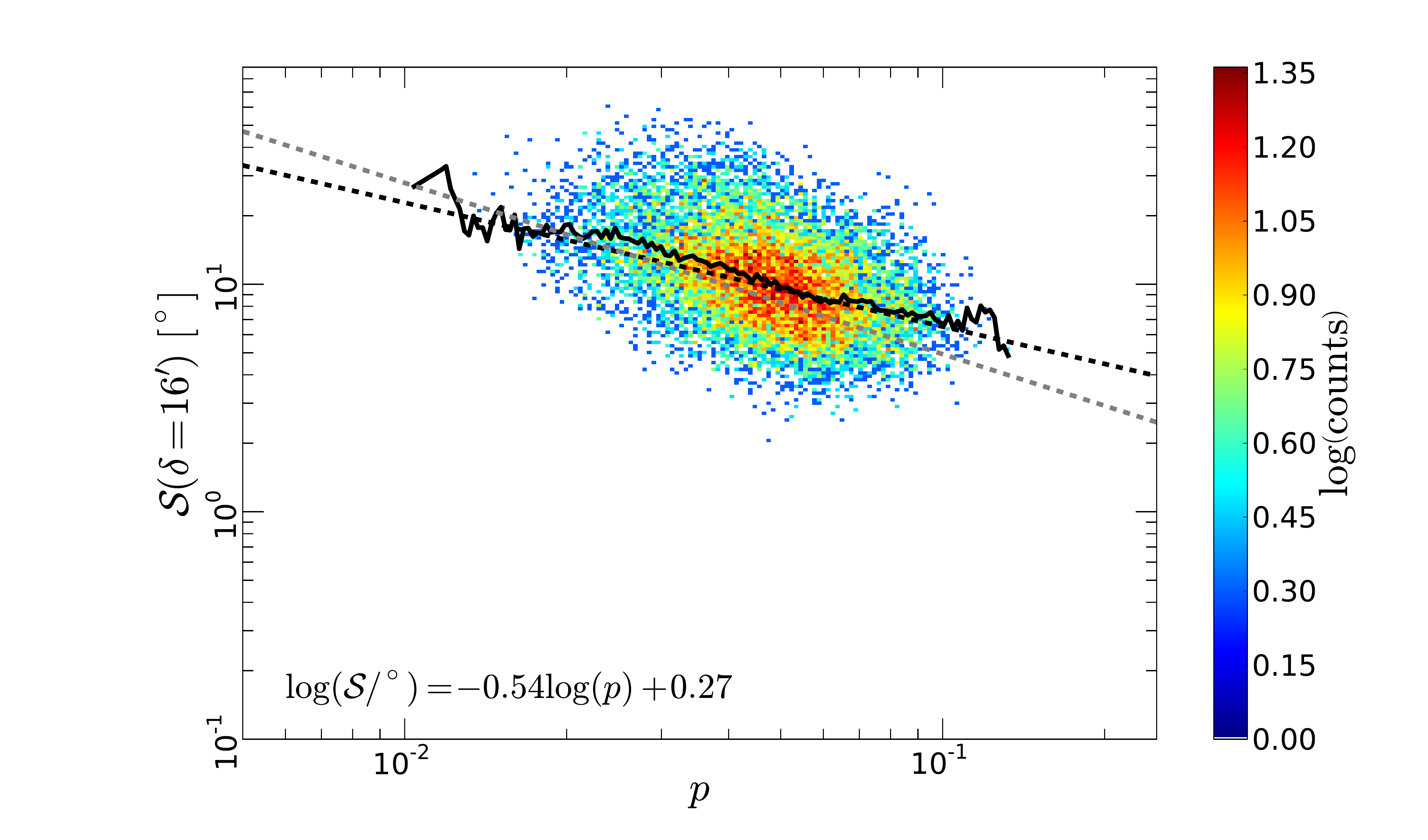}}
\caption{Same as Fig.~\ref{fig:PI-dpsi-data}, but for the Polaris Flare field. Two-dimensional distribution function of $\DeltaAng$ and polarization fraction $\polfrac$. The {\DeltaAngName} $\DeltaAng$ is computed at a lag $\delta=16\arcm$. Only pixels for which $\polfrac/\sigpolfrac>3$ are retained. The dashed grey line is the large-scale fit (with $\mathrm{FWHM}=1\deg$ and $\delta=1\pdeg07$) $\log{\left(\DeltaAng\right)}=-0.75\log\left({\polfrac}\right)-0.06$, the solid black line shows the mean $\DeltaAng$ for each bin in $\polfrac$ (the bin size is $\Delta\log(\polfrac)=0.008$) and the dashed black line is a linear fit of that curve in log-log space, restricted to bins in $\polfrac$ which contain at least 1\% of the total number of points (so about 150 points per bin).}
\label{fig:PI-dpsi-data-Polaris}
\end{figure}

\begin{figure}[htbp]
\centerline{\includegraphics[width=8.8cm,trim=120 0 60 0,clip=true]{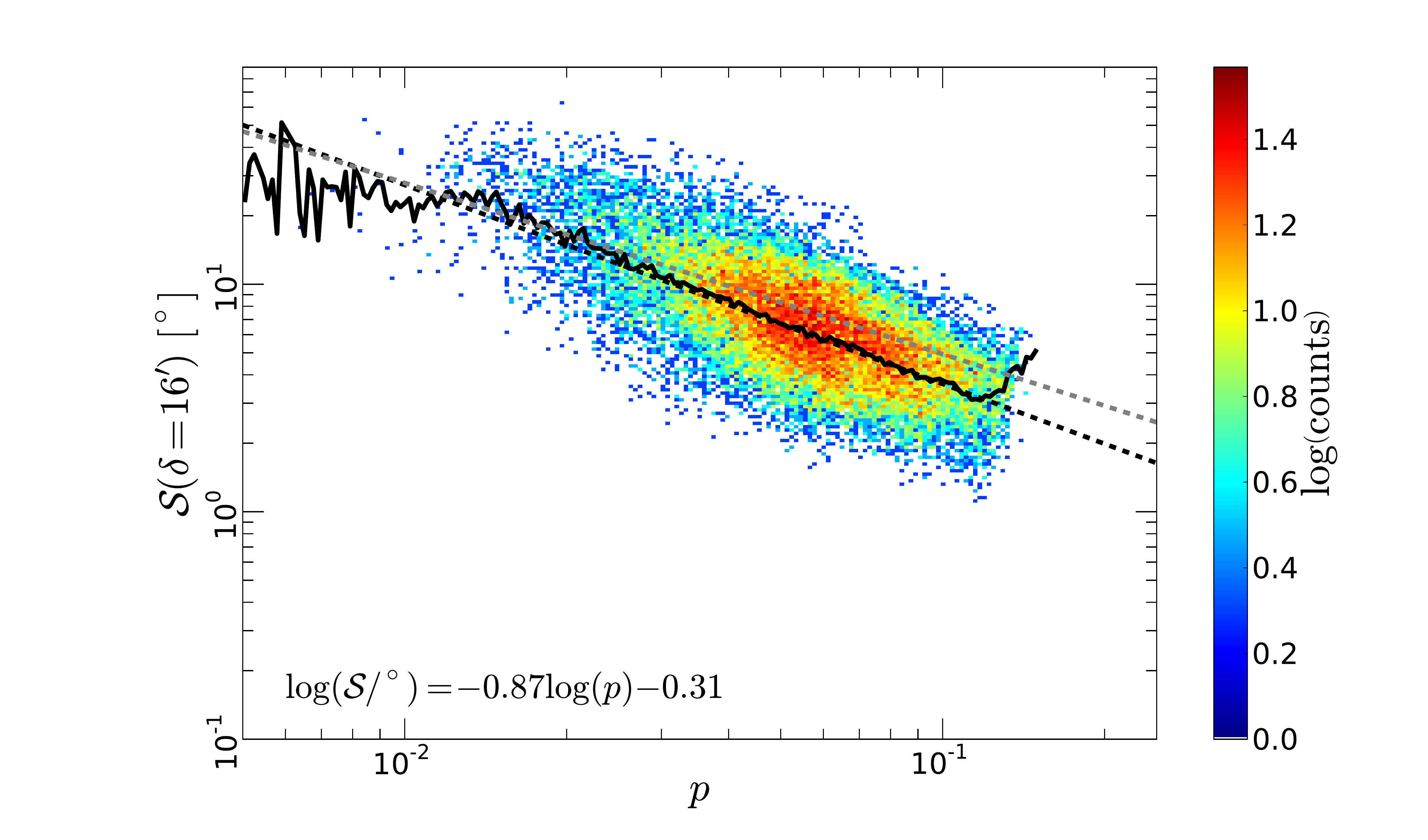}}
\caption{Same as Fig.~\ref{fig:PI-dpsi-data}, but for the Taurus field.}
\label{fig:PI-dpsi-data-Taurus}
\end{figure}

\begin{figure}[htbp]
\centerline{\includegraphics[width=8.8cm,trim=120 0 60 0,clip=true]{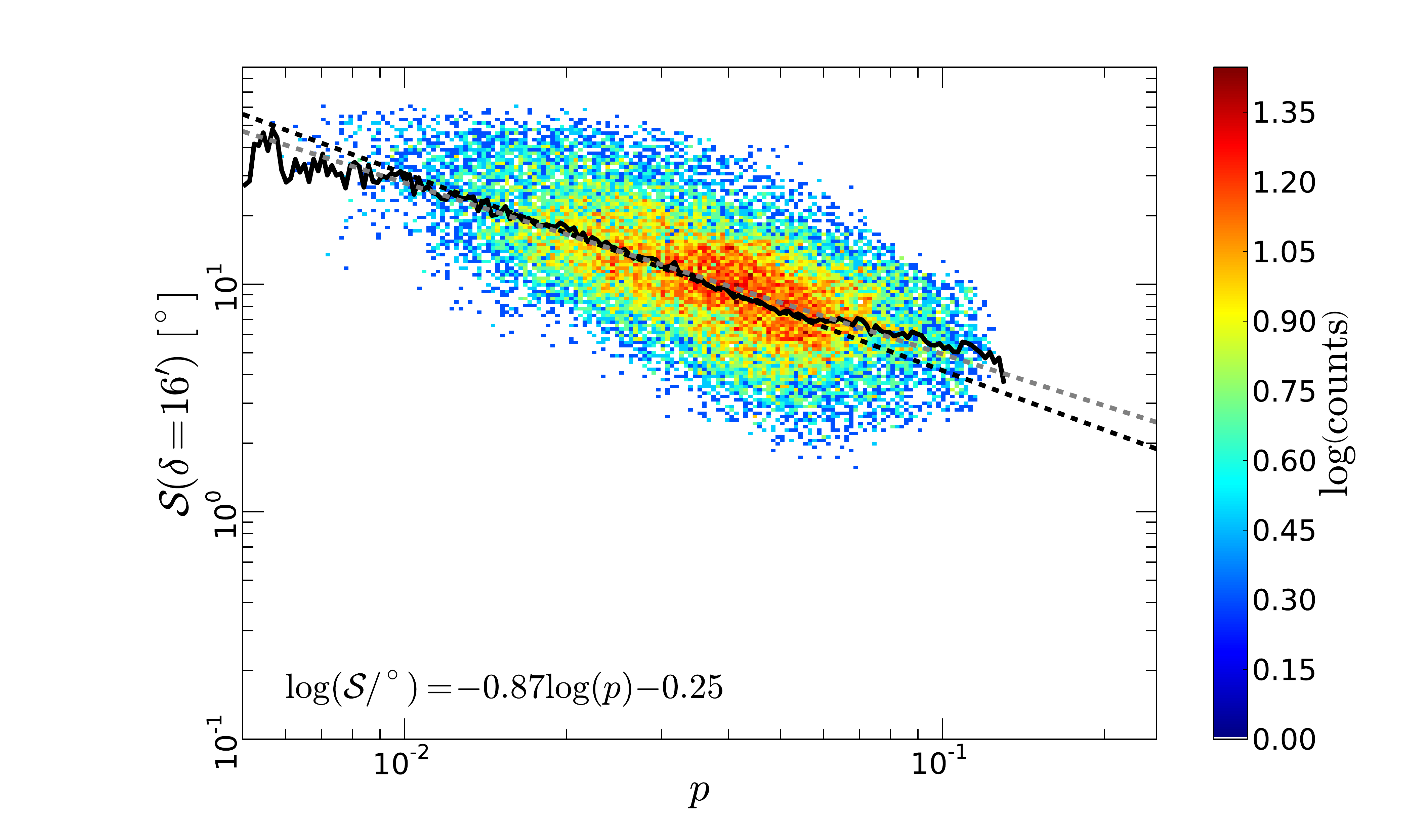}}
\caption{Same as Fig.~\ref{fig:PI-dpsi-data}, but for the Orion field.}
\label{fig:PI-dpsi-data-Orion}
\end{figure}

\begin{figure}[htbp]
\centerline{\includegraphics[width=8.8cm,trim=120 0 60 0,clip=true]{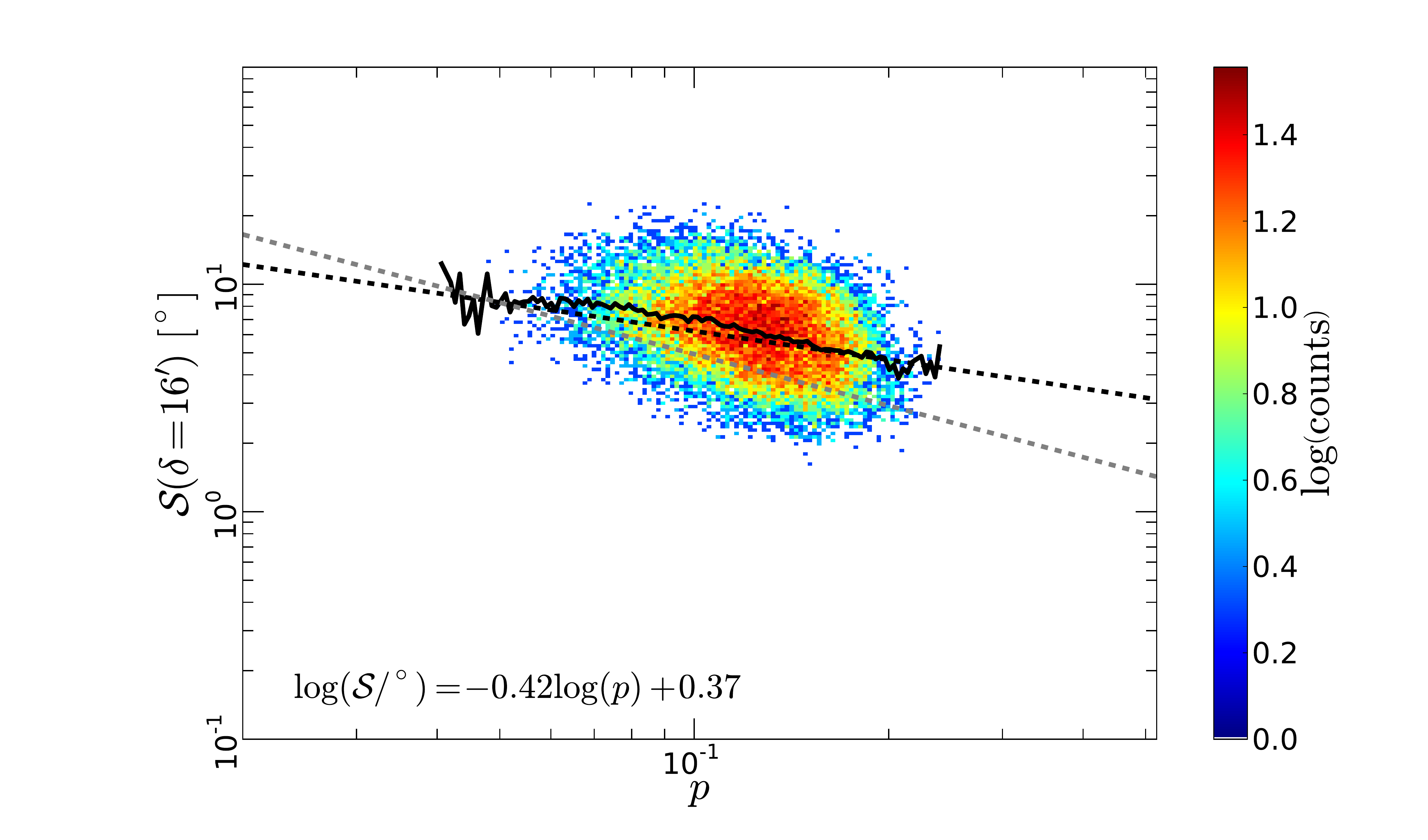}}
\caption{Same as Fig.~\ref{fig:PI-dpsi-data}, but for the Microscopium field. Note that the range in $\polfrac$ is different from Fig.~\ref{fig:PI-dpsi-data}.}
\label{fig:PI-dpsi-data-Microscopium}
\end{figure}

\begin{figure}[htbp]
\centerline{\includegraphics[width=8.8cm,trim=120 0 60 0,clip=true]{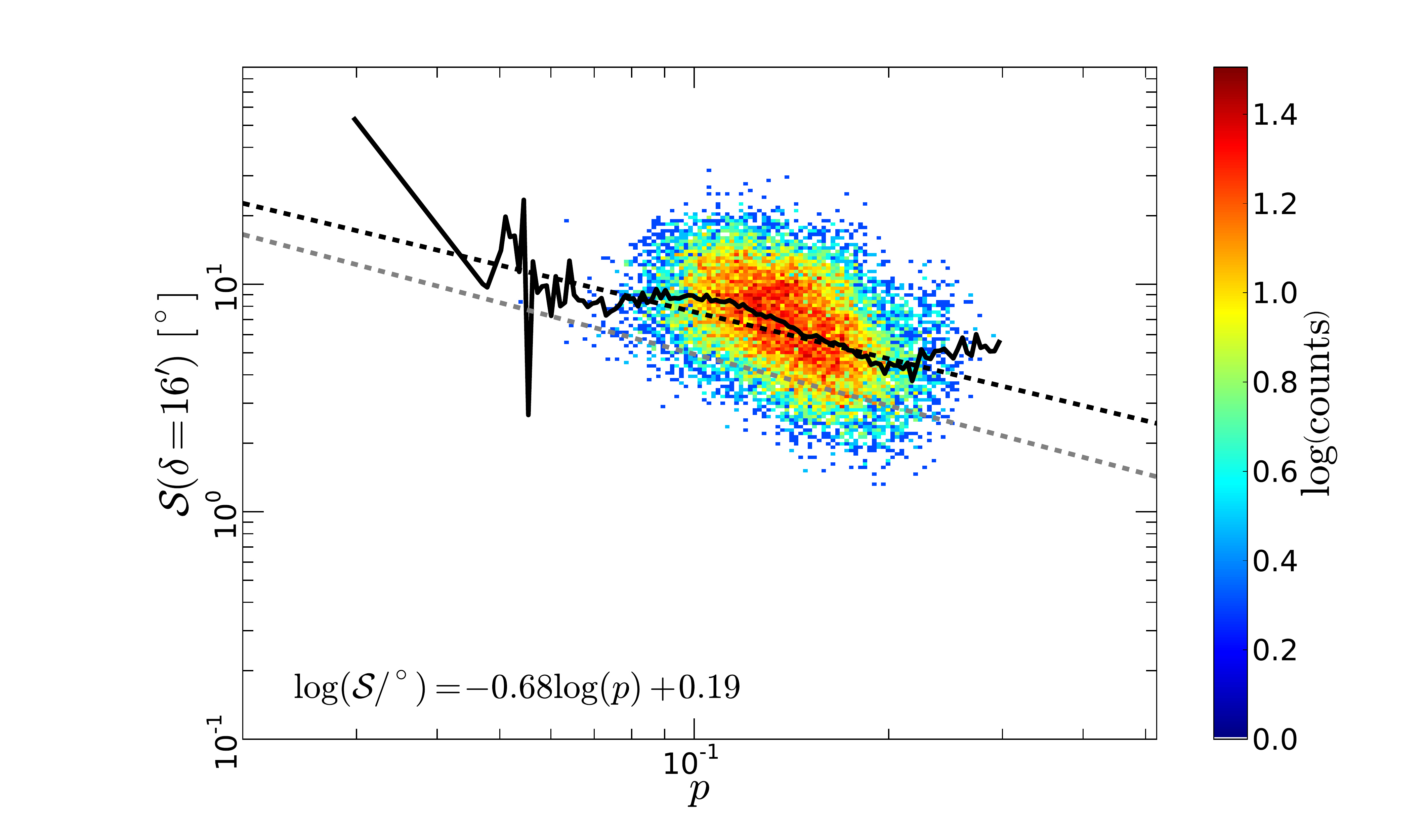}}
\caption{Same as Fig.~\ref{fig:PI-dpsi-data}, but for the Pisces field. Note that the range in $\polfrac$ is different from Fig.~\ref{fig:PI-dpsi-data}.}
\label{fig:PI-dpsi-data-Pisces}
\end{figure}

\begin{figure}[htbp]
\centerline{\includegraphics[width=8.8cm,trim=120 0 60 0,clip=true]{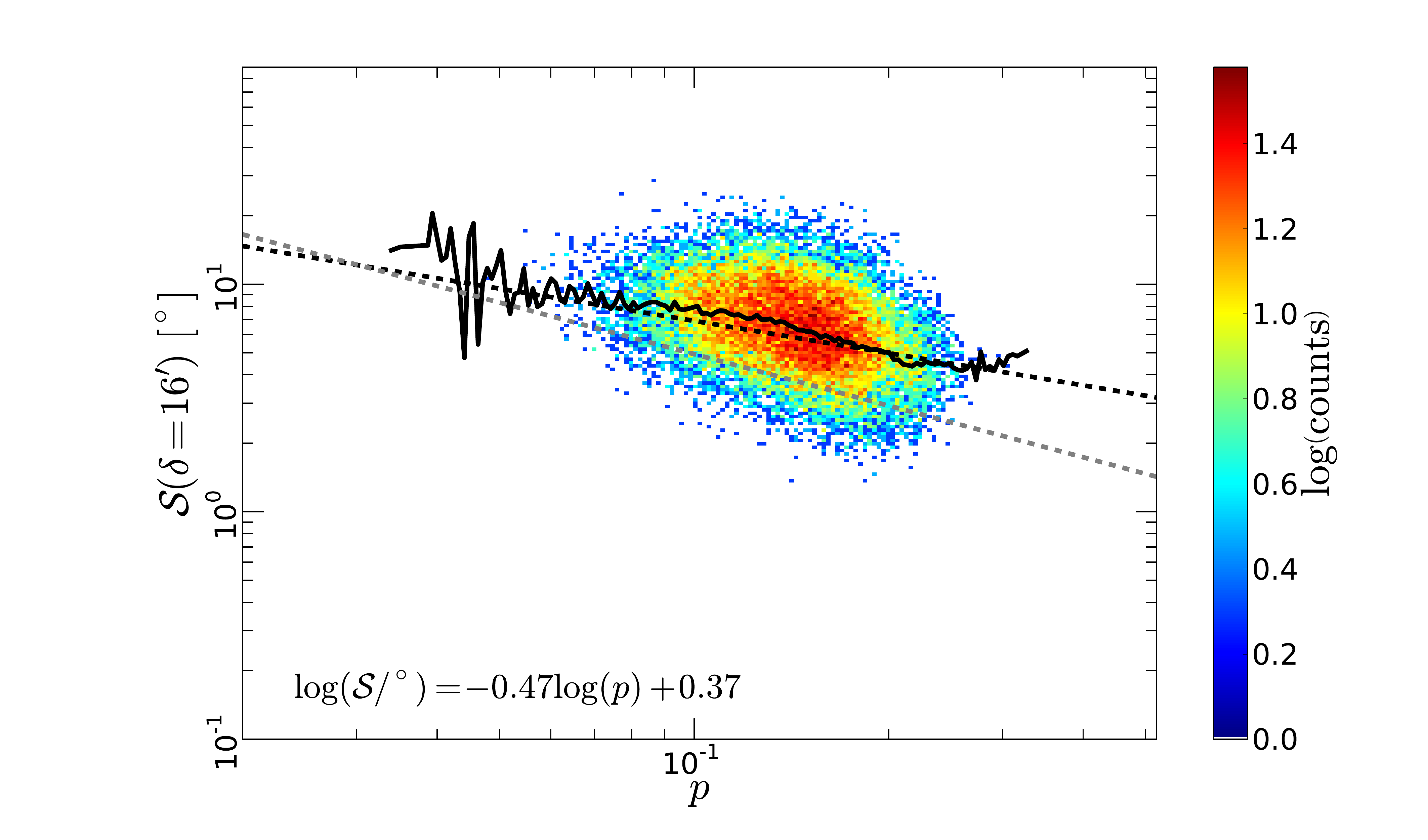}}
\caption{Same as Fig.~\ref{fig:PI-dpsi-data}, but for the Perseus field. Note that the range in $\polfrac$ is different from Fig.~\ref{fig:PI-dpsi-data}.}
\label{fig:PI-dpsi-data-Perseus}
\end{figure}

\begin{figure}[htbp]
\centerline{\includegraphics[width=8.8cm,trim=120 0 60 0,clip=true]{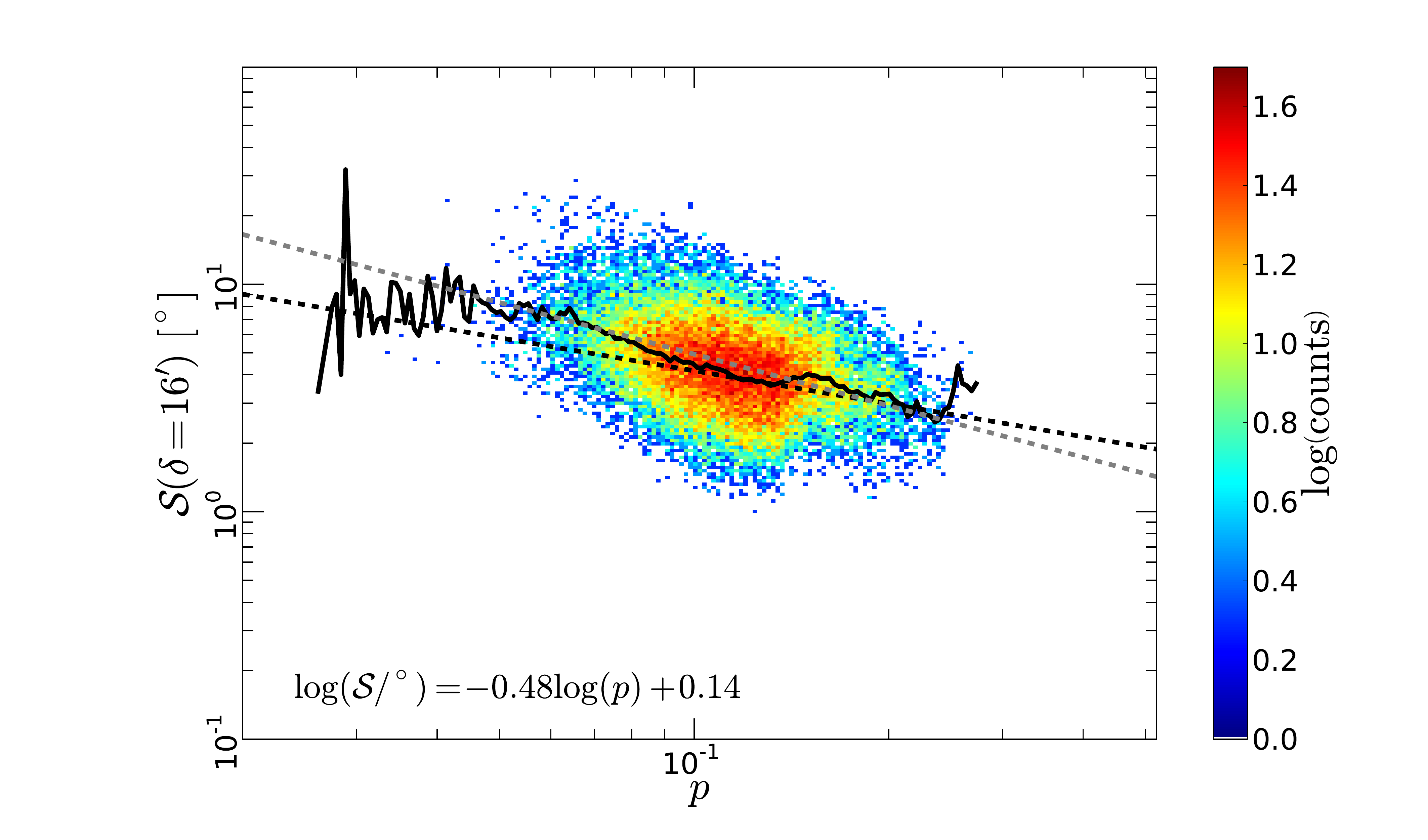}}
\caption{Same as Fig.~\ref{fig:PI-dpsi-data}, but for the Ara field. Note that the range in $\polfrac$ is different from Fig.~\ref{fig:PI-dpsi-data}.}
\label{fig:PI-dpsi-data-Ara}
\end{figure}

\begin{figure}[htbp]
\centerline{\includegraphics[width=8.8cm,trim=120 0 60 0,clip=true]{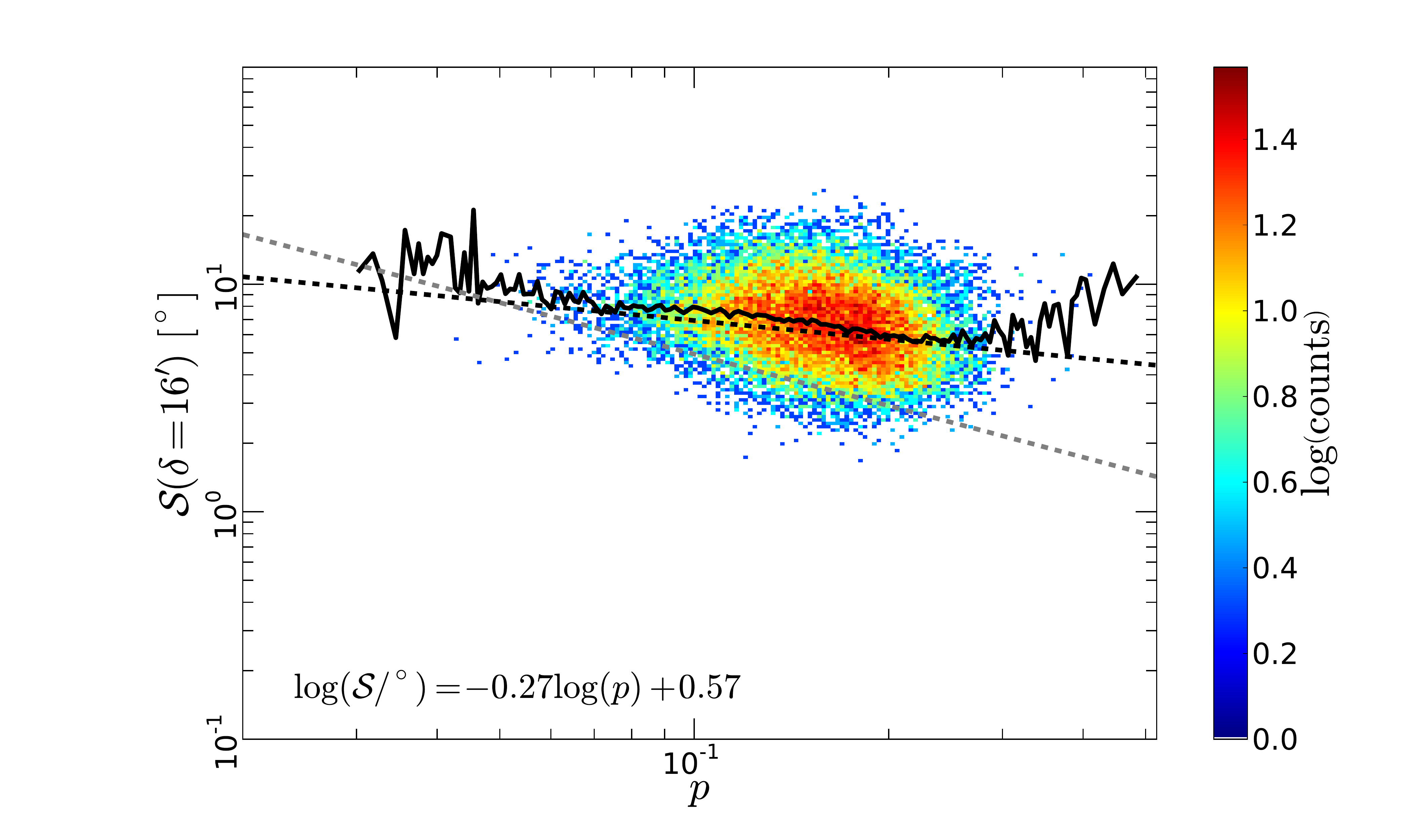}}
\caption{Same as Fig.~\ref{fig:PI-dpsi-data}, but for the Pavo field. Note that the range in $\polfrac$ is different from Fig.~\ref{fig:PI-dpsi-data}.}
\label{fig:PI-dpsi-data-Pavo}
\end{figure}

In the main body of the paper, we showed maps and plots for the Chamaeleon-Musca and Ophiuchus fields. In this appendix we show similar figures for the remaining eight fields, in the same order as in Tables~\ref{table-fields} and~\ref{table-fields-properties}. We first show maps similar to Fig.~\ref{fig:PI-B-NH_Ophiuchus-Chamaeleon} (Figs.~\ref{fig:PI-B-NH_Polaris} to \ref{fig:PI-B-NH_Pavo}), then distribution functions of $\polfrac$ and $N_\mathrm{H}$ similar to Fig.~\ref{fig:PI-vs-NH} (Figs.~\ref{fig:PI-vs-NH-Polaris} to~\ref{fig:PI-vs-NH-Pavo}), and finally distribution functions of $\DeltaAng\left(\delta=16\arcm\right)$ and $\polfrac$ similar to Fig.~\ref{fig:PI-dpsi-data} (Figs.~\ref{fig:PI-dpsi-data-Polaris} to~\ref{fig:PI-dpsi-data-Pavo}).

\clearpage

\section{Derivation of the Stokes parameters for emission}
\label{sec:Stokes-LD85}

The derivation of the Stokes equations Eqs.~\ref{eq:simulated_I}, \ref{eq:simulated_Q}, and \ref{eq:simulated_U}, as presented by \cite{wardle_90} based upon \cite{lee_85}, considers the extinction cross sections $C_\parallel$ and $C_\perp$ for light that is polarized parallel or perpendicular to the grain symmetry axis, and distinguishes oblate and prolate grains. Say that at each point $M$ on the line of sight we define a reference frame $\left(Mx_0y_0z_0\right)$ such that $z_0$ points to the observer, and the local magnetic field $\boldsymbol{B}$ is in the $\left(My_0z_0\right)$ plane. With $\beta$ the angle between $\boldsymbol{B}$ and the angular momentum $\boldsymbol{J}$ of a rotating grain at $M$, and $\gamma$ the angle between $\boldsymbol{B}$ and the plane of the sky, as defined in Fig.~\ref{fig:anglesdefinition}, \cite{lee_85} give, for oblate grains
\begin{equation}
C_{x_0}=C_\perp-\frac{C_\perp-C_\parallel}{2}\sin^2\beta\\
\end{equation}
\begin{equation}
C_{y_0}=C_\perp-\frac{C_\perp-C_\parallel}{2}\left[\sin^2\beta+\cos^2\gamma\left(3\cos^2\beta-1\right)\right]\\
\end{equation}
and for prolate grains
\begin{equation}
C_{x_0}=C_\perp+\frac{C_\parallel-C_\perp}{4}\left(1+\cos^2\beta\right)\\
\end{equation}
\begin{equation}
C_{y_0}=C_\perp+\frac{C_\parallel-C_\perp}{4}\left[1+\cos^2\beta-\cos^2\gamma\left(3\cos^2\beta-1\right)\right]\\
\end{equation}
For spherical grains, all these cross-sections are of course equal, $C_{x_0}=C_{y_0}=C_\perp=C_\parallel$. The expressions for the Stokes parameters in terms of the cross-sections are
\begin{equation}
I=\int n_\mathrm{d}B_\nu\left(T_\mathrm{d}\right)\frac{\langle{C_{x_0}+C_{y_0}}\rangle}{2}\mathrm{d}s
\end{equation}
\begin{equation}
Q=\int n_\mathrm{d}B_\nu\left(T_\mathrm{d}\right)\frac{\langle{C_{x_0}-C_{y_0}}\rangle}{2}\cos\left(2\phi\right)\mathrm{d}s
\end{equation}
\begin{equation}
U=\int n_\mathrm{d}B_\nu\left(T_\mathrm{d}\right)\frac{\langle{C_{x_0}-C_{y_0}}\rangle}{2}\sin\left(2\phi\right)\mathrm{d}s
\end{equation}
where the average $\langle\ldots\rangle$ is performed on the possible angles $\beta$. The equivalent expressions given by \cite{wardle_90} are incorrect in omitting the factor $1/2$ (it is easily checked that our expressions match the expected form of $I$ in the case of spherical grains, and of $P/I$ in the case of fully polarizing grains: 100\% polarization when $C_{y_0}=0$. 

Computation of the sums and differences of $C_{x_0}$ and $C_{y_0}$ for both grain geometries lead to the same expressions for the Stokes parameters
\begin{equation}
I=\int n_\mathrm{d}B_\nu\left(T_\mathrm{d}\right)C_\mathrm{avg}\left[1-p_0\left(\cos^2\gamma-\frac{2}{3}\right)\right]\mathrm{d}s
\end{equation}
\begin{equation}
Q=\int n_\mathrm{d}B_\nu\left(T_\mathrm{d}\right)C_\mathrm{avg}p_0\cos\left(2\phi\right)\cos^2\gamma\mathrm{d}s
\end{equation}
\begin{equation}
U=\int n_\mathrm{d}B_\nu\left(T_\mathrm{d}\right)C_\mathrm{avg}p_0\sin\left(2\phi\right)\cos^2\gamma\mathrm{d}s
\end{equation}
where we have introduced the average cross-section
\begin{equation}
C_\mathrm{avg}=\frac{1}{3}\left(2C_\perp+C_\parallel\right),
\end{equation}
and the polarization cross section
\begin{equation}
C_\mathrm{pol}=\frac{C_\perp-C_\parallel}{2}\qquad\textrm{(for oblate grains)}
\end{equation}
\begin{equation}
C_\mathrm{pol}=\frac{C_\parallel-C_\perp}{4}\qquad\textrm{(for prolate grains).}
\end{equation}

These expressions match those in~\cite{martin_72}, \cite{martin_74}, \cite{martin_75}, and \cite{draine_fraisse_09}; those adopted by \cite{lee_85} are a factor 2 larger. The parameter $p_0$ is then given by
\begin{equation}
p_0=\frac{C_\mathrm{pol}}{C_\mathrm{avg}}\frac{3}{2}\left(\langle\cos^2\beta\rangle-\frac{1}{3}\right)=\frac{C_\mathrm{pol}}{C_\mathrm{avg}}R
\end{equation}
with $R$ a Rayleigh reduction factor accounting for the chosen form of imperfect alignment~\citep{lee_85}. Writing the equations for $I$, $Q$ and $U$ using the optical depth $\tau_\nu$ (which is small in the submillimetre) in place of the physical position $s$ on the line of sight, one is lead to Eqs.~\ref{eq:simulated_I}, \ref{eq:simulated_Q}, and \ref{eq:simulated_U}.

The \IntrinsicpName~is easily computed for both grain geometries:
\begin{equation}
p_\mathrm{i}=\frac{C_\perp-C_\parallel}{C_\perp+C_\parallel}\qquad\textrm{(for oblate grains)}
\end{equation}
\begin{equation}
p_\mathrm{i}=\frac{C_\parallel-C_\perp}{3C_\perp+C_\parallel}\qquad\textrm{(for prolate grains).}
\end{equation}

\end{document}